\documentclass{aa}
\usepackage{amsmath}
\usepackage{txfonts}
\usepackage[T1]{fontenc}
\usepackage[utf8]{inputenc}
\usepackage{natbib}
\usepackage{graphicx}
\usepackage{booktabs}

\usepackage[caption=false, farskip=2pt,labelformat=empty]{subfig}

\bibpunct{(}{)}{;}{a}{}{,} 

\begin{document}

\title{Large scale IRAM 30m CO-observations\\ in the giant molecular cloud complex W43}
\titlerunning{IRAM 30m CO-observations in W43}

\author{P. Carlhoff\inst{1}
  \and Q. Nguyen Luong\inst{2}
  \and P. Schilke\inst{1}
  \and F. Motte\inst{3}
  \and N. Schneider\inst{4,5}
  \and H. Beuther\inst{6}
  \and S. Bontemps\inst{4,5}
  \and F. Heitsch\inst{7}
  \and T. Hill\inst{3}
  \and C. Kramer\inst{8}
  \and V. Ossenkopf\inst{1}
  \and F. Schuller\inst{9}
  \and R. Simon\inst{1}
  \and F. Wyrowski\inst{9}}

\offprints{P. Carlhoff, \email{carlhoff@ph1.uni-koeln.de}}

\institute{1. Physikalisches Institut, Universit\"at zu K\"oln, Z\"ulpicher Str. 77, D-50937 K\"oln, Germany
  \and Canadian Institute for Theoretical Astrophysics -- CITA, University of Toronto, Toronto, Ontario,
  M5S 3H8, Canada
  \and Laboratoire AIM, CEA/IRFU -- CNRS/INSU -- Universit\'{e} Paris Diderot, Service d'Astrophysique, 
  B\^{a}t. 709, CEA-Saclay, F-91191 Gif-sur-Yvette Cedex, France
  \and Univ. Bordeaux, LAB, UMR 5804, F-33270, Floirac, France
  \and CNRS, LAB, UMR 5804, F-33270, Floirac, France
  \and Max-Planck-Institut f\"ur Astronomie, K\"onigsstuhl 17, 69117 Heidelberg, Germany
  \and Department of Physics and Astonomy, University of North Carolina Chapel Hill, CB 3255, Phillips Hall, Chapel Hill, NC 27599, USA
  \and Instituto Radioastronom\'{\i}a Milim\'{e}trica, Av. Divina Pastora 7, Nucleo Central, 18012 Granada, Spain
  \and Max-Planck-Institut f\"ur Radioastronomie, Auf dem H\"ugel 69, 53121 Bonn, Germany}

\date{Received March 28 2013; accepted August 29, 2013}

\abstract 
{
We aim to fully describe the distribution and location of dense molecular clouds in the
giant molecular cloud complex W43. It was previously identified as one of the most massive star-forming regions
in our Galaxy.
To trace the moderately dense molecular clouds in the W43 region, we initiated W43-HERO, a large program using the IRAM 30m telescope, which covers a wide
dynamic range of scales from 0.3 to 140 pc. We obtained on-the-fly-maps
in $^{13}$CO~(2--1) and C$^{18}$O~(2--1) with a high spectral resolution of 0.1~km\,s$^{-1}$ and a spatial resolution of 12\arcsec. These maps cover an area
of $\sim$1.5 square degrees and include the two main clouds of W43 and the lower density gas surrounding them. 
A comparison to Galactic models and previous distance calculations confirms the location of W43 near 
the tangential point of the Scutum arm at approximately 6 kpc from the Sun.
The resulting intensity cubes of the observed region are separated into subcubes, which are centered on single clouds and then analyzed in detail. 
The optical depth, excitation temperature, and H$_2$ column density maps are derived out of the $^{13}$CO and C$^{18}$O data.
These results are then compared to those derived from Herschel dust maps.
The mass of a typical cloud is several $10^4\,M_{\sun}$ while 
the total mass in the dense molecular gas (>~$10^2$~cm$^{-3}$) in W43 is found
to be $\sim 1.9 \times 10^{6} M_{\sun}$.
Probability distribution functions obtained from column density maps derived from molecular line data
and Herschel imaging show a log-normal distribution for low column densities and a power-law tail for high densities. 
A flatter slope for the molecular line data probability distribution function may imply that those selectively show the gravitationally collapsing
gas.
}

\maketitle

\section{Introduction}

The formation of high-mass stars is still not fully understood, although they play an important role in the cycle of star-formation and in the balance of the interstellar medium.
What we do know is that these stars form in giant molecular clouds (GMCs) \citep[see][]{mckee2007}. To understand high-mass star-formation, it is crucial to understand these GMCs. 
One of the most important points to be studied is their formation process.

The region around 30\degr\, Galactic longitude was identified as one of the most active star-forming regions in the Galaxy about 10 years ago \citep{motte2003}. 
It was shown to be heated by a cluster of Wolf-Rayet and OB stars \citep{lester1985, blum1999}.
Although the object W43 \citep{westerhout1958} was previously known, \citet{motte2003} were the first to consider it as a Galactic ministarburst region.

In the past, the name W43 was used for the single cloud \mbox{(G030.8+0.02)} that is known today as W43-Main. \citet{nguyenluong2011} characterized the complex by analyzing
atomic hydrogen continuum emission \citep{stil2006} from the Very Large Array (VLA) and the $^{12}$CO~(1--0) \citep{dame2001} and $^{13}$CO~(1--0) \citep{jackson2006} Galactic plane surveys. 
They concluded that W43-Main and \mbox{G29.96-0.02} (now called W43-South) should be considered as a single connected complex.

\begin{figure*}[htb]
\centering
\begin{minipage}{0.516\textwidth}
\resizebox{\hsize}{!}{\includegraphics{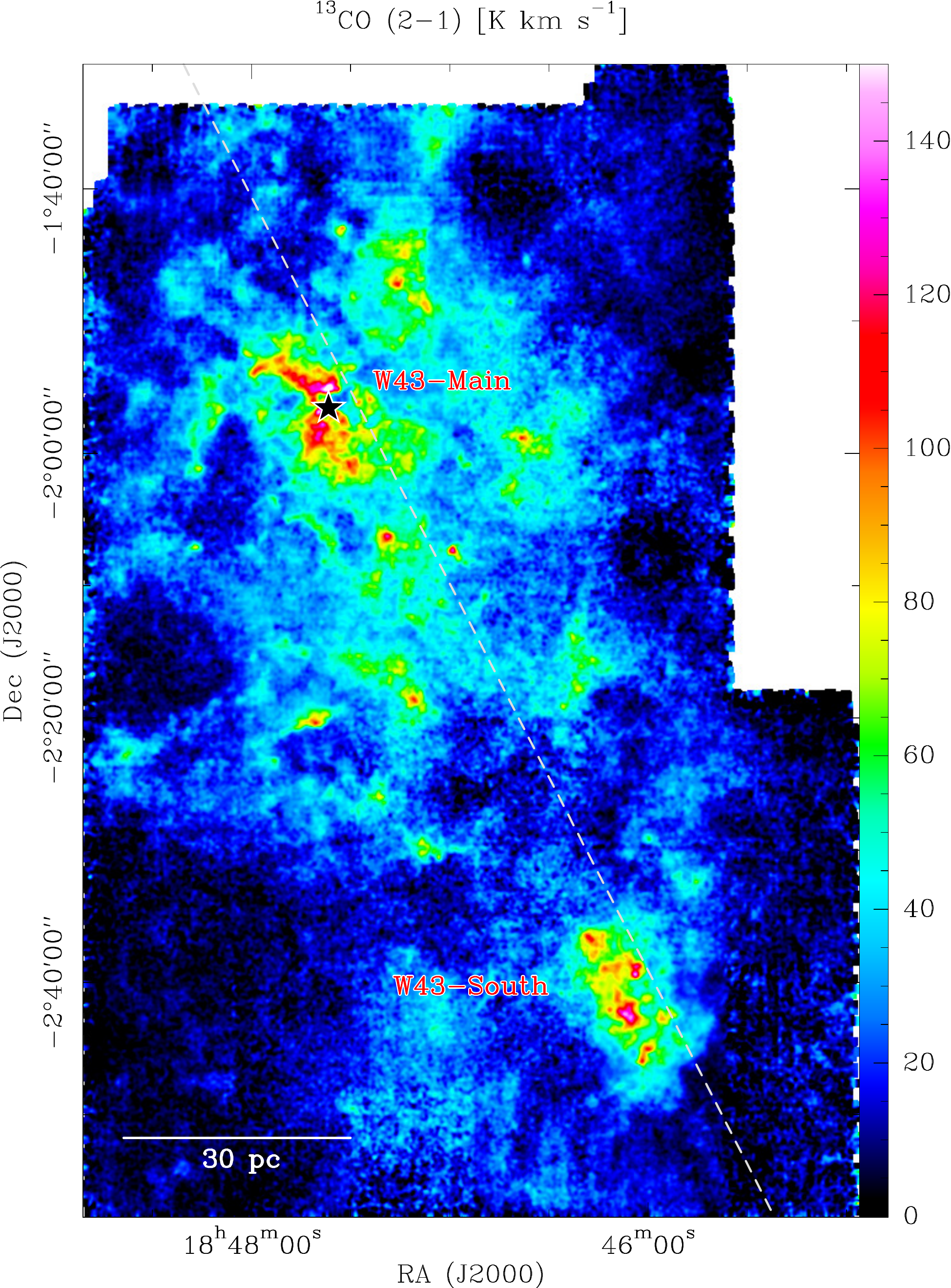}}
\end{minipage}
\hfill
\begin{minipage}{0.464\textwidth}
\resizebox{\hsize}{!}{\includegraphics{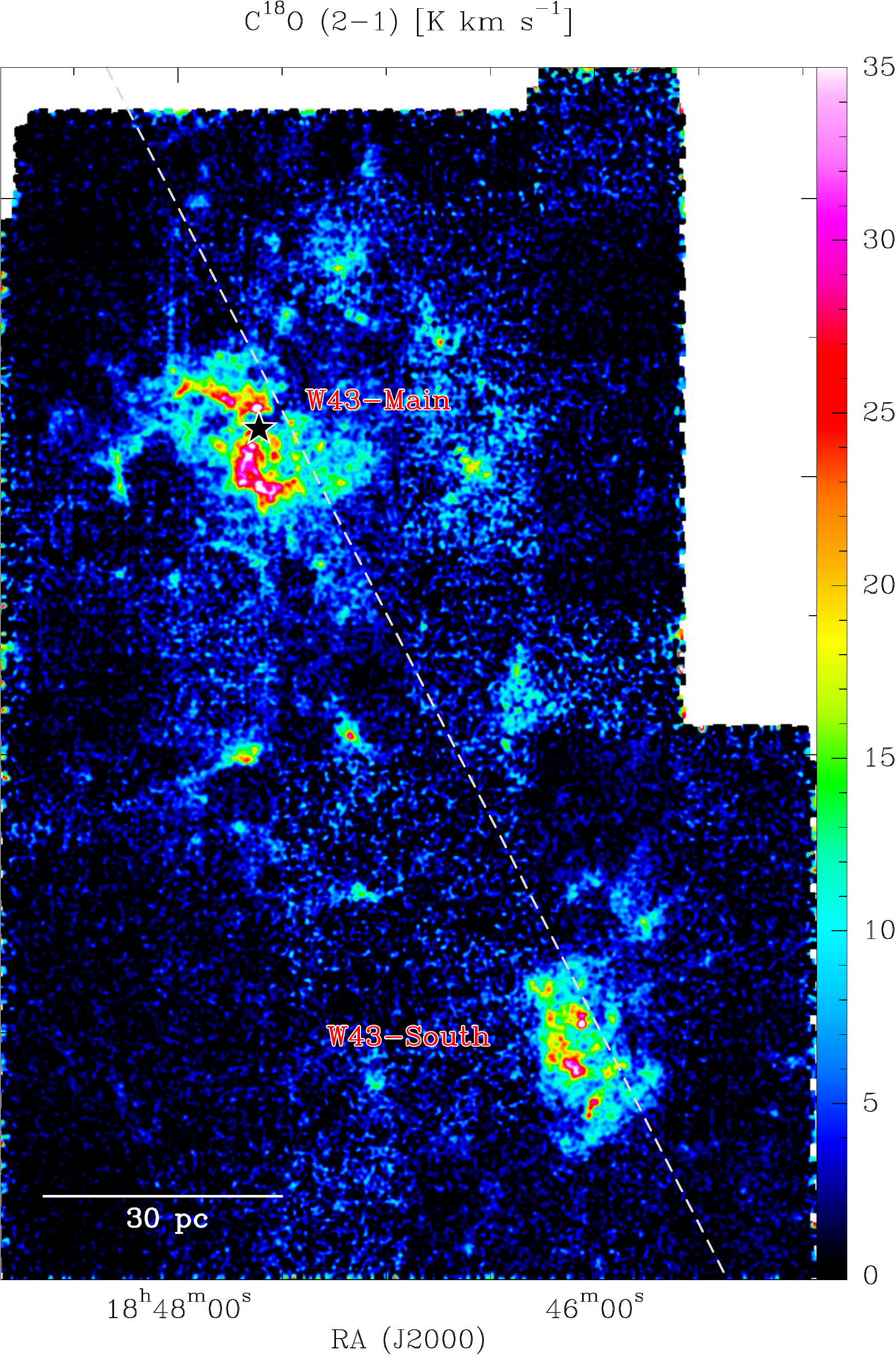}}
\end{minipage}
\caption{Integrated intensity maps created from the resulting data cubes of the complete W43 field in units of K\,km\,s$^{-1}$. The complete spectral range of 30 - 130~km\,s$^{-1}$
has been used to create these maps. The dashed lines denote the Galactic plane, while the star in W43-Main marks the OB star cluster.  
The left plot shows $^{13}$CO~(2--1), while the the right side plots the  C$^{18}$O~(2--1) map.}
\label{fig:completemaps}
\end{figure*}

From the position in the Galactic plane and its radial velocity, \citet{nguyenluong2011} concluded that W43 is located at the junction point of the Galactic long bar 
\citep{churchwell2009} and the Scutum spiral arm at 6~kpc relative to the Sun. The kinematic distance ambiguity, arising from the Galactic rotation curve, gives relative distances for 
W43 of $\sim$6 and $\sim$8.5~kpc for the near and the far kinematic distance, respectively. 
Although there have been other distances adopted by other authors \citep[e.g.][]{pandian2008}, most publications \citep{pratap1999,anderson2009,russeil2011}
favor the near kinematic distance.

This position in the Galaxy makes W43 a very interesting object for studying the formation of molecular clouds. Despite its distance, it is possible to analyze the details of this cloud due 
to its large spatial scale of $\sim$150~pc and the large amount of gas at high density \citep[see][]{nguyenluong2011}.

In this paper, we present initial results of the large IRAM\footnote{IRAM is supported by INSU/CNRS (France), MPG (Germany), and IGN (Spain).} 
30m project entitled ``W43 Hera/EmiR Observations'' (W43 HERO, PIs P. Schilke and F. Motte). 
By observing the kinematic structure of this complex, the program aims to draw conclusions about the formation processes of both molecular clouds from atomic gas
and high-mass stars from massive clouds and so-called ridges.

One part of the project aimed at mapping the large scale mid-density molecular gas ($\sim10^2-10^3$~cm$^{-3}$) in the complete W43 region in the $^{13}$CO~(2--1) and C$^{18}$O~(2--1) emission lines. 
The resulting dataset and a first analysis is presented in this paper. 
The second part of the project observed several high-density tracers in the densest parts of W43. This data and its analysis will be published in
a separate article (Nguyen Luong et al. in prep).

This paper is structured as follows. We first give an overview of the observations and the technical details in Sect.~\ref{sec:observations}.
In Sect.~\ref{sec:results}, we present the resulting line-\citep{stil2006}integrated and PV-maps and a list of clouds that were separated using the Duchamp Sourcefinder software \citep{whiting2012}.
We also show the velocity structure of the complex and determine its position in the Milky Way in Sect.~\ref{subsec:distdeterminationW43}.
Section~\ref{sec:analysis} describes the calculations that we conducted, including optical depth, excitation temperature, and column density of the gas.
We then systematically compare our data to other datasets in Sect.~\ref{sec:comparison} to further characterize the sources we identified.
In Sect.~\ref{sec:sourcedescription}, we give a more detailed description of the main clumps.
The summary and conclusions are given in Sect.~\ref{sec:conclusion}

\section{Observations}\label{sec:observations}

The following data has been observed with the IRAM 30m telescope on Pico Veleta, Spain between
November 2009 and March 2011. We simultaneously observed the molecular emission lines $^{13}$CO~(2--1) and 
C$^{18}$O~(2--1) at 220.398684~GHz and 219.560358~GHz, respectively.
Smaller regions around the two main cloud complexes were additionally observed in high-density tracers, such as
HCO$^+$~(3--2), H$^{13}$CO$^+$~(2--1), N$_2$H$^+$~(1--0), and C$^{34}$S~(2--1) (Nguyen Luong et al. in prep.).

This survey spans the whole W43 region, which includes the two main clouds, W43-Main and W43-South, and several smaller 
clouds in their vicinity. It covers a rectangular map with a size of $\sim$1.4$\times$1.0 degrees.
This translates to spatial dimensions of $\sim$145$\times$105~pc, given an estimated distance of about 6~kpc to the source (see Sect.~\ref{subsec:distdeterminationW43}).
The center of the map lies at \mbox{18:46:54.4 -02:14:11 (EQ J2000)}.
The beam size of the $^{13}$CO and C$^{18}$O observations is 11.7$\arcsec$\footnote{http://www.iram.es/IRAMES/mainWiki/Iram30mEfficiencies},
which corresponds to 0.34~pc at this distance.

For the observations, we used the heterodyne receiver array (HERA) of the IRAM 30m \citep{schuster2004}. It consists of 3$\times$3 pixels
separated by 24$\arcsec$ and has two polarizations, which point at the same location on the sky. This gave us the 
possibility of observing both CO isotopologes in one pass, where we observed one line per polarization. 
The instrument HERA can be tuned in the range of 215 to 272~GHz and has a receiver noise temperature of about 100~K at 220~GHz.
Typically, the system temperature of the telescope was in the range of 300 to 400~K during our observations.

We used the versatile spectrometer assembly (VESPA) autocorrelator as a backend, which was set to a spectral resolution of 80~kHz per channel with a bandwidth of 80~MHz. 
This translates to a resolution of 0.15~km\,s$^{-1}$ and a bandwidth of $\sim$100~km\,s$^{-1}$, which was set to cover the velocity range of 30~km\,s$^{-1}$ to 
130~km\,s$^{-1}$ to cover the complete W43 complex.

\begin{figure}[htb]
 \centering
 \begin{minipage}{0.48\textwidth}
  \resizebox{0.98\hsize}{!}{\includegraphics{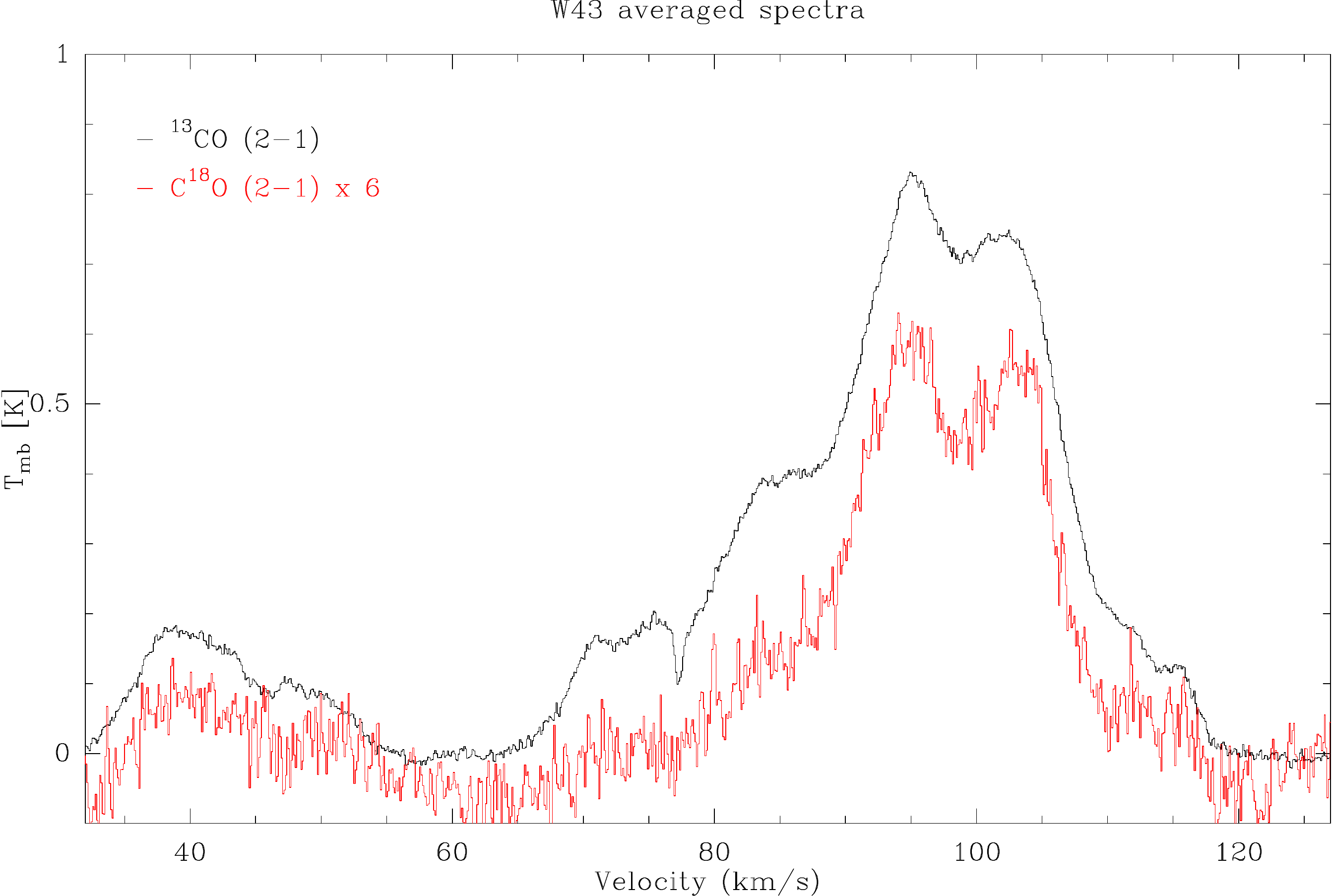}}\vspace{0.2cm}\\
  \resizebox{0.98\hsize}{!}{\includegraphics{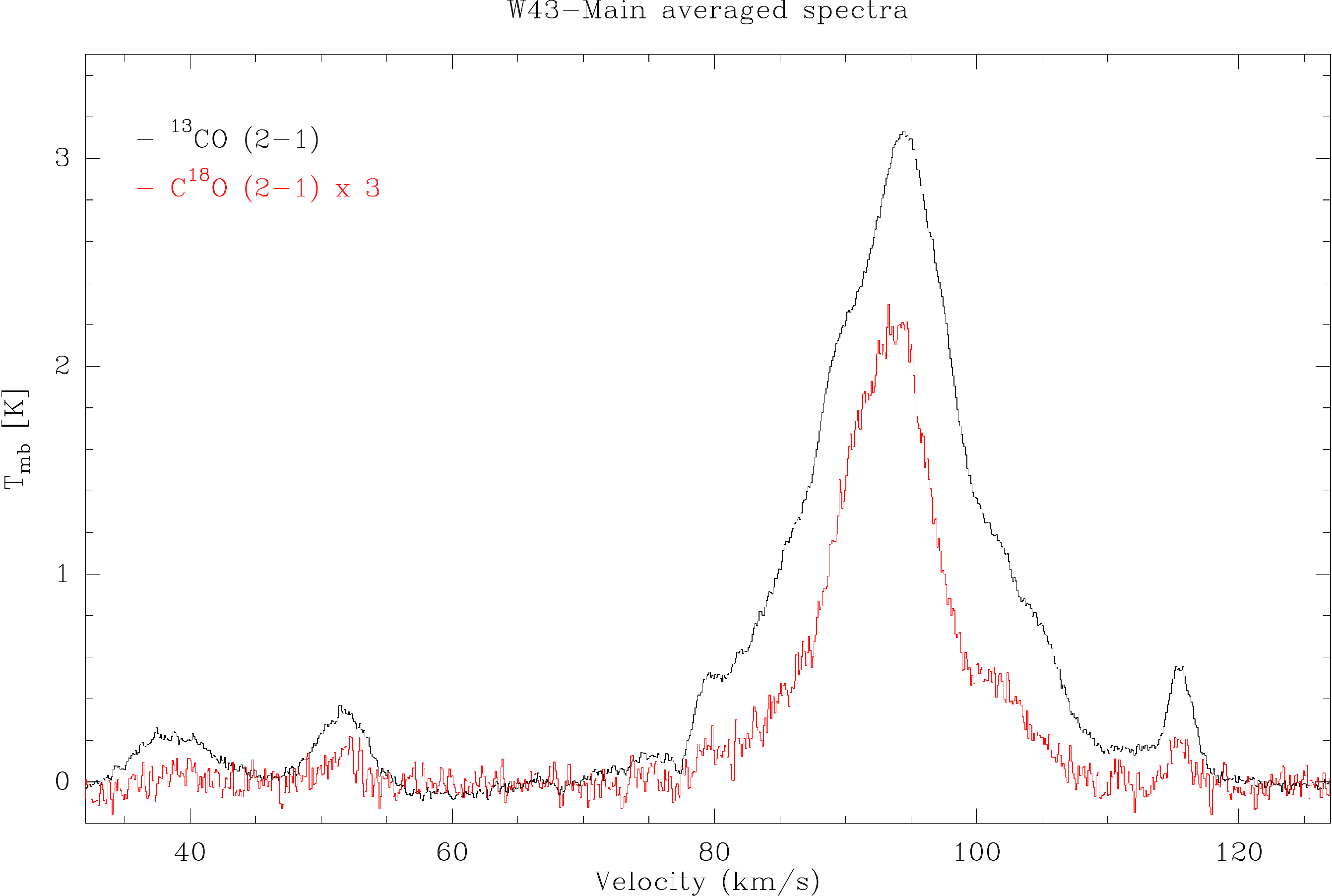}}\vspace{0.2cm}\\
  \resizebox{0.98\hsize}{!}{\includegraphics{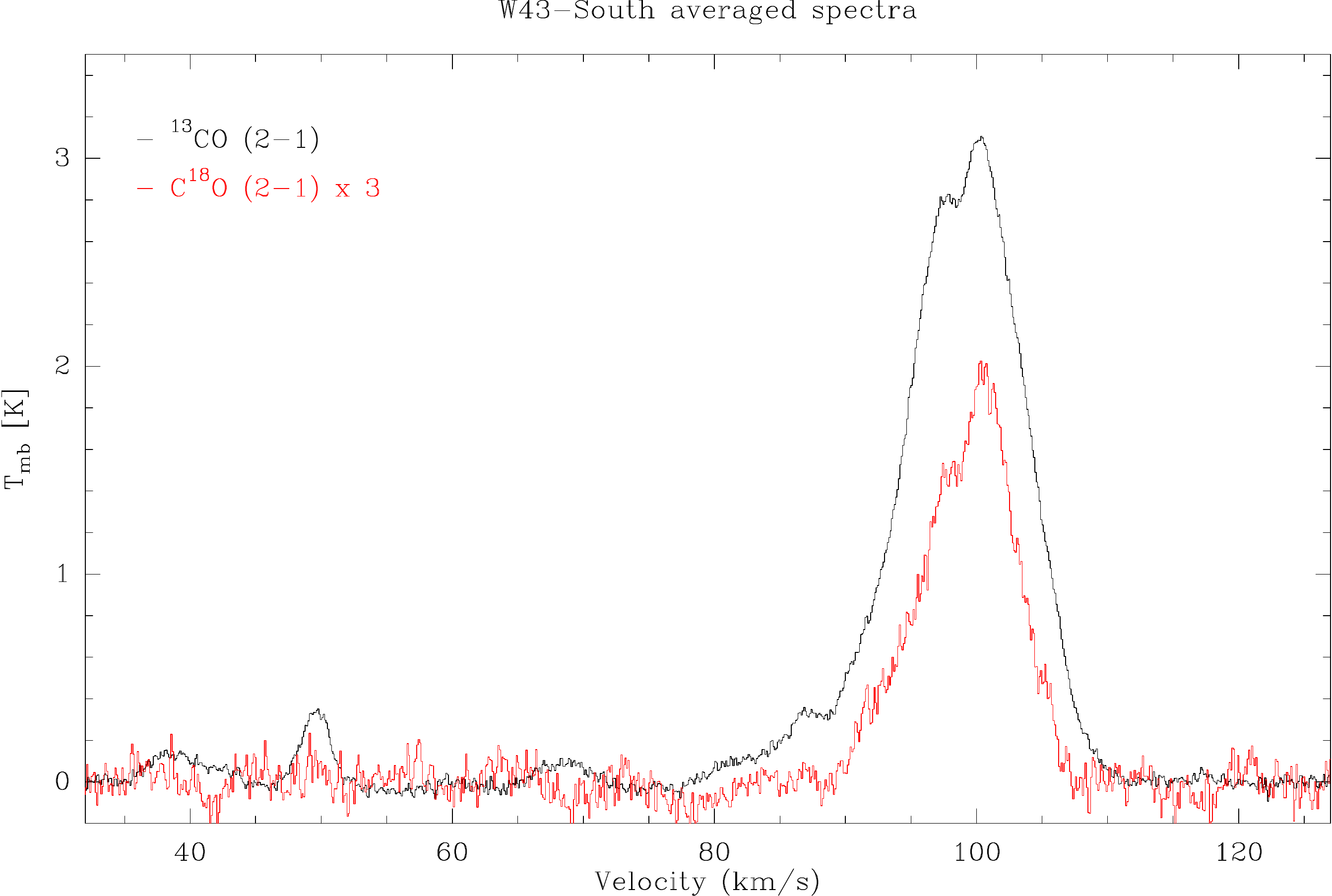}}
 \end{minipage}
 \caption{{\bf Top:} Spectra of $^{13}$CO~(2--1) and C$^{18}$O~(2--1) averaged over the complete data cubes. 
 The C$^{18}$O line has been scaled to highlight the features in the line. {\bf Center:} Averaged spectra of W43-Main.
 {\bf Bottom:} Averaged spectra of W43-South.}
 \label{fig:average_spectra}
\end{figure}

A Nyquist-sampled on-the-fly mapping mode was used to cover 10$\arcmin \times$10$\arcmin$ tiles, taking
about 20 minutes each. Each tile was observed in two orthogonal scanning directions to reduce striping
in the results. The tiles uniformly cover the whole region.
A total of $\sim$3 million spectra in both CO lines was received that way, taking a total observation time
of nearly 80 hours.

Calibration scans, pointing, and focus were done on a regular basis to assure a correct calibration later.
Calibration scans were done every 10 minutes and a pointing every 60 to 90 minutes. A focus scan was done every
few hours with more scans performed around sunset and sunrise as the atmosphere was less stable then. For the pointing, we used \mbox{G34.3}, a strong nearby ultracompact 
\ion{H}{ii} region. The calibration was conducted with the MIRA package, which is part of the GILDAS\footnote{This 
software is developed and maintained by IRAM. http://www.iram.fr/IRAMFR/GILDAS} software.
We expect the flux calibration to be accurate within error limits of $\sim10\%$.

\subsection{Data reduction}

\begin{figure*}[htb]
 \centering
 \resizebox{0.98\hsize}{!}{\includegraphics{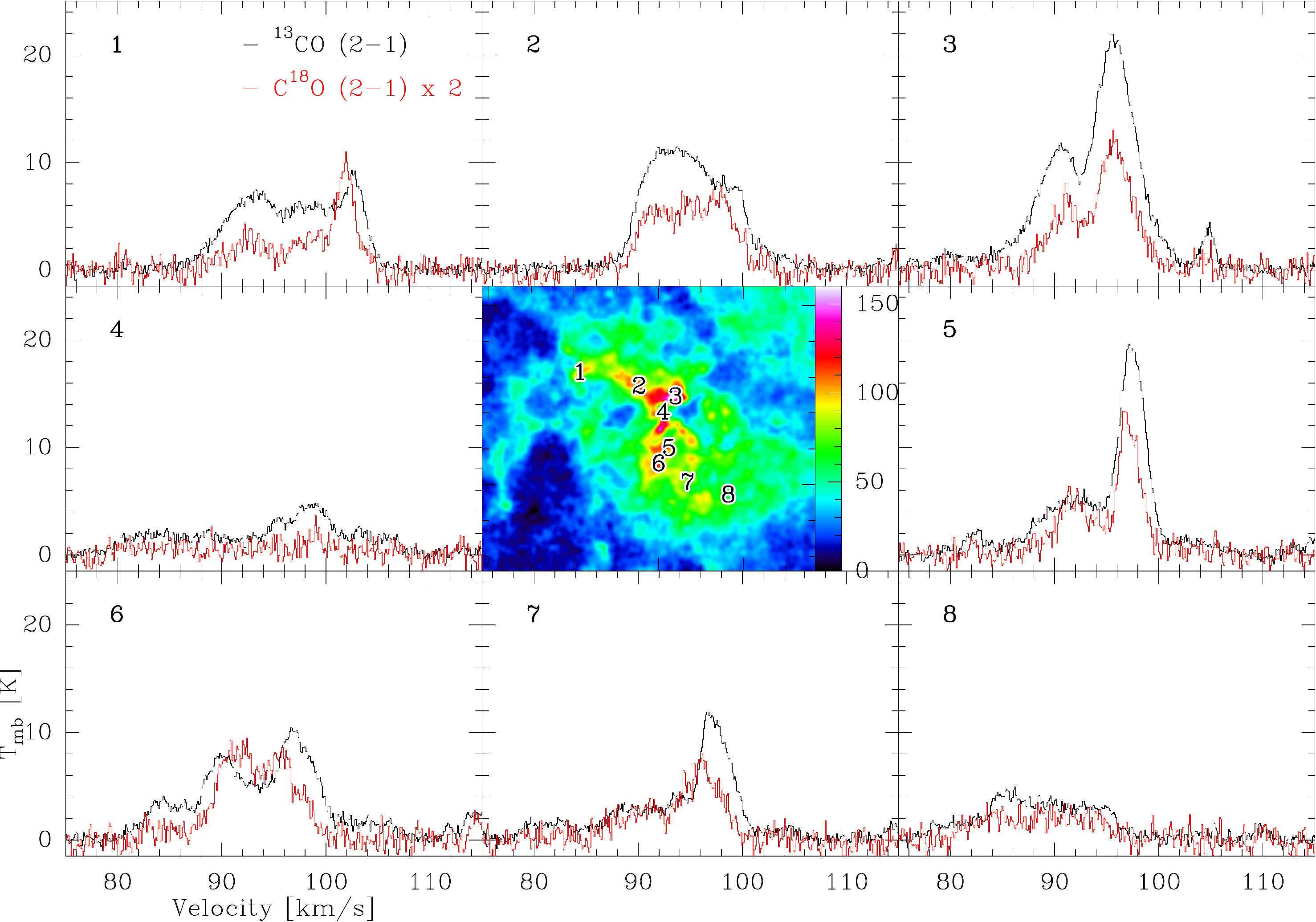}}
 \caption{$^{13}$CO~(2--1) and C$^{18}$O~(2--1) spectra of several points in W43-Main. The central map is a $^{13}$CO intensity map integrated over the velocity range of
 78 to 110~km\,s$^{-1}$ in units of K\,km\,s$^{-1}$.}
 \label{fig:w43main_spectra}
\end{figure*}

The raw data were processed using the GILDAS$^3$ software package. 
The steps taken for data reduction included flagging of bad data (e.g., noise level that are too high or platforms that could not be removed), platform removal in the spectra, baseline subtraction, 
and gridding to create three dimensional data cubes. About 10 percent of the data had to be flagged due to
excessive platforming or strong noise. Platforming, which is an intensity jump in the spectra, is sometimes produced by
the VESPA backend and occurs only in specific pixels of the array that correspond to fixed frequencies. We
calculated the intensity offsets by taking several baselines on each side of the jump to remove these effects on the spectra.

Baseline subtraction turned out to be complicated in some regions that are crowded with emission over a large part of the
band. A first order baseline fit was usually adequate, but a second order baseline was 
needed for a small number of pixels and scans. We then corrected for the main beam efficiency via $T_{\mathrm{mb}} = (F_{\mathrm{eff}}/B_{\mathrm{eff}})\,T_{\mathrm{A}}^*$, where
$F_{\mathrm{eff}}=0.94$ was the forward efficiency of the IRAM 30m telescope and $B_{\mathrm{eff}}=0.63$ was the main beam efficiency at 210~GHz
(No efficiency measurements had been carried out for the IRAM 30m at 220~GHz, but the values should not deviate much from those at 210~GHz).

Finally, the single spectra were gridded to two data cubes with one for each line. The pixels are separated by half
beam steps, 5.9$\arcsec$ in spatial dimension, and have channel widths of 0.15~km\,s$^{-1}$. This step includes the convolution with a Gaussian of beam width. The 
final cubes have dimensions of 631$\times$937$\times$917 data points\footnote{The readily reduced data cubes can be
obtained from: http://www.iram-institute.org/EN/content-page-292-7-158-240-292-0.html} (RA--DEC--velocity).

The noise of single spectra varied with the weather and also with the pixel of the HERA-array. Maps that show the noise
level for each spatial point for both lines are shown in Fig.~\ref{fig:noisemaps} in the Appendix. Typical values are $\sim$1~K, which are corrected for the main beam
efficiency, while several parts of the southern map that are observed during worse weather conditions have rms values of up to 3~K.
In general, the noise level of the C$^{18}$O is higher than that of the $^{13}$CO line.
Despite our dedicated reduction process, scanning effects are still visible in the resulting maps. 
They appear as stripes (See upper part of $^{13}$CO map in Fig.~\ref{fig:completemaps}.)
and tiling patterns (See noise difference of diffuse parts of C$^{18}$O in Fig.~\ref{fig:completemaps}.). 

\section{Results}\label{sec:results}

Integrated intensity maps of the whole W43 region in both $^{13}$CO~\mbox{(2--1)} and C$^{18}$O~(2--1) lines are shown in Fig.~\ref{fig:completemaps}. 
The maps use the entire velocity range from 30~km\,s$^{-1}$ to 130~km\,s$^{-1}$ and show a variety of clouds and filaments. The two main cloud complexes, W43-Main in the upper left part 
of the maps and W43-South in the lower right part, are clearly visible.

In Fig.~\ref{fig:average_spectra}, we show several spectra taken from the data. The upper plot shows the spectra of $^{13}$CO~(2--1) and C$^{18}$O~(2--1) 
averaged over the complete complex; the center and bottom plots show averaged spectra of the W43-Main and W43-South clouds.
The spectra of the complete cubes show emission nearly across the whole velocity range in  $^{13}$CO. Only the components at $55-65$~km\,s$^{-1}$ and velocities
higher than 120~km\,s$^{-1}$ do not show any emission. The C$^{18}$O follows that distribution, although it is not as broad. We thus can already distinguish
two separated velocity components with one between 35 and 55~km\,s$^{-1}$. Most of the emission is concentrated in the velocity range between 65 and 120~km\,s$^{-1}$.
To give an impression of the complexity of some sources, we plot several spectra of the W43-Main cloud in Fig.~\ref{fig:w43main_spectra}.

\subsection{Decomposition into subcubes}\label{subsec:decomposition}

\begin{figure}[htb]
 \centering
 \begin{minipage}{0.48\textwidth}
  \resizebox{0.98\hsize}{!}{\includegraphics{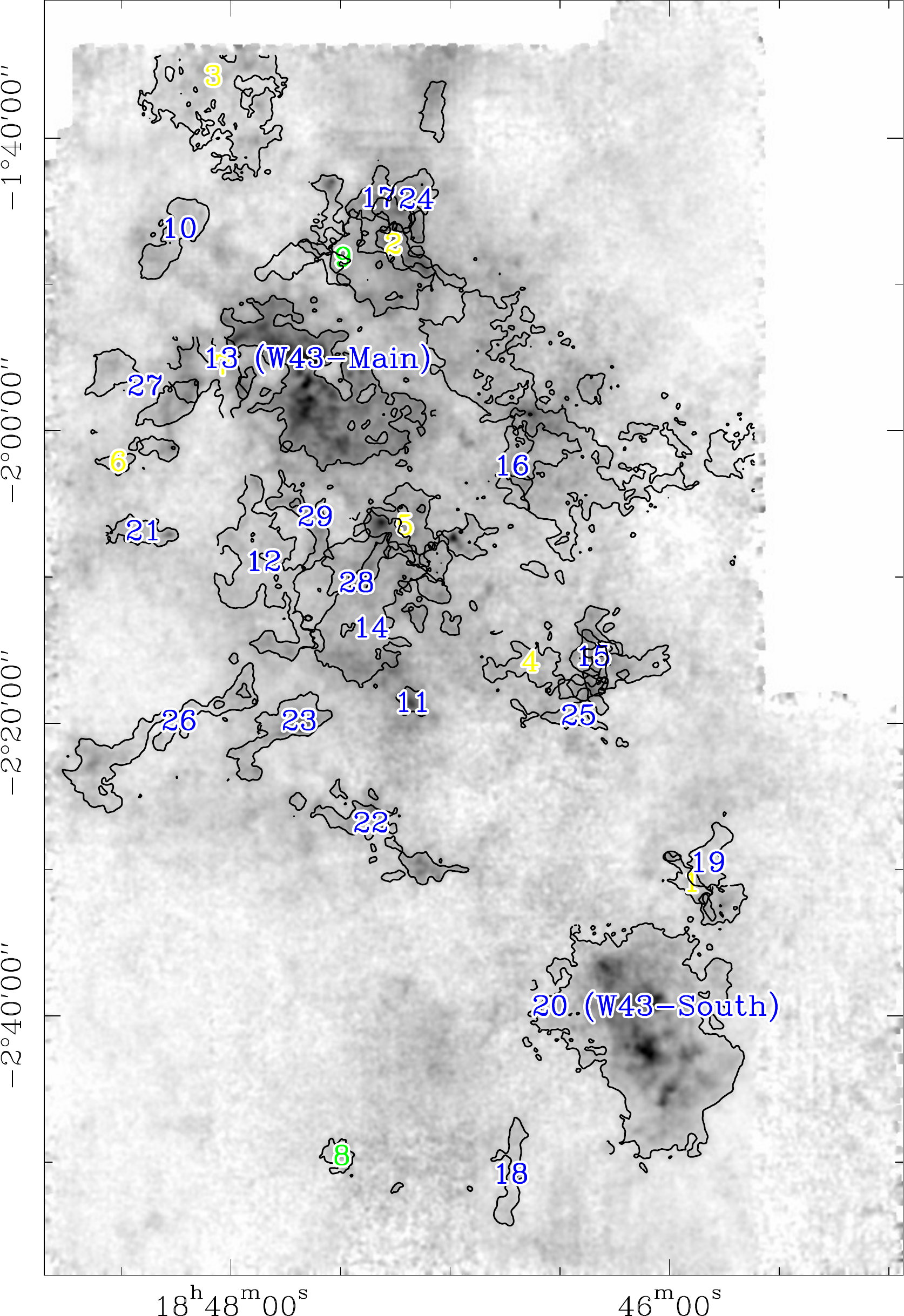}}
 \end{minipage}
 \caption{Detection map of the Duchamp Sourcefinder. Detected clouds are overlaid on a gray-scale map of $^{13}$CO~(2--1).
 Sources are ordered by their peak velocities.
 The color-coding relates to the distance of identified clouds: blue for 6~kpc, green for 4.5~kpc, and yellow for the 4 and 12~kpc component.}
 \label{fig:duchampdetectionmap}
\end{figure}

The multitude of sources found in the W43 region complicates the analysis of the complete data cube. Details get 
lost when integrating over a range in frequency that is too large. We want to examine each source separately, so we need
to decompose the data cube into subcubes that only contain one single source each. This is only done on the 
$^{13}$CO cube, as this is the stronger molecular line. This breakdown is then copied to the C$^{18}$O cube.

We use the Duchamp Sourcefinder software package to automatically find a decomposition. See \citet{whiting2012} for a detailed description of this software.
The algorithm finds connected structures in three-dimensional ppv-data cubes by searching for emission that lies 
above a certain threshold. The value of this threshold is crucial for the success of the process and needs to
be carefully adjusted by hand. For the decomposition of the $^{13}$CO cube, we use two different cutoffs. 
The lower cutoff of $5\,\sigma$ per channel is used to identify weaker sources. 
A higher cutoff of $10\,\sigma$ per channel is needed to distinguish sources in the central part of the complex. 

We identify a total of 29 clouds (see Table~\ref{table:clumplist}), 20 in the W43 complex itself and 9 in the fore-/background (see Sect.~\ref{subsec:distdeterminationW43} for details).
The outcome of this method is not trivial, as it is not always clear which parts are still to be considered
associated and which are separate structures. It still needs some correction by hand in some of the very
weak sources and the strong complexes.
A few weak sources that have been identified by eye are manually added to our list (e.g. sources 18 and 28).
These are clearly coherent separate structures but are not identified by the algorithm. 
On the other hand, a few sources are merged by hand (e.g. source 26) that clearly belong together but are divided into several subsources by the software.
Some of these changes are open to interpretation, but they show that some adjustment of the software result is needed.
However, the algorithm works well and identifies 25 out of 29 clouds on its own.

The resulting detection map is shown in Fig.~\ref{fig:duchampdetectionmap}. The numbers shown there are color-coded to
show the distance of each detected cloud (See Sect.~\ref{subsec:distdeterminationW43} for details.).
Sources are sorted by their peak velocities.
See Table~\ref{table:clumplist} for positions and dimensions of the clouds, Table~\ref{table:clumplist_derived} for derived properties,
and Sect.~\ref{sec:sourcedescription} for a detailed description of the main complexes, while plots of all clouds can be found in Appendix~\ref{app:sourcesplots}.
We see a number of different sizes and shapes that range from small spherical clouds to expanded filaments and more complex structures.

The resulting data cubes show clouds of different shapes, while the typical spatial scales lie in the range of 10 to 20~pc (see Table~\ref{table:clumplist}).
We also show the area that the $^{13}$CO emission of each source covers, which was determined
by defining a polygon for each source that contains the $^{13}$CO emission. Thus, it accounts for shapes that deviate from spheres or rectangles, which is true for most of our clouds.
Hereafter we define clouds as objects with a size on the order of 10 pc, while we define clumps as objects on the parsec scale. The whole W43 region is considered a 
cloud complex.

\subsection{PV-diagram of the region}

\begin{figure*}[htb]
 \centering
 \begin{minipage}{\textwidth}
  \subfloat[{\bf (a)} PV-diagram of $^{13}$CO~(2--1) in \ensuremath{[}K\ensuremath{]} that is averaged along the Galactic latitude, showing several 
  velocity components. The y-axes show the velocity and distance, according to the rotational model seen in (b).]{\resizebox{0.485\hsize}{!}{\includegraphics{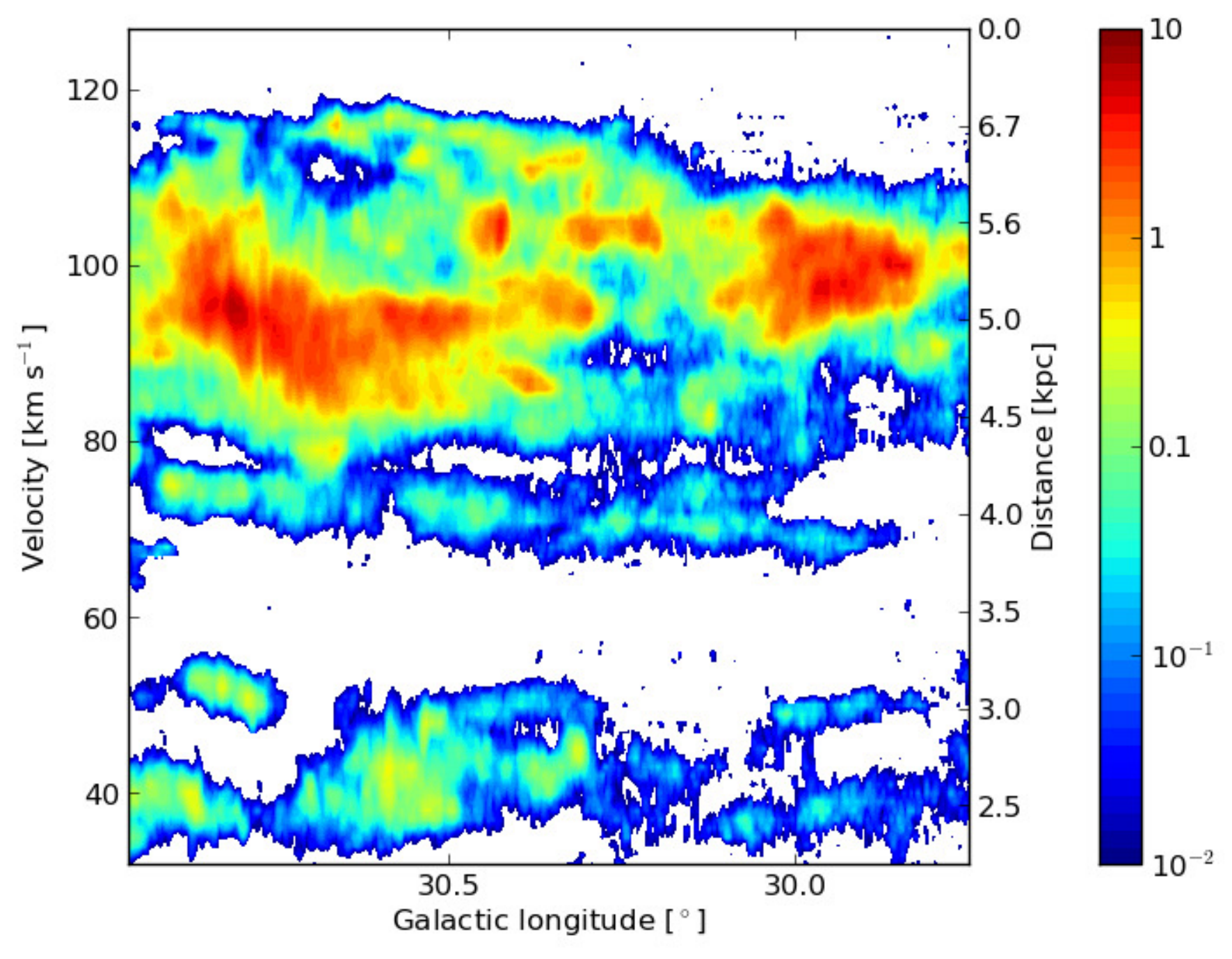}}}\hfill
  \subfloat[{\bf (b)} Simple rotational model of the Milky Way relating the relative velocity and the distance to the Sun for 30$^{\circ}$ Galactic longitude. 
  The stripes denote the different velocity components  found in (a).]{\resizebox{0.495\hsize}{!}{\includegraphics{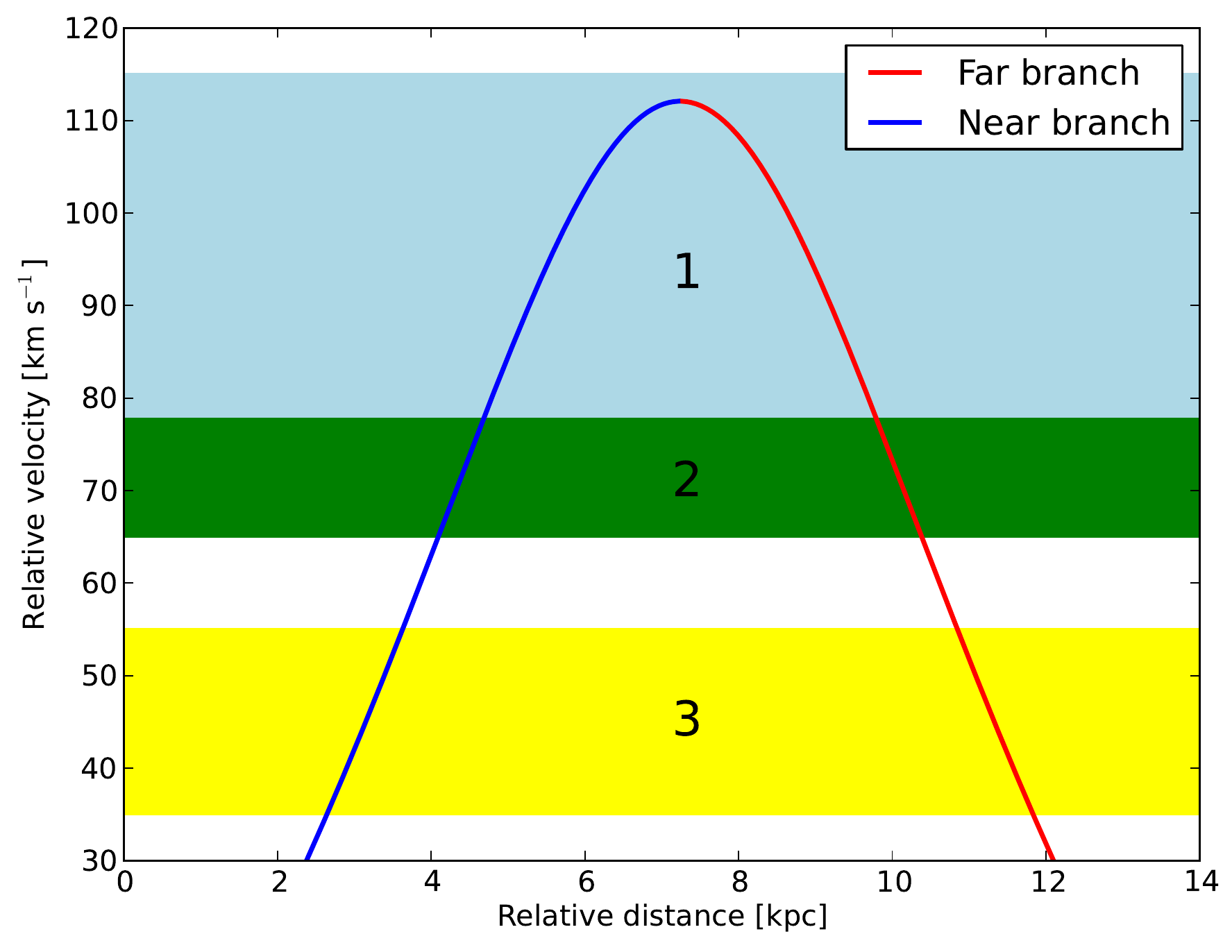}}}\\
  \subfloat[ {\bf (c)} Model of the spiral arm structure of the Milky Way from \citet{vallee2008}. The position of the W43 complex is marked at (1),
  the fore-/background complexes at (2), (3), and (3$^{\prime}$).]{\resizebox{0.465\hsize}{!}{\includegraphics{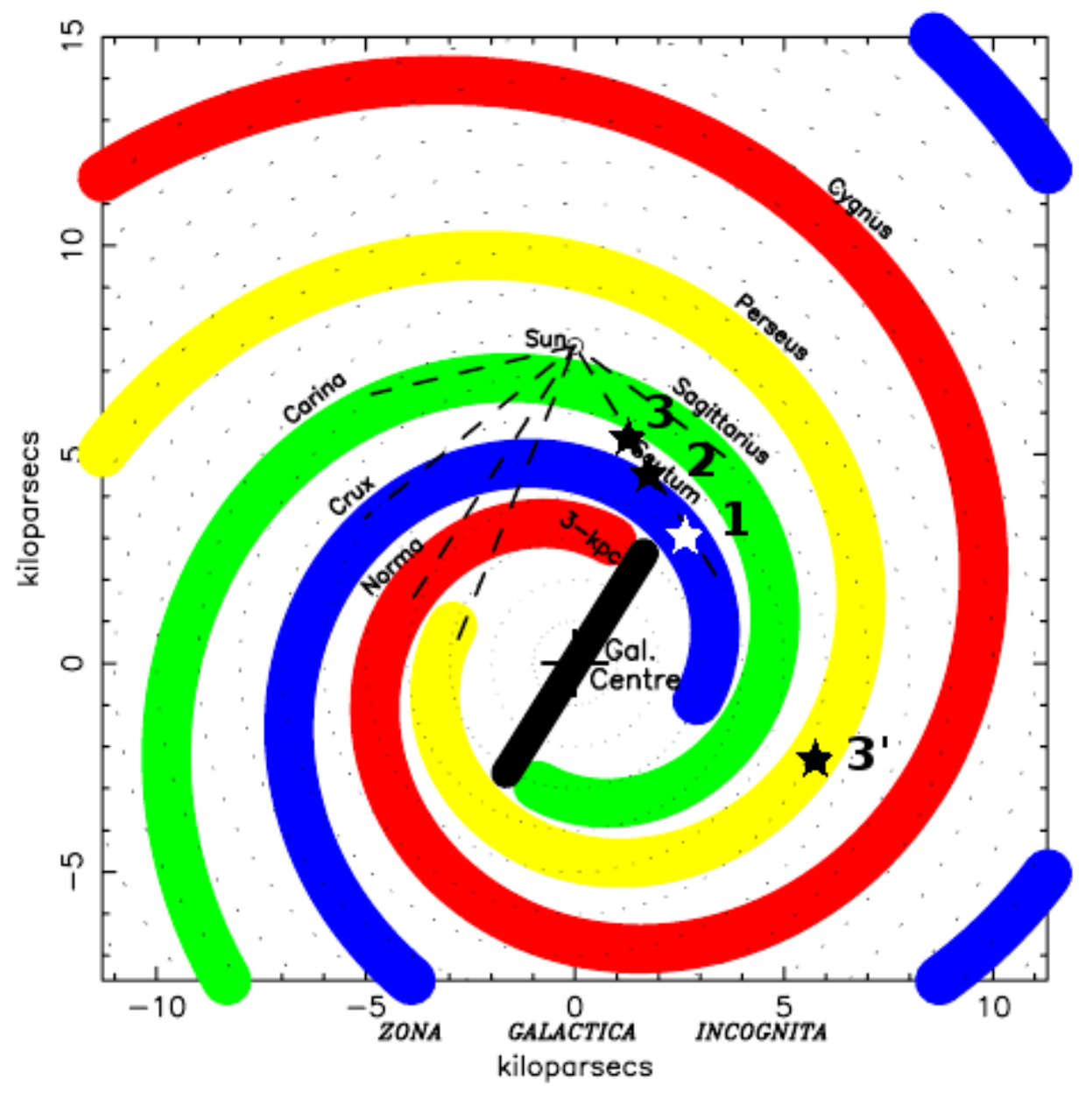}}}\hfill
  \subfloat[ {\bf (d)} PV-diagram of the model from \citep{vallee2008} showing relative velocities of spiral arms in the first quadrant.
  The interesting part is around 30$^{\circ}$ Galactic longitude (gray box), 
  where W43 is located.]{\resizebox{0.515\hsize}{!}{\includegraphics{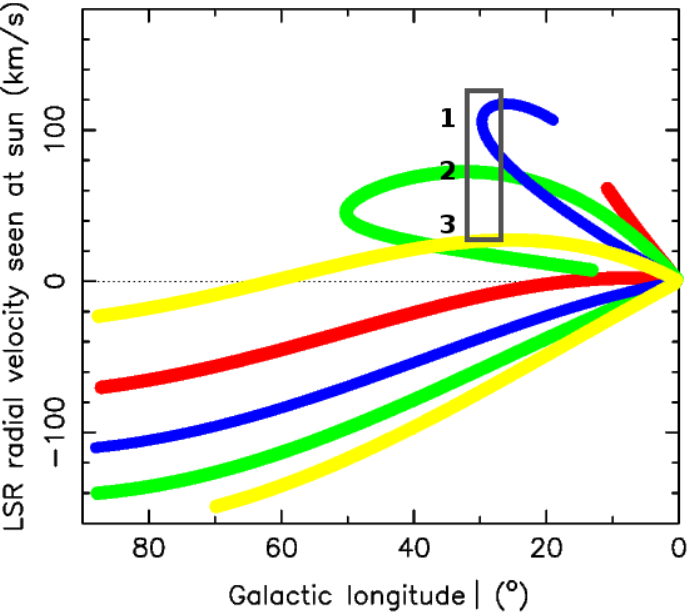}}}
 \end{minipage}
 \caption{Set of plots, illustrating the distance determination of W43 and its position in the Galaxy.}
 \label{fig:pvsum}
\end{figure*}

For a more advanced analysis of the velocity structure, we create a position velocity diagram of the $^{13}$CO~(2--1) line that is averaged along the Galactic latitude (see Fig.~\ref{fig:pvsum}~(a)).
The distribution of emission across the velocity range that is seen in the averaged spectra in Fig.~\ref{fig:average_spectra} can also be identified here but with 
additional spatial information along the Galactic longitude.
We note the similarity of this figure to the plot of $^{13}$CO~(1--0) displayed in \citet{nguyenluong2011}.
However, we see more details in our plot due to the higher angular resolution of our data.

We further analyze the position velocity diagram, as shown in Fig.~\ref{fig:pvsum}~(a), to separate our cube into several velocity components. We assume that these different components
are also spatially separated.

Two main velocity complexes can be distinguished: one between
35 and 55~km\,s$^{-1}$, the other between 65 and 120~km\,s$^{-1}$. Both complexes are clearly separated from each other, indicating that they are situated at different positions in the Galaxy.
On second sight, it becomes clear that the complex from 65 to 120~km\,s$^{-1}$ breaks down into a narrow component, spanning the 
range from 65 to 78~km\,s$^{-1}$ and a broad component between 78 and 120~km\,s$^{-1}$. Analyzing the channel maps of the cube verifies
that these structures are indeed separated. 
See Fig.~\ref{fig:velocitycomplexmaps} for integrated maps of each velocity component.

On the other hand, it is clearly visible that all three velocity components span the complete spatial dimension along the Galactic longitude. The
broadest complex at 78-120~km\,s$^{-1}$ shows two major components at 29.9$\degr$ and 30.8$\degr$ that coincide with W43-South and W43-Main, respectively.
The gap between both complexes is bridged by a smaller clump, and all three clumps are surrounded by diffuse gas, which forms an envelope around the whole complex. 
It is thus suggested to consider W43-Main and W43-South as one giant connected molecular cloud complex.
This connection becomes more clear in the PV-plot than in the spatial map in Fig.~\ref{fig:completemaps} \citep[see also][]{nguyenluong2011}, 
although the averaging of values causes blurring which might merge structures.

The lower velocity complex between 35 and 55~km\,s$^{-1}$ is a bit more fragmented than the other components. One central object at 30.6$\degr$ spans the whole velocity range, 
but it splits into two subcomponents to the edges of the map. 
It is hard to tell if we actually see one or two components. 

\subsection{Determination of the distance of W43}\label{subsec:distdeterminationW43}

\begin{figure*}[htb]
 \centering
 \begin{minipage}{\textwidth}
  \subfloat[{\bf (a)} 35 to 55~km\,s$^{-1}$ component at a distance of 4 and 12~kpc.]{\resizebox{0.317\textwidth}{!}{\includegraphics[scale=1]{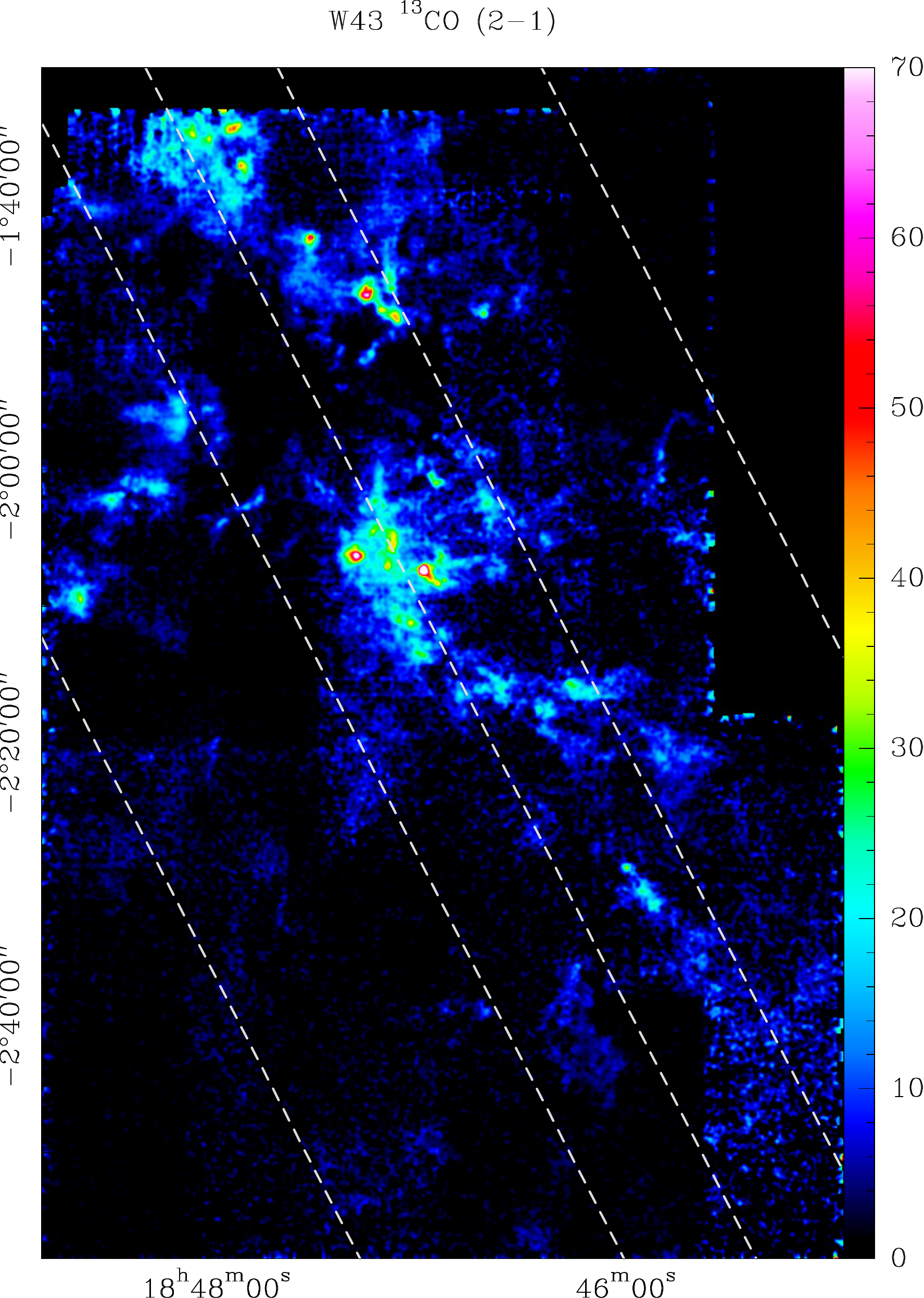}}}\hfill
  \subfloat[{\bf (c)} 65 to 78~km\,s$^{-1}$ component at a distance of 4.5~kpc.]{\resizebox{0.317\textwidth}{!}{\includegraphics[scale=1]{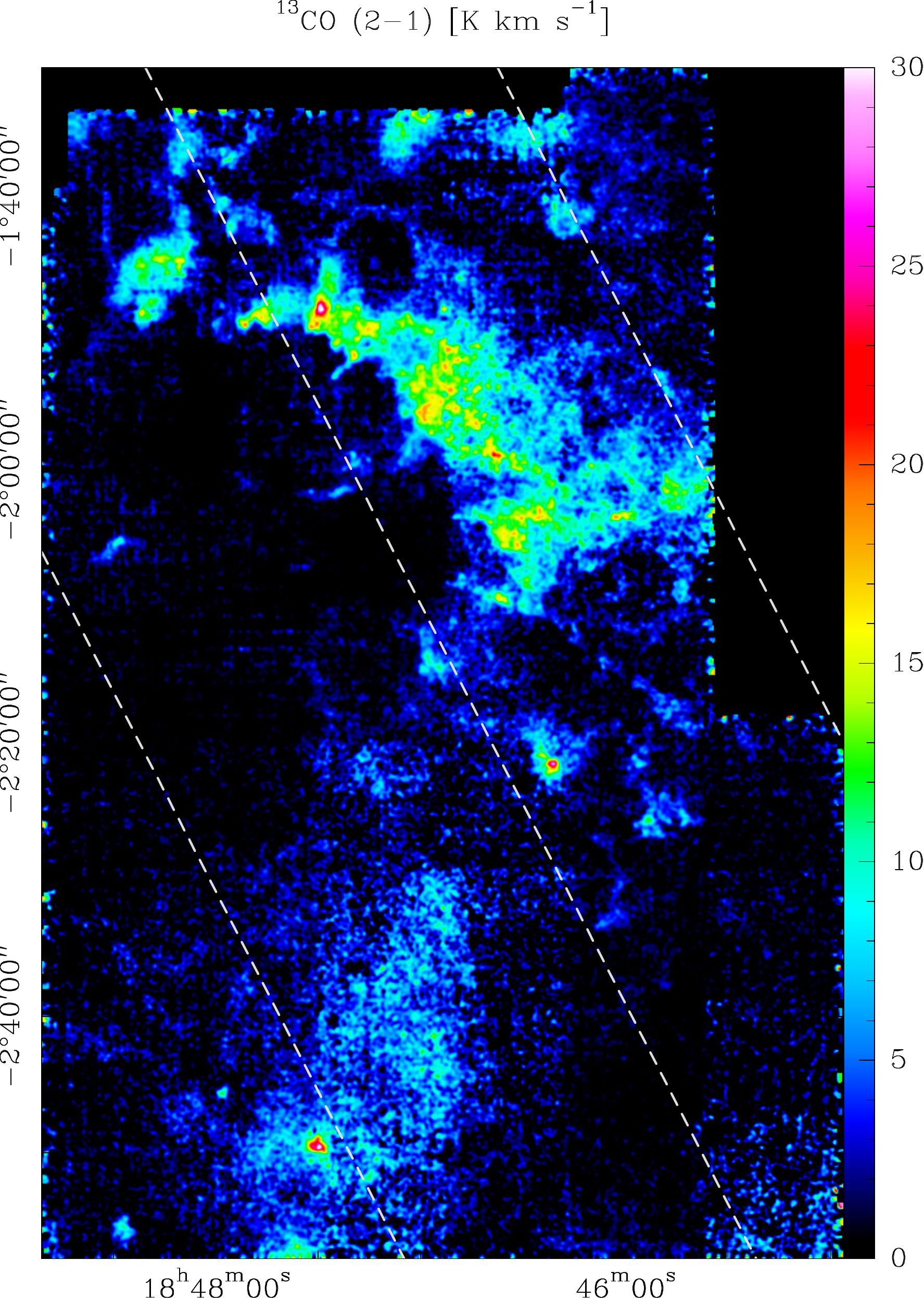}}}\hfill
  \subfloat[{\bf (d)} 78 to 120~km\,s$^{-1}$ component at a distance of 5 to 7~kpc.]{\resizebox{0.324\textwidth}{!}{\includegraphics[scale=1]{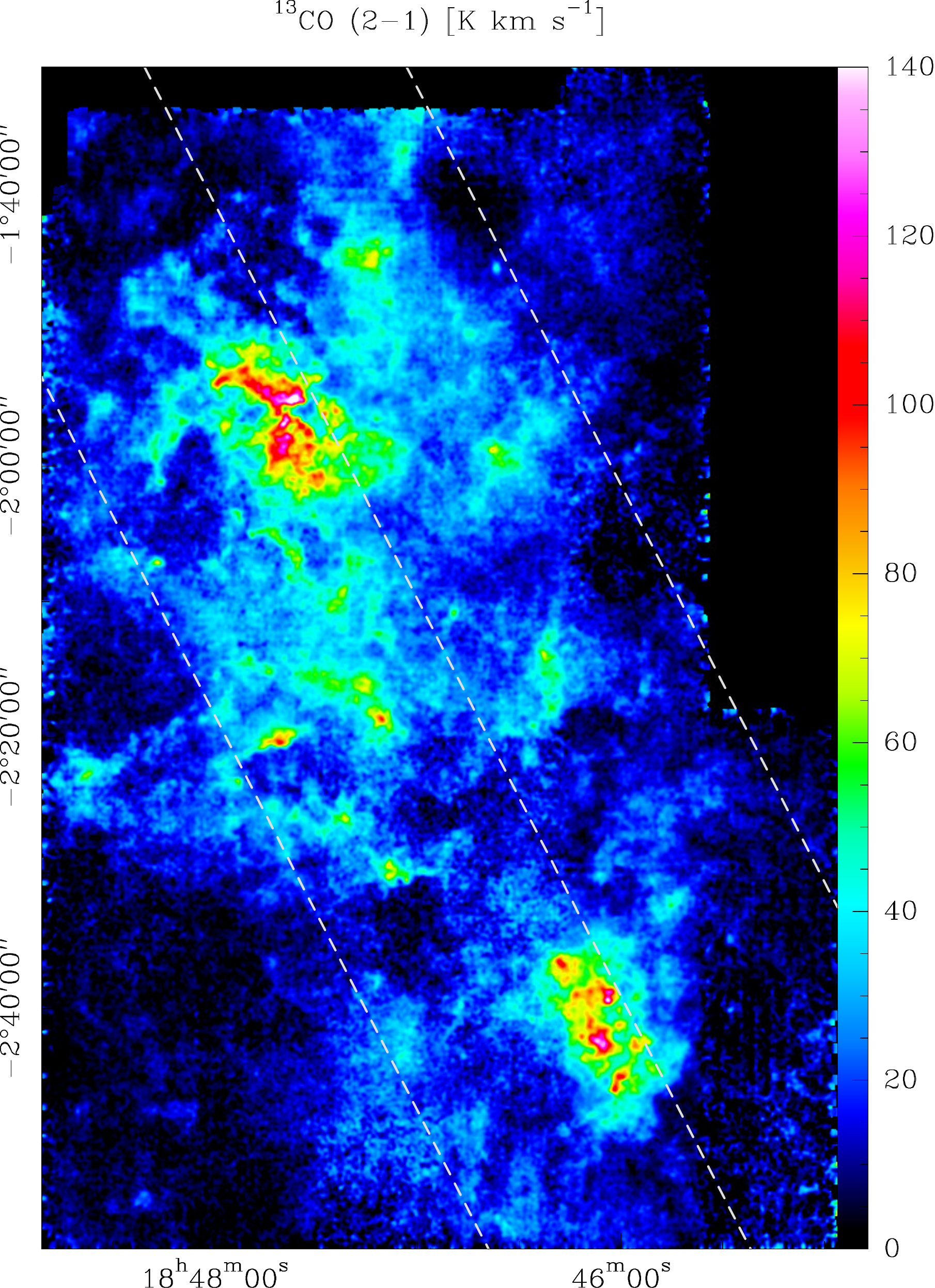}}}
 \end{minipage}
 \caption{Integrated $^{13}$CO~(2--1) maps of the separated velocity complexes. The white stripes mark the Galactic plane and the planes 30~pc above and below it.
 Figure (a) shows sources at two different distances and therefore two different scales.}
 \label{fig:velocitycomplexmaps}
\end{figure*}

We can analyze our data by using a simple rotational model of the Milky Way. For this model, we assume a rotational curve that increases linearly in the inner
3~kpc of the Galaxy, where the bar is situated. For radii larger than that, we assume a rotation curve that has a value of 254~km\,s$^{-1}$ at $R_{\odot}$
and slightly rises with the radius at a rate of 2.3~km\,s$^{-1}$\,kpc$^{-1}$ \citep[see][]{reid2009}.
The Galactocentric radius of a cloud with a certain relative velocity can be calculated using the formula

\begin{equation}
 r = R_{\odot}\, \mathrm{sin}(l)\frac{V(r)}{V_{\mathrm{LSR}} + V_{\odot}\,\mathrm{sin}(l)},
\end{equation}

\noindent as used in \citet{roman_duval2009}. The parameter $R_{\odot}$ is the Galactocentric radius of the Sun, which is assumed to be 8.4~kpc \citep[see][]{reid2009}, 
$l$ is the Galactic longitude of the source (30$^{\circ}$), $V_{\odot}$ the radial velocity of the Sun (254~km\,s$^{-1}$), and $V(r)$ the radial velocity of the source. 
The parameter $V_{\mathrm{LSR}}$ is the measured relative velocity between the source and the Sun. With the knowledge of the radius $r$, we can then compute the relative distance to the source
by the equation:

\begin{equation}
 d = R_{\odot}\, \mathrm{cos}(l)\pm\sqrt{r^2-R_{\odot}^2\mathrm{sin^2}(l)}.
\end{equation}

\noindent Up to the tangent point, two different possible distances exist for each measured velocity: one in front of and one behind the tangent point.
However, this calculation is not entirely accurate as the assumptions of the geometry of the Galaxy bear large errors. 
\citet{reid2009} state that the errors of the kinematic distance can sum up to a factor as high as 2.
One main reason for uncertainties are the streaming motions of molecular clouds relative to the motion of the spiral arms \citep{reid2009}.
Figure~\ref{fig:pvsum}~(b) shows the kinematic distance curve for our case at 30$^{\circ}$ Galactic longitude.
This works only for the circular orbits in the spiral arms and not the elliptical orbits in the Galactic bar \citep[see][]{rodriguezfernandez2008,rodriguezfernandez2011}.

To determine the location of each velocity complex, we need to break the kinematic distance ambiguity;
that is we need to decide for each complex if we assume it to be on the near or the far side of the tangent point.
Here, we use the distance estimations by \citet{roman_duval2009}, who utilized H{\sc I} self absorption from the VGPS project \citep{stil2006}.
It is possible to associate several entries of their extensive catalog with clouds we found in our dataset.
Thus, we are able to remove the distance ambiguity and attribute distances to these clouds.
In combination with the detailed model of the Milky Way of \citet{vallee2008}, we are able to fix their position in our Galaxy.
We cannot assign a distance to each single cloud in our dataset, as each was not analyzed by \citeauthor{roman_duval2009}
We assume the missing sources to have the same distances as sources nearby. 
This may not be exact in all cases, but the only unclear assignments are two clouds in the 35 to 55~km\,s$^{-1}$ velocity component (sources 1 and 3).
The distance of the W43 complex is unambiguous as it is well determined by the calculations found in \citet{roman_duval2009}.

The object W43 (78 to 120~km\,s$^{-1}$) is found to lie on the near side of the tangential point with distances from 5 to 7.3~kpc, which increases with radial velocity.
This places it near the tangential point of the Scutum arm at a Galactocentric radius of $R_{\mathrm{GC}}= 4$~kpc (marker 1 in Fig.~\ref{fig:pvsum}~(c)).
For our analysis, we use an average distance of 6~kpc for the whole complex, since this is where the mass center is located.

The second velocity component (65 to 78~km\,s$^{-1}$) lies in the foreground of the first one at a distance of 4.5~kpc to the Sun and $R_{\mathrm{GC}}=4.8$~kpc (marker 2 in Fig.~\ref{fig:pvsum}~(c)).
Another indication that these sources are located on the near side of the tangential point is their position above the Galactic plane, as seen in Fig.~\ref{fig:velocitycomplexmaps}~(b).
The larger the distance is from the Sun, the further above the Galactic plane it would be positioned. 
This would be difficult to explain, since high-mass star-forming regions are typically located within the plane.
It is unclear if this cloud is still situated in the Scutum arm or if it is located between spiral arms. 
According to the model, it would be placed at the edge of the Scutum arm.
In light of the previously discussed uncertainties, it is still possible that this cloud is part of the spiral arm.

The third component between 35 and 55~km\,s$^{-1}$ is more complex than the others, since we find sources to be located on both the near and far side of the tangential point.
The brightest sources in the center of our map are in the background of W43 in the Perseus arm with a distance of 11 to 12~kpc to the Sun and of $R_{\mathrm{GC}}=6$~kpc to the 
(marker 3 in Fig.~\ref{fig:pvsum}~(c)).
However, several other sources in the north and south are found by \citeauthor{roman_duval2009} to be near the Sun at 3.5 to 4~kpc (marker 3$^{\prime}$ in Fig.~\ref{fig:pvsum}~(c)). 
These sources also have a Galactocentric radius of 6~kpc.
Table~\ref{table:clumplist} gives an overview of the distance of each source, while Fig.~\ref{fig:velocitycomplexmaps} shows integrated intensity plots of the individual velocity components.



We can now apply this calculation to our data in Fig.~\ref{fig:pvsum}~(a) by changing the velocity scale into a distance scale. 
The distance scale is inaccurate for the parts of the lowest component that lie on the far side of the tangent point.
Although it is not possible to disentangle clouds that are nearby and far away in this plot, we still show this scale.
These values, taken from the rotation curve, are smaller than the actual distances found when compared to \citet{roman_duval2009}, since we used the newer rotation curve of \citet{reid2009}.
\citeauthor{roman_duval2009} use the older values from \citet{clemens1985}, which explains the discrepancy.
However, this axis still gives us an idea of the distribution of the clouds.
We note that no distance can be assigned for velocities larger than 112~km\,s$^{-1}$, hence the zero for the 120~km\,s$^{-1}$ tick in Fig.~\ref{fig:pvsum}~(a).
Subplot (d) shows the related modeled PV-diagram from \citet{vallee2008}. Our dataset is indicated by the gray box.


Figure~\ref{fig:pvsum}~(c) summarizes our determination of distance in a plot taken from \citet{vallee2008}. 
The W43 complex (78-120~km\,s$^{-1}$, marker 1) lies between 5 and 7~kpc, where the distance increases with velocity, which we found to be located on the near branch.
The complex 65 to 78~km\,s$^{-1}$ (marker 2) is located at the near edge of the Scutum arm, while the 35 to 55~km\,s$^{-1}$ component is marked by 3 and 3$^{\prime}$ on both
sides of the tangent point. The far component is located in the Perseus arm at 12~kpc distance and the near component at a distance of 4.5~kpc between the Scutum and the Sagittarius arm.

It may be a bit surprising that no emission from the local part of the Sagittarius arm is seen in our dataset. The reason is that our observed velocity range only goes
to 30~km\,s$^{-1}$. Possible molecular clouds nearby would have even lower relative velocities of $\sim$20~km\,s$^{-1}$, which can be seen in the model in Fig.~\ref{fig:pvsum}~(d).
In \citet{nguyenluong2011}, the $^{13}$CO~(1--0) spectrum, which is averaged over the W43 complex, shows an additional velocity component at 5 to 15~km\,s$^{-1}$ which fits to 
this spiral arm. 

\subsection{Peak velocity and line width}\label{subsec:peakvelocity}

\begin{figure*}[htb]
\centering
\begin{minipage}{0.48\textwidth}
\subfloat[{\bf (a)} Peak velocity map.]{\resizebox{\hsize}{!}{\includegraphics{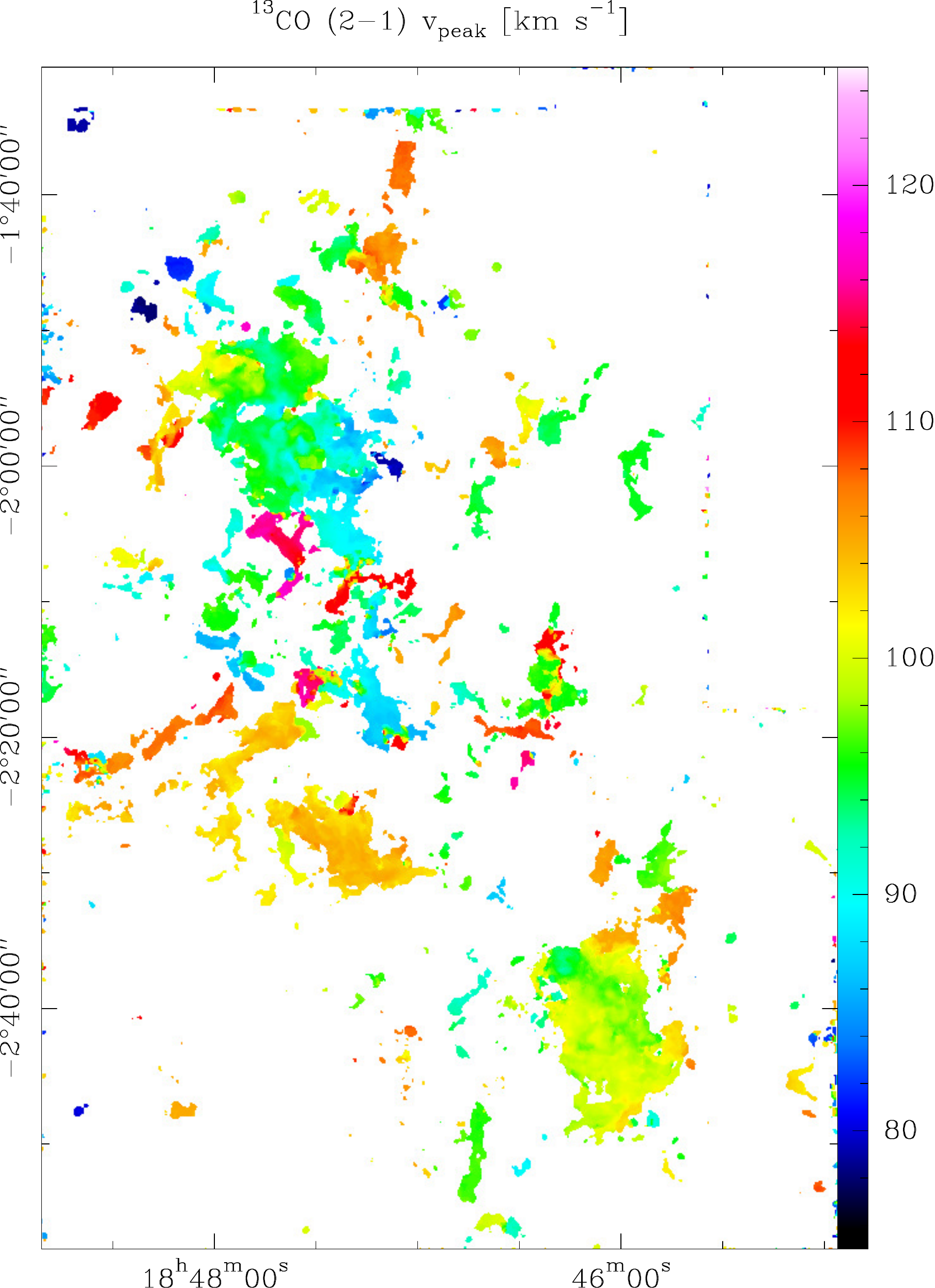}}}
\end{minipage}
\hfill
\begin{minipage}{0.47\textwidth}
\subfloat[{\bf (b)} Line width (FWHM) map.]{\resizebox{\hsize}{!}{\includegraphics{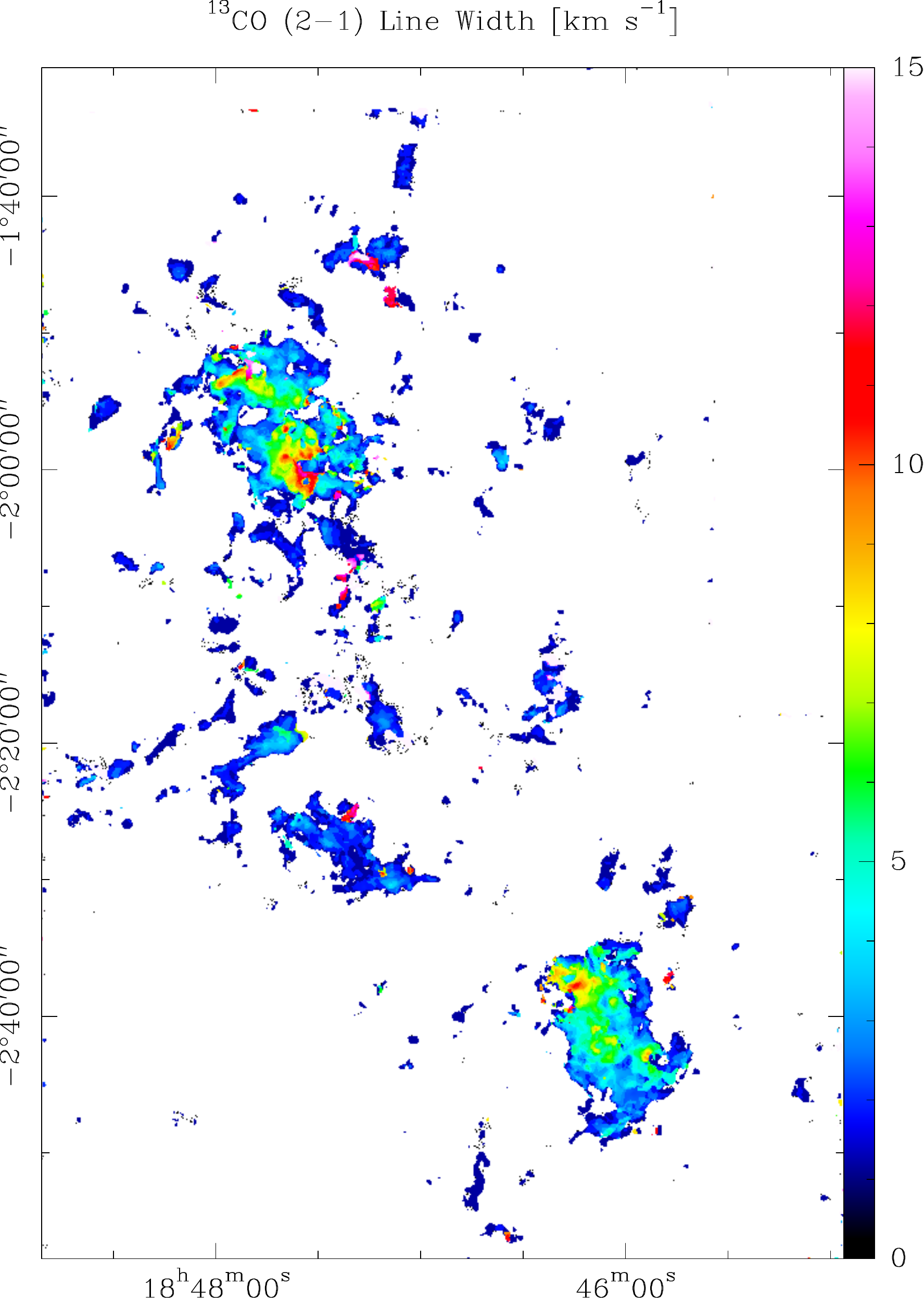}}}
\end{minipage}
\caption{Moment maps of the W43 complex derived from the $^{13}$CO~(2--1) cube. A cutoff of 4~K per channel was used.}
\label{fig:velocitylinewidthmaps}
\end{figure*}

After separating the different velocity components, we created moment maps of each component. For each spectra of the $^{13}$CO data cube, a Gaussian line profile was fitted. The first moment resembles
the peak velocity, the position of the line peak. The second moment is the width of the line. The maps of the W43 complex are shown in Fig.~\ref{fig:velocitylinewidthmaps}, while
the plots of the background components can be seen in the Appendix in Fig.~\ref{fig:appendixpeakvelocitymaps}.
Care should be taken in interpretation of these maps. 
As some parts of the maps show complex spectra (see Fig.~\ref{fig:w43main_spectra} for some examples), a Gaussian profile is not always a good approximation.
In regions where several velocity components are found, the maps only give information about the strongest component.
In case of self-absorbed lines, the maps may even be misleading. 
This especially concerns the southern ridge of W43-Main, called W43-MM2 as defined in \citep{nguyenluong2013}.

The line peak velocity map in Fig.~\ref{fig:velocitylinewidthmaps}~(a) traces a variety of coherent structures. Most of these correspond to the sources we identified
with the Duchamp software. However, some structures, as mentioned above, overlap and cannot be defined simply from using this velocity map.

The two main clouds, W43-Main and W43-South, are again located in the upper left and lower right part of map, respectively. 
As in the PV-diagram in Fig.~\ref{fig:pvsum}~(a), we note that both clouds are slightly shifted 
in velocity. While W43-Main lies in the range of 85 to 100~km\,s$^{-1}$, W43-South spans velocities from 95 to 105~km\,s$^{-1}$. Several smaller sources bridge the gap
between the two clouds, especially in the higher velocities. This structure is also seen in the PV-diagram.

In comparison to the PV-diagram, this plot shows the peak velocity distribution in both spatial dimensions.
On the other hand, we lose information of the shape of the lines.
Here, we see that the velocity of W43-South is rather homogeneous across the whole cloud. 
In contrast, W43-Main shows strong velocity gradients from west to east and from south to north, which are already seen in \citet{motte2003}.
The velocity changes by at least 30~km\,s$^{-1}$ on a scale of 25~pc.
We can interpret this as mass flows across the cloud, which makes it kinematically much more active than W43-South.

Figure~\ref{fig:velocitylinewidthmaps}~(b) shows a map of the FWHM line width of each pixel. Some parts in W43-Main show unrealistically large values of more than 10~km\,s$^{-1}$. This is a line-of-sight
effect and originates in several velocity components located at the same point on the sky. Therefore, it is more accurate to analyze the line width of each source separately.
From these single sources, we determine the mean line width, which is given in Table~\ref{table:clumplist}~(7).

\section{Analysis}\label{sec:analysis}

\begin{table*}[htb]
 \centering
 \begin{minipage}{\textwidth}
 \resizebox{\hsize}{!}{
 \begin{tabular}[c]{l l c c c r@{ -- }l c c c c c c}
  \toprule
        &        & (1)  & (2)  & (3)            & \multicolumn{2}{ c }{(4)}      & (5)       & (6)       & (7)            & (8)      & (9)             & (10) \\
        & Number & RA   & DEC  & Peak $^{13}$CO & \multicolumn{2}{ c }{Velocity} & $^{13}$CO & C$^{18}$O & Mean $^{13}$CO line & Assumed  & Dimensions      & Area \\
        &        & peak & peak & velocity       & \multicolumn{2}{ c }{extent}   & peak      & peak      & width (FWHM)     & distance & RA $\times$ DEC & of $^{13}$CO \\
  \midrule
        &  & & & [km\,s$^{-1}$] & \multicolumn{2}{ c }{[km\,s$^{-1}$]} & [K] & [K] & [km\,s$^{-1}$] & [kpc] & [pc] & [pc$^2$] $\times10^2$ \\
  \midrule
        & 1  & 18:45:59.5 & -02:29:08.3 & 35.8  & 33 & 40   & 9.0  & 4.5  & 1.5 & 12  & 12.9 $\times$ 13.3 & 1.6 \\
        & 2  & 18:47:15.5 & -01:47:17.3 & 39.9  & 33 & 45   & 12.7 & 4.0  & 1.6 & 4   & 9.7 $\times$ 8.3 & 0.7 \\
        & 3  & 18:47:54.5 & -01:35:02.5 & 41.9  & 33 & 48   & 14.5 & 4.0  & 1.7 & 4   & 11.3 $\times$ 9.3 & 1.0 \\
 Fore-  & 4  & 18:46:16.3 & -02:15:39.3 & 45.9  & 37 & 47   & 8.5  & 3.3  & 1.2 & 12  & 47.8 $\times$ 22.0 & 8.7 \\
 ground & 5  & 18:46:59.1 & -02:07:22.5 & 47.8  & 38 & 51   & 21.0 & 5.7  & 1.4 & 12  & 27.9 $\times$ 43.3 & 7.0 \\
        & 6  & 18:48:30.7 & -02:02:09.8 & 49.0  & 46 & 54   & 8.4  & 2.7  & 1.2 & 3.5 & 6.7 $\times$ 4.4 & 0.2 \\
        & 7  & 18:48:02.0 & -01:55:34.3 & 51.8  & 48 & 57   & 8.4  & 4.0  & 1.4 & 3.5 & 5.7 $\times$ 5.6 & 0.2 \\
        & 8  & 18:47:29.8 & -02:49:38.1 & 72.5  & 69 & 75   & 12.2 & 5.8  & 0.8 & 4.5 & 13.1 $\times$ 9.2 & 0.8 \\
        & 9  & 18:47:29.5 & -01:48:11.9 & 75.7  & 69 & 78   & 10.9 & 5.4  & 0.9 & 4.5 & 44.9 $\times$ 35.4 & 9.5 \\
  \midrule
        & 10 & 18:48:10.6 & -01:45:37.7 & 80.9  & 74 & 85   & 15.7 & 4.9  & 1.6 & 6  & 10.3 $\times$ 9.9  & 1.0 \\
        & 11 & 18:47:10.2 & -02:18:37.7 & 86.9  & 84 & 92   & 18.5 & 10.0 & 1.4 & 6  & 7.3  $\times$ 8.4  & 0.8 \\
        & 12 & 18:47:52.8 & -02:03:33.5 & 91.8  & 88 & 98   & 15.5 & 4.5  & 2.0 & 6  & 11.5 $\times$ 22.2 & 2.3 \\
        & 13\footnote{W43-Main} & 18:47:36.4 & -01:55:06.6 & 93.8  & 78 & 108  & 21.9 & 9.0  & 5.2 & 6  & 29.0 $\times$ 20.9 & 6.4 \\
        & 14 & 18:47:22.6 & -02:12:01.8 & 94.6  & 91 & 97   & 8.2  & 3.3  & 1.9 & 6  & 18.0 $\times$ 23.4 & 3.5 \\
        & 15 & 18:46:22.6 & -02:14:04.2 & 94.8  & 90 & 100  & 13.6 & 5.2  & 2.0 & 6  & 18.0 $\times$ 10.3 & 1.2 \\
        & 16 & 18:46:43.3 & -02:02:28.1 & 94.8  & 91 & 97   & 9.0  & 3.1  & 1.0 & 6  & 26.4 $\times$ 23.9 & 3.2 \\
        & 17 & 18:47:27.1 & -01:44:59.1 & 94.9  & 89 & 102  & 12.8 & 3.7  & 2.2 & 6  & 17.5 $\times$ 18.7 & 2.3 \\
        & 18 & 18:46:43.9 & -02:53:59.6 & 96.5  & 94 & 98   & 11.2 & 3.9  & 1.0 & 6  & 4.4  $\times$ 14.7 & 0.7 \\
 W43    & 19 & 18:45:50.9 & -02:30:31:9 & 96.7  & 93 & 102  & 10.7 & 4.5  & 1.1 & 6  & 9.9  $\times$ 15.4 & 1.3 \\
        & 20\footnote{W43-South} & 18:46:04.0 & -02:39:22.2 & 97.6  & 89 & 107  & 27.5 & 12.8 & 4.8 & 6  & 23.7 $\times$ 31.1 & 6.0 \\
        & 21 & 18:48:16.6 & -02:07:06.0 & 103.2 & 97 & 107  & 11.8 & 4.5  & 1.5 & 6  & 8.2  $\times$ 3.7  & 0.4 \\
        & 22 & 18:47:08.0 & -02:29:32.3 & 103.9 & 98 & 108  & 19.5 & 9.7  & 2.2 & 6  & 25.0 $\times$ 15.4 & 3.3 \\
        & 23 & 18:47:40.1 & -02:20:24.3 & 104.9 & 100 & 108 & 20.6 & 9.0  & 2.0 & 6  & 13.1 $\times$ 11.3 & 1.1 \\
        & 24 & 18:47:09.5 & -01:44:08.4 & 105.5 & 102 & 110 & 11.5 & 5.7  & 1.6 & 6  & 11.2 $\times$ 26.5 & 2.9 \\
        & 25 & 18:46:26.1 & -02:19:11.5 & 108.1 & 106 & 117 & 13.7 & 6.3  & 1.0 & 6  & 11.2 $\times$ 14.8 & 1.2 \\
        & 26 & 18:48:38.1 & -02:22:49.1 & 108.2 & 104 & 112 & 11.9 & 4.8  & 1.5 & 6  & 22.0 $\times$ 17.6 & 2.7 \\
        & 27 & 18:48:32.2 & -01:55:28.8 & 111.1 & 107 & 116 & 10.4 & 4.6  & 1.5 & 6  & 14.1 $\times$ 10.5 & 1.4 \\
        & 28 & 18:47:22.2 & -02:09:25.2 & 112.5 & 110 & 116 & 10.7 & 4.2  & 1.0 & 6  & 12.0 $\times$ 7.7  & 0.9 \\
        & 29 & 18:47:48.6 & -02:05:03.7 & 115.9 & 110 & 120 & 15.2 & 5.8  & 1.2 & 6  & 8.2  $\times$ 12.7 & 1.2 \\
  \bottomrule
 \end{tabular}
 }
 \end{minipage}
\caption{Clouds found by the Duchamp Sourcefinder and their characteristics derived from the CO datasets.}
\label{table:clumplist}
\end{table*}

\subsection{Calculations}\label{subsec:calculations}

\begin{figure*}[htb]
\centering
\subfloat[{\bf (a)} $^{13}$CO (2--1)]{\resizebox{0.19\textwidth}{!}{\includegraphics{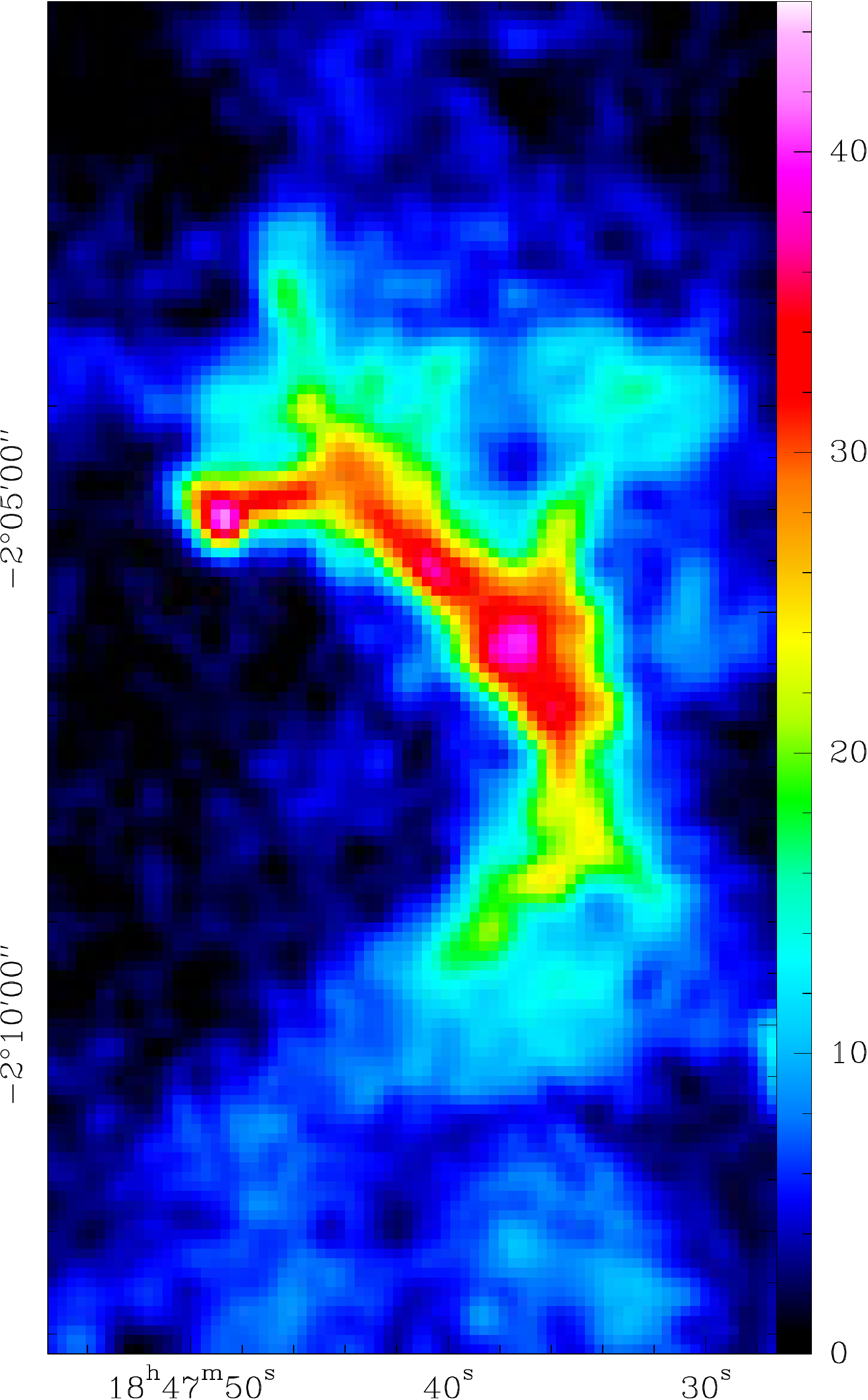}}}
\hfill
\subfloat[{\bf (b)} C$^{18}$O (2--1)]{\resizebox{0.19\textwidth}{!}{\includegraphics{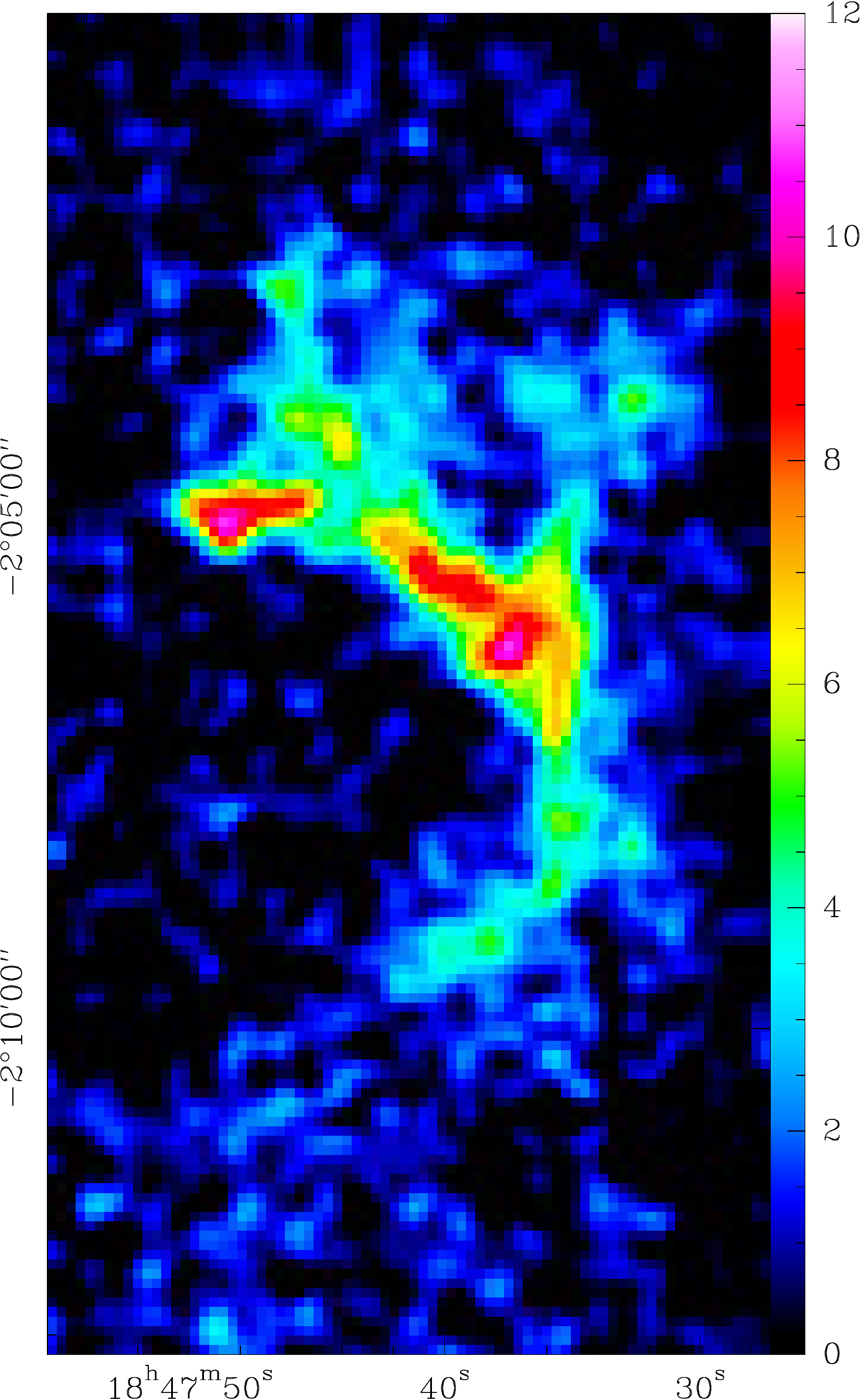}}}
\hfill
\subfloat[{\bf (c)} Optical depth]{\resizebox{0.185\textwidth}{!}{\includegraphics{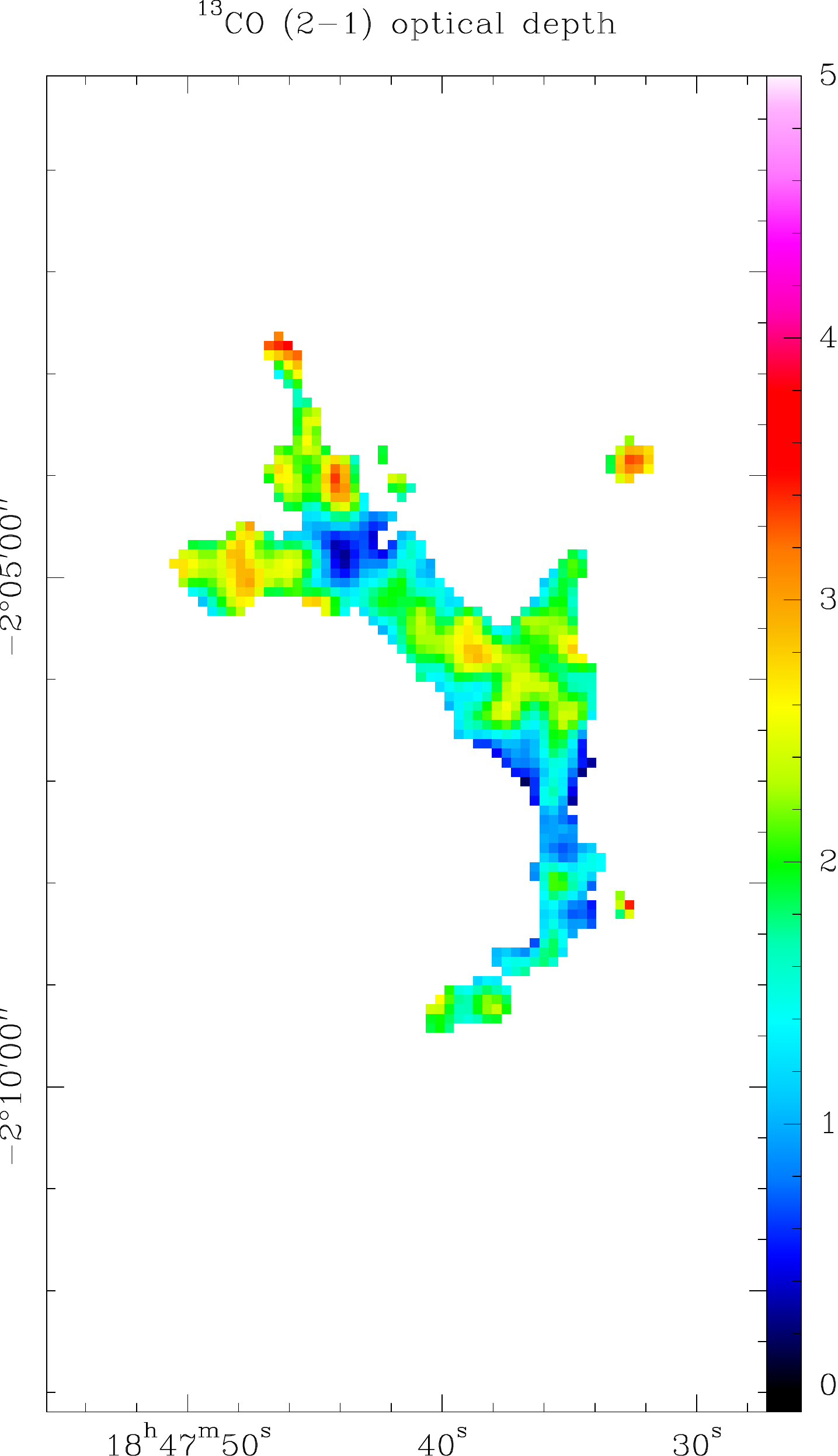}}}
\hfill
\subfloat[{\bf (d)} Excitation temperature]{\resizebox{0.19\textwidth}{!}{\includegraphics{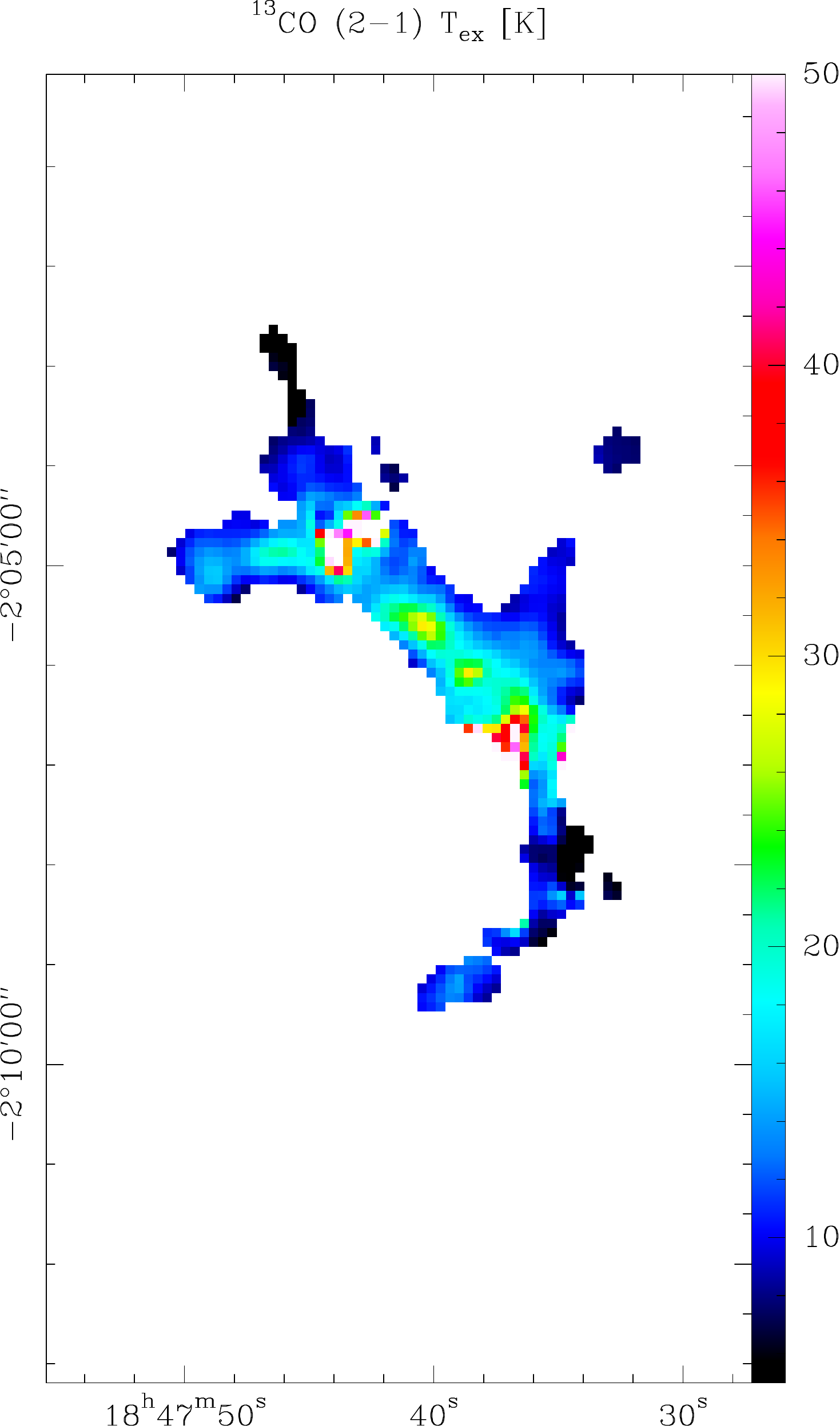}}}
\hfill
\subfloat[{\bf (e)} H$_2$ column density]{\resizebox{0.195\textwidth}{!}{\includegraphics{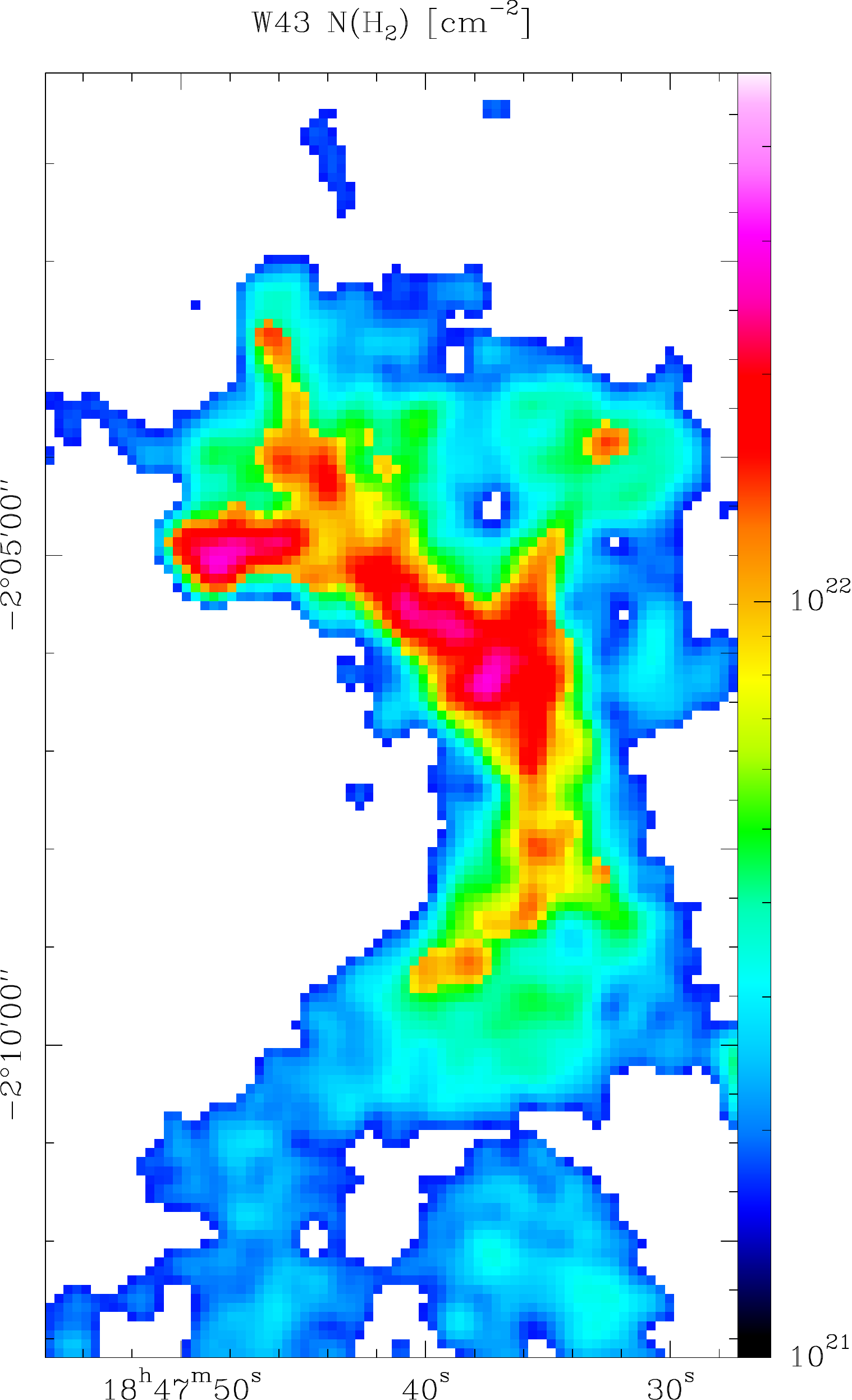}}}
\caption{Series of plots showing the different steps of calculations carried out for each source. The example shows source 29, which is located in the center of the
W43 complex at a velocity of 115~km\,s$^{-1}$.
From left to right: {\bf (a)} 
$^{13}$CO~(2--1) in [K\,km\,s$^{-1}$], {\bf (b)} C$^{18}$O~(2--1) in [K\,km\,s$^{-1}$], {\bf (c)} optical depth $\tau$ of $^{13}$CO~(2--1), 
{\bf (d)} excitation temperature in [K], and {\bf (e)} H$_2$ column density map in [cm$^{-2}$] derived from the CO lines as described in Sect.~\ref{subsubsec:column_density}. 
The column density has been calculated from an assumed constant excitation temperature.}
\label{fig:examplecalc}
\end{figure*}

\subsubsection{Optical depth}

For each identified source, we conducted a series of calculations to determine its physical properties.
We did this on a pixel by pixel basis, using maps integrated over the velocity range that is covered by the 
specific source. 
The optical depth of the $^{13}$CO gas was calculated from the ratio of the intensities of $^{13}$CO (2--1) and
C$^{18}$O (2--1), assuming that C$^{18}$O is optically thin. 
(This assumption holds for H$_2$ column densities up to \mbox{$\sim10^{23}$}~cm$^{-2}$, but a clear threshold cannot be given.)
Then we computed the excitation temperature of this gas and the H$_2$ column
density, which was then used to estimate the total mass along the line-of-sight. 
All these calculations are explained in detail in Appendix~\ref{app:calculations}.
Example maps for a small filament (source 29) can be seen in Fig.~\ref{fig:examplecalc}.

We first calculated the $^{13}$CO~\mbox{(2--1)} optical depth from the ratio of the two CO lines. We note, that the intrinsic ratios of the different CO isotopologes
used for this calculation are dependent on the Galactocentric radius, so we have to use different values for W43 and the fore-/background sources.
An example map of $\tau$ of $^{13}$CO~\mbox{(2--1)} is shown in Fig.~\ref{fig:examplecalc}~(c). 
Typical clouds have optical depths of a fraction of 1 in the outer parts and up to 4 at most in the central cores. The extreme case is the
W43-main cloud, where the $^{13}$CO optical depth goes up to 8.
This means that most parts of the clouds are optically thin and we can see through them. Even at positions where $^{13}$CO become optically thick,
C$^{18}$O still remains optically thin. Only for the extreme case of the densest part of W43-Main, C$^{18}$O starts to become optically thick.
This means that the combination of the two isotopologes reveals most of the information about the medium density CO gas in the W43 complex.

\subsubsection{Excitation temperature}\label{subsubsec:ex_temp}

\begin{figure}[htb]
 \centering
 \begin{minipage}{0.48\textwidth}
  \resizebox{\hsize}{!}{\includegraphics{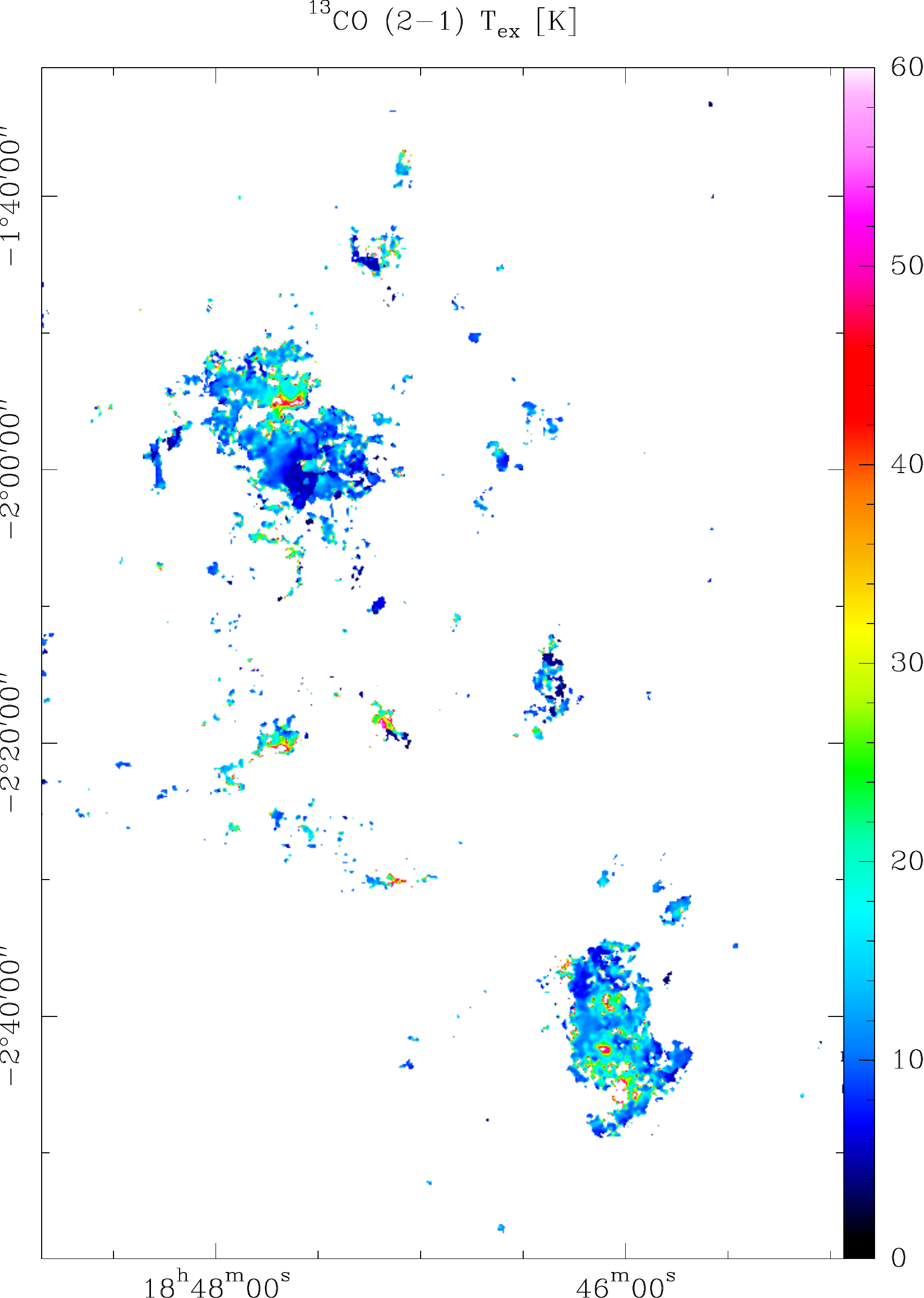}}
 \end{minipage} 
 \caption{Map of the derived excitation temperature in the W43 complex. This map shows unrealistic low temperatures of $\sim5$\,K in several regions 
 (cp. Sect.~\ref{subsubsec:ex_temp} for a discussion of error sources).
 The beam size is identical to those of the CO maps (11\arcsec).}
 \label{fig:tex_map_total}
\end{figure}

The formula used for the computation of the excitation temperatures is explained in Appendix~\ref{app:calculations}. 
The resulting map is shown in Fig.~\ref{fig:tex_map_total}.
Certain assumptions are made. First, we assumed that $T_{\mathrm{ex}}$ is the same for the $^{13}$CO and the 
C$^{18}$O gas. This method becomes unrealistic when the temperature distribution along the line-of-sight is not uniform anymore. If there was 
a temperature gradient, we would miss the real ratio of the $^{13}$CO to C$^{18}$O line intensities and thus either over- or underestimate the temperature. Thus the calculated temperature
might be incorrect for very large cloud structures
that show a complex temperature distribution along the line-of-sight.
This problem is partially circumvented by using the spectral information of our observations,
but we use intensity maps integrated over at least several km\,s$^{-1}$ for our calculation of the excitation temperature, which still leaves room for uncertainties.
This means we do not confuse different clouds, but we still average the temperature along the line-of-sight over the complete clouds.

The centers of the two main clouds in the W43 region are candidates for an underestimated excitation temperature. 
In case these regions were internally heated, a decreasing gradient in the excitation temperature would appear from the inside of the cloud to the outside.
As $^{13}$CO is rather optically thick, only the cool outside of the cloud would be seen by the observer. 
In contrast, C$^{18}$O would be optically thin; thus, the hot center of the cloud would also be observed.
Averaging along the line-of-sight, $I$(C$^{18}$O) would be increased relative to $I$($^{13}$CO), which would lead to an overestimated 
optical depth. This would then lead to an underestimated calculated excitation temperature.
External heating, on the other hand, would result in an overestimated excitation temperature. 
However, we find the first case with regard to low excitation temperatures is more likely in some clouds.

Another effect, which leads to a reduced excitation temperature, is the beam-filling factor $\eta$. In our calculations, we assume it to be 1. This corresponds to extended clouds that
completely fill the telescope beam. This is not a true representation of molecular clouds, as they are structured on the subparsec scale and we would have to use a factor $\eta <1$. 
Technically speaking, we calculate the value of $T_{\mathrm{ex}}\times\eta$, which is smaller than $T_{\mathrm{ex}}$. 

As the C$^{18}$O line is much weaker than the $^{13}$CO line,
we cannot use the ratio of them for those pixels where no C$^{18}$O is detected, even if $^{13}$CO is present. 
We find typical temperatures to be between 6 and 25~K; in some rare cases, it is up to 50~K with a median of 12~K.

Due to the sparsely covered maps (see Fig.~\ref{fig:tex_map_total}) and the uncertainties described above, we concluded that is was best not to use the excitation
temperature maps for the following calculations of the H$_2$ column density.
Instead, we assumed a constant excitation temperature for the complete W43 region. We chose the value to be 12~K, since this was the median temperature found in the W43 complex.
Assuming a constant temperature value across the cloud is likely not a true representation of the cloud; in particular, it does not distinguish between star forming cores and the ambient background.
However, such an assumption is a good first approximation to the temperature in the cloud and is more representative of star-forming cores than the aforementioned unrealistically low values.

\subsubsection{H$_2$ column density}\label{subsubsec:column_density}

We also calculated the column density along the line-of-sight of the $^{13}$CO gas from the assumed constant excitation temperature, the $^{13}$CO integrated emission,
and a correction for the opacity (See Appendix~\ref{app:calculations} for details.).
Assuming a constant ratio between $^{13}$CO and H$_2$, it was then possible to find the H$_2$ column density.
Please note that H$_2$ column densities derived by assuming a constant temperature are also subject to the same caveats and accuracies.

\begin{figure}[htb]
 \centering
 \begin{minipage}{0.48\textwidth}
  \resizebox{\hsize}{!}{\includegraphics{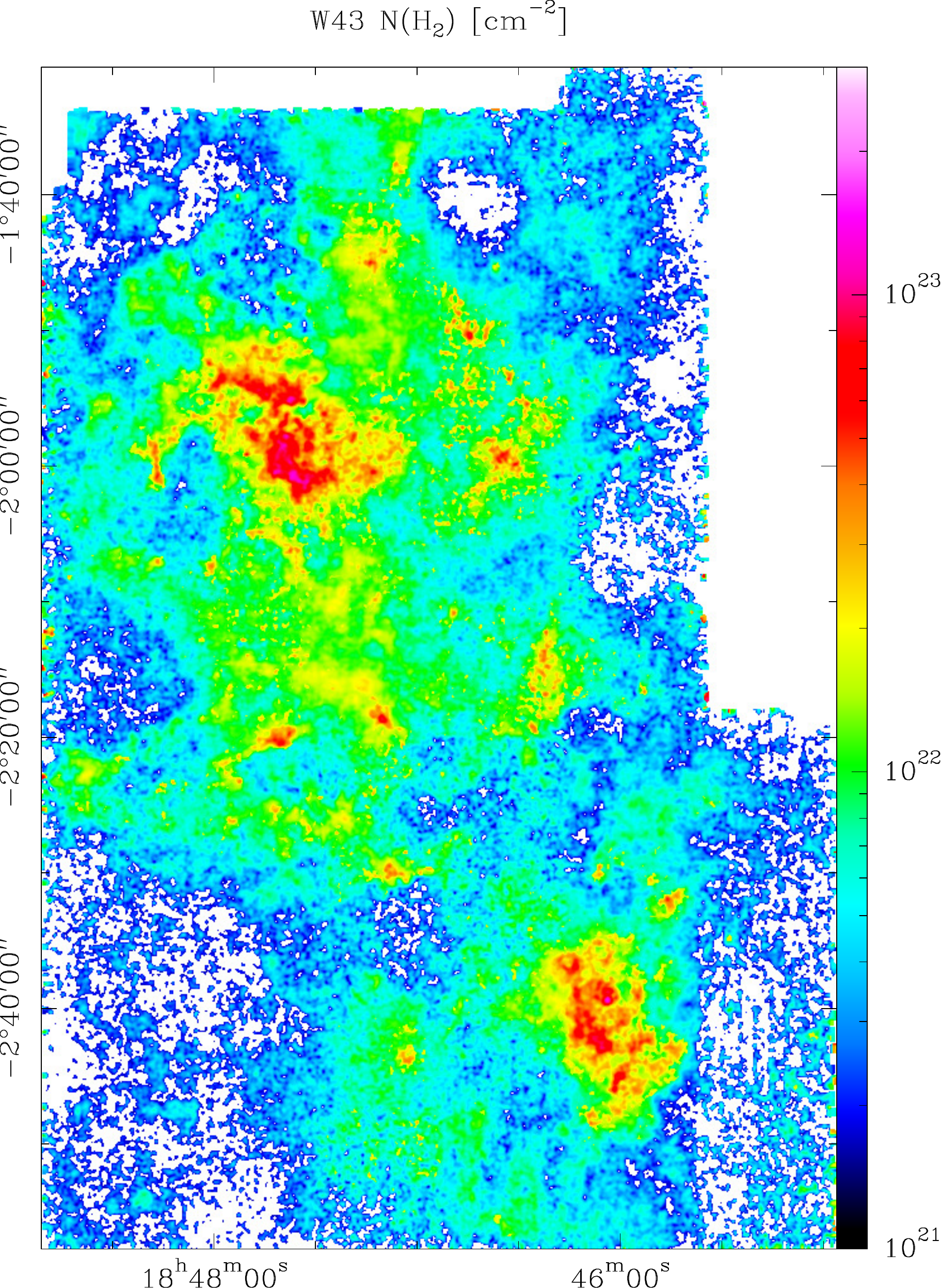}}
 \end{minipage}
 \caption{Map of the H$_2$ column density of the complete W43 complex derived from the IRAM 30m $^{13}$CO and C$^{18}$O maps. The velocity range between 78 and 120~km\,s$^{-1}$ was used.
 The beam size is identical to those of the CO maps (11\arcsec).}
 \label{fig:h2_map_total}
\end{figure}

All ratios between H$_2$ and CO isotopologes bear errors, since they depend on the Galactocentric radius. 
These errors add up with the uncertainty on the assumed excitation temperature.
The final results for column densities and masses must be taken with caution, because there is at least an
uncertainty of a factor of 2. 
Fig.~\ref{fig:h2_map_total} shows the calculated H$_2$ column density map of the full W43 complex. The column density has been calculated at those points, where the $^{13}$CO integrated
intensity is higher than 5~K\,km\,s$^{-1}$.
The resulting values range from values that are a few times of $10^{21}$~cm$^{-2}$ in the diffuse surrounding gas up to $\sim2\times10^{23}$~cm$^{-2}$ in the center of W43-Main.

The southern ridge of W43-Main, where we calculate high column densities, is the most problematic part of our dataset. The spectra reveal that $^{13}$CO is self-absorbed
in this part of the cloud. We use the integrated intensity ratio to calculate the opacity at each point, which is strongly overestimated in this case. This leads to both low
excitation temperatures and high column densities. The results for this part of the cloud should be used with caution.

\subsubsection{Total mass}\label{subsubsec:totalmass}

From the H$_2$ column density, we then determine the total mass of our sources, given in Table~\ref{table:clumplist}.
We find that the total mass of a typical cloud is in the range of a few $\times 10^4$ solar masses. Of course, this is only the mass
seen in the mid- to high-density sources. The very extended diffuse molecular gas cannot be seen with $^{13}$CO; it is generally traced by $^{12}$CO lines and accounts for 
a major fraction of the gas mass \citep[][]{nguyenluong2011}.

The total H$_2$ mass as derived from our $^{13}$CO~\mbox{(2--1)} and C$^{18}$O~\mbox{(2--1)} observations is found to be 

\begin{equation}
 M(\mathrm{H}_2)_{\mathrm{CO}} = 1.9 \times 10^6\, M_{\sun}
\end{equation}

\noindent for the W43 complex with about 50\% within the clouds that we have identified and the rest in the diffuse surrounding gas.
Here, we have excluded the foreground sources and only considered the W43 complex itself.
\citet{nguyenluong2011} used similar areas and velocity ranges ($\sim$80$\times$190~pc, 80-120~km\,s$^{-1}$) and determined a molecular gas mass in W43 clouds 
from the Galactic Ring Survey \citep{jackson2006} of $4.2\times 10^6$~$M_{\sun}$.
A different estimation of the H$_2$ column density in W43 was done by \citet{nguyenluong2013} using Herschel dust emission maps. 
Using these maps, we find a value of $2.6 \times 10^6$~$M_{\sun}$.
See Sect.~\ref{subsec:higal_comparison} for a discussion of this difference.

We underestimate the real mass where $^{12}$CO exists but no $^{13}$CO is seen, where C$^{18}$O might become optically thick, and where our assumption for the excitation temperature is too high.
On the other hand, we overestimate the real mass, where our assumption for the excitation temperature is too low. 
In extreme cases of very hot gas, the gas mass can be overestimated by 40\% at most (cp. Fig~\ref{fig:appendix_13co_excitation_plot} in Appendix~\ref{app:subsec:masscalc}),
while very cold cores can be underestimated by a factor of nearly 10.
Both effects partly cancel out each other, when integrating over the whole region; however, we estimate that the effects, which underestimate the real mass, are stronger.
Therefore, the mass we calculated should be seen as a lower limit of the real molecular gas mass in the W43 complex.

\begin{table*}[htb]
 \centering
 \begin{minipage}{0.98\textwidth}
 \resizebox{\hsize}{!}{
 \begin{tabular}[c]{l l c c c c c c c c}
  \toprule
         &        & (1)        		 & (2)     	     	& (3)               & (4)                      	& (5)			& (6)    	 & (7)       	& (8) \\
         & Number & Integrated 		 & Mean   	     	& Maximum 	    & Median $T_{\mathrm{ex}}$ 	& mean H$_2$     	& max H$_2$ 	 & Total 	& Virial \\
         &        & $^{13}$CO intensity  & $\tau$($^{13}$CO) 	& $\tau$($^{13}$CO) &                          	& column density 	& column density & mass 	& mass   \\
  \midrule
         & & [K\,km\,s$^{-1}$\,pc$^2$]$\times10^3$ & & & [K] & [cm$^{-2}$] $\times\,10^{21}$ & [cm$^{-2}$] $\times\,10^{22}$ & [$M_{\sun}$] $\times\,10^{3}$ 	& [$M_{\sun}$] $\times\,10^{3}$ \\
  \midrule
         & 1      		  & 1.3  & 1.33   		& 3.4 		    & 7.0  			& 10.5		 	& 3.4  		 & 21.5   	& 18.4  \\
         & 2      		  & 0.9  & 0.87   		& 3.5 		    & 11.1 			& 8.7   	 	& 5.4  		 & 13.2	  	& 14.1  \\
         & 3      		  & 1.3  & 0.88   		& 4.1 		    & 9.2  			& 10.2 		 	& 3.8  		 & 18.6	  	& 19.3 \\
  Fore-  & 4      		  & 5.9  & 1.01   		& 3.9 		    & 7.1  			& 7.3 		 	& 3.2 		 & 73.3  	& 27.9  \\
  ground & 5      		  & 15.2 & 0.52   		& 2.2 		    & 10.1 			& 9.2 		 	& 8.1  		 & 202.5  	& 33.9  \\
         & 6      		  & 0.2  & 0.81   		& 2.8 		    & 7.7  			& 7.4 		 	& 2.0  		 & 2.0   	& 8.1  \\
         & 7      		  & 0.3  & 1.27   		& 4.5 		    & 7.7  			& 10.2 		 	& 5.7  		 & 4.8	  	& 11.6  \\
         & 8      		  & 0.5  & 1.41   		& 2.6 		    & 10.4 			& 3.7 		 	& 1.8  		 & 2.2   	& 3.7  \\
         & 9      		  & 5.0  & 1.22   		& 4.4 		    & 7.5  			& 4.7 		 	& 5.1  		 & 43.2  	& 16.3  \\
  \midrule                           
         & 10     		  & 1.3  & 0.97   		& 3.6 		    & 8.8  			& 5.1 		 	& 2.8  		 & 10.5   	& 17.1  \\
         & 11     		  & 1.2  & 2.15   		& 5.2 		    & 15.0 			& 7.1 		 	& 8.1  		 & 11.3  	& 11.5  \\
         & 12     		  & 3.5  & 1.05   		& 4.4 		    & 6.9  			& 5.2 		 	& 3.4  		 & 26.8  	& 39.5  \\
         & 13\footnote{W43-Main}  & 30.8 & 1.70   		& 8.0 		    & 10.0 			& 20.9 		 	& 12.7 		 & 306.3 	& 448.4 \\
         & 14     		  & 5.8  & 0.61   		& 3.3 		    & 8.3  			& 4.6 		 	& 2.0  		 & 38.3  	& 44.5  \\
         & 15     		  & 2.7  & 1.09   		& 3.0 		    & 8.9  			& 5.6 		 	& 2.8  		 & 22.0  	& 28.5  \\
         & 16     		  & 3.4  & 1.50   		& 5.9 		    & 7.9  			& 4.3 		 	& 2.9  		 & 25.0  	& 11.7  \\
         & 17     		  & 3.9  & 1.15   		& 4.2 		    & 8.2  			& 6.2 		 	& 3.2  		 & 30.2  	& 47.6  \\
         & 18     		  & 0.5  & 1.03   		& 4.2 		    & 13.9  			& 3.7 		 	& 0.9  		 & 3.0   	& 5.6   \\
  W43    & 19     		  & 1.2  & 1.37   		& 5.2 		    & 8.3  			& 4.2 		 	& 2.0  		 & 9.7   	& 9.0  \\
         & 20\footnote{W43-South} & 23.2 & 1.46   		& 5.3 		    & 11.3 			& 14.8 		 	& 14.9 		 & 205.8 	& 369.2 \\
         & 21     		  & 0.5  & 1.24   		& 4.8 		    & 14.8  			& 6.0 		 	& 4.3  		 & 4.1   	& 8.8  \\
         & 22     		  & 4.6  & 1.11   		& 5.0 		    & 11.0  			& 5.5 		 	& 3.1  		 & 33.9  	& 57.1  \\
         & 23     		  & 3.2  & 1.45   		& 3.3 		    & 13.2  			& 8.1 		 	& 6.9  		 & 27.3  	& 27.1  \\
         & 24     		  & 2.9  & 1.76   		& 6.5 		    & 8.5  			& 5.0 		 	& 3.1  		 & 24.5  	& 28.4  \\
         & 25     		  & 1.2  & 1.80   		& 6.4 		    & 9.3  			& 4.6 		 	& 4.1  		 & 6.8   	& 7.3  \\
         & 26     		  & 2.3  & 1.51   		& 6.1 		    & 8.5  			& 4.7 		 	& 2.9  		 & 16.8  	& 24.3  \\
         & 27     		  & 1.7  & 1.34   		& 3.7 		    & 7.0  			& 4.6 		 	& 2.6  		 & 14.0  	& 17.4  \\
         & 28     		  & 0.7  & 1.28   		& 4.1 		    & 9.9  			& 4.1 		 	& 1.8  		 & 4.3   	& 6.3  \\
         & 29     		  & 1.5  & 1.66   		& 5.4 		    & 12.6  			& 5.1 		 	& 2.8  		 & 11.9  	& 10.3  \\
  \bottomrule
 \end{tabular}
 }
 \end{minipage}
\caption{Physical properties of the W43 clouds and foreground structures as described in Table~\ref{table:clumplist}.}
\label{table:clumplist_derived}
\end{table*}

\subsection{Shear parameter}

Investigating the motion of gas streams in the Galaxy is important to explain how large molecular clouds like W43 can be accumulated. 
While Motte et al. in prep. will investigate streams of $^{12}$CO and H{\sc I} gas in W43 in detail, we only consider here the aspect of radial shear in this field.
The shear that is created by the differential rotation of the Galaxy at different Galactocentric radii can prevent the formation of dense clouds if it is too strong.
It is possible to calculate a shear parameter $S_{\mathrm{g}}$ as described in \citet{dib2012} by considering the Galactocentric radius of a region, its spatial and velocity extent, and 
its mass. For values of $S_{\mathrm{g}}$ higher than 1, the shear is so strong that clouds get ripped apart, while they are able to form for values below 1.

The values we use to calculate $S_{\mathrm{g}}$ are the total mass, calculated above, of $1.9 \times 10^6\, M_{\sun}$, a velocity extent of 40~km\,s$^{-1}$,
a Galactocentric radius of 4~kpc, and an area of $8\times10^3\,$pc$^2$.
This is the area that is covered by emission and is smaller than the total size of our map.
These values yield a shear parameter $S_{\mathrm{g}}=0.77$. 
Accordingly, shear forces are not strong enough to disrupt the W43 cloud. 
However, we have to keep in mind that we probably underestimate the total gas mass, as described in Sect.~\ref{subsec:calculations}. A higher mass would lead to a lower shear parameter.
This calculation is only valid for an axial symmetric potential (i.e. orbits outside the Galactic bar). 
Shear forces inside the Galactic bar could be stronger due to the different shape of orbits there.
As we, however, located W43 at the tip of the bar, the calculation still holds.

We can also conduct this calculation for the larger gas mass derived from $^{13}$CO (1--0) by \citet{nguyenluong2011}. 
They find a gas mass of $4.2 \times 10^6\, M_{\sun}$ that is spread out over an area of $1.5\times10^4\,$pc$^2$.
These values lead to a shear parameter of $S_{\mathrm{g}}=0.66$, which is lower than our value above.

\subsection{Virial masses}

In Table~\ref{table:clumplist}~(7) and (10), we have given the mean line width and the area of our sources. 
This allows us to calculate virial masses by defining an effective region radius by $R=(A/\pi)^{1/2}$, where $A$ is the area of the cloud.
This area cannot be determined exactly, because the extent of a cloud depends on the used molecular line. 
Here, we use the area of $^{13}$CO emission above a certain threshold (20\% of the peak intensity).

Virial masses can then be computed using the relation

\begin{equation}
 M_{\mathrm{V}} = 5 \frac{R\,\sigma^2}{G}(M_{\odot}),
\end{equation}

\noindent where $\sigma$ is the Gaussian velocity dispersion averaged over the area $A$ and $G$ is the gravitational constant.

The resulting virial masses are shown in Table~\ref{table:clumplist_derived}~(8).
We notice that most sources in W43 have masses derived from $^{13}$CO that are smaller than their virial masses.
Sources 4 and 5 show much larger molecular than virial masses, which might indicate that their distance was overestimated.
If the sources would be completely virialized, we would need bigger masses to produce the observed line widths. 
On the other hand, systematic motion of the gas, apart from turbulence, like infall, outflows, or colliding flows would also broaden the lines.
This could be an explanation for the observed large line widths.

\citet{ballesteros-paredes2006} stated that usually turbulent molecular clouds are not in actual virial equilibrium, since there is a flux of mass, momentum, and energy between the clouds
and their environment. What is normally viewed as virial equilibrium is an energy equipartition between self-gravity, kinetic, and magnetic energy.
This energy equipartition is found for most clouds due to observational limitations. 
Clouds out of equilibrium are either not observed due to their short lifetime or not considered clouds at all.
 
Of course, we also need to consider the shape of our sources. 
Non-spherical sources have a more complicated gravitational behavior than spheres. 
Therefore, one has to be extremely careful using these results.
In addition, we neglect here the influence of external pressure and magnetic fields on the virial masses.
What we observe agrees with \citet{ballesteros-paredes2006} in that most of our detected clouds show a molecular mass in the order of their virial mass, or up to a factor of 2 higher.

\section{Comparison to other projects}\label{sec:comparison}

To gather more information about the W43 complex, we compare the IRAM 30m CO data to other existing datasets. We pay special 
attention to three large-scale surveys in this section: the Spitzer GLIMPSE and MIPSGAL projects, the Herschel Hi-GAL survey, and the Galactic
plane program ATLASGAL as observed with the APEX telescope.

All of these datasets consist of total power maps over certain bands. They naturally do not contain spectral information, so line-of-sight confusion is considerable, since 
the W43 region is a complex accumulation of different sources. It can sometimes be complicated to assign the emission of these maps to single sources. Nevertheless,
the additional information is very valuable.

\subsection{Spitzer GLIMPSE and MIPSGAL}

The Spitzer Space Telescope program, Galactic Legacy Infrared Mid-Plane Survey Extraordinaire (GLIMPSE) \citep{benjamin2003,churchwell2009}, observed the Galactic plane at several IR wavelengths 
between 3.6 and 8~$\mu \mathrm{m}$. It spans the Galactic plane from $-65 \degr$ to $65 \degr$ Galactic longitude.

Here, we concentrated on the 8~$\mu \mathrm{m}$ band. It is dominated by UV-excited PAH emission \citep{peeters2004}. 
These photon dominated regions (PDRs) \citep{hollenbach-tielens1997} are heated by young OB stars. 
By studying this band in comparison
to the IRAM 30m CO maps, we can determine which parts of the molecular clouds contain UV-heated dust. This is seen as extended emission in the Spitzer maps.
We can also identify nearby UV-heating sources, as seen as point sources. Finally, some parts of specific clouds appear in absorption against the background.
These so-called infrared dark clouds (IRDCs) \citep[See][]{egan1998,simon2006,peretto_fuller2009} show denser dust clouds that are not heated by UV-radiation. In this way, we are able to determine which sources
heat part of the gas and which parts are shielded from UV radiation. We can also tell if YSOs have already formed inside the clumps that we have observed and thus estimate the evolutionary stage of the clouds.
\citet{nguyenluong2011} used this tracer to estimate the star-formation rate \citep[SFR][]{wu2005} of the W43 complex.

\begin{figure}[htb]
 \centering
 \begin{minipage}{0.48\textwidth}
  \subfloat[{\bf (a)} IRAM 30m]{\resizebox{0.46\textwidth}{!}{\includegraphics[scale=1]{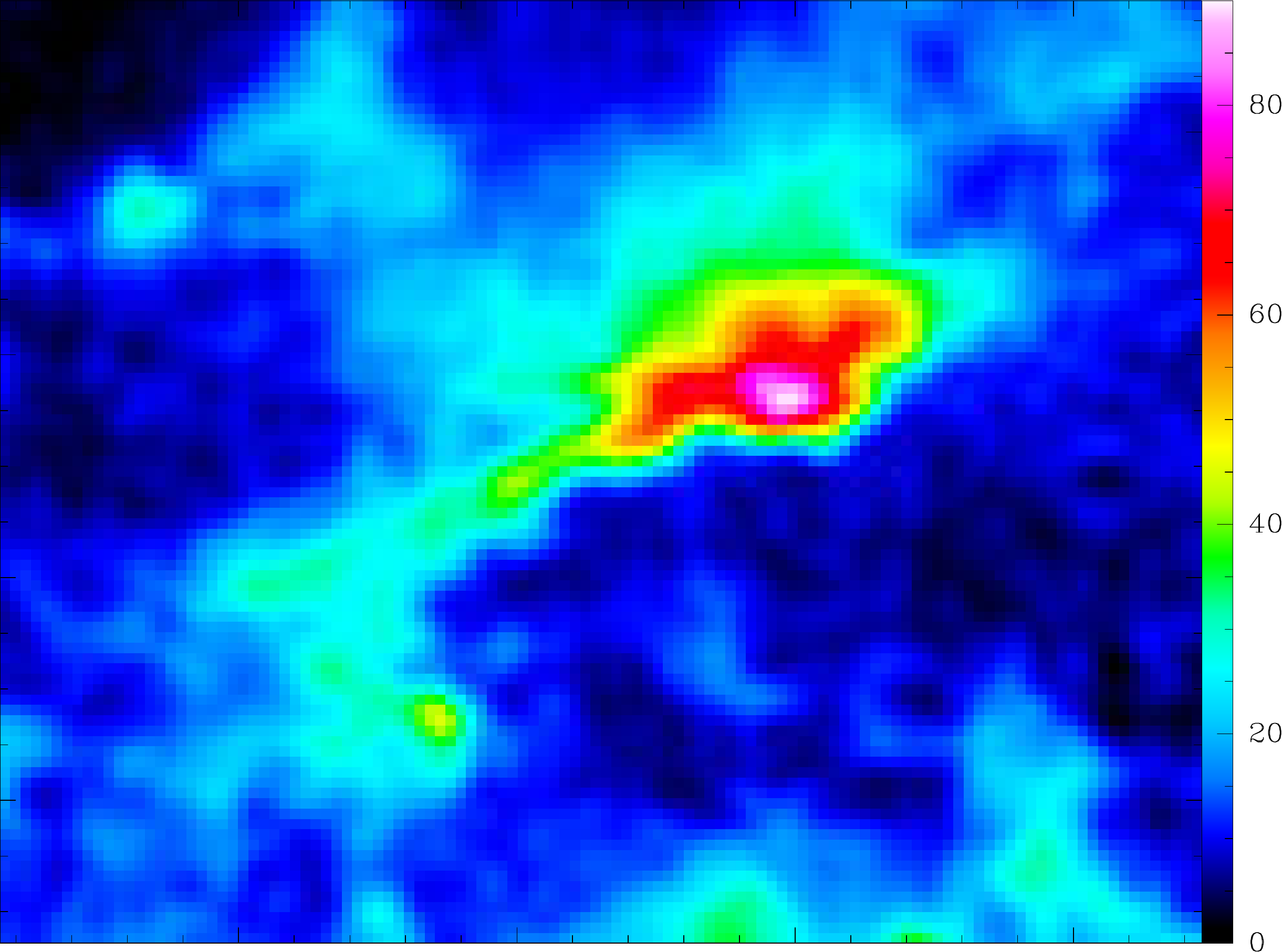}}}\hspace{0.03\textwidth}
  \subfloat[{\bf (b)} ATLASGAL]{\resizebox{0.46\textwidth}{!}{\includegraphics[scale=1]{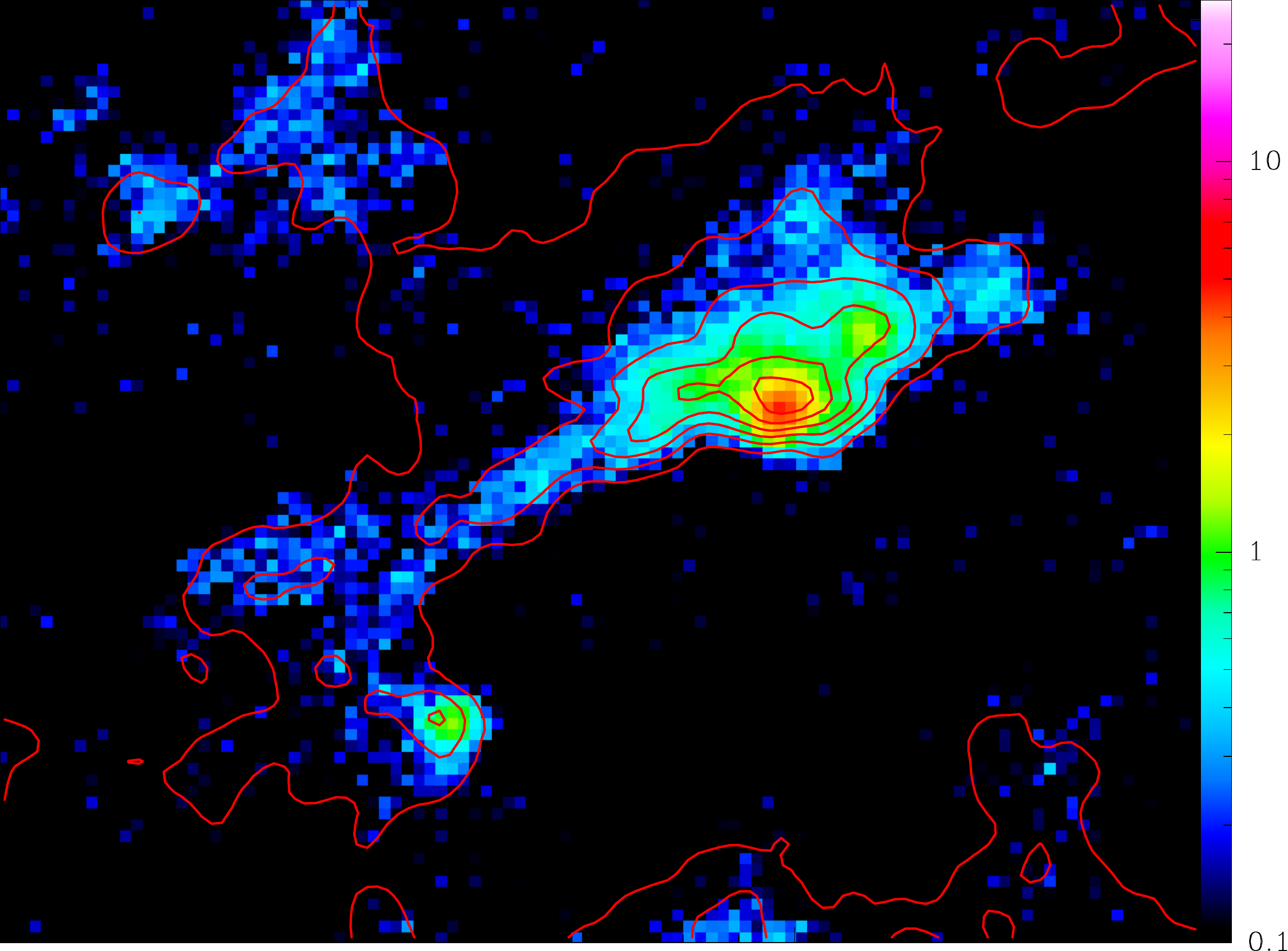}}}\hfill\\
  \subfloat[{\bf (c)} Hi-GAL]{\resizebox{0.455\textwidth}{!}{\includegraphics[scale=1]{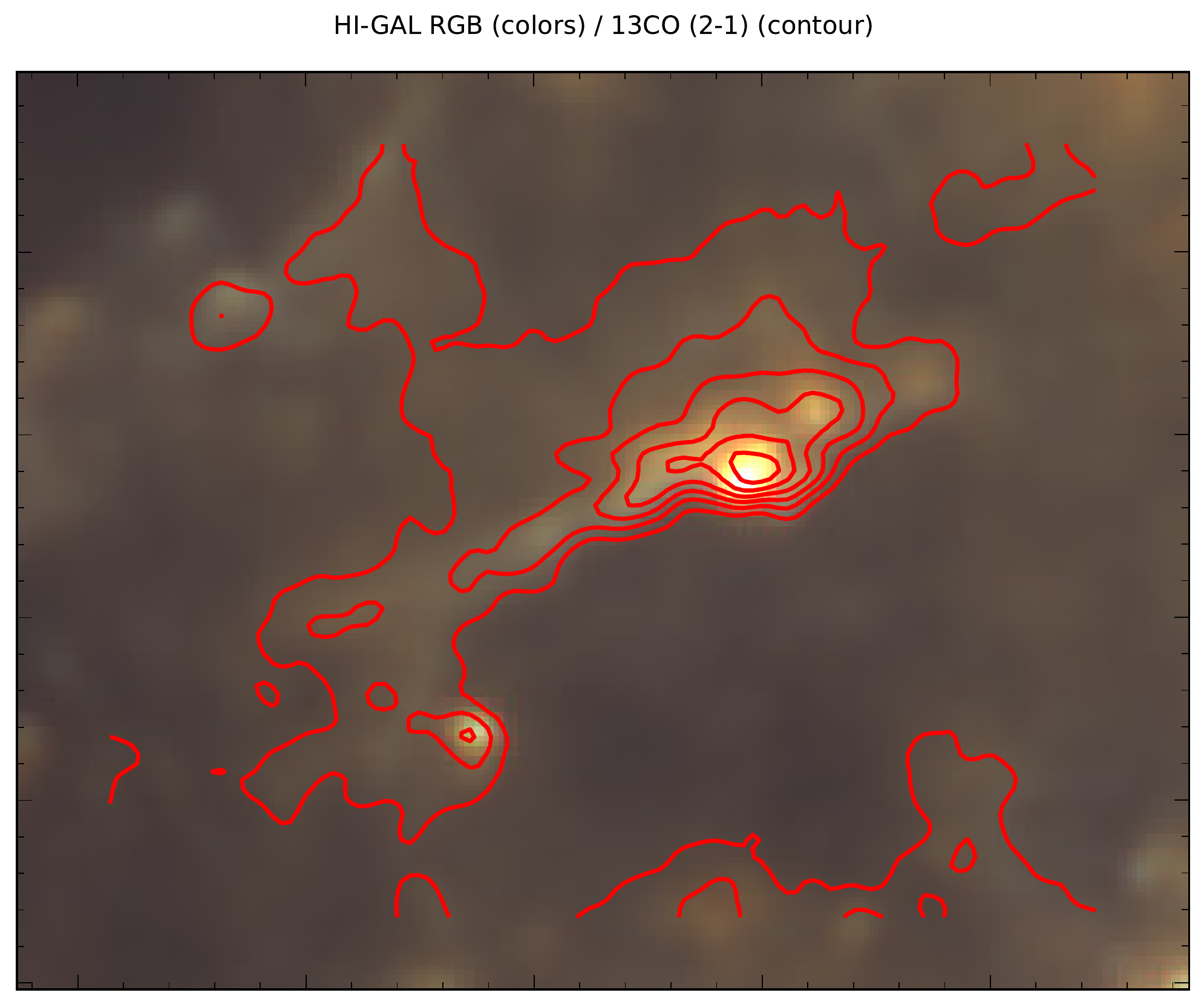}}}\hspace{0.03\textwidth}
  \subfloat[{\bf (d)} GLIMPSE]{\resizebox{0.445\textwidth}{!}{\includegraphics[scale=1]{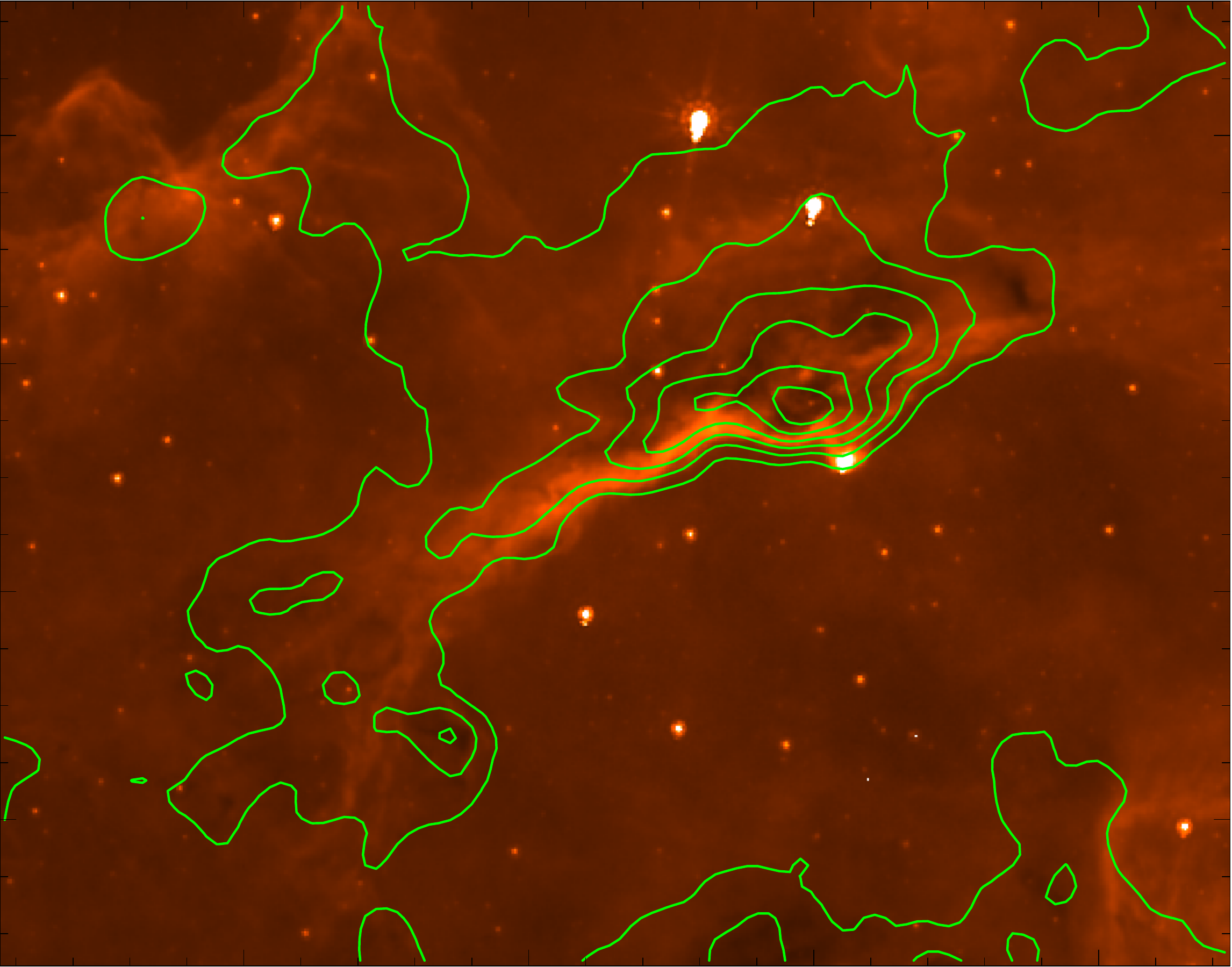}}}\hfill
 \end{minipage}
 \caption{Series of plots showing the comparison of different datasets of source 23 (see Table~\ref{table:clumplist}). From upper left to lower right: 
 {\bf (a)} IRAM 30m integrated intensity map of $^{13}$CO~(2--1),
 {\bf (b)} IRAM 30m $^{13}$CO map contours on ATLASGAL 870~$\mu \mathrm{m}$ map,
 {\bf (c)} IRAM 30m $^{13}$CO map contours on Hi-GAL RGB map composed of the 350, 250 and 160~$\mu \mathrm{m}$ maps, and
 {\bf (d)} IRAM 30m $^{13}$CO map contours on GLIMPSE 8~$\mu \mathrm{m}$ map
 }
 \label{fig:projectcomparison23}
\end{figure}

MIPSGAL \citep{carey2009} is a Galactic plane survey using the MIPS instrument onboard Spitzer and has created maps at 24 and 70~$\mu$m. 
Here, we inspect the 24~$\mu$m bandm which is dominated by the emission of
small dust grains. The MIPS instrument also detects proto-stellar cores, although these cores are usually too small to be resolved at a distance of 6~kpc or more.

\begin{table}[htb]
 \centering
 \begin{minipage}{0.49\textwidth}
 \resizebox{\hsize}{!}{
 \begin{tabular}[c]{c l l l c }
  \toprule
         &        & (1)		   	& (2)       				& (3)		   \\
         & Number & $^{13}$CO	   	& Spitzer   				& Mass ratio 	   \\
         &        & structure	   	& structure				& Hi-GAL/$^{13}$CO \\
  \midrule
         &        &       	  	&           				& \\
  \midrule
         & 1  & Filament		& Single core 		      		& \\
         & 2  & Ensemble of cores 	& Two cores + extended emission   	& \\
         & 3  & Ensemble of cores 	& Single core + extended emission 	& \\
  Fore-  & 4  & Joining filaments 	& No related emission 			& \\
  ground & 5  & Two cores 	   	& Two cores + extended emission  	& \\
         & 6  & Complex 	   	& Extended emission 			& \\
         & 7  & Complex 	   	& Single core + extended emission 	& \\
         & 8  & Single core 	   	& Single core + absorption 		& \\
         & 9  & Ensemble of cores 	& Single core + absorption 		& \\
  \midrule
         & 10 & Two cores 	   	& Two cores + absorption 		& 1.8 \\
         & 11 & Single core 	   	& Single core + absorption 		& 1.0 \\
         &    &			   	& + extended emission 			&     \\
         & 12 & Complex 	   	& Several cores 			& 1.1 \\
         & 13\footnote{W43-Main} & Complex 	   & Bright emission + absorption, 	& 1.2 \\
         &    &			   	& Bubble around star cluster		&     \\
         & 14 & Ensemble of cores 	& Absorption 				& 1.3 \\
         & 15 & Complex 	   	& One extended core + absorption 	& 1.2 \\
         & 16 & Filament 	   	& Absorption 				& 1.5 \\
         & 17 & Complex 	   	& Several cores + absorption 		& 1.9 \\
         & 18 & Filament 	   	& No related emission 			& 1.1 \\
  W43    & 19 & Complex 	   	& Two extended cores + absorption 	& 1.0 \\
         & 20\footnote{W43-South} & Ensemble of cores & Bright extended emission 		& 0.8 \\
         &    &			   	&  + absorption 			&     \\
         & 21 & Filament 	   	& Single core + extended emission 	& 1.9 \\
         &    &			   	&  					&     \\
         & 22 & Complex 	   	& Several cores + absorption 		& 1.1 \\
         &    &			   	& + extended emission 			&     \\
         & 23 & Filament 	   	& Single core + absorption 		& 0.9 \\
         &    &			   	& + extended emission 			&     \\
         & 24 & Joining Filaments 	& Two cores + absorption 		& 1.7 \\
         & 25 & Filament 	   	& Two cores + extended emission 	& 1.2 \\
         & 26 & Filament 	   	& Single core + extended emission 	& 0.9 \\
         & 27 & Two cores 	   	& Single core + extended emission 	& 1.2 \\
         & 28 & Filament 	   	& No related emission 			& 1.6 \\
         & 29 & Filament 	  	& Several cores 			& 1.0 \\
  \bottomrule
 \end{tabular}
 }
 \end{minipage}
\caption{List of sources, as described in Table \ref{table:clumplist} and a description of their structure.}
\label{table:clumplist_characterization}
\end{table}

\subsection{APEX ATLASGAL}

The APEX telescope large area survey of the galaxy (ATLASGAL) \citep{schuller2009} used the LABOCA camera, which is installed at the APEX telescope. It observed 
the Galactic plane from a Galactic longitude of $-60 \degr$ to $60 \degr$ at 870~$\mu \mathrm{m}$. This wavelength traces cold dust and is therefore also a good indicator of dense molecular
cloud structures, especially of high-mass star-forming clumps. \citeauthor{schuller2009} also identified hot cores, proto-stars, compact H{\sc II} regions, and young embedded stars by combining 
their map with other data.

Our project is a direct follow-up of ATLASGAL, from which the idea to observe the W43 region in more detail was born.


\subsection{Herschel Hi-GAL}\label{subsec:higal_comparison}

The Hi-GAL project \citep{molinari2010} utilizes Herschel's PACS and SPIRE instruments to observe the Galactic plane from a Galactic longitude of $-60 \degr$ to $60 \degr$
at five wavelengths between 70 and 500~$\mu \mathrm{m}$.
A part of the Hi-GAL maps of the W43 complex are presented in \citet{bally2010}.

Figure~\ref{fig:projectcomparison23} shows one example of the comparison of the different datasets that we carried out for all identified sources. It shows source 23, sticking to the notation of
Table~\ref{table:clumplist}. See Appendix~\ref{app:sec:important_sources} for an in depth description of the different sources.

In Table~\ref{table:clumplist_characterization}, we categorize our sources, whether they have a filamentary shape, consist of cores, or show a more complex structure. We also list the structure
of the Spitzer 8 and 24~$\mu$m maps here. Usually, the two wavelengths are similar.

\begin{figure*}[htb]
\centering
\begin{minipage}{0.47\textwidth}
\subfloat[{\bf (a)} Dust temperature derived from three Herschel channels \mbox{(160-350~$\mu$m)} in color and $^{13}$CO~(2--1) contours.]{\resizebox{\hsize}{!}{\includegraphics{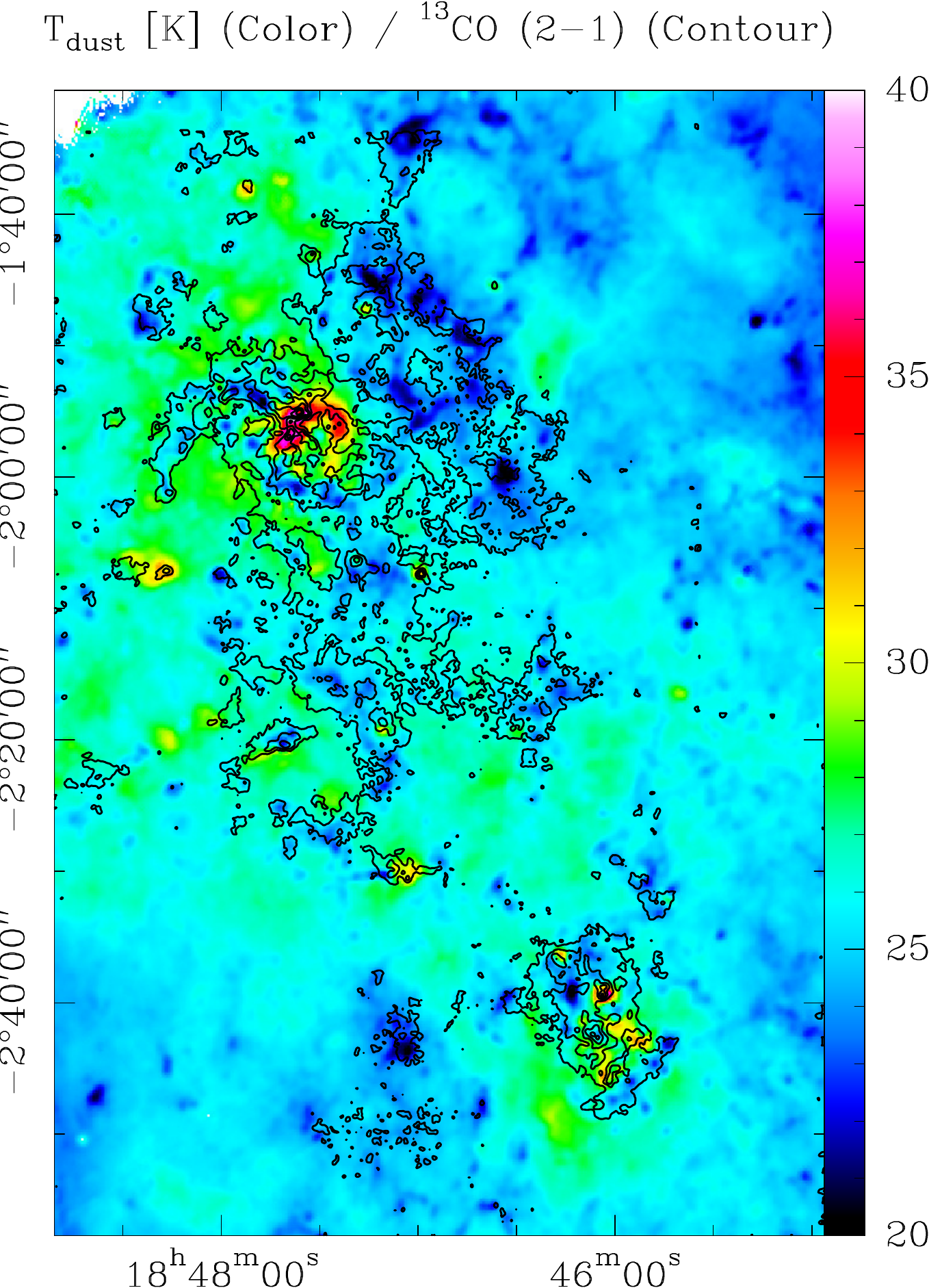}}}
\end{minipage}
\hfill
\begin{minipage}{0.49\textwidth}
\subfloat[{\bf (b)} H$_2$ column density map derived from three Herschel channels \mbox{(160-350~$\mu$m)} in color and $^{13}$CO~(2--1) contours.]{\resizebox{\hsize}{!}{\includegraphics{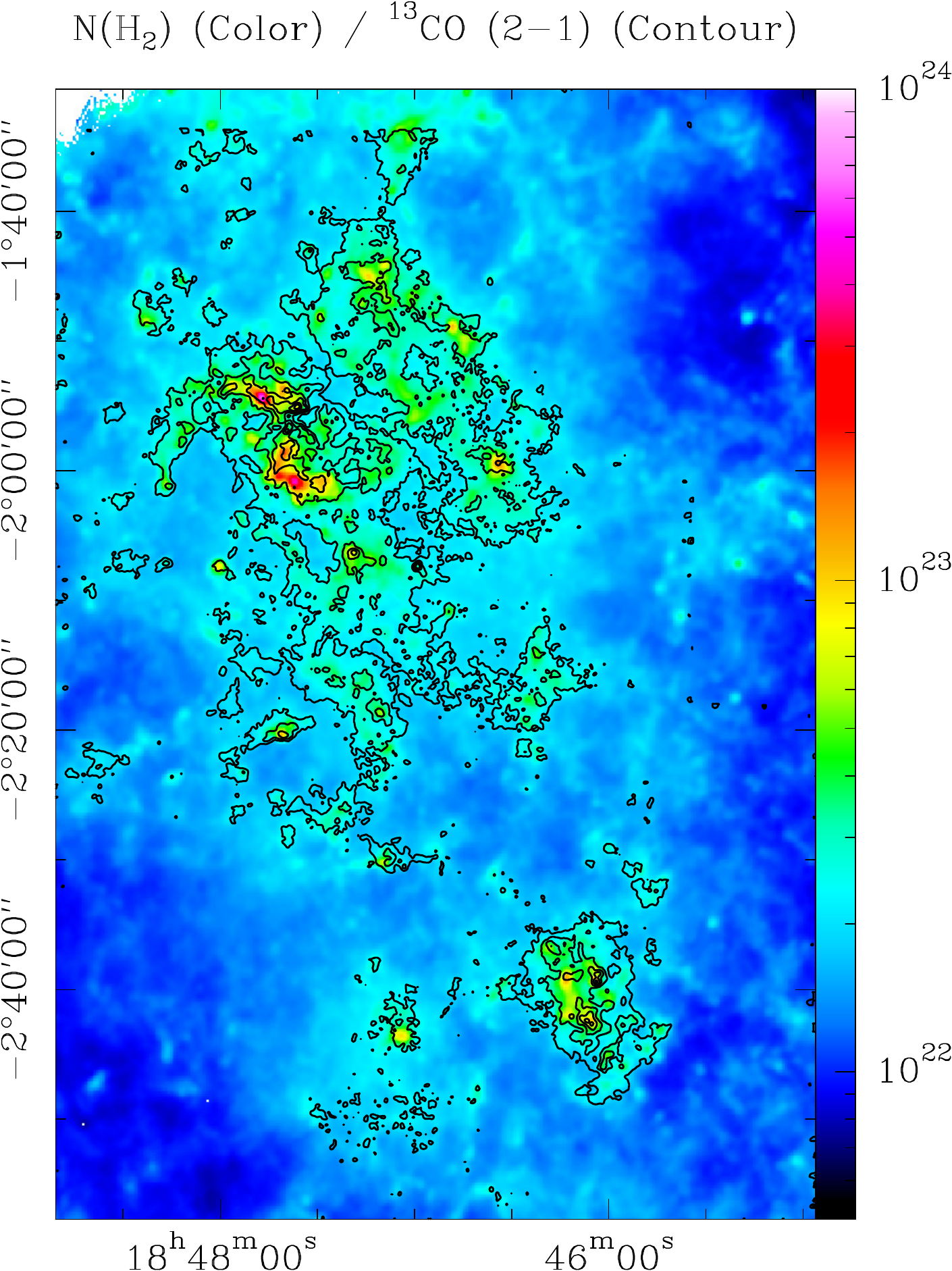}}}
\end{minipage}
\caption{Maps derived from Hi-GAL and HOBYS data \citep[see][]{nguyenluong2013}.}
\label{fig:higal_temp_nh2}
\end{figure*}

From the Hi-GAL dust emission, it is possible to derive a temperature and a total (gas + dust) H$_2$ column density map \citep[e.g.,][]{battersby2011}.
These calculations were conducted for the W43 region by \citet{nguyenluong2013}, following the fitting routine detailed in \citet{hill2009,hill2010} and adapted and applied to Herschel data as in
\citet{hill2011,hill2012}, \citet{molinari2010}, and \citet{motte2010}. 
This approach uses Hi-GAL data for the calculations where possible. 
As on very bright positions, Hi-GAL data become saturated or enter the nonlinear response regime, HOBYS data was used to fill the missing data points.
The idea is to fit a modified black body curve to the different wavelengths (in this case, the 160 to 350~$\mu$m channels) for each pixel using a dust opacity law
of $\kappa=0.1\times(300\,\mu \mathrm{m}/\lambda)^{\beta}\,$cm$^2\,$g$^{-1}$ with $\beta=2$.
The final angular resolution of the calculated maps (25\arcsec\, in this case) results from the resolution of the longest wavelength used.
This is the reason that the 500~$\mu$m channel has been omitted.
Planck and IRAS offsets were added before calculating the temperature and H$_2$ column density.

This approach assumes the temperature distribution along the line-of-sight to be constant.
As discussed above in Sect.~\ref{subsubsec:ex_temp}, this is not necessarily the case in reality. 
Due to temperature gradients along the line-of-sight, the calculated H$_2$ column density might deviate from the real value.
This error is found in the Herschel and the CO calculations.

Gas and dust temperatures can deviate in optically thick regions, because the volume densities play a key role here.
Only at fairly high densities of more than about $10^5\,$cm$^{-3}$ does the gas couple to the dust temperature.
At lower densities, the dust grains are an excellent coolant, in contrast to the gas. 
Therefore, the dust usually shows lower temperatures than the gas.
This is in contrast to our findings and may indeed indicate that we underestimate the $T_{\mathrm{ex}}$ of the gas, as discussed in Sect.~\ref{subsubsec:ex_temp}.
Even in optically thin regions,
the kinetic temperature and the values derived from an SED-fit do not correspond perfectly \citep[see][]{hill2010}.
Figure~\ref{fig:higal_temp_nh2} shows the results of these calculations for the W43 region. We show the temperature and H$_2$ column density maps from the photometry data and overlay $^{13}$CO contours
to indicate the location of the molecular gas clouds.

The dust temperature derived by \citet{nguyenluong2013} (Fig.~\ref{fig:higal_temp_nh2}~(a)) lies in the range from 20 to 40~K; the outer parts of the complex show temperatures between 25 and 30~K. 
The regions where dense molecular clouds are found are colder (about 20~K) than their surroundings; for example the dense ridges in W43-Main are clearly visible. Some places, where the dust
is heated by embedded UV-sources, are hotter (up to 40~K), especially for the OB-star cluster in W43 and one core in W43-South that catches the eye.

If we compare these results with the excitation temperatures that we calculated above in Sect.~\ref{subsec:calculations}, we note that the temperatures derived from CO are
lower ($\sim7\,-\,20$~K) than those using Herschel images. 
It is possible that gas and dust are not mixed well and that both could have a different temperature. 
Another possibility is that the CO gas is subthermally excited and that the excitation temperature is lower than its kinetic temperature. 
The Herschel dust temperature map is showing the averaged temperature along the line-of-sight. 
Thus, lower temperatures of the dense clouds are seen with hotter diffuse gas around it, which leads to higher averaged temperatures.
As discussed in Sect.~\ref{subsubsec:ex_temp}, we most probably underestimate the excitation temperature in our calculations above due to subbeam clumpiness. 
Also, we cannot neglect that the temperatures calculated from Herschel bear large errors on their own. 

The H$_2$ column density map derived from Herschel data (Fig.~\ref{fig:higal_temp_nh2}~(b)) nicely traces the distribution of molecular gas, as indicated by the $^{13}$CO~(2--1) contours.
There is a background level of $\sim1-2\times10^{22}$~cm$^{-2}$ that is found outside the complex. 
The column density rises in the molecular clouds up to a value of $\sim6\times10^{23}$~cm$^{-2}$ in the ridges of W43-Main.

A comparison to the column density values derived from CO (See Sect.~\ref{subsec:calculations} and the plot in Fig.~\ref{fig:h2_map_total}.) reveal certain differences.
In the mid-density regions, calculations of both the medium-sized clouds and most parts of the two large clouds are comparable after subtracting the background level from the Herschel maps.
These are still systematically higher, but the difference does rarely exceed a factor of 2.
The same is true for the extended gas between the denser clouds. The typical values are around a few $10^{21}$~cm$^{-2}$ for the Herschel and CO maps.
However, the Herschel H$_2$ column density reaches values of several $10^{23}$~cm$^{-2}$ in the densest parts of W43-Main, with a maximum of $\sim6\times10^{23}$~cm$^{-2}$, while the CO derived values
peak at $\sim2\times10^{23}$~cm$^{-2}$.

The offset of $\sim1\times10^{22}$~cm$^{-2}$ (which could be a bit higher but we chose the lower limit) that has to be subtracted from the Herschel map can partly be explained 
by diffuse cirrus emission along the line-of-sight, which is not associated with W43. 
This statistical error of the background brightness has been described for the Hi-GAL project by \citet{martin2010}.
\citeauthor{nguyenluong2013} find this offset agrees with \citet{battersby2011}.



The total mass that is found for the W43 complex still deviates between both calculations.
The exact value depends on the specific region that we integrated over.
We find a total mass of W43 from $^{13}$CO of $1.9 \times 10^6$~$M_{\sun}$.
The Herschel map gives $2.6 \times 10^6$~$M_{\sun}$, if we use the total map with an area of $1.5\times10^4$~pc$^2$
and consider the diffuse cirrus emission that is included by the Herschel data.
This is a factor of 1.4 higher than the CO result, although we did not calculate a value for every single pixel for the CO H$_2$ column density map.
The mean H$_2$ column density is $7.5\times10^{21}$~cm$^{-2}$ ($^{13}$CO) and 
\mbox{$7.4\times10^{21}$~cm$^{-2}$} (Herschel), respectively. 
The difference in total mass can be explained when we consider that the Herschel map covers more points
(the difference is reduced to a factor of 1.2, when comparing a smaller region which is covered in both maps, although this might be too small and biased toward the larger clouds).
For a comparison of each Duchamp cloud, see Table~\ref{table:clumplist_characterization}~(3). A comparison of the foreground clouds would be complicated due to line-of-sight effects, so we only give numbers
for the W43 sources. Most values lie in the range 1 to 1.5, which affirms
that both calculations deviate by about a factor of 1.4. The difference in the H$_2$ column density again depends on the examined region. 

As stated in Sect.~\ref{subsubsec:totalmass}, the mass derived from CO is a lower limit to the real molecular gas mass. 
As we used the lower limit of the Herschel column density offset subtracted in W43, these values are thus an upper limit.
Taking this and the errors still included in both calculations into account, the values are nearly consistent.

\subsection{Column density histogram - PDF}  
  
A detailed investigation of the H$_2$ column density structure of W43 
is done by determining a histogram of the H$_2$ column density, 
which is normalized to the average column density. These probability 
distribution functions (PDFs) are a useful tool to scrutinize 
between the different physical processes that determine the density 
structure of a molecular cloud, such as turbulence, gravity, feedback, 
and magnetic fields. Theoretically, it was shown that isothermal 
turbulence leads to a log-normal PDF \citep[e.g.][]{federrath2010}, 
while gravity \citep{klessen2000} and non-isothermality \citep{passot1998} 
provoke power-law tails at higher densities. 
Observationally, power-law tails seen in PDFs that are obtained from column 
density maps of visual extinction or Herschel imaging were 
attributed to self-gravity for low-mass star-forming regions \citep[e.g.,][]{kainulainen2009,schneider2013} and high-mass 
star-forming regions \citep[e.g.,][Russeil et al. in press]{hill2011,schneider2012}. 
Recently, it was shown \citep[][Tremblin et al. in prep.]{schneider2012} that feedback processes, such as the compression of 
an expanding ionization front, lead to a characteristic `double-peaked' 
PDF and a significant broadening. 
 
The determination of PDFs from molecular line data was attempted by \citet{goodman2009} and \citet{wong2008}, but it turned 
out to be problematic when these lines become optically thick and thus 
do not correctly reflect the molecular cloud spatial and density 
structure. In addition, uncertainties in the abundance can complicate 
conversion into H$_2$ column density.  On the other hand, molecular 
lines allow us to significantly reduce line-of-sight confusion, 
because PDFs can be determined for selected velocity ranges. In 
addition, using molecules with different critical densities in 
selected velocity ranges allows us to make dedicated PDFs that focus
on a particular subregion like a dense filament. 
 
\begin{figure}[htb]  
 \centering  
 \begin{minipage}{0.48\textwidth}  
  \resizebox{\hsize}{!}{\includegraphics{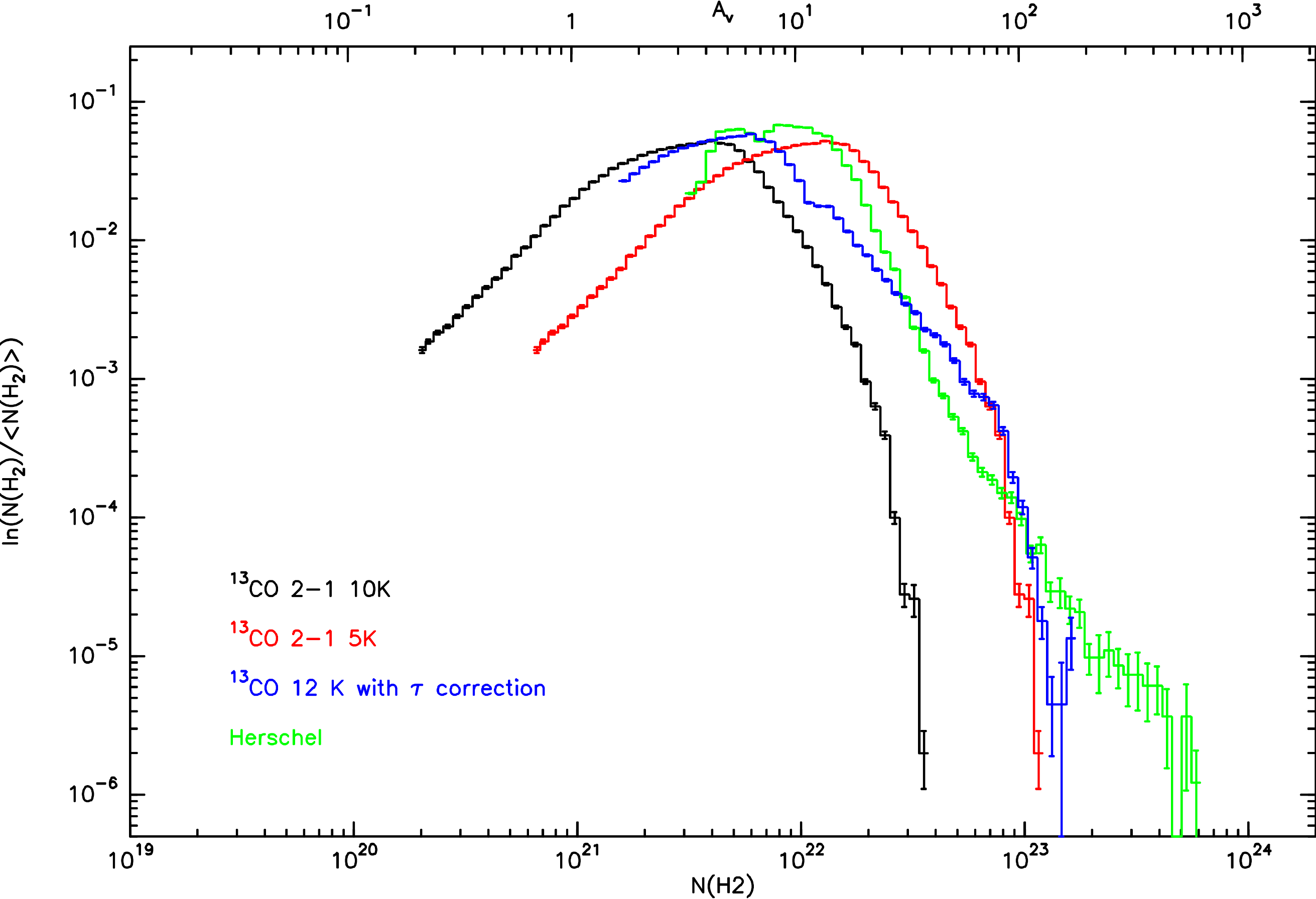}}  
 \end{minipage}  
 \caption{Probability distribution functions of the H$_2$ column 
   density derived for the whole W43 complex from different data 
   sets and methods. The y-axis is normalized to the mean value of N(H$_2$).
   Isothermal CO column densities (5 and 10~K) are plotted in red and black; CO derived values with optical depth correction are in blue, and Herschel results are in green.} 
 \label{fig:h2_pdf}  
\end{figure}

In this study, we determined the PDFs of W43 in three ways: (i) from a 
simple conversion of the $^{13}$CO~(2--1) map into H$_2$ column 
density by using a constant conversion factor and one 
temperature (5 or 10~K); (ii) from the H$_2$ column density map 
derived from the $^{13}$CO~(2--1) emission, which includes a correction for the optical depth that is derived from both CO lines
(see Sect.~\ref{subsec:calculations});
and (iii) from the column density map 
obtained with Herschel using SPIRE and PACS photometry. Figure~\ref{fig:h2_pdf} shows the resulting 
distributions. 
For simplicity, we used the conversion $N(\mathrm{H}_2+\mathrm{H})/A_\mathrm{V} = 9.4\times 10^{20}\mathrm{cm}^{-2}$ \citep{bohlin1978} for all maps, though the 
Herschel column density map is a mixture of HI and H$_2$ while the CO derived map is most likely dominated by H$_2$.

The `isothermal' PDFs from $^{13}$CO without and optical depth correction (in black and 
red) clearly show that there is a cut-off in the PDF at high column 
densities where the lines become optically thick (A$_{\rm v} \sim$20~mag 
for 10~K and A$_{\rm v} \sim$70~mag for 5~K). Obviously, there is also a 
strong temperature dependence that shifts the peak of the PDF to lower 
column densities with increasing temperature. 
The assumed gas temperature (see discussion in Sect.~\ref{subsec:calculations}) thus has a strong impact 
on the resulting PDFs positions, but not their shape (not considering the uncertainty in the conversion factor). 

With the more sophisticated approach to create a column density map out of the 
$^{13}$CO emission and to include the information provided by the optically thinner C$^{18}$O which is corrected for the optical depth $\tau$, 
the PDF (in blue) is more reliable. 
It does not show the cut-off at high column densities, because this effect is compensated by using the optically thin C$^{18}$O. 
Only in those regions where this line becomes optically thick, the method gives lower limits for the column density. 
Therefore, it drops below the Herschel PDF for high column densities, because the molecular lines underestimate the H$_2$ column densities for very hot gas.

It is remarkable that the PDF derived from $^{13}$CO and C$^{18}$O shows a log-normal distribution for low 
column densities and a power-law tail for higher densities. This feature is also observed in the Herschel PDF. 
The approach to determine a PDF from the cloud/clump distribution by correcting the opacity using C$^{18}$O appears to be the right way to 
get a clearer picture of the distribution of higher column densities.  
Note that the absolute scaling in column density for both PDFs -- from CO or Herschel -- remains problematic due to the 
uncertainty in the conversion factor (and temperature) for the CO data, the line-of-sight confusion, and opacity variations for the Herschel maps.

We observe that the slope of the Herschel power-law tail is steeper than the one obtained from the CO-data and shows a
`double-peak' feature (The low column density component is not strictly log-normally shaped but shows two subpeaks.) as seen in other regions with stellar feedback 
(Tremblin et al. in prep.). The column density structure of W43 could thus be explained in a scenario where gravity is the dominating 
process for the high density range, leading to global cloud collapse and individual core collapse; compression by expanding 
ionization fronts from embedded HII-regions may lead to an increase in column density, and the lower-density extended emission follows a turbulence 
dominated log-normal distribution. 
This scenario is consistent with what was proposed for other high-mass star-forming regions, such as 
Rosette \citep{schneider2012}, RCW36, M16 (Tremblin et al. in prep.), and W3 \citep{rivera-ingraham2013b}. 
   
 

Though the overall shape of the PDFs is similar, there are significant 
differences in the slope of the power-law tails.
The power-law tail of the Herschel PDF is steeper than the CO-PDF.
A possible explanation is that the Herschel PDF 
contains atomic hydrogen in addition to molecular hydrogen (which 
constitutes mainly the CO PDF), which is less `participating' in the 
global collapse of the region and individual clump/core collapse. In 
this case, our method to derive the H$_2$ column density from $^{13}$CO 
and C$^{18}$O 
turns out to be an efficient tool to identify only the collapsing gas that ends up into a proto-star.


\section{Description of W43-Main and W43-South}\label{sec:sourcedescription}

Here, we give a detailed description of the two most important sources in the W43 complex, W43-Main and W43-South. 
Several other interesting sources are described in Appendix~\ref{app:sec:important_sources}.

\subsection{W43-Main, Source 13}

Source 13 (see Fig.~\ref{fig:appendixsourcemaps}~(m)), or W43-Main, is the largest and most prominent of all sources in the W43 complex. 
Located in the upper central region of the map with an extent of roughly 
30 by 20~pc, it shows a remarkable Z-shape of connected, elongated ridges. The upper part of this cloud extrudes far to the east with a strong emitting filament, where it curves down south
in a weaker extension of this filament. This structure is especially clear in ATLASGAL and Hi-GAL dust emission.

\begin{figure}[htb]
 \centering
 \begin{minipage}{0.48\textwidth}
  \subfloat[{\bf (a)} Shell structure]{\resizebox{0.48\hsize}{!}{\includegraphics{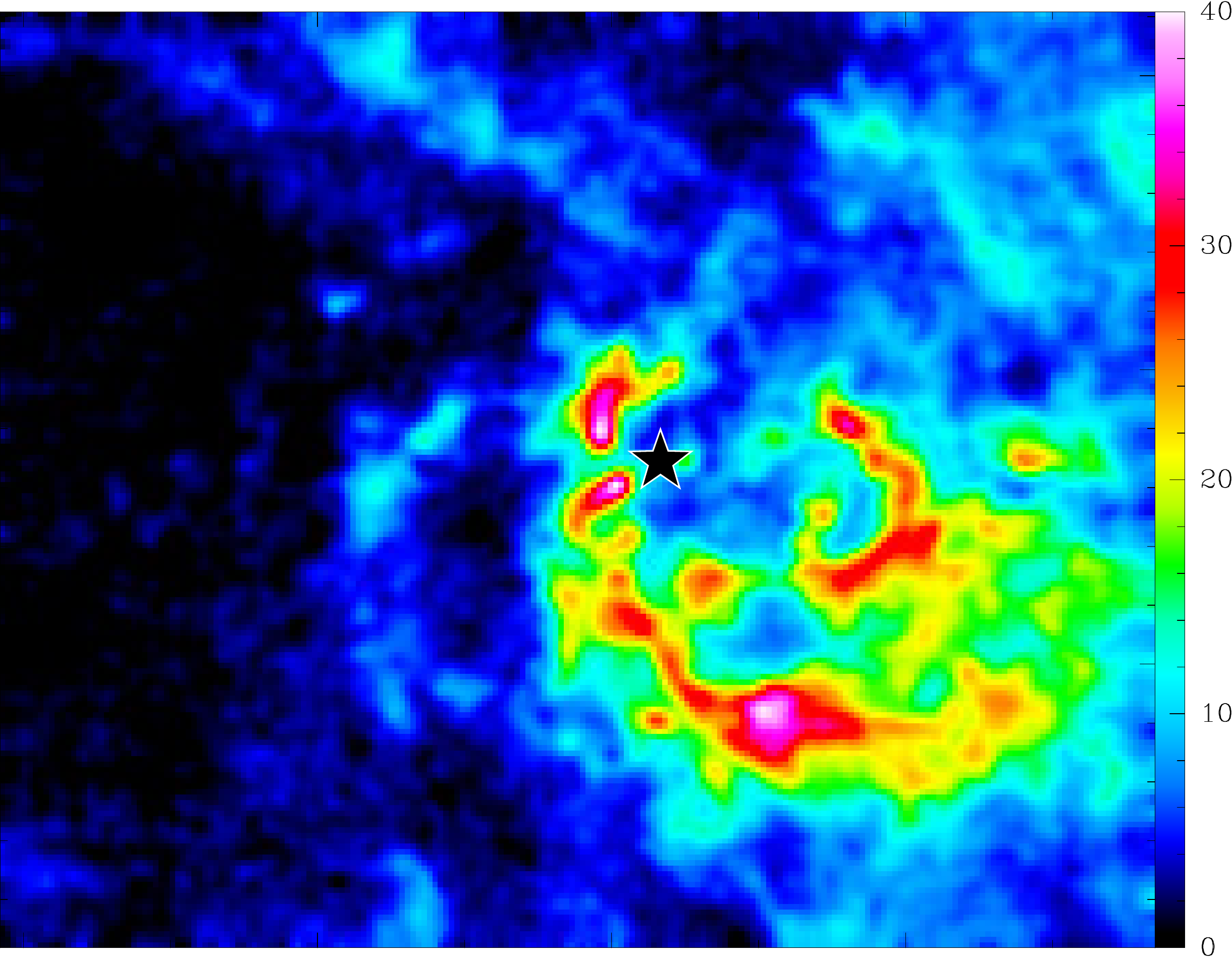}}}\hfill
  \subfloat[{\bf (b)} Canal structure]{\resizebox{0.48\hsize}{!}{\includegraphics{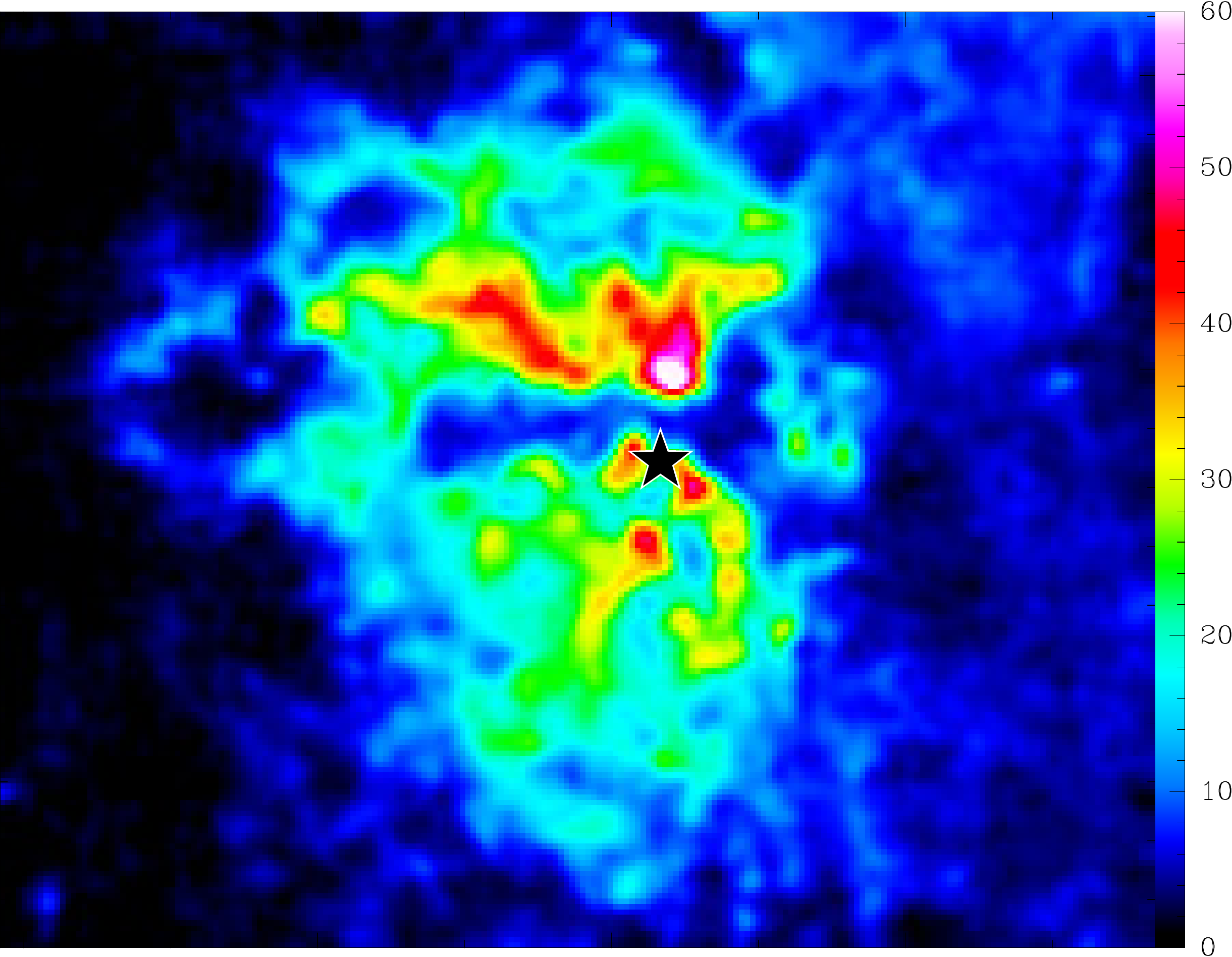}}}
 \end{minipage}
 \caption{Selected velocity ranges of W43-Main (source 13). The location and size of the figures match those of Fig.~\ref{fig:w43mainvelocitymap}.
 {\bf (a)} Shell-like structure in the $^{13}$CO map integrated over the channels from 80 to 90~km\,s$^{-1}$. 
 {\bf (b)} Northern and southern parts of the cloud are separated by an abyss.
 The black star marks the position of the embedded OB star cluster.
 The plot is integrated over channels between 94 and 98~km\,s$^{-1}$.}
 \label{fig:source13detailmaps}
\end{figure}

There are some details hidden in this cloud that cannot be seen clearly in the complete integrated map.
In the velocity range of 80 to 90~km\,s$^{-1}$, which are located in the southwest of the source, we see a circular, shell-like structure surrounding an empty bubble (Fig.~\ref{fig:source13detailmaps}~(a)).
This bubble is elliptically
shaped with dimensions of 10$\times$6~pc, while the molecular shell is about 1.5~pc thick. It is located where a cluster of young OB stars is situated. Possibly, this cavity is formed by the radiation 
of this cluster. See \citet{motte2003} for a description of the expansion of clouds at the periphery of this (H{\sc II}) bubble.

\begin{figure}[htb]
 \centering
 \begin{minipage}{0.48\textwidth}
  \resizebox{0.96\hsize}{!}{\includegraphics{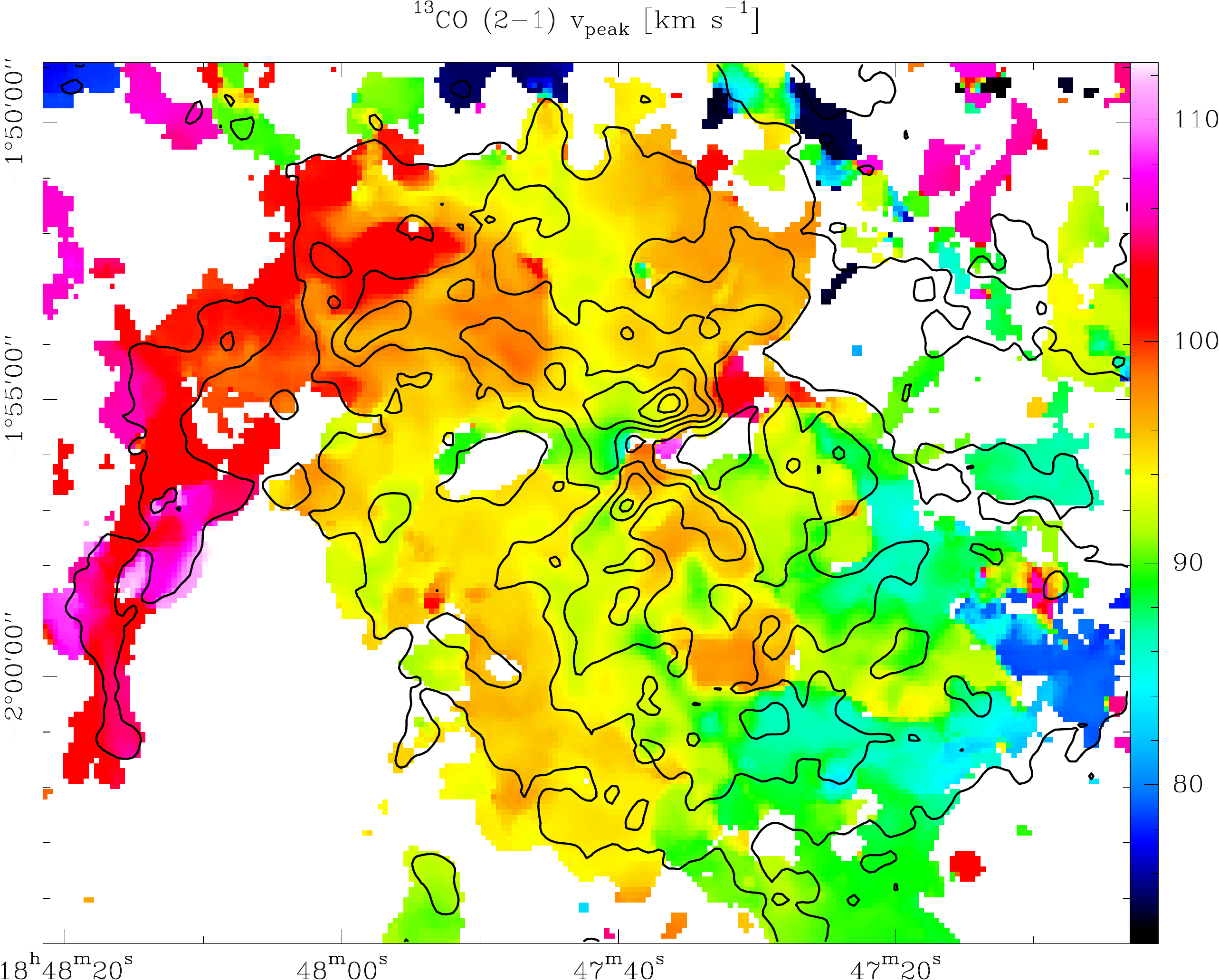}}
 \end{minipage}
 \caption{Peak velocity map of W43-Main derived from the $^{13}$CO data cube in colors; contours show $^{13}$CO integrated intensities, where levels range from 36 to 162~K\,km\,s$^{-1}$
 (20 and 90\% of peak intensity).}
 \label{fig:w43mainvelocitymap}
\end{figure}

The central Z-shape of this cloud appears to be monolithic on the first sight. 
However, Fig.~\ref{fig:source13detailmaps}~(b) shows that the northern and southern parts are separated by a gap in the channel maps between
94 and 98~km\,s$^{-1}$. Both parts are still connected in channels higher and lower than those velocities. This chasm that we see is narrow in the center of the cloud, where it has a width of 1 to 2~pc,
and opens up to both sides. 
The origin of this structure is the cluster of WR and OB stars situated in the very center that blows the surrounding material out along a plane perpendicular to the line-of-sight.
This agrees with the presence of 4 HCO$^+$ clouds in the 25~km\,s$^{-1}$ range along the line-of-sight of the Wolf-Rayet cluster \citep{motte2003}.

W43-Main is the most luminous source in our set with integrated $^{13}$CO~(2--1) emission of up to 170~K\,km\,s$^{-1}$ and line peaks of up to 23~K at the peak in the northern filament. Most inner parts
of this cloud show integrated intensities of $\sim$90 to 120~K\,km\,s$^{-1}$ and 45 to 60~K\,km\,s$^{-1}$ in the outer parts. It is also the most optically thick
with an optical depth of $^{13}$CO up to 8 in the southern arm, while the bulk of the cloud with 2 to 3 is not exceptionally opaque. 
It is possible that we overestimated the opacity in the south of this cloud. 
The spectra show that $^{13}$CO is self-absorbed here, while C$^{18}$O is not. 
This would lead to an unrealistic ratio of the two isotopologes and an overestimated optical depth.

The cloud shows a velocity gradient across its complete structure (see Fig.~\ref{fig:w43mainvelocitymap}). 
Beginning at a relative velocity of $\sim 80$~km\,s$^{-1}$ at the most southwestern tip, it winds through the Z shape and ends in the
eastern extension filament at $\sim 110$~km\,s$^{-1}$. This velocity difference of $\sim 30$~km\,s$^{-1}$, already described in \citet{motte2003} is huge and the largest discovered in the W43 complex.
It could be the sign of a rotation of the cloud.
Comparably impressive are the line widths of the central parts of this clouds. We find them to be up to 15~km\,s$^{-1}$ especially in the central parts, indicating very turbulent gas or global motions.

In the most luminous parts in the south, we calculate the highest H$_2$ column densities of about
$2\times10^{23}$~cm$^{-2}$. This is due to the high opacities that have been calculated here.
The other central parts of this cloud show H$_2$ column densities of $5\,-\,10\times10^{22}$~cm$^{-2}$.
\citet{liszt1995} derived the cloud mass of W43-Main from $^{13}$CO~(1--0) and find a value of $10^6$~M$_{\odot}$.
We find a total mass in this source of $\sim 3 \times 10^5$~$M_{\sun}$ and thus a large fraction of the mass of the complete W43 complex (~20\%).

\begin{figure}[htb]
 \centering
 \begin{minipage}{0.48\textwidth}
  \resizebox{0.96\hsize}{!}{\includegraphics[angle=-0]{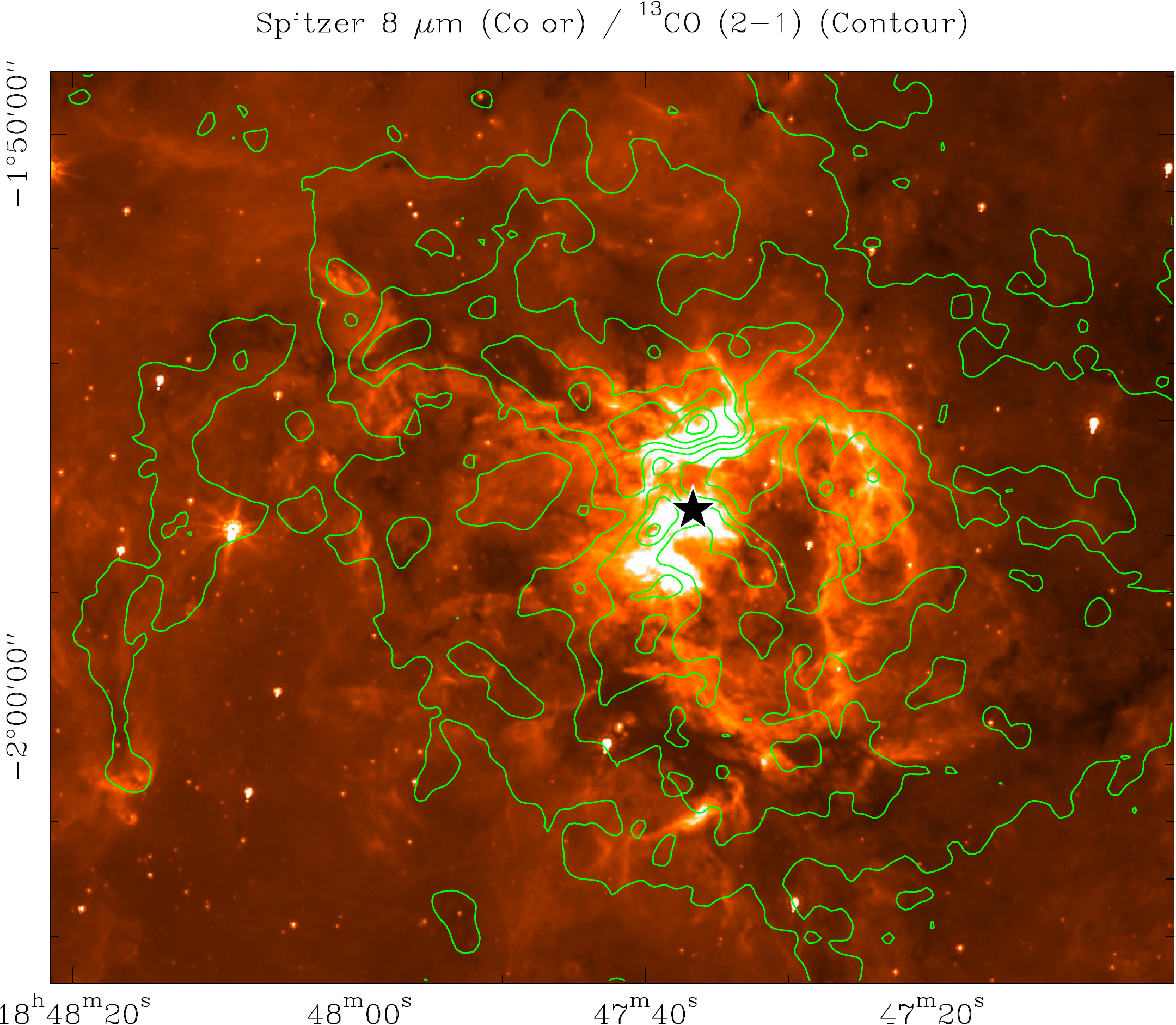}}
 \end{minipage}
 \caption{Spitzer GLIMPSE 8~$\mu$m map of the W43-Main cloud with IRAM 30m $^{13}$CO~(2--1) contours (levels ranging from 36 to 162~K\,km\,s$^{-1}$). 
 The black star marks the location of the embedded OB star cluster.}
 \label{fig:w43mainglimpse}
\end{figure}

In the GLIMPSE maps at this position, we see strong extended IR emission along the walls of a cavity, which are just west of the central Z-shape, 
formed by a cluster of young OB stars that is located here (see Fig.~\ref{fig:w43mainglimpse}). 
This cluster is probably the first result of the star-formation going on here.
The cavity seen in infrared is not the same as the one in CO described above, but both appear to be connected. The CO channel may be due to the strong radiation of the nearby stars.
The northern and southern arm of the Z-shape are seen in absorption in the 8~$\mu \mathrm{m}$ band of GLIMPSE against the infrared background. 
This indicates dense cold dust and molecular clouds that are shielded from the UV radiation of the stars, which is verified by strong emission of cold dust, as seen in the ATLASGAL and Hi-GAL maps.

\subsection{W43-South, Source 20}

Source 20, also known as W43-South and \mbox{G029.96--0.02} \citep[see e.g][]{beltran2013}, is the second largest source in the W43 complex and dominates the southwestern part our map. 
Figure~\ref{fig:appendixsourcemaps}~(t) shows a plot of this source.
It has approximately the shape of a tilted ellipse with 
the dimensions of about 24$\times$31 parsec with several smaller clumps scattered across the cloud. These clumps emit strongly in $^{13}$CO~(2--1) up to 150~K\,km\,s$^{-1}$ and are surrounded by less
luminous gas, where we see emission between 60 to 90~K\,km\,s$^{-1}$ and down to 30~K\,km\,s$^{-1}$ in the outer parts of the cloud. Maximum line peaks are 30~K. This source is less optically
thick than W43-Main; the opacity is around 2 to 3 for most parts of the cloud and does not exceed 4 in the dense clumps.

\begin{figure}[htb]
 \centering
 \begin{minipage}{0.48\textwidth}
  \includegraphics[angle=-90,width=0.96\textwidth]{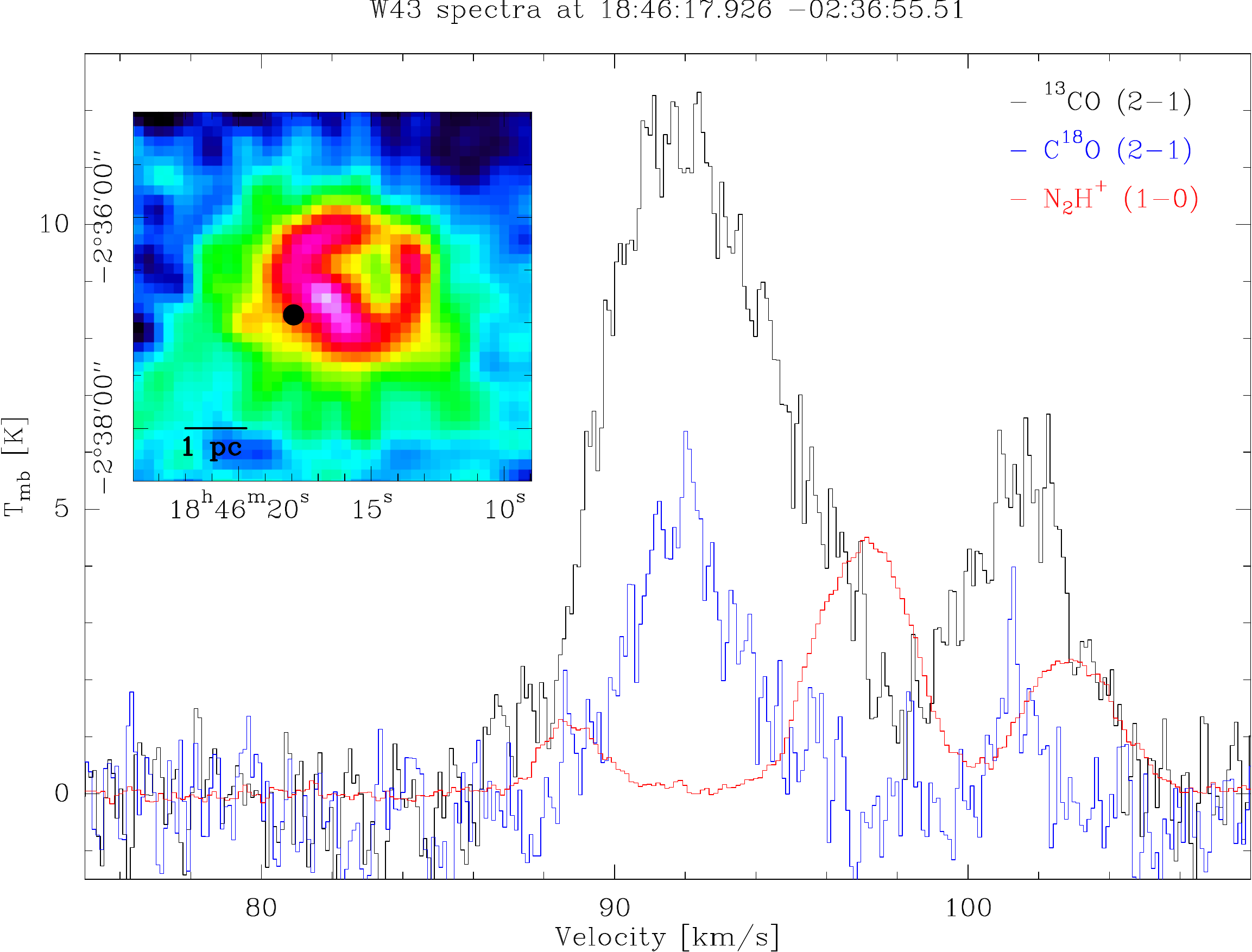}
 \end{minipage}
 \caption{Ring-like structure, as seen in W43-south. Examples of corresponding spectra that show typical infall signatures. The inlay shows the $^{13}$CO channel map at 92.7~km\,s$^{-1}$.
 The black circle marks the position where the spectra were taken and their beam size.}
 \label{fig:source20detailmaps}
\end{figure}

Studying the details of this source, we see that several of the dense clumps are actually shells of gas. The ringlike structures are clearly recognizable in some channel maps.
Figure~\ref{fig:source20detailmaps} shows the most intriguing example.
The spectra across the whole ring show infall signatures, as discussed in \citet{walker1994} in the optically thick $^{13}$CO lines. 
However, C$^{18}$O, which usually is optically thin shows the same signature as it is still optically thick at this position. 
To really trace infall, the optically thin line should show a single peak at the position of the
absorption feature in the optically thick line. This is the case for the N$_2$H$^+$~(1--0) line, which is also taken at the IRAM 30m telescope during the second part of our program,
which is optically thin (although it shows a hyper-fine structure, the strongest peak is centered on the correct velocity). 
This could be interpreted as a bubble of gas which is heated from the inside by an embedded UV source,
although it is not associated with any ultracompact H{\sc II} region
identified by the CORNISH survey \citep{purcell2013}.
The Spitzer 8~$\mu$m also show a heated ring of dust, which indicates an embedded heating source.

The relative radial velocity of the gas is more or less constant around 100~km\,s$^{-1}$ across the whole cloud W43-South. 
Small parts in the east are slower at 95~km\,s$^{-1}$, while a tip in the northwest is faster with velocities
around 105~km\,s$^{-1}$. However, the velocity gradient is not very pronounced. 
FWHM line widths show typical values of 5 to 10~km\,s$^{-1}$ with broad lines that are found mostly in the eastern part of the cloud.
The bright clumps do not show distinct broad lines.

The calculated H$_2$ column density is on the order of around a few times of $10^{22}$~cm$^{-2}$ for most of the cloud, but parts in the northwest and the center peak at about $1.5 \times 10^{23}$~cm$^{-2}$. 
The total mass of this cloud is $\sim 2 \times 10^5$~$M_{\sun}$, and it is thus the second most massive source in the W43 complex.

All bright clumps, except for the one in the central north, are seen in bright emission in the GLIMPSE 8~$\mu$m map. 
They are obviously heated from the inside; no external UV sources can be identified. 
Apparently, YSOs have formed in the dense but separated clumps that are seen in CO emission and in dust emission. 
In the northeast of the cloud, one slab of cold gas is seen in absorption in
the GLIMPSE map, separating some of the bright clumps.

\section{Conclusions}\label{sec:conclusion}

Following the investigations of \citet{motte2003}, ATLASGAL, and \citet{nguyenluong2011}, we observed the W43 region in the $^{13}$CO~(2--1) and C$^{18}$O~(2--1) emission lines with the IRAM 30m telescope. 
We have presented integrated maps of the resulting position-position-velocity cubes in which we identified numerous clouds. 

\begin{enumerate}
 \item We have confirmed that W43 is indeed one connected complex, as described in \citet{nguyenluong2011}. The connection between the two main clouds W43-main and W43-south is shown in the PV-diagram in
 Fig.~\ref{fig:pvsum}~(a).
 
 \item An analysis of the velocity distribution of our dataset and comparison to the Galactic model of \citet{vallee2008} reveals emission not only from the W43 complex itself but also from fore-/background
sources. According to this model, W43 is situated near the tangential point of the Scutum arm, where it meets the Galactic bar. The low velocity sources are located in the Perseus arm and the space
between the Sagittarius and Scutum arms (see Fig.~\ref{fig:pvsum}~(c)). 
The separation of the different components lets us avoid the confusion and line-of-sight effects in our analysis.

\item We decomposed the data cubes into subclouds, using Duchamp Sourcefinder. 
We identified a total of 29 clouds whith 20 were located in the W43 complex, while 9 were found to be foreground and background sources.
 
 \item We have derived physical properties like excitation temperature, H$_2$ column density, and total mass, of each subcloud of our dataset (see Table~\ref{table:clumplist_derived}). 
Typical smaller sources have spatial scales of 10 to 20~pc and masses
of several $10^4$~$M_{\sun}$. 
The two most outstanding sources are W43-Main (source 13) and W43-South (source 20) that have masses of a few $10^5\,M_{\sun}$.

 \item We have determined the total mass of dense clouds 
(>$10^2$~cm$^{-3}$) in the W43 complex to be $1.9\times10^6$~$M_{\sun}$. This is a factor of 1.4 lower than the mass derived from Herschel dust emission maps,
which is a discrepancy that can be explained by the details of both calculations.

 \item The shear parameter of W43 ($S_{\mathrm{g}}=0.77$) shows that the accumulation of mass in molecular clouds in this region is not disrupted by shear forces of the Galactic motion.

 \item We have created probability distribution functions obtained from column density maps. We use both the molecular line maps
and Herschel imaging data (Fig.~\ref{fig:h2_pdf}).
Both show a log-normal distribution for low column densities and a power-law tail for high densities. 
Still, there are differences seen in peak position and a power-law slope.
Possibly, the flatter slope of the molecular line data PDFs imply that those could be used to selectively show the gravitationally collapsing
gas.

\end{enumerate}

\begin{acknowledgements}
This project is carried out within the Collaborative Research Council 956, subproject A4, funded by the Deutsche For\-schungs\-gemein\-schaft (DFG).
Part of this work was supported by the \mbox{ANR-11-BS56-010} project ``STARFICH''.
Part of this work was supported by the French National Agency for Research (ANR) project ``PROBeS'', number ANR-08-BLAN-0241.
We thank the referee for constructive and valuable comments and ideas.
\end{acknowledgements}

\bibliographystyle{aa} 
\bibliography{paperrefs.bib} 

\appendix

\section{Noise maps}

\begin{figure}[htb]
\centering
\begin{minipage}{0.24\textwidth}
\resizebox{\hsize}{!}{\includegraphics{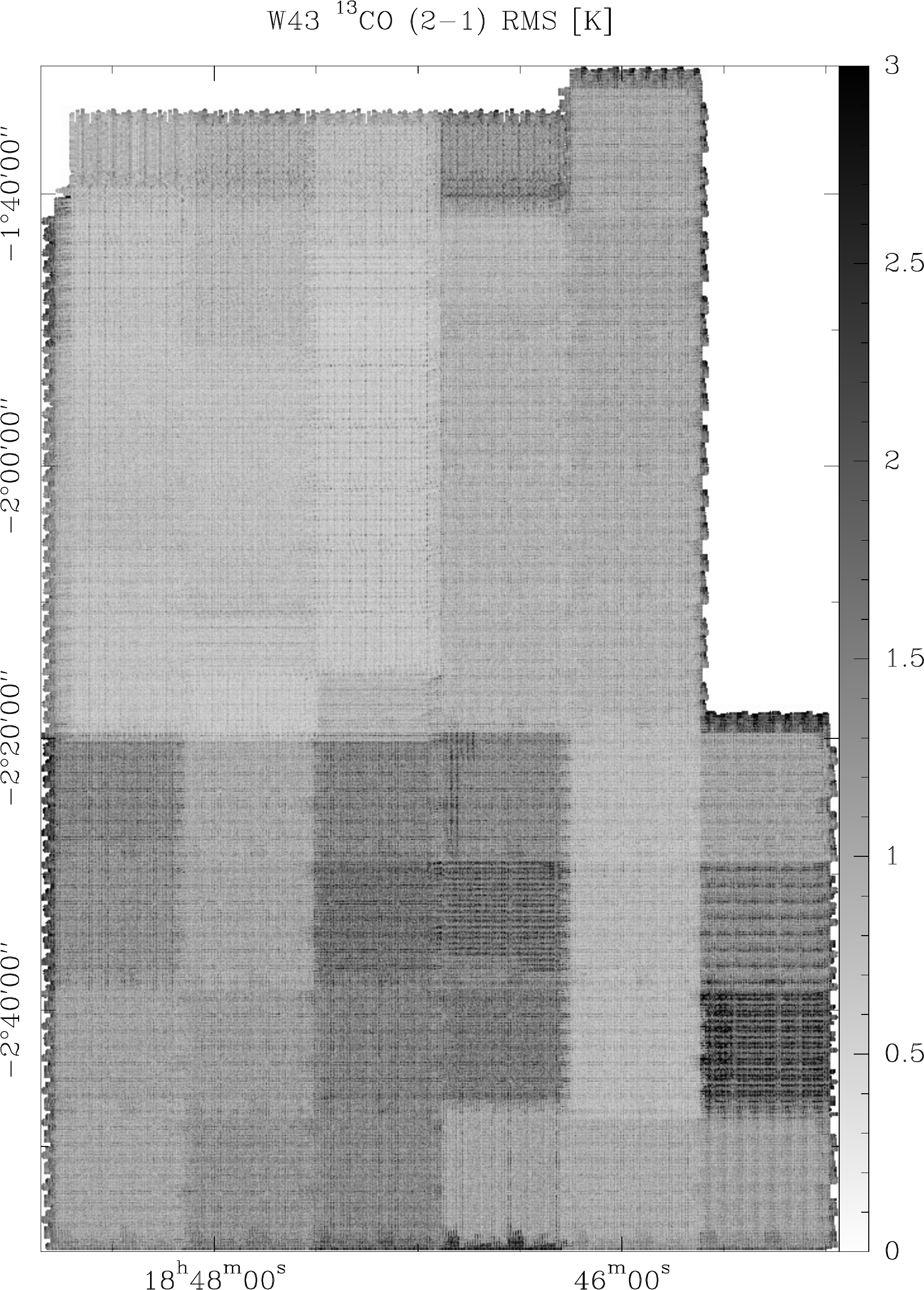}}
\end{minipage}
\hfill
\begin{minipage}{0.24\textwidth}
\resizebox{\hsize}{!}{\includegraphics{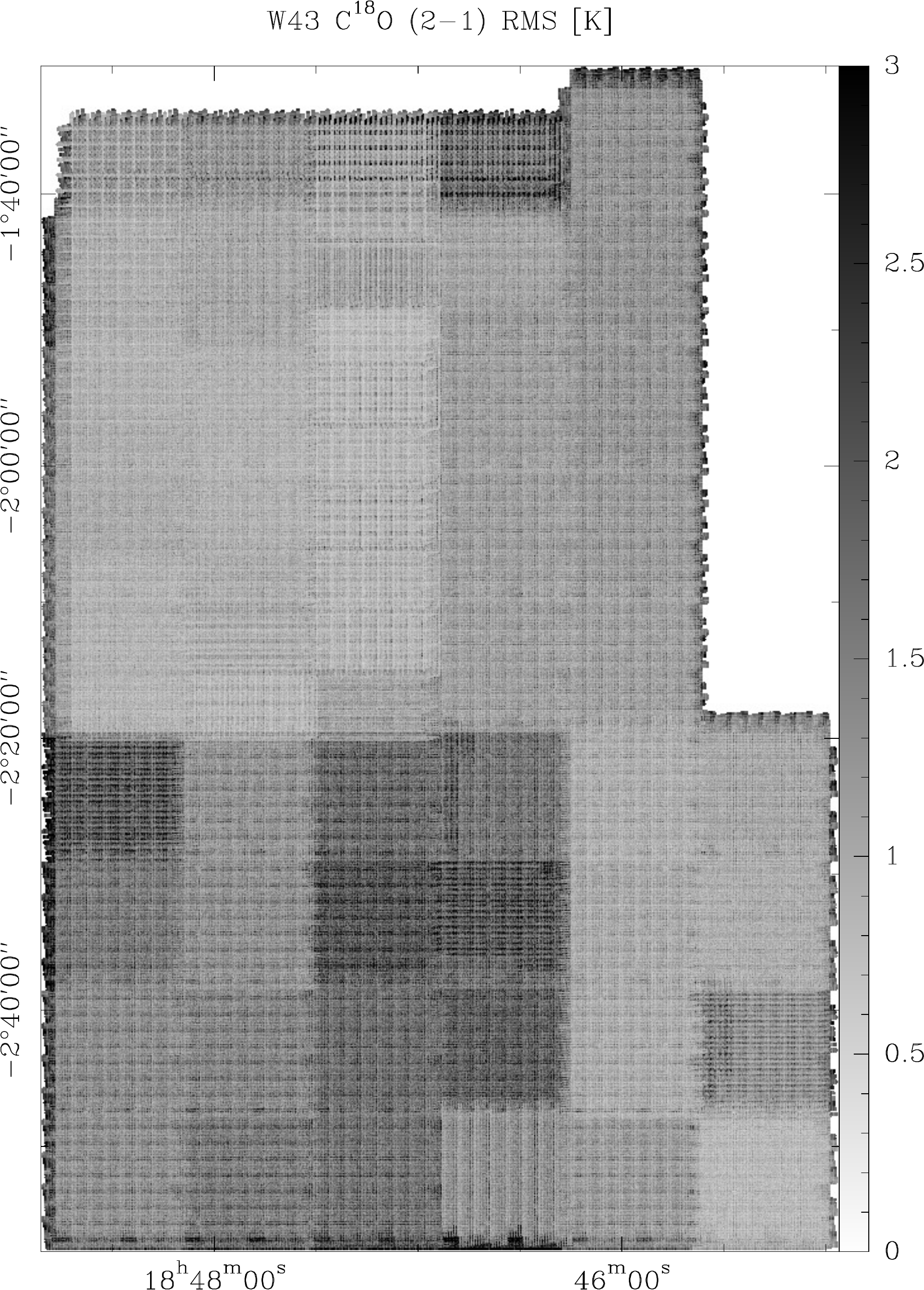}}
\end{minipage}
\caption{Noise level maps in $T_{\mathrm{mb}}$ with scale in [K]. {\bf Left:} $^{13}$CO~(2--1) map. {\bf Right:} C$^{18}$O~(2--1) map.}
\label{fig:noisemaps}
\end{figure}

We create noise maps from both the $^{13}$CO~(2--1) and the C$^{18}$O~\mbox{(2--1)} data cubes. 
For each spectrum, we determine the rms and create maps from these values.
For this purpose, we need to calculate the root-mean-square (rms) from parts of the spectra that are emission-free.
We use the velocity range between 120 and 130~km\,s$^{-1}$, because it is free of emission for
the complete region that we mapped. 
Typical values are found to be around 1~K or even less, while some parts in the south show values of up to 3~K.
All values given here are in $T_{\mathrm{mb}}$, which are corrected for main beam efficiency. The results can be seen in Fig.~\ref{fig:noisemaps}. 

We find that the structure of the noise is similar for both lines. 
The largest differences arise from weather conditions and time of day.
This is seen in the squarish pattern, as each square shows single observations that have been carried out in a small time window. 
Still, there is a striped pattern visible that overlays the whole map. 
This stems from the nine different pixels that make up the HERA receiver.
These pixels have different receiver temperatures, hence the different noise levels.
We also note that observations in the northern part of the map are usually less noisy than those in the south. 
This probably results from the different weather in which the observations have been carried out.
Last, we find that the C$^{18}$O (mean rms of 1.3~K, maximum of 6.7~K) data shows a little increase in noise temperature in general compared to the $^{13}$CO line (mean rms of 1.1~K, maximum of 3.7~K). 
The latter has been observed with the HERA1 polarization of the HERA receiver, whereas the first has been observed with HERA2, which has an overall higher receiver temperature.

\section{Peak velocity and line width of foreground components}

\begin{figure}[htb]
 \centering
 \begin{minipage}{0.48\textwidth}
  \subfloat[{\bf (a)} 35 to 55~km\,s$^{-1}$ component peak velocity]{\resizebox{0.485\textwidth}{!}{\includegraphics[scale=1]{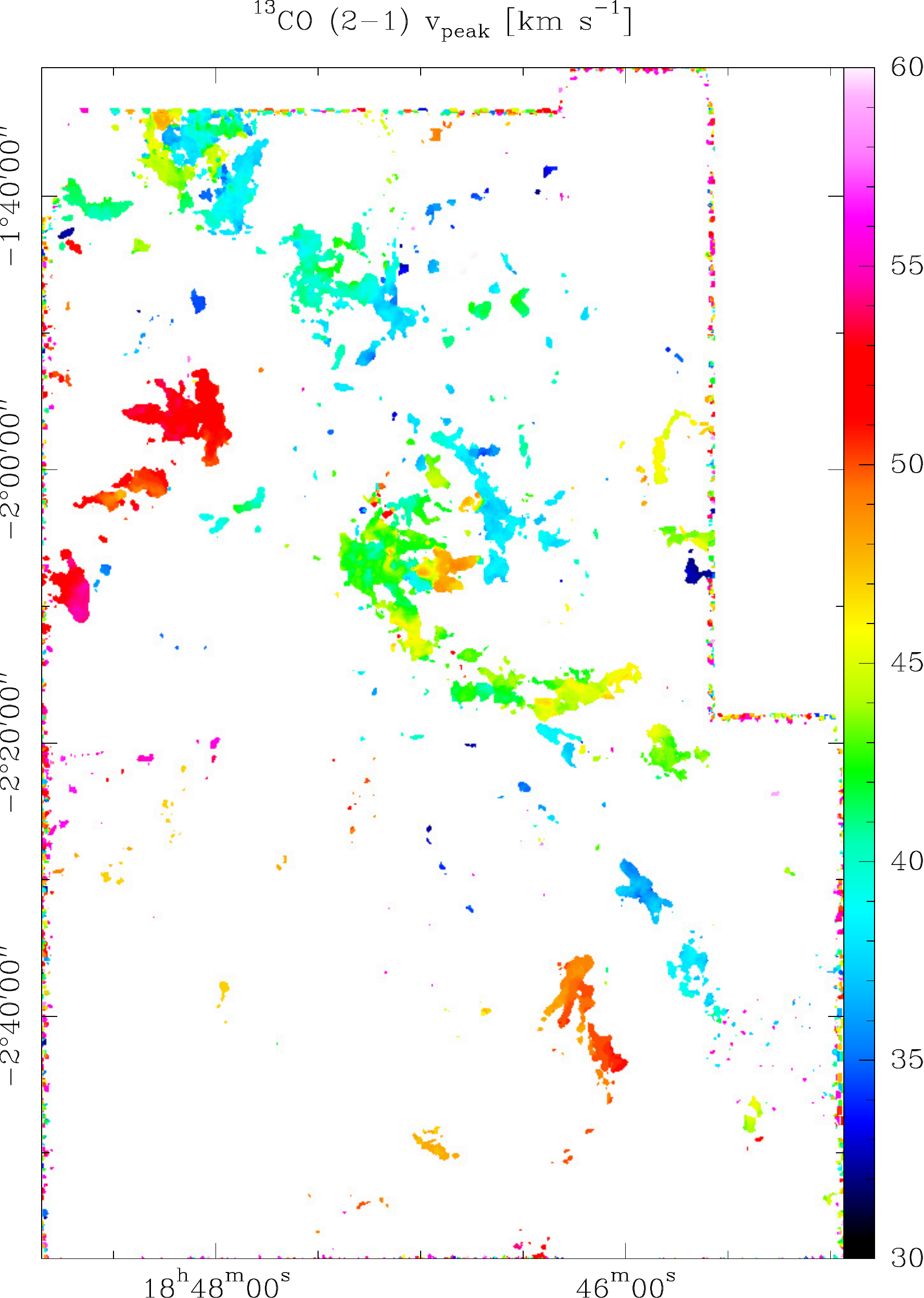}}}\hfill
  \subfloat[{\bf (b) 35} to 55~km\,s$^{-1}$ component line width]{\resizebox{0.475\textwidth}{!}{\includegraphics[scale=1]{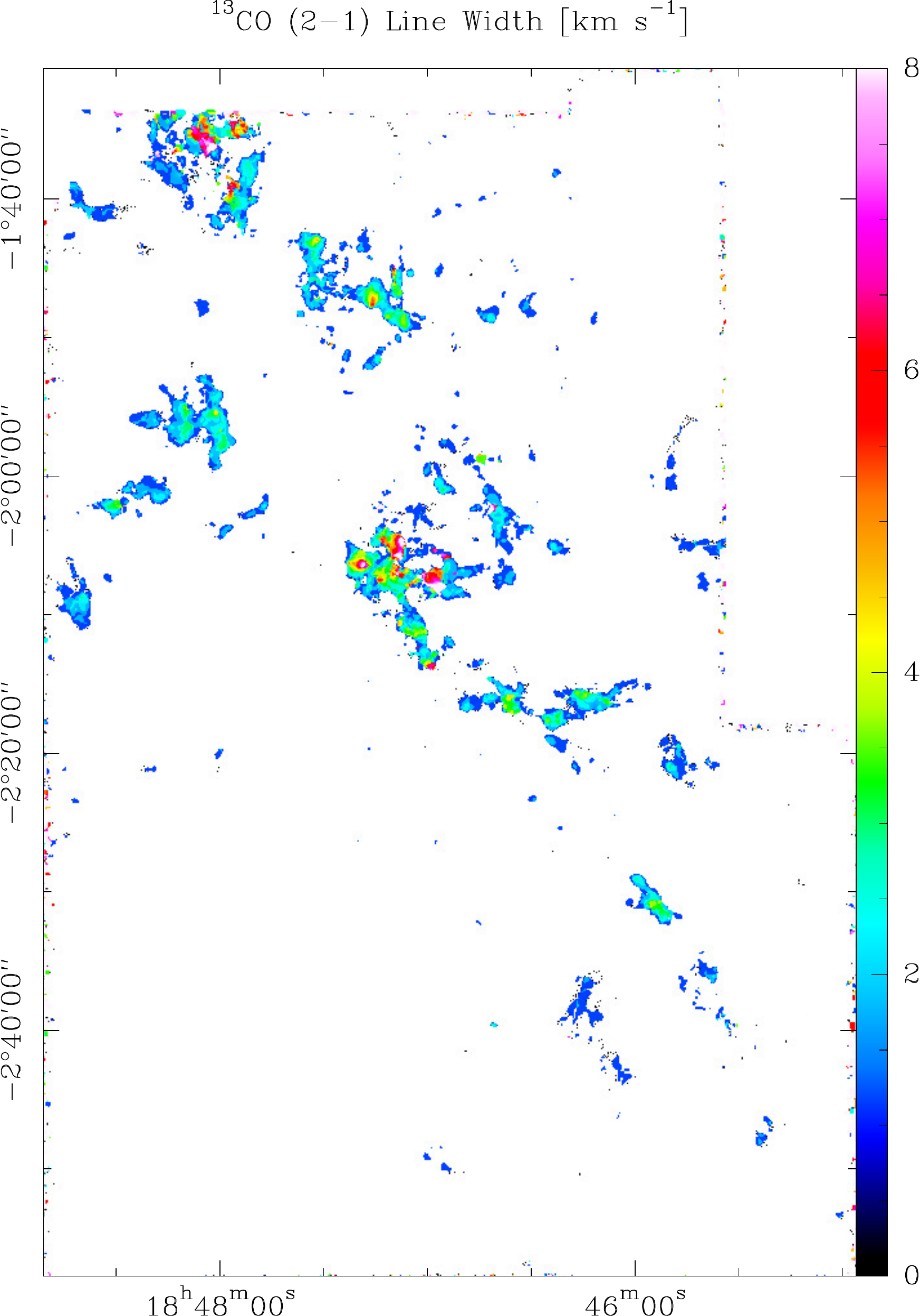}}}\\
  \subfloat[{\bf (c) 65} to 78~km\,s$^{-1}$ component peak velocity]{\resizebox{0.485\textwidth}{!}{\includegraphics[scale=1]{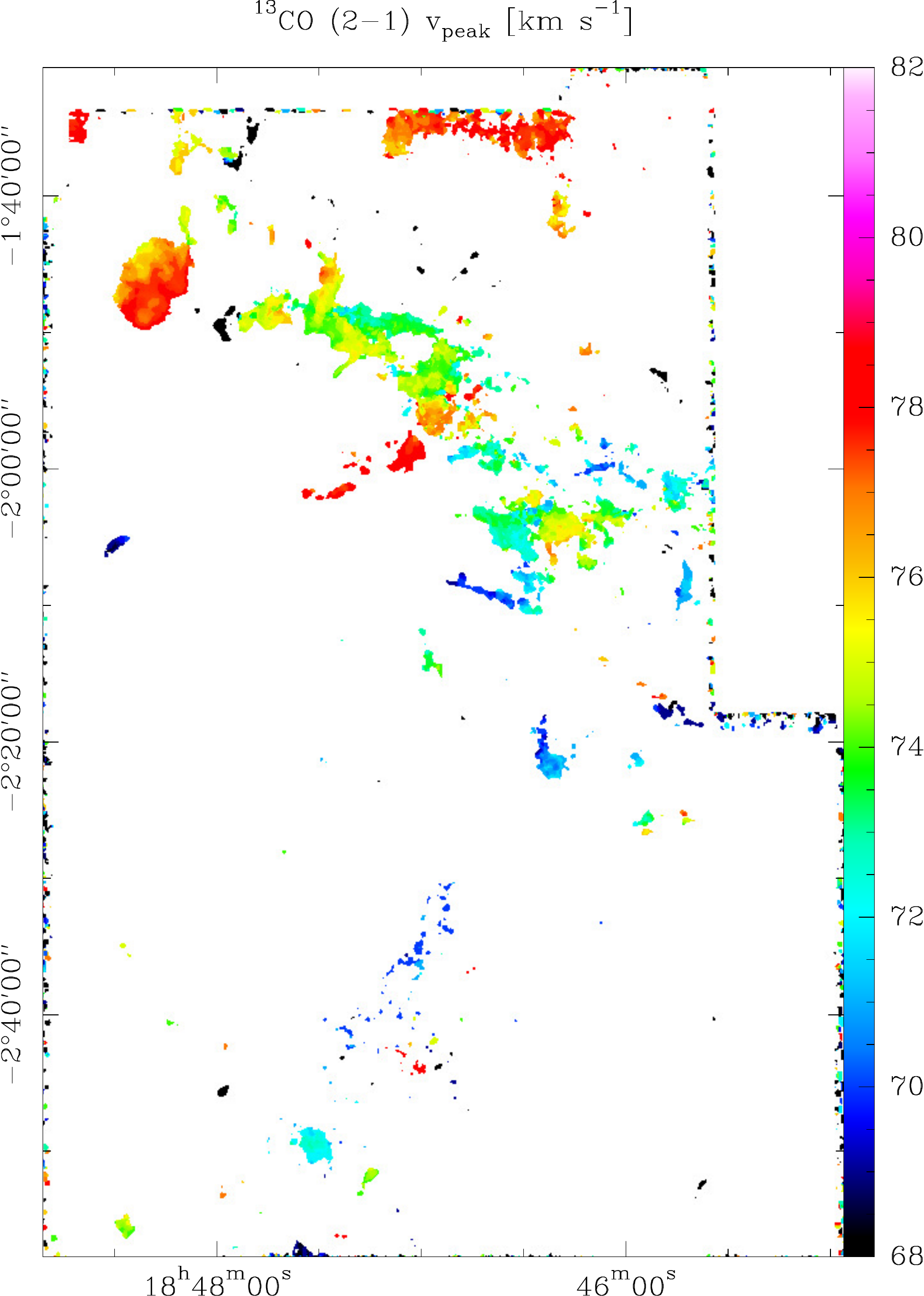}}}\hfill
  \subfloat[{\bf (d) 65} to 78~km\,s$^{-1}$ component line width]{\resizebox{0.475\textwidth}{!}{\includegraphics[scale=1]{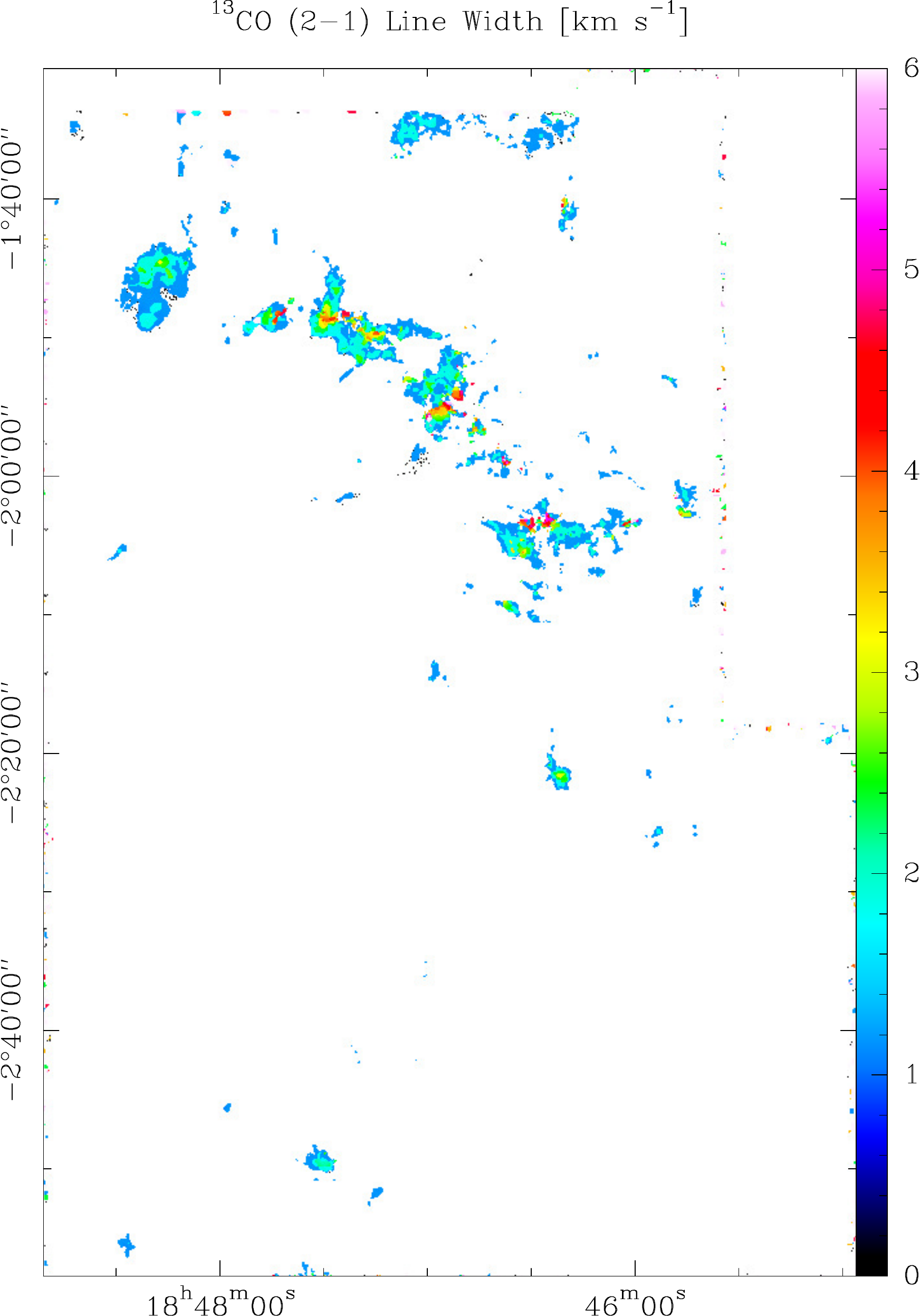}}}
 \end{minipage}
 \caption{Maps of the first (line peak velocity) and second (line width FWHM) moment maps of the two fore-/background components.}
 \label{fig:appendixpeakvelocitymaps}
\end{figure}

Figure~\ref{fig:appendixpeakvelocitymaps} shows plots of the peak velocity position and the FWHM line width of the two lower velocity components. The maps of the W43 complex itself are shown in 
Fig.~\ref{fig:velocitylinewidthmaps} and are described in Sect.~\ref{subsec:peakvelocity}.

\section{Calculations}\label{app:calculations}

\subsection{$^{13}$CO Optical depth}

Assuming a constant abundance ratio of \mbox{$^{12}$CO:$^{13}$CO:C$^{18}$O}, we can estimate the optical depth of the $^{13}$CO gas \citep[see e.g.][]{myers1983,ladd1998}.
We compare the intensities of the $^{13}$CO and C$^{18}$O line emission integrated over the analyzed cloud and solve the equation,

\begin{equation}
 \frac{I(^{13}\mathrm{CO})}{I(\mathrm{C}^{18}\mathrm{O})} \approx \frac{1-\exp(-\tau_{13})}{1-\exp(-\tau_{13} / R)},
\end{equation}

\noindent for $\tau_{13}$, where $R$ is the intrinsic ratio of of the two mapped CO isotopologes. 

The isotopic abundance of C and O in the Milky Way is known to depend on the Galactocentric radius.
Often cited values are found in \citet{wilson_rood1994}. They find the ratio of $^{16}$O and $^{18}$O to be 272 at 4~kpc radius, 302 at 4.5~kpc,
and 390 at 6~kpc, so we take those numbers for the \mbox{$^{12}$CO:C$^{18}$O} ratio.
Recent values for the \mbox{C/$^{13}$C} abundance are given in \citet{milam2005}. Here, we take values derived from CO observations and get 
a \mbox{$^{12}$CO:$^{13}$CO} ratio of 31 at a 
Galactocentric radius of 4~kpc for the main component and a ratio of 43 and 52 for the foreground components at a radius of 4.5 and 6~kpc, respectively. 
In total, we use an intrinsic ratio of
\mbox{$^{12}$CO:$^{13}$CO:C$^{18}$O} of \mbox{1:1/31:1/272} for sources in the main complex and ratios of \mbox{1:1/43:1/302} and \mbox{1:1/52:1/390} respectively for foreground sources.

\subsection{Excitation temperature}

Once we know the optical depth, we can determine the excitation temperature of the CO gas.

For this, we use \citet{ladd1998}:

\begin{equation}
  T_{\mathrm{mb}} = \frac{1}{\eta} \left( J(T_{\mathrm{ex}}) - J(T_{\mathrm{BG}}) \right)  \left( 1-\exp(-\tau) \right). 
\end{equation}

\noindent With 

\begin{equation}
  J(T) = \frac{h\nu}{k_\mathrm{B}\, \exp \left( \frac{h\nu}{k_\mathrm{B} T} -1 \right)},
\end{equation}

\noindent we use the line peak intensity $T_{\mathrm{peak}}$ for $T_{\mathrm{mb}}$. 
The parameter $T_{\mathrm{BG}}$ is the cosmic background radiation of 2.7~K, $\tau$ the $^{13}$CO optical depth, and $\eta$ the beam filling factor.
This expression can then be solved for $T_{\mathrm{ex}}$.

Here, we assume that the excitation temperature of $^{13}$CO and C$^{18}$O is the same.
We then assume that the beam filling factor $\eta$ is always 1.
It is likely that there is some substructure that we cannot resolve with our beam size. 
This would mean that the real $\eta$ is lower than 1, and we calculate $T_{\mathrm{ex}} \times \eta$ rather than just $T_{\mathrm{ex}}$.
Thus, we underestimate the temperature in cases where there is indeed substructure.
The line peak $T_{\mathrm{peak}}$ is calculated from fitting a Gaussian to all spectra in the $^{13}$CO line emission cube. 

We cannot calculate a $T_{\mathrm{ex}}$ in this way for all pixels, even if $^{13}$CO is present, as the C$^{18}$O is much weaker. 
As the ratio of $^{13}$CO and C$^{18}$O is needed, a self-consistent temperature can only be computed for points where C$^{18}$O is present. 
The main uncertainty of the calculation itself is the assumption that the beam filling factor is always 1,
and the real value can only be correctly derived where this is true.
This led to the decision not to use these T$_{\mathrm{ex}}$ maps for the further calculation of the H$_2$ column density. 
We used the calculations to get an idea of the gas temperature and then used a constant value for all further steps of analysis.
We chose this value to be 12 K, as this was the median temperature found across the whole W43 region.

\subsection{Column density and mass}\label{app:subsec:masscalc}


We then compute the H$_2$ column density by using the assumed excitation temperature $T_{\mathrm{ex}}$ of 12 K
and the integrated molecular line emission from our observation through

\begin{equation}
 N(^{13}\mathrm{CO}) = \frac{\tau}{1-\exp(-\tau)} \, f(T_{\mathrm{ex}}) \int T_{\mathrm{mb}}\, \mathrm{d} v ,
\end{equation}

\noindent where the factor containing $\tau$ accounts for the effect that the full gas is not seen for optically thicker clouds and
 
\begin{equation}
  f(T_{\mathrm{ex}}) = \frac{3h}{8\mathrm{\pi}^3 \mu^2} \frac{Z}
 {\left( J\left( T_{\mathrm{ex}} \right) - J\left( T_{\mathrm{BG}}  \right) \right) \left(1 - \exp \left( -\frac{h\nu}{k_\mathrm{B} T_{\mathrm{ex}}} \right)\right)}
\end{equation}

\noindent with the function

\begin{equation}
 Z = \exp\left(\frac{ B J_\mathrm{u} (J_\mathrm{u}+1) h}{ k_\mathrm{B} T_{\mathrm{ex}}} \right)\frac{k_\mathrm{B} T_{\mathrm{ex}}}{J_\mathrm{u} B h} ,
\end{equation}

\noindent where $B=5.5099671\times10^{10}\,$s$^{-1}$ is the rotational constant for $^{13}$CO, $\mu=0.112\,$D is its dipole moment, and $J_\mathrm{u}$ is the upper level of 
our transition (2 in this case). 
We correct all those points for the optical depth where we find an opacity larger than 0.5. 
We assume this is the minimum value we can determine correctly as we might confuse emission with noise for lower opacities.

\begin{figure}[htb]
 \centering
 \begin{minipage}{0.48\textwidth}
  \resizebox{0.96\hsize}{!}{\includegraphics[angle=-0]{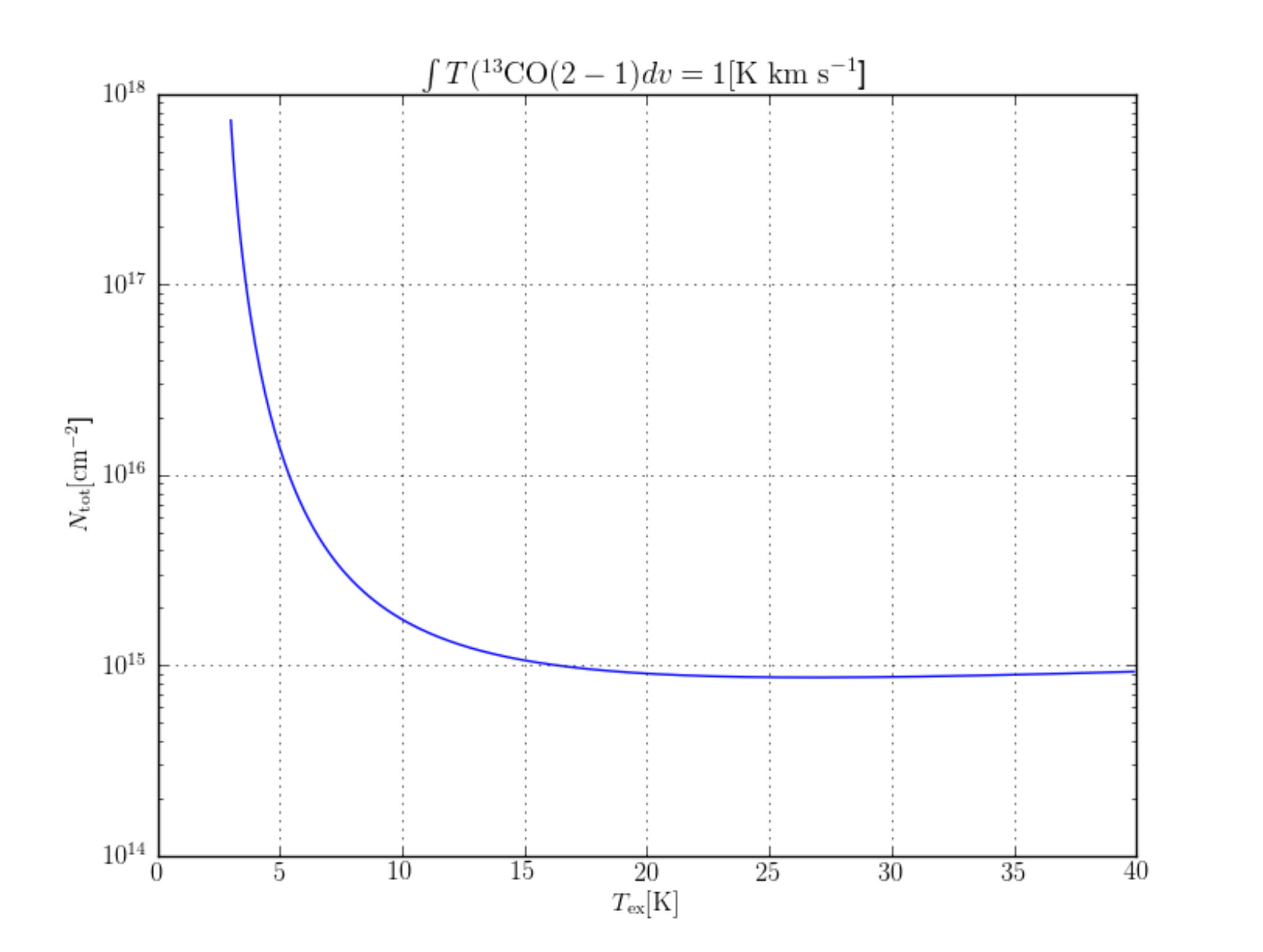}}
 \end{minipage}
 \caption{Plot of $^{13}$CO column density depending on the excitation temperature.}
 \label{fig:appendix_13co_excitation_plot}
\end{figure}

In Fig.~\ref{fig:appendix_13co_excitation_plot} we plot the dependency of the $^{13}$CO column density and the excitation temperature. 
We note that the column density for values of $> 15$~K is nearly independent of the excitation temperature.
On the contrary, it rises steeply for temperatures below 10~K. 
This is important as we most probably would underestimate the excitation temperature for most clouds as discussed above, if we actually used the calculated T$_{\mathrm{ex}}$. 
We would thus overestimate the column density. 
Therefore, we can assume that using an excitation temperature of 12 K for our calculation results in a lower limit for the actual column density.

As we want to calculate the H$_2$ column density, we need to translate $N(^{13}$CO) into $N(\mathrm{H}_2)$. 
The standard factor of $N(\mathrm{H}_2)$:$N(^{12}\mathrm{CO})$ is $10^4$ for local molecular clouds, but the ratio varies with sources and also with the Galactocentric radius.
\citet{fontani2012} derives a radius dependent formula for \mbox{$N(\mathrm{H}_2)$:$N(\mathrm{CO})$}, using values of \mbox{$^{12}$C/H} and \mbox{$^{16}$O/H} from \citet{wilson1992}.
This formula gives a ratio of 6550 for a Galactic radius of 4~kpc 7000 at 4.5~kpc and 8500 at 6~kpc.
Using our ratios of \mbox{$^{12}$CO:$^{13}$CO} from above we get $N(\mathrm{H}_2) = 2.0\times10^5\, N(^{13}\mathrm{CO})$ for the 
W43 complex with a radius of 4~kpc and the factors of $3.0\times 10^5$ and $4.4 \times 10^5$ for the fore-/background complexes with radii of 4.5 and 6~kpc components, respectively.
These factors are prone to large errors of at least a factor of 2.

To calculate the mass of the observed source from the H$_2$ column density it is necessary to consider the relative distance from the Sun to the source.
See Sect.~\ref{subsec:distdeterminationW43} for the distance determination.
We assume the main complex clouds are 6~kpc away, while the foreground clouds have distances of 3.5, 4, 3.5, and 12~kpc.
Then, we just have to count the number of H$_2$ molecules per pixel to receive the mass per pixel in solar masses.

\begin{equation}
 M_{\mathrm{tot}} = N(\mathrm{H}_2) \, \frac{m(\mathrm{H}_2)}{M_{\sun}} \, d^2 \, \alpha^2 \, \mu.
\end{equation}

\noindent Here, $d$ is the distance toward the source, $\alpha$ the angular extent of one pixel on the sky.
The value of $\mu=1.36$ accounts for higher masses of molecules, apart from H$_2$ \citep[compare][]{schneider2010}.

The resulting H$_2$ column density map of the full W43 complex is shown in Fig.~\ref{fig:h2_map_total}, which was calculated using the velocity range between
78 and 120~km\,s$^{-1}$. It has been calculated for all points, where the integrated $^{13}$CO shows an intensity of more than 5~K\,km\,s$^{-1}$. 
The integrated $^{13}$CO map in Fig.~\ref{fig:completemaps}~(a) shows diffuse emission in-between the brighter sources that were isolated with the Duchamp sourcefinder.
This diffuse emission accounts for about 50\% of the the total mass in W43.

%

\section{Description of important sources}\label{app:sec:important_sources}

In the following, we want to give a description of several important and interesting sources of the W43 complex ($d=6$~kpc) found in our datasets.
Information on these sources are listed in Table~\ref{table:clumplist}, while their location is indicated in Fig.~\ref{fig:duchampdetectionmap}. 
The corresponding maps can be found in Fig.~\ref{fig:appendixsourcemaps}. 
We discuss the shape, topology, and intensity of the maps and fundamental properties like velocity gradients, FWHM line widths, temperature, and column density.
We also mention conclusions from the comparison to different datasets.
Sources are ordered by their peak velocities.

\subsection{Source 23}

Source 23 (plots can be seen in Fig.~\ref{fig:projectcomparison23} and also Fig.~\ref{fig:appendixsourcemaps}~(w)) consists of one central elliptical clump with one elongated thin extension, 
protruding from the southeast, that ends in a hook-like tip and is curved to the south.
The central clump is elliptically shaped with a length of 5 and 3 parsec on the major axes and is bound sharply at the southern edge, while it is much more diffuse and more 
extended in the north.
The extension has a length of 7.5~pc. 
We find the maximum integrated intensity of the $^{13}$CO~(2--1) line to be  about 90~K\,km\,s$^{-1}$ at the peak of the clump, while the filamentary extension lies around 30 to 45~K\,km\,s$^{-1}$.
The line peak intensity rises from 12~K in the filament to 24~K in the clump.
The opacity has typical values of 1 to 2.5 with higher values in the central clump.

We see a gradient in the radial velocity of the cloud from the filament to the center of the source of $\sim 3$~km\,s$^{-1}$, which can be interpreted as a flow of gas along the outrigger onto the clump.
The line width (FWHM) changes between 4.5~km\,s$^{-1}$ in the inner clump and 2 to 2.5~km\,s$^{-1}$ in the outer parts of the cloud.

The H$_2$ column density that we calculated rises from $\sim 5 \times 10^{21}$~cm$^{-2}$ in the edges of the cloud to $\sim 7 \times 10^{22}$~cm$^{-2}$ in the center. 
The total mass is calculated to be $2.7 \times 10^4$~$M_{\sun}$ and thus resembles a typical total mass of our set of sources.

The CO emission that we measure in our maps is nearly exactly matched by the dust emission maps of ATLASGAL and Hi-GAL (see Fig.~\ref{fig:projectcomparison23}~(b) and (c), respectively). 
Both show the strong peak in the central clump and the weaker filament in the southeast, including that of the curved tip. 
The GLIMPSE map shows very interesting features (see Fig.~\ref{fig:projectcomparison23}~(d)). There is one strong UV point source less than 1~pc off to the south of the CO clump.

\subsection{Source 25}

This filament, as seen in Fig.~\ref{fig:appendixsourcemaps}~(y), resides in the central western part of the W43 complex, which is half-way between W43-Main and W43-South. It is shaped like an inverted L 
with two branches and connected by an orthogonal angle. The vertical branch has a length of 14~pc; the horizontal one is 10~pc long. The typical width of both branches is between 2 and 3~pc.
One strong clump is seen in the southern part with an integrated line intensity of the $^{13}$CO~(2--1) line of 40~K\,km\,s$^{-1}$, while the rest of the filament backbone only reaches 18 to 
22~K\,km\,s$^{-1}$. Line peak intensities range from a few K in the outer parts of the filament up to 15~K in the strong southern clump.

Investigating the line peak velocity map, we realize that the two branches of this source are actually separated. The horizontal branch has a constant radial velocity of 108~km\,s$^{-1}$ across,
while the vertical branch shows a gradient from 110~km\,s$^{-1}$ in the north to 115~km\,s$^{-1}$ in the south. Line widths range between 1 and 2~km\,s$^{-1}$ in the whole source.

The H$_2$ column density varies between $2 \times 10^{21}$~cm$^{-2}$ in the outer parts and $4 \times 10^{22}$~cm$^{-2}$ around the southern core. 
The total mass is $\sim 6.8\times 10^3$~$M_{\sun}$.
Comparing this source to the complementary projects is complicated, since the source 17 is located at the same place and overlaps this source. Most emission that is seen in the northern part of the
source is presumably part of source 17. Only the embedded core in the south is clearly seen in dust emission and as a compact Spitzer source.

\subsection{Source 26}

Located in the easternmost central part of the W43 complex lies this filamentary shaped source, whose plot is found in Fig.~\ref{fig:appendixsourcemaps}~(z). It stretches over a range of 26~pc 
from southeast to northwest. The filament consists of three subsections that contain several clumps and has a typical width of 5~pc.
The integrated emission map of the $^{13}$CO~(2--1) line shows values of up to 35~K\,km\,s$^{-1}$ in the clumps, which is surrounded by weaker gas. The strongest clump lies in the southeastern end of the filament
while the highest line peak intensities are found in the northwest with up to 13~K.

The velocity structure of this filament is nearly symmetrical, starting around 106~km\,s$^{-1}$ in the middle of the filament and increasing toward its tips up to 110~km\,s$^{-1}$ in the west and 
112~km\,s$^{-1}$ in the east. The width of the lines has nearly homogeneous values around 2~km\,s$^{-1}$ in the center and western part of the filament but shows broad lines with a width of more 
than 5~km\,s$^{-1}$ in the eastern clump.

This filament shows a typical distribution of its H$_2$ column density. Several denser clumps are embedded along the filament. Column densities vary from a few $10^{21}$~cm$^{-2}$
in the outer parts of the filament up to a maximum of $3\times10^{22}$~cm$^{-2}$ in one clump. 
The total mass of this source is $1.7\times 10^4$~$M_{\sun}$.

The CO emission of this source matches nicely with the dust emission of ALTASGAL and Hi-GAL. However, the eastern part of the filament is stronger in the 850~$\mu$m map than the 
shorter wavelengths of Hi-GAL and vice versa in the western part. Only the clump in the west of this filament can be seen in the 8~$\mu$m Spitzer data, the east is not traced. There is one 
extended source seen in emission in the center part of this source, but this is probably unrelated.

\subsection{Source 28}

Source 28 (see Fig.~\ref{fig:appendixsourcemaps}~(ab)) is located in the southwest of W43-Main in the central region of the complex. It is not visible in the total integrated maps of the region, 
as it is confused with sources of a different relative
velocity. It becomes visible by investigating the channel maps between 110 and 115~km\,s$^{-1}$ radial velocity. The source has dimensions of 12~pc in the east-west direction and 8~pc in 
the north-south direction. Its shape is that of a two-armed filament, whose two parts join at an angle of $\sim 135\degr$ where the eastern arm runs from southeast to northwest and the
western arm from east to west. We see two stronger clumps in the eastern filament: one in the center of it and one in the southeastern tip.
This is a relatively weak source with an integrated $^{13}$CO~(2--1) emission that peaks at only 25~K\,km\,s$^{-1}$ in the center of the eastern filament, where the maximum line peak is around 13~K. 
The integrated intensity goes down to 8~K\,km\,s$^{-1}$ in the outskirts of the filament.
Yet, it is a valuable source, due to its pronounced filamentary structure and the embedded clumps, which is
a good candidate for more investigations in filament formation \citep[See][for additional observations and analysis of this cloud.]{carlhoff2013b}.

The eastern arm is especially interesting. It has a length of 6~pc and a typical width of 1.5~pc. 
We note two embedded clumps embedded in it and a velocity gradient along the filament, which starts at
110~km\,s$^{-1}$ in the north and increases to 114~km\,s$^{-1}$ in the southeastern part of this arm. 
The typical line width varies between 1 and 2~km\,s$^{-1}$,
increasing toward the inside of the cloud and reaching the maximum width at the clumps.

As this source is rather weak, we also find H$_2$ column densities to be only around $2 \times 10^{22}$~cm$^{-2}$ at the maximum around the embedded clumps. 
The total derived mass of the molecular gas is $\sim 4300$~$M_{\sun}$, which makes it one of the less massive sources identified.

Surprisingly, this source is one of the few that is not traced at all in the GLIMPSE 8~$\mu$m map. It appears that there are no nearby UV sources that could heat the gas. 
The two nearest sources
were identified to be related to background sources in another Galactic spiral arm. Also, the gas and the related dust is obviously too faint to appear in absorption. This is also verified by the 
ATLASGAL an Hi-GAL maps, which show only weak dust emission in the filament.

\subsection{Source 29}

\begin{figure*}[htb]
\centering
\begin{minipage}{18cm}
 \subfloat[(a) Source 1]{\resizebox{!}{0.105\textheight}{\includegraphics[scale=1]{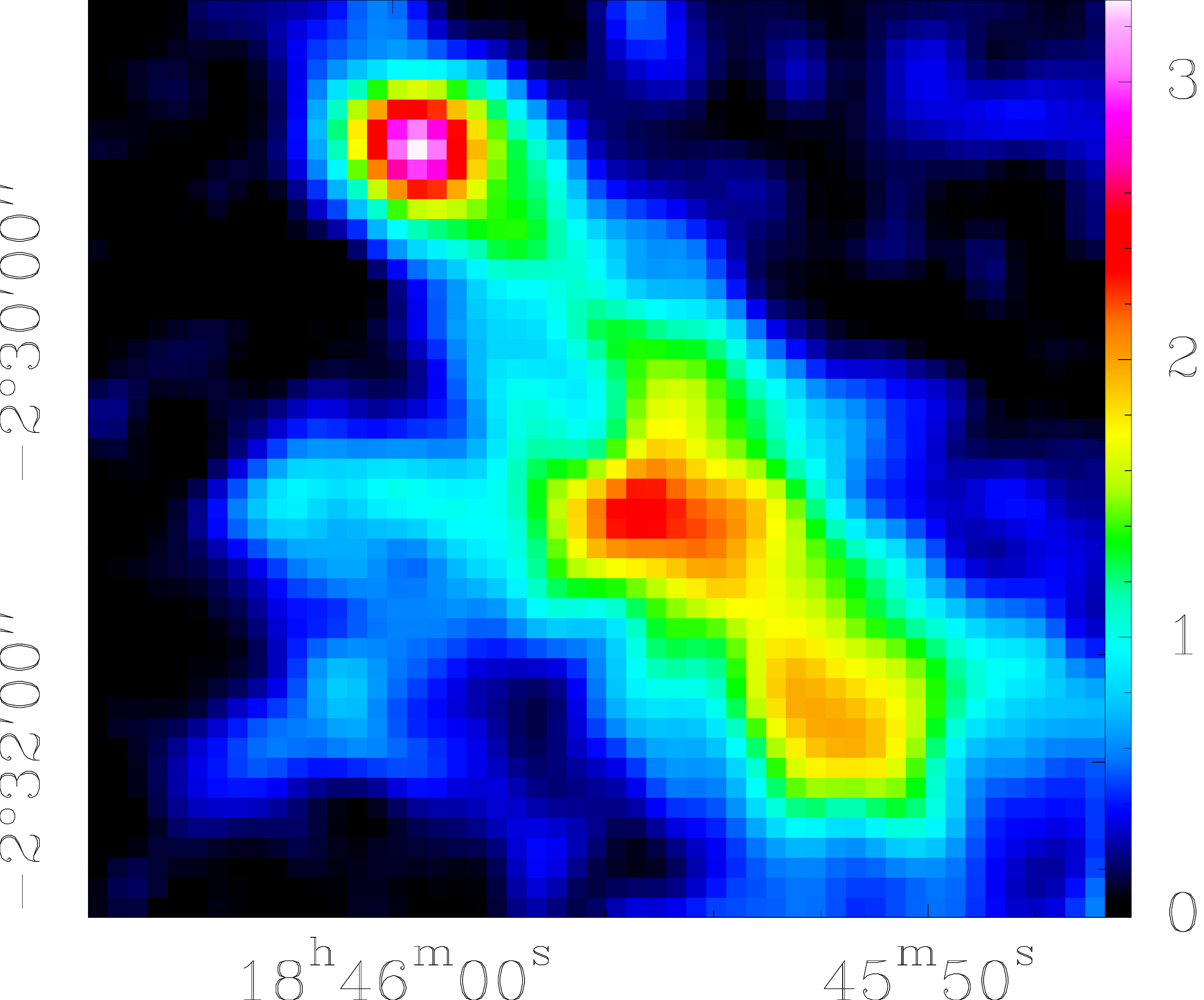}}}\hfill
 \subfloat[(b) Source 2]{\resizebox{!}{0.105\textheight}{\includegraphics[scale=1]{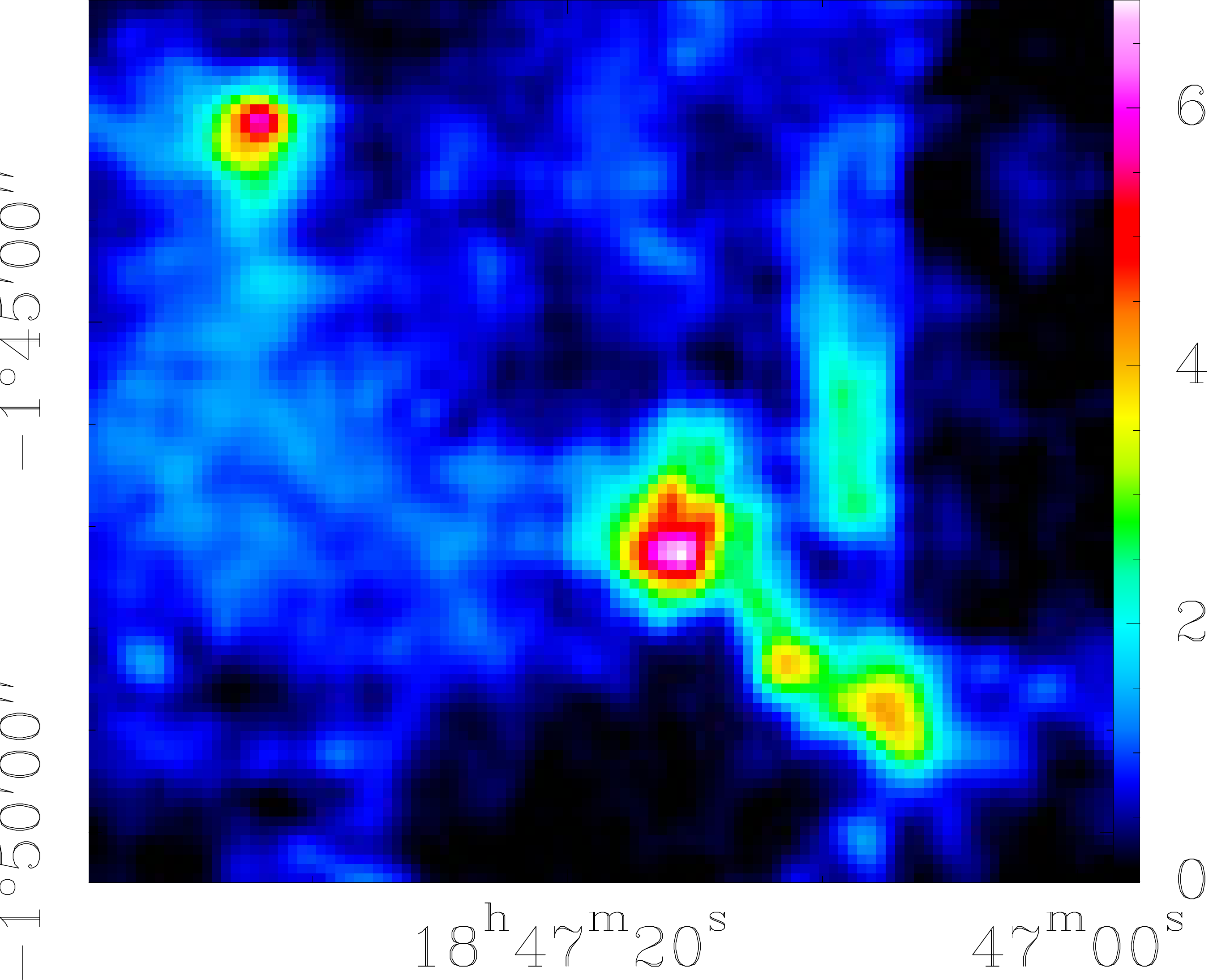}}}\hfill
 \subfloat[(c) Source 3]{\resizebox{!}{0.105\textheight}{\includegraphics[scale=1]{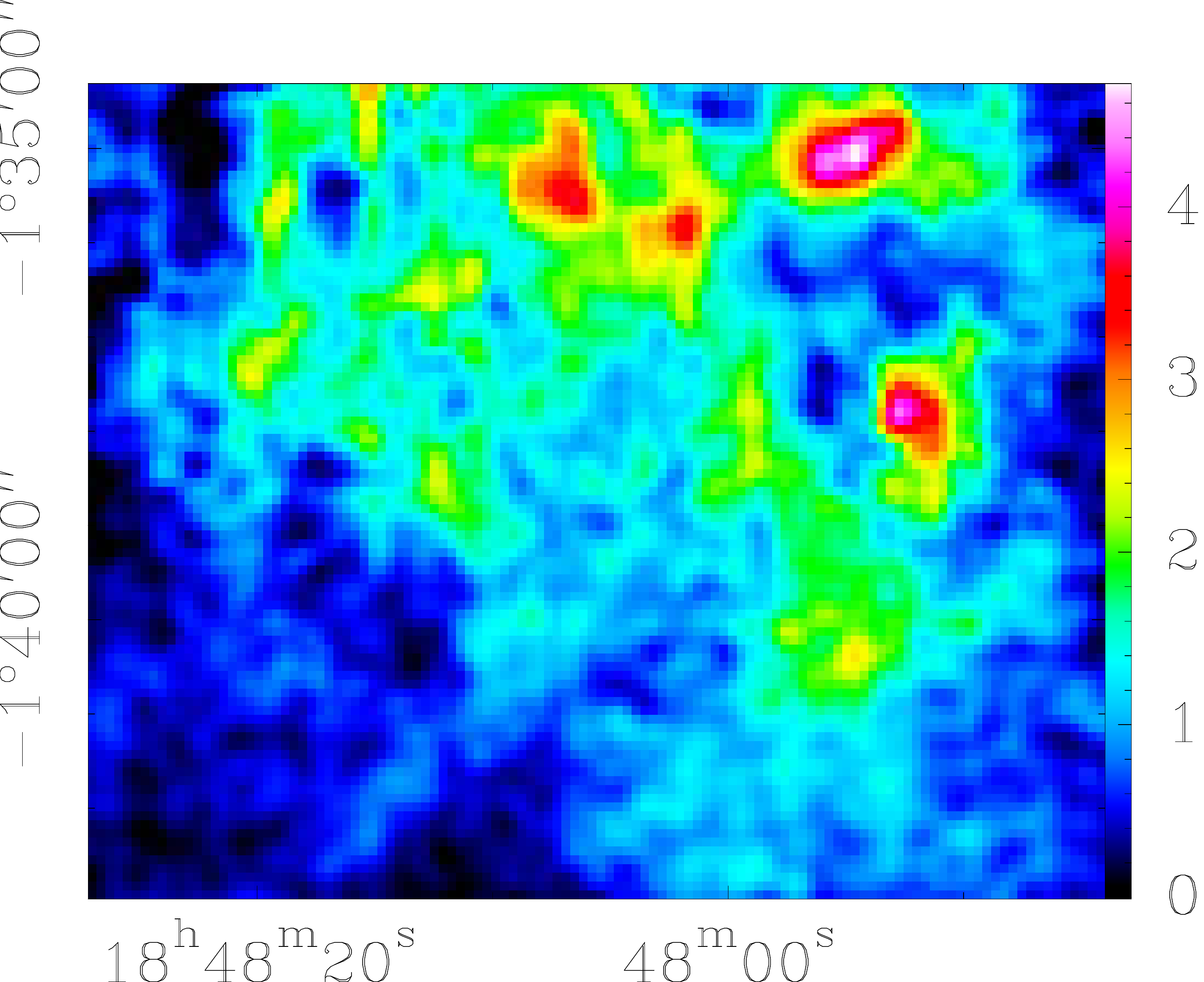}}}\hfill
 \subfloat[(d) Source 4]{\resizebox{!}{0.105\textheight}{\includegraphics[scale=1]{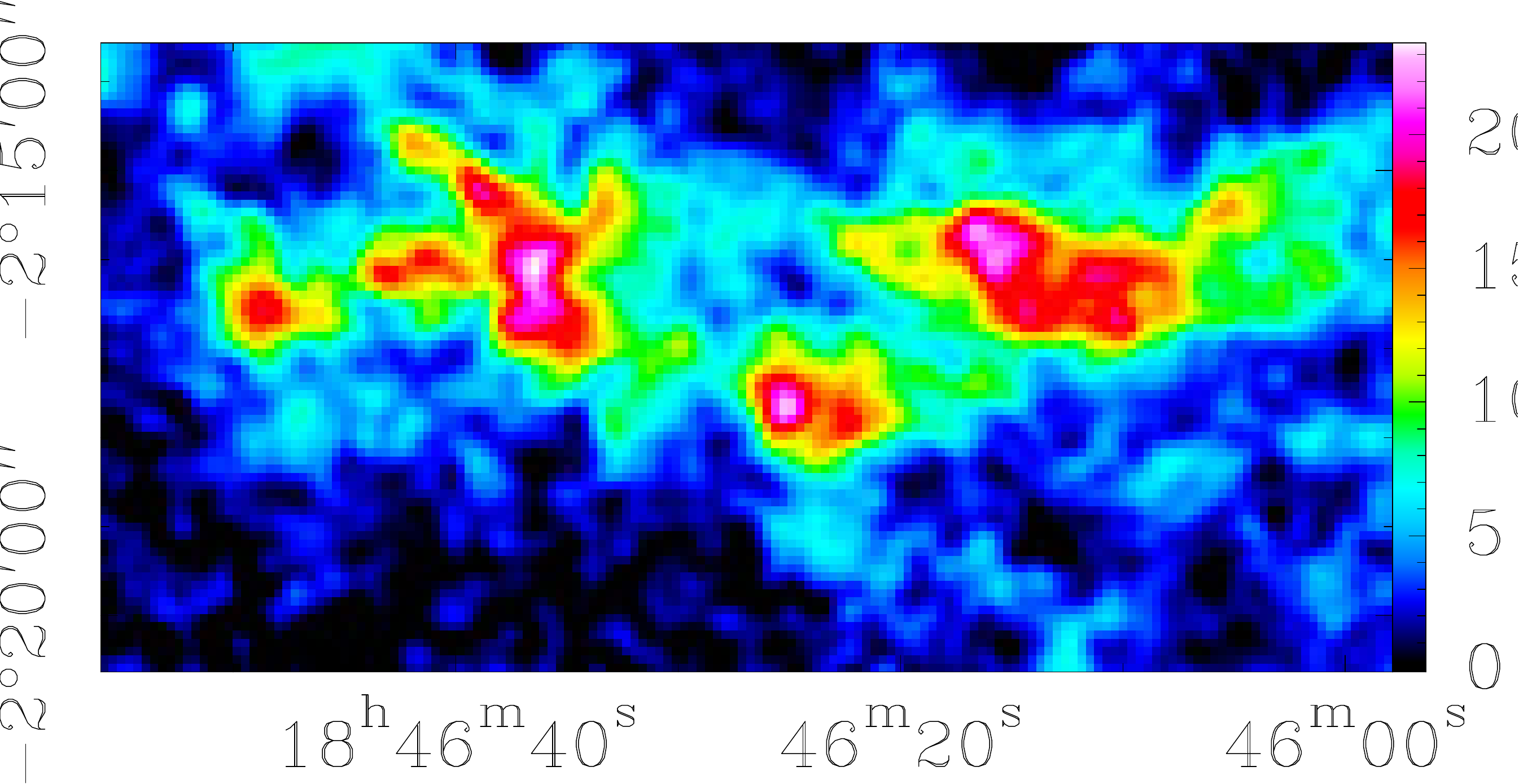}}}\hfill
 \subfloat[(e) Source 5]{\resizebox{!}{0.105\textheight}{\includegraphics[scale=1]{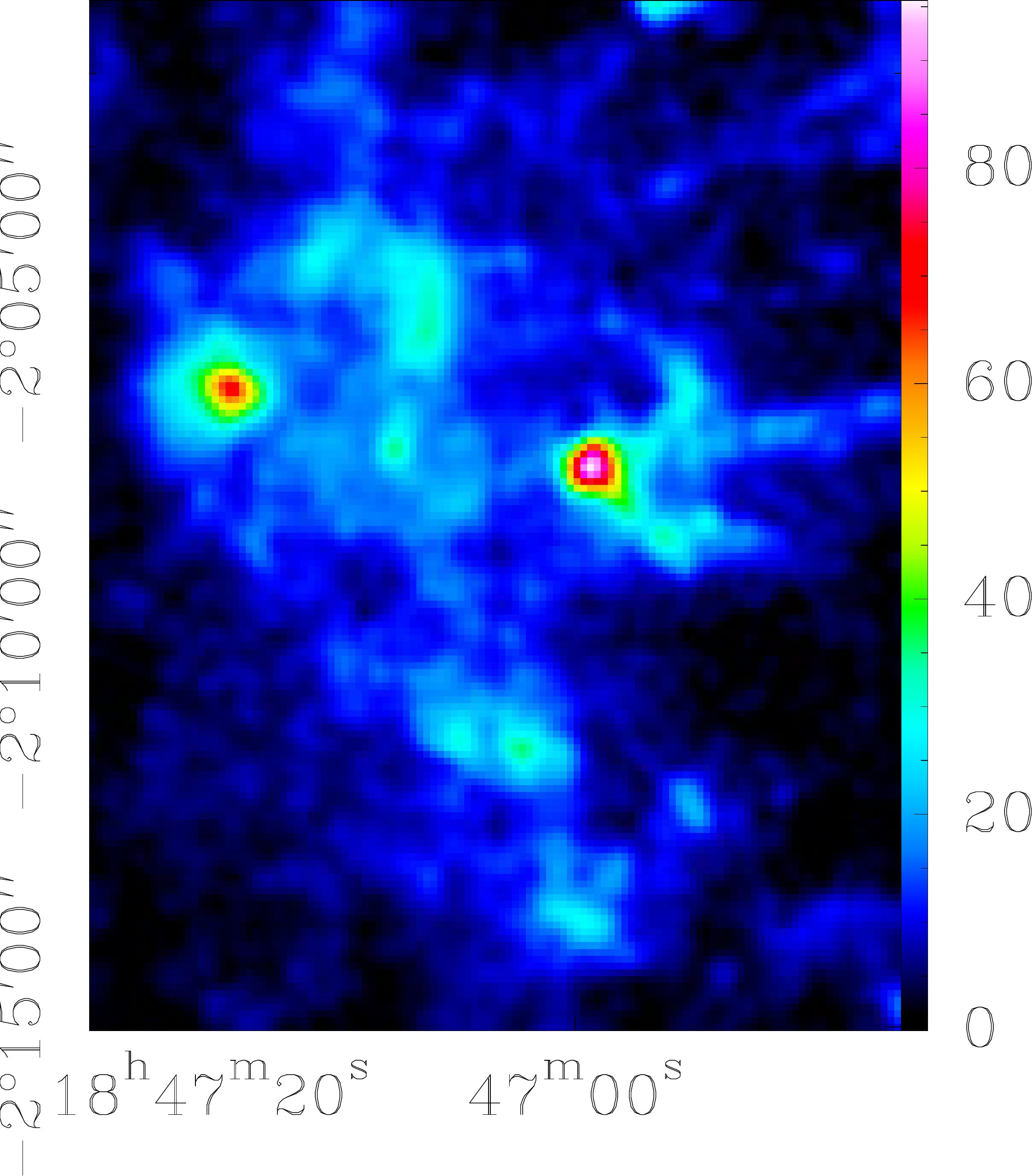}}}\\
 \subfloat[(f) Source 6]{\resizebox{!}{0.125\textheight}{\includegraphics[scale=1]{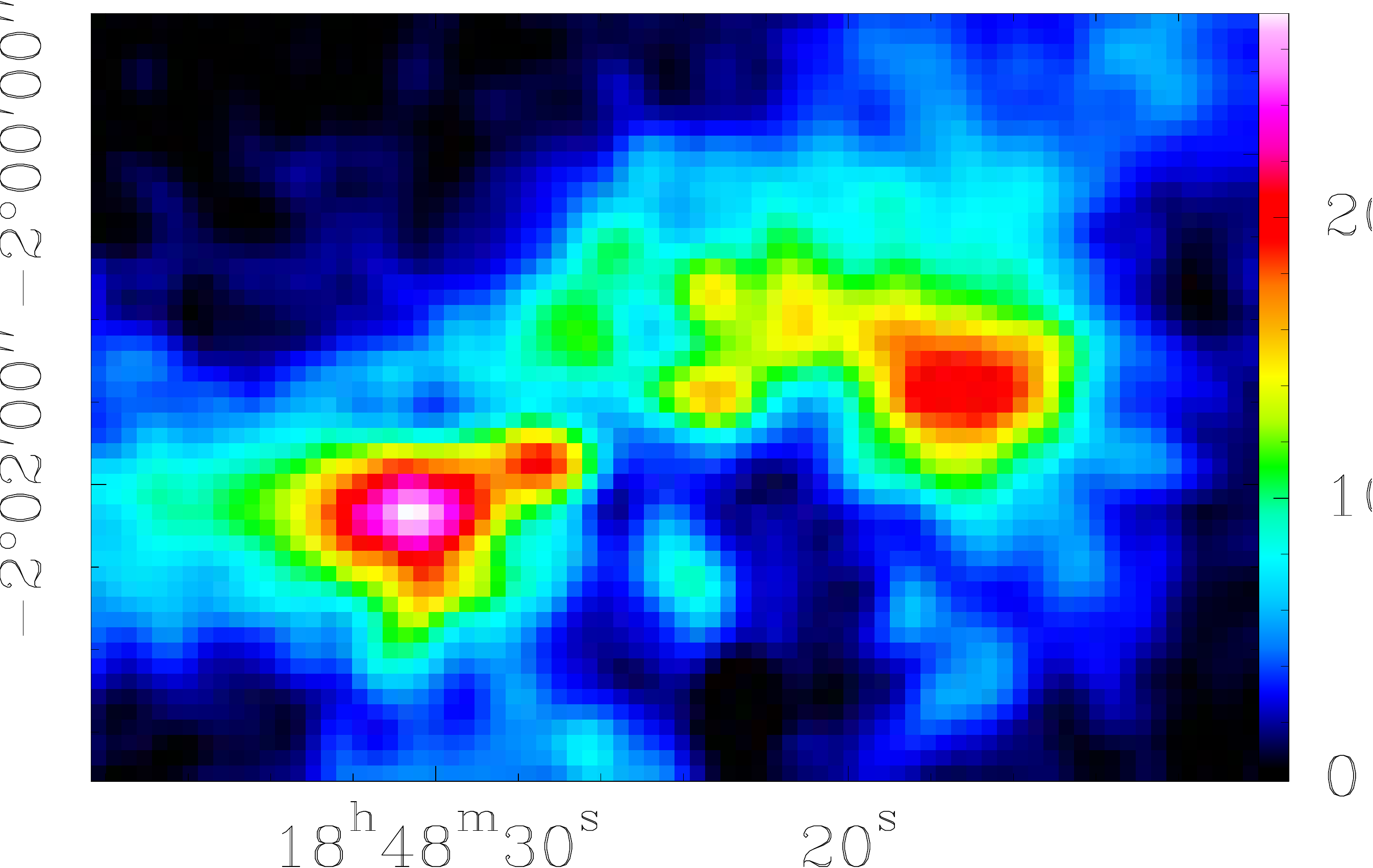}}}\hfill
 \subfloat[(g) Source 7]{\resizebox{!}{0.125\textheight}{\includegraphics[scale=1]{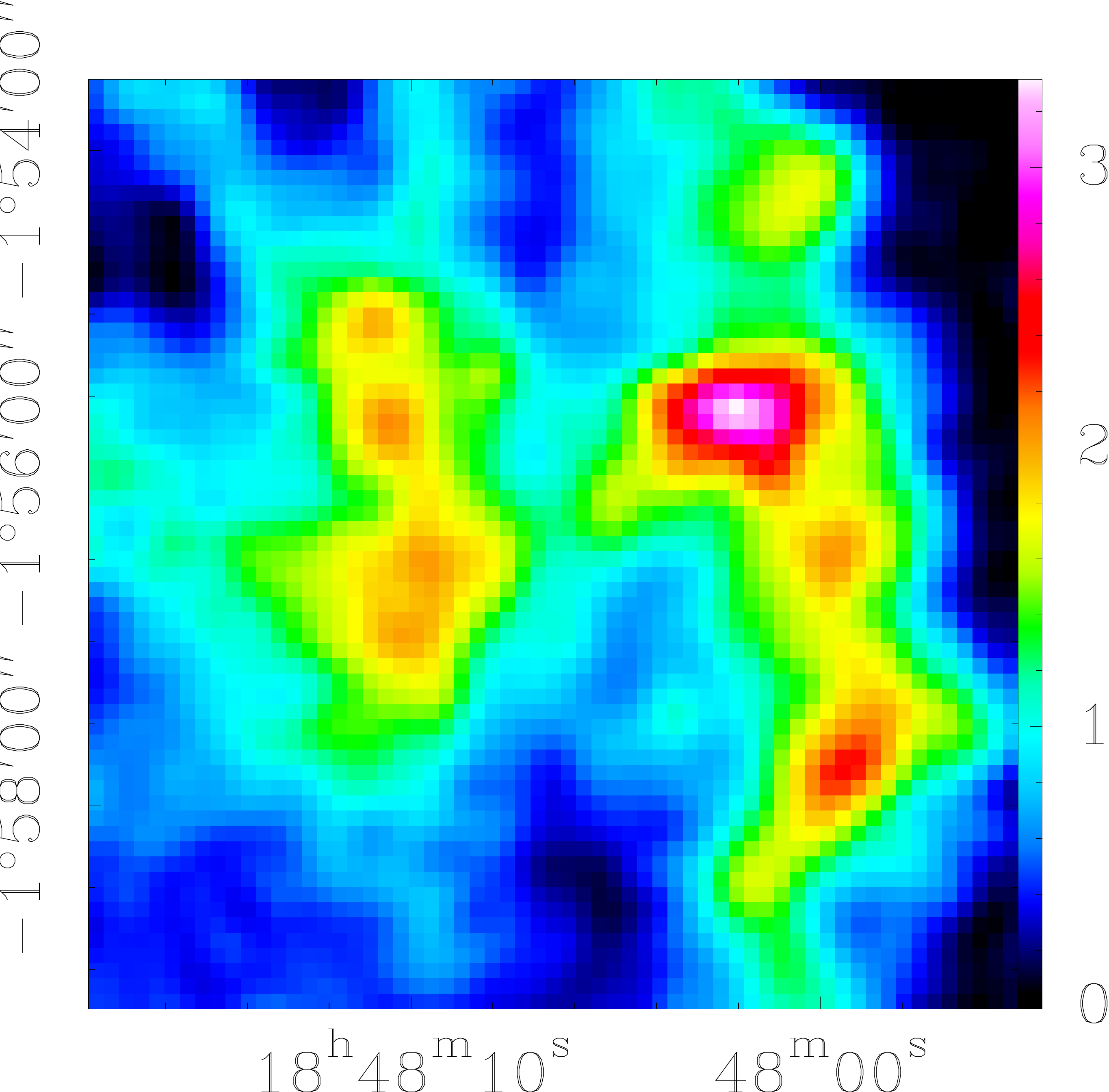}}}\hfill
 \subfloat[(h) Source 8]{\resizebox{!}{0.125\textheight}{\includegraphics[scale=1]{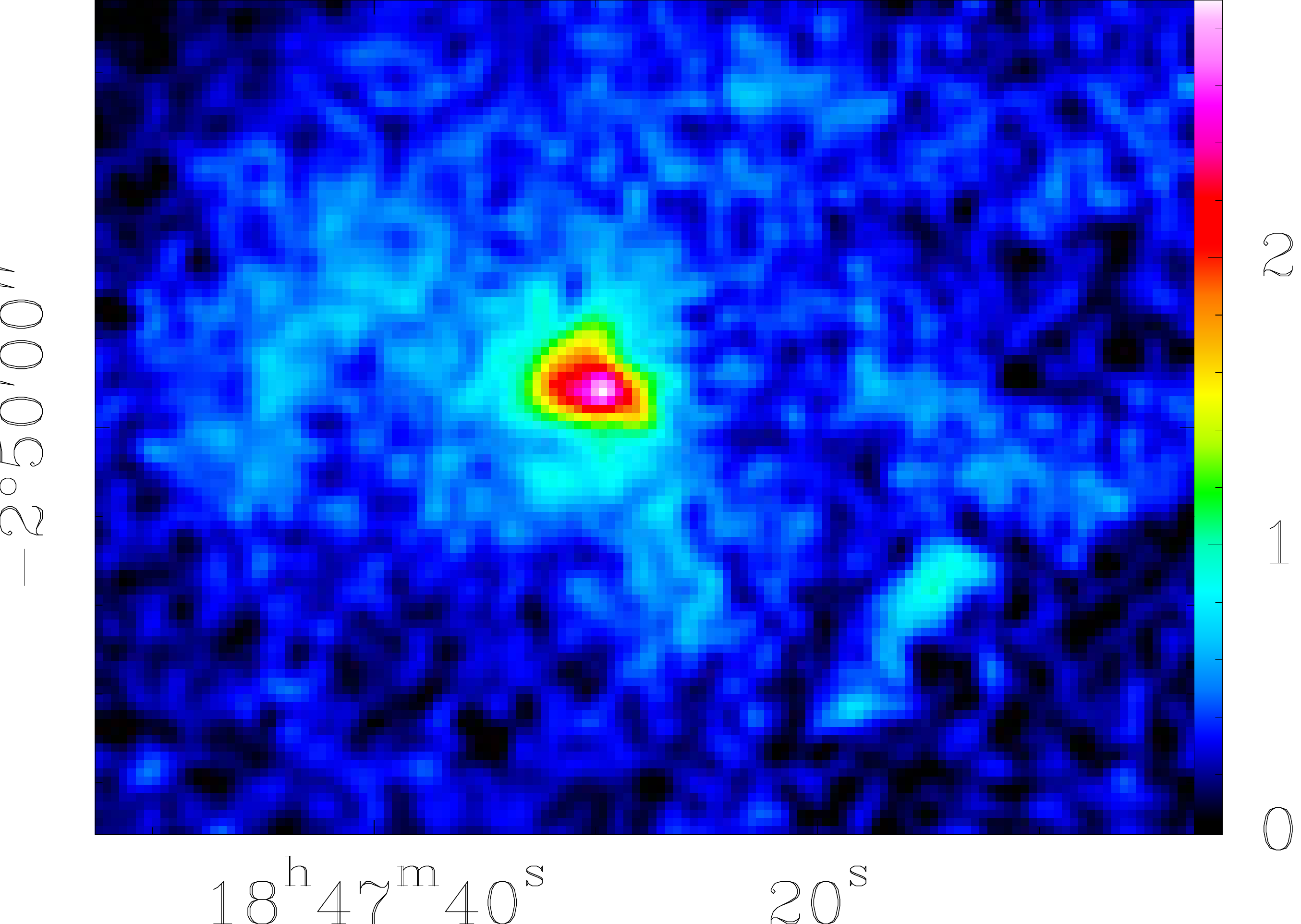}}}\hfill
 \subfloat[(i) Source 9]{\resizebox{!}{0.125\textheight}{\includegraphics[scale=1]{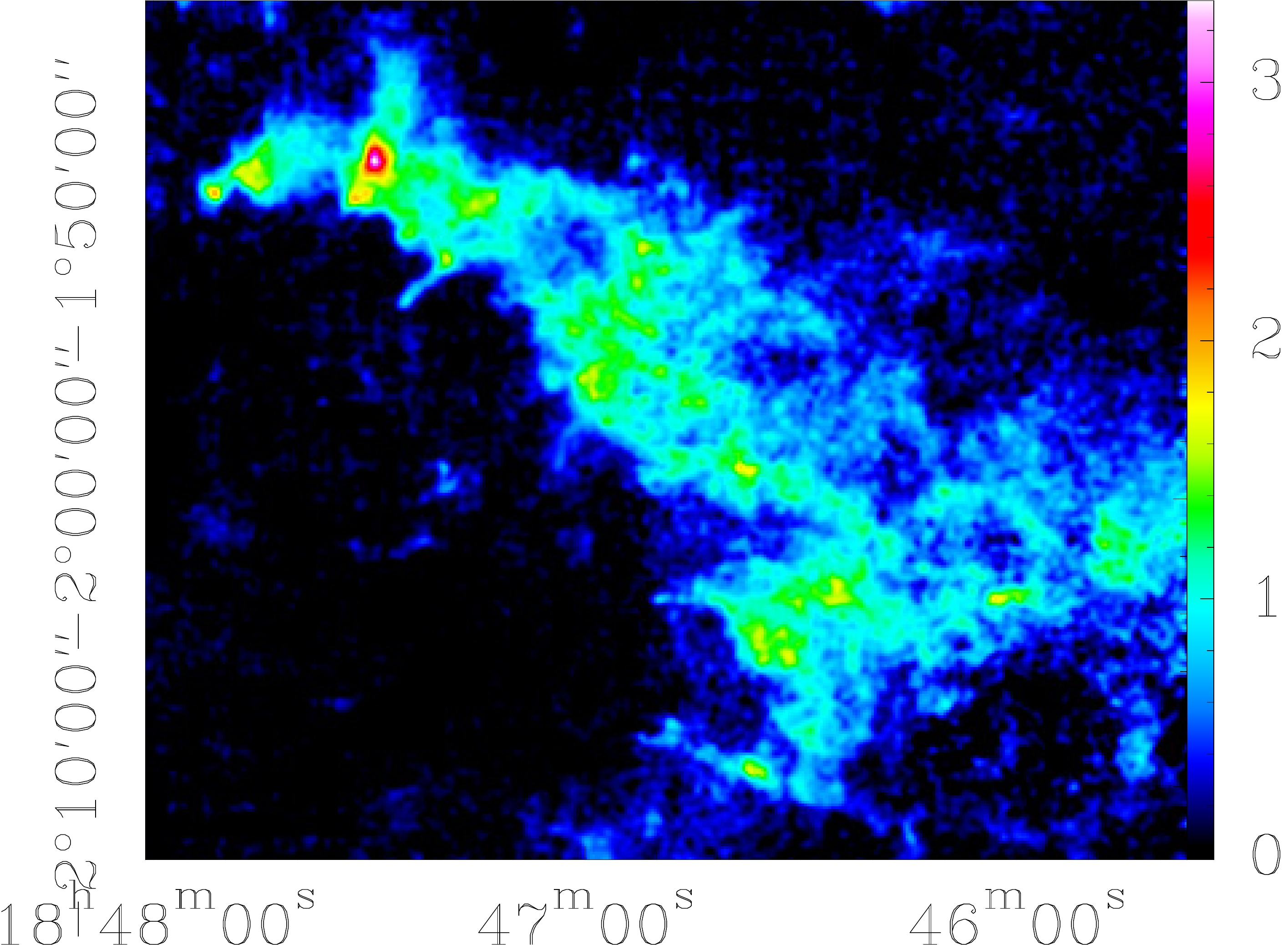}}}\\
 \subfloat[(j) Source 10]{\resizebox{!}{0.125\textheight}{\includegraphics[scale=1]{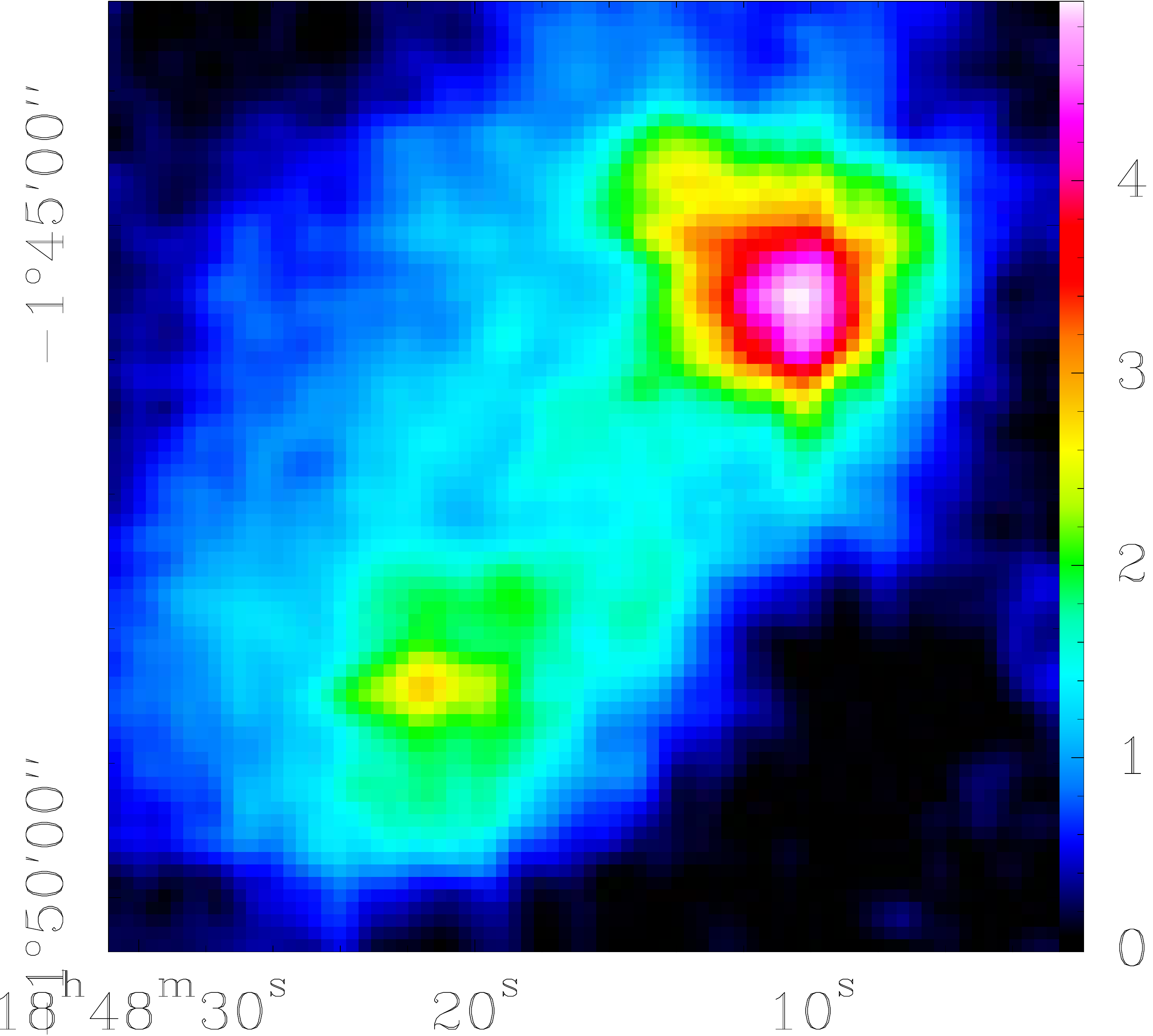}}}\hfill
 \subfloat[(k) Source 11]{\resizebox{!}{0.125\textheight}{\includegraphics[scale=1]{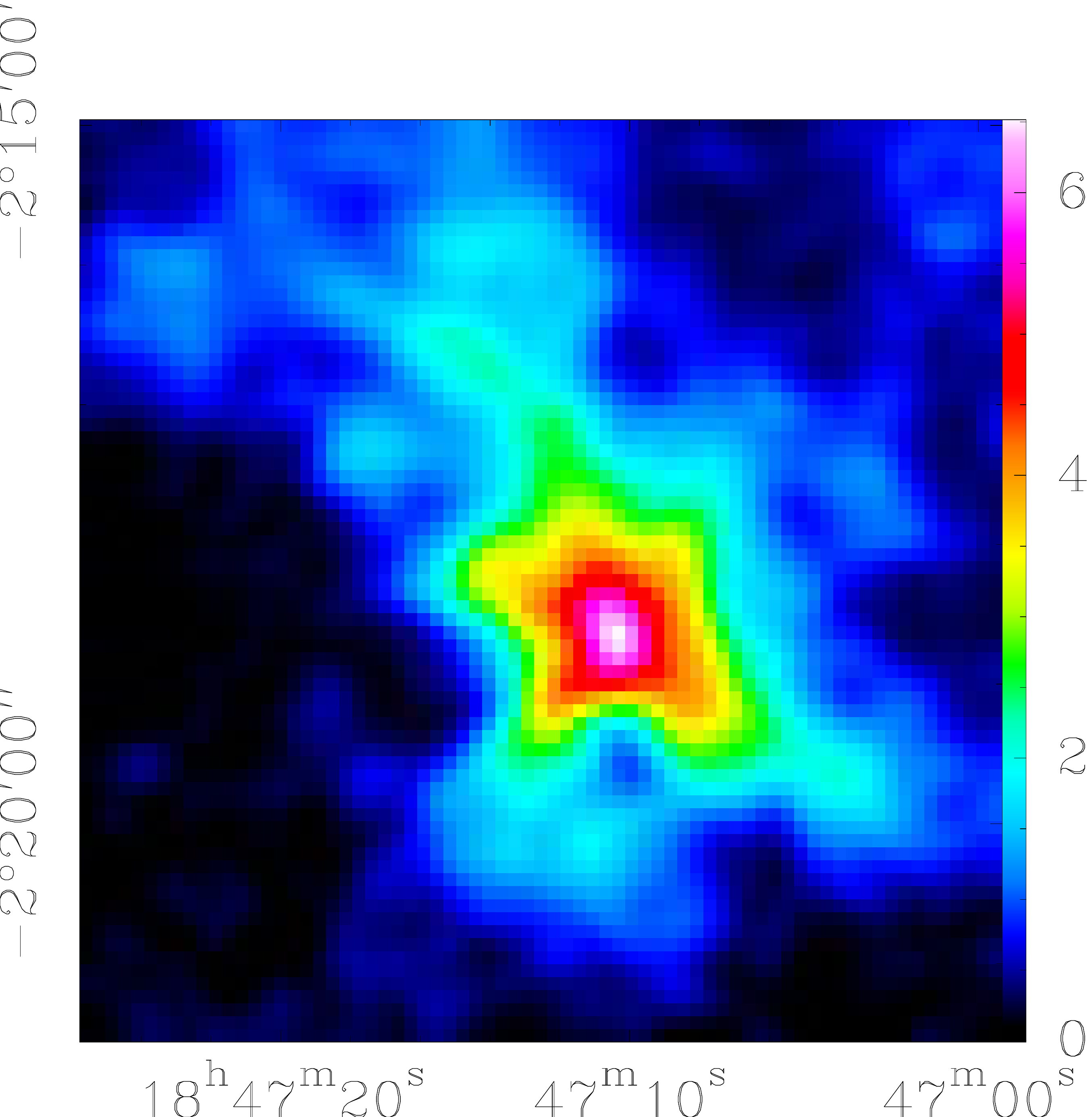}}}\hfill
 \subfloat[(l) Source 12]{\resizebox{!}{0.125\textheight}{\includegraphics[scale=1]{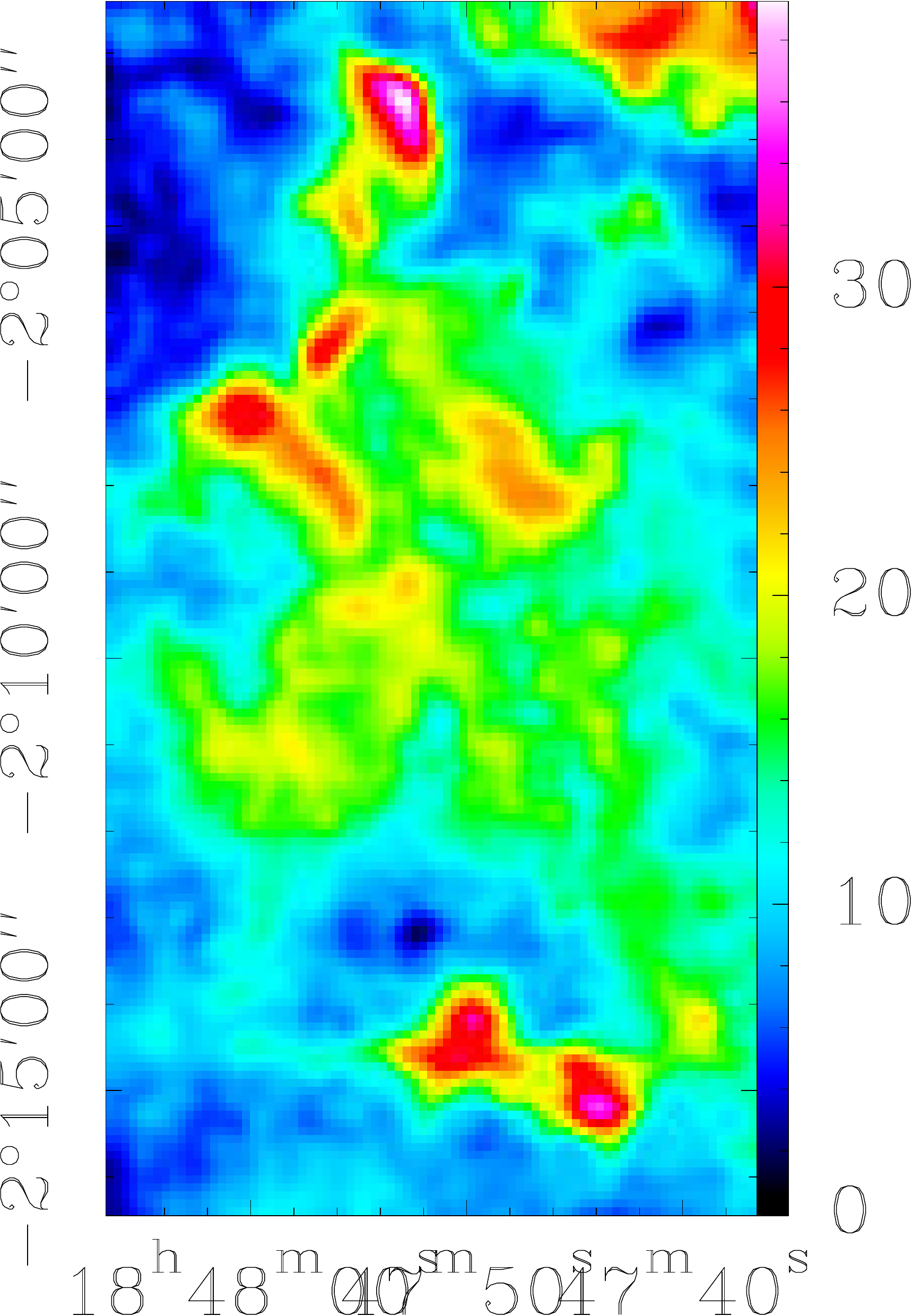}}}\hfill
 \subfloat[(m) Source 13]{\resizebox{!}{0.125\textheight}{\includegraphics[scale=1]{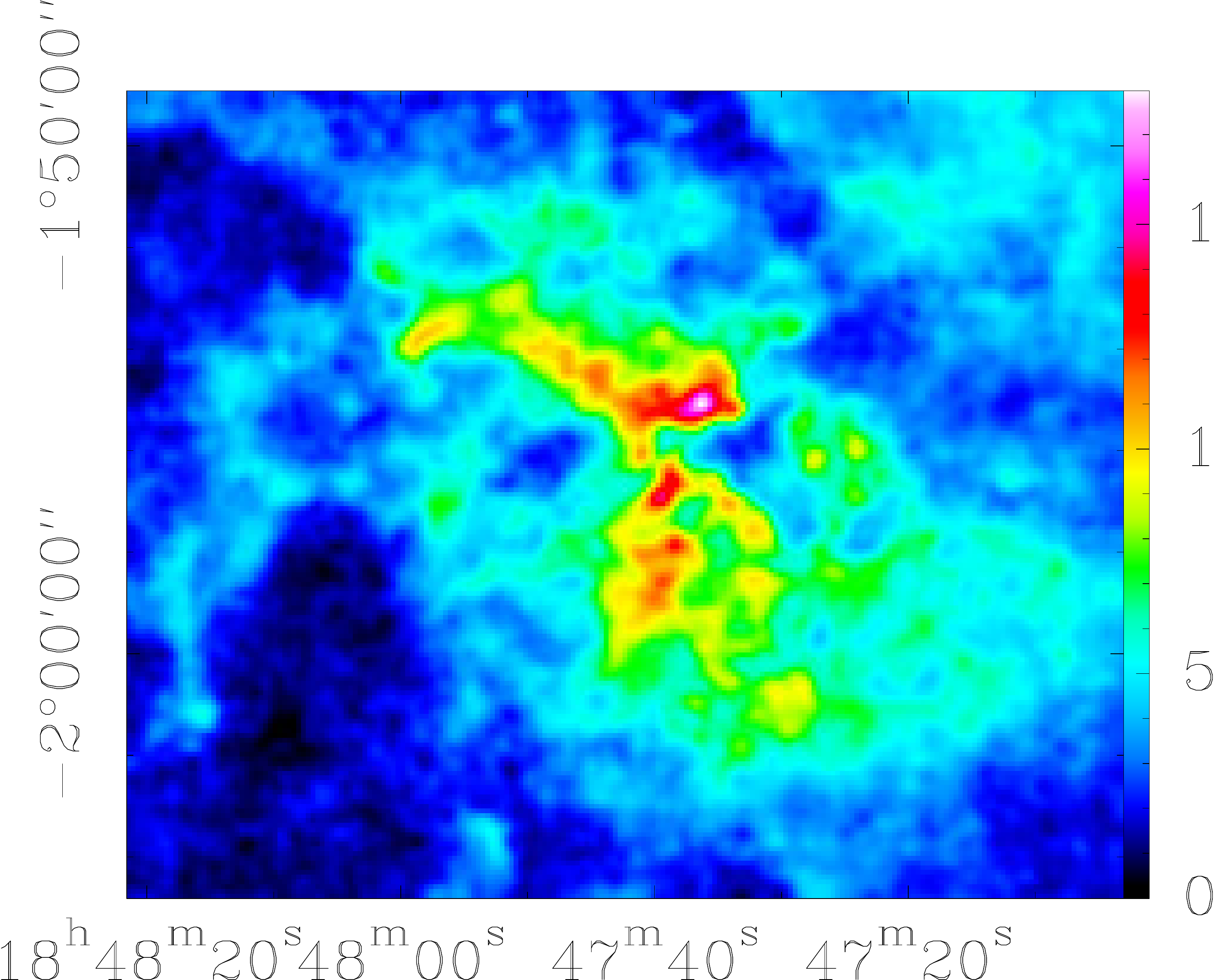}}}\hfill
 \subfloat[(n) Source 14]{\resizebox{!}{0.125\textheight}{\includegraphics[scale=1]{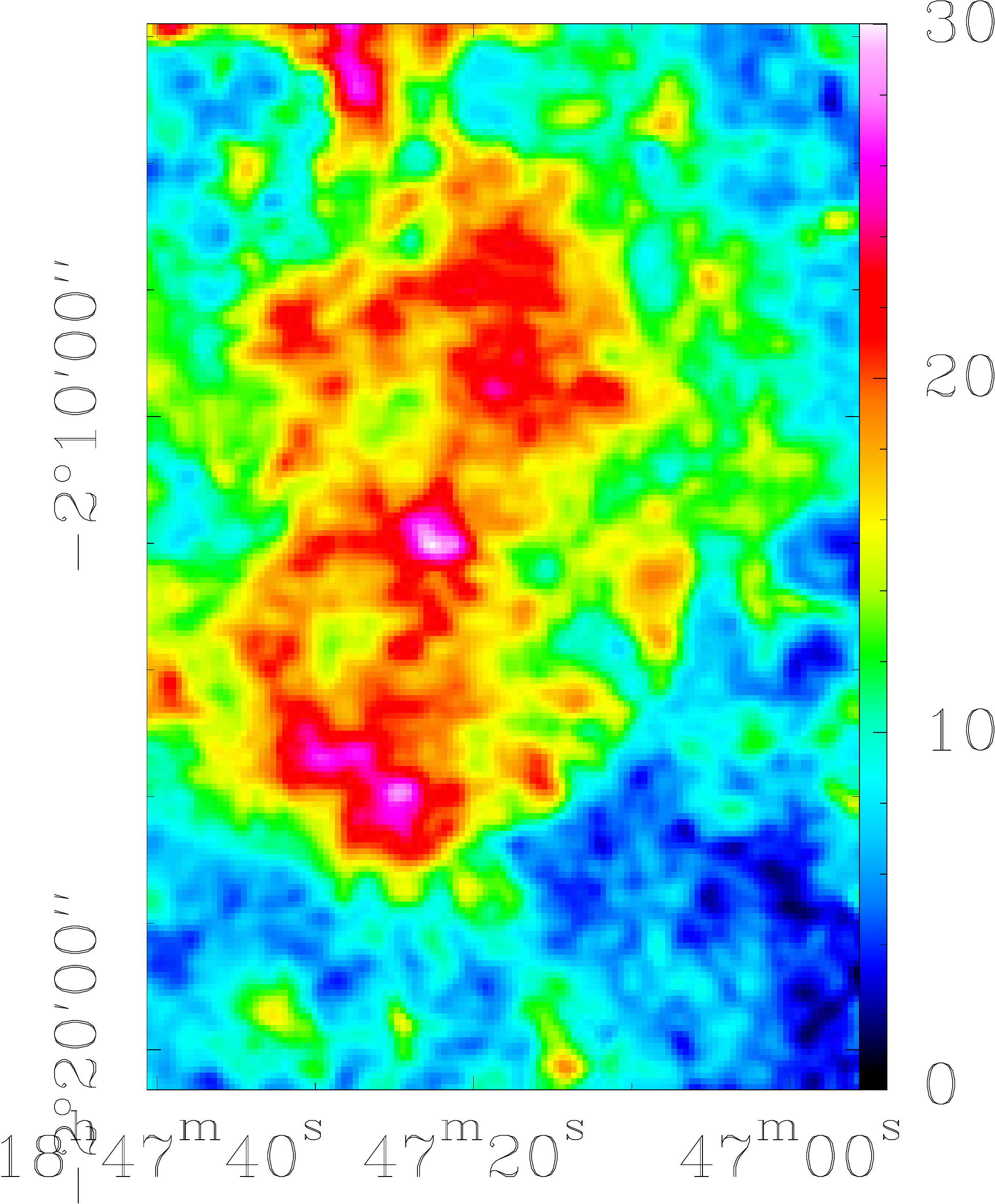}}}\\
 \subfloat[(o) Source 15]{\resizebox{!}{0.125\textheight}{\includegraphics[scale=1]{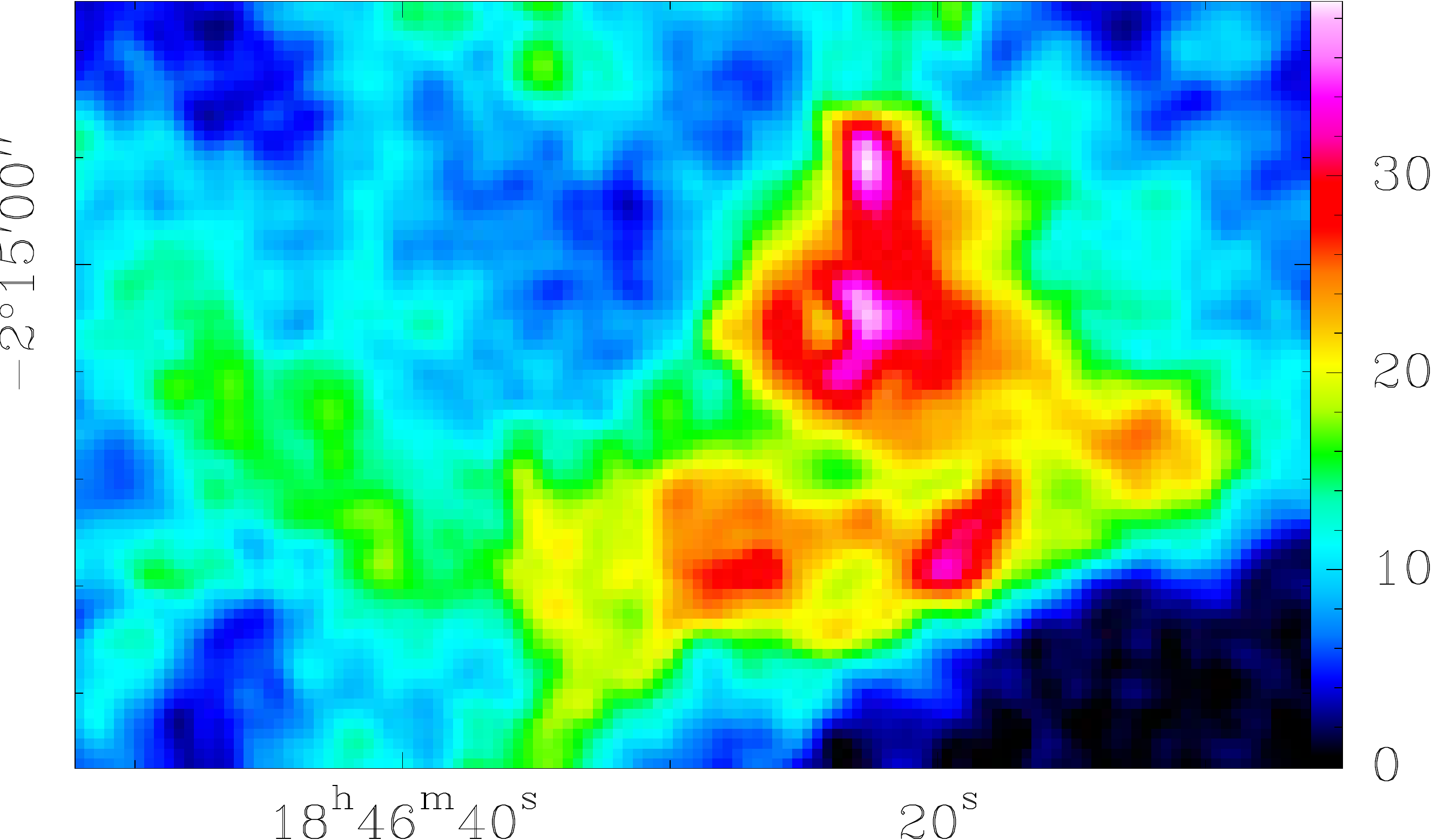}}}\hfill
 \subfloat[(p) Source 16]{\resizebox{!}{0.125\textheight}{\includegraphics[scale=1]{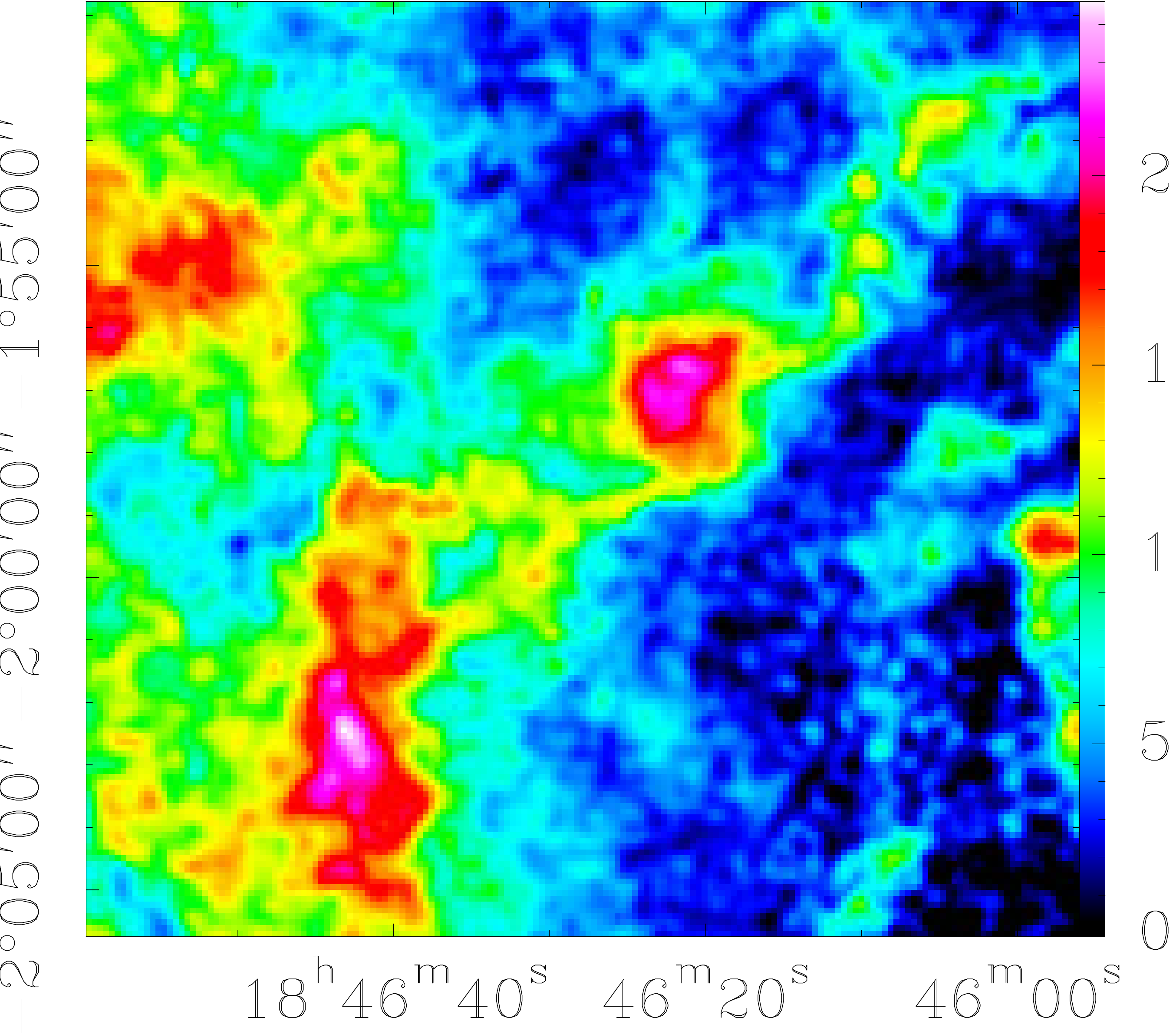}}}\hfill
 \subfloat[(q) Source 17]{\resizebox{!}{0.125\textheight}{\includegraphics[scale=1]{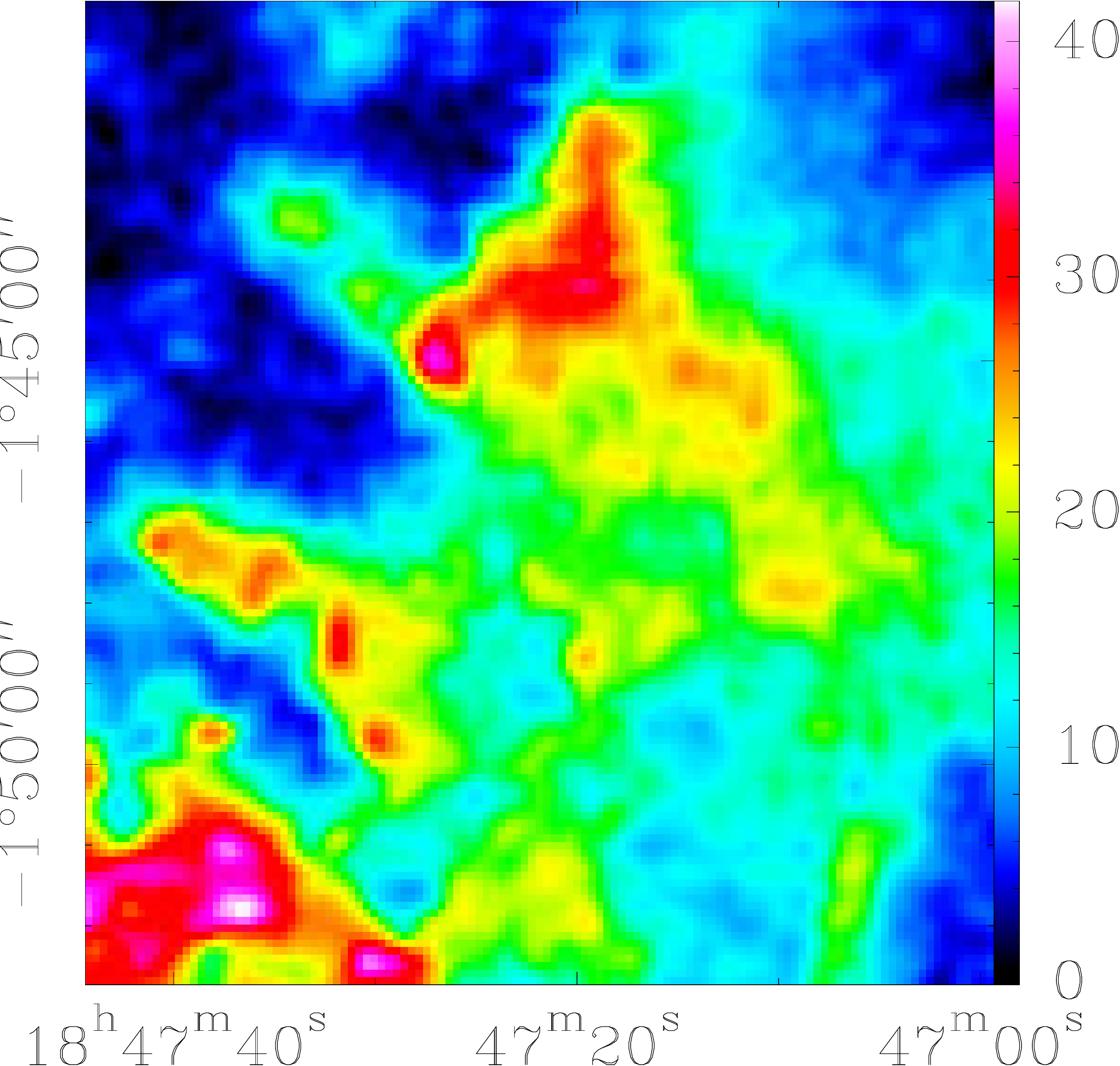}}}\hfill
 \subfloat[(r) Source 18]{\resizebox{!}{0.125\textheight}{\includegraphics[scale=1]{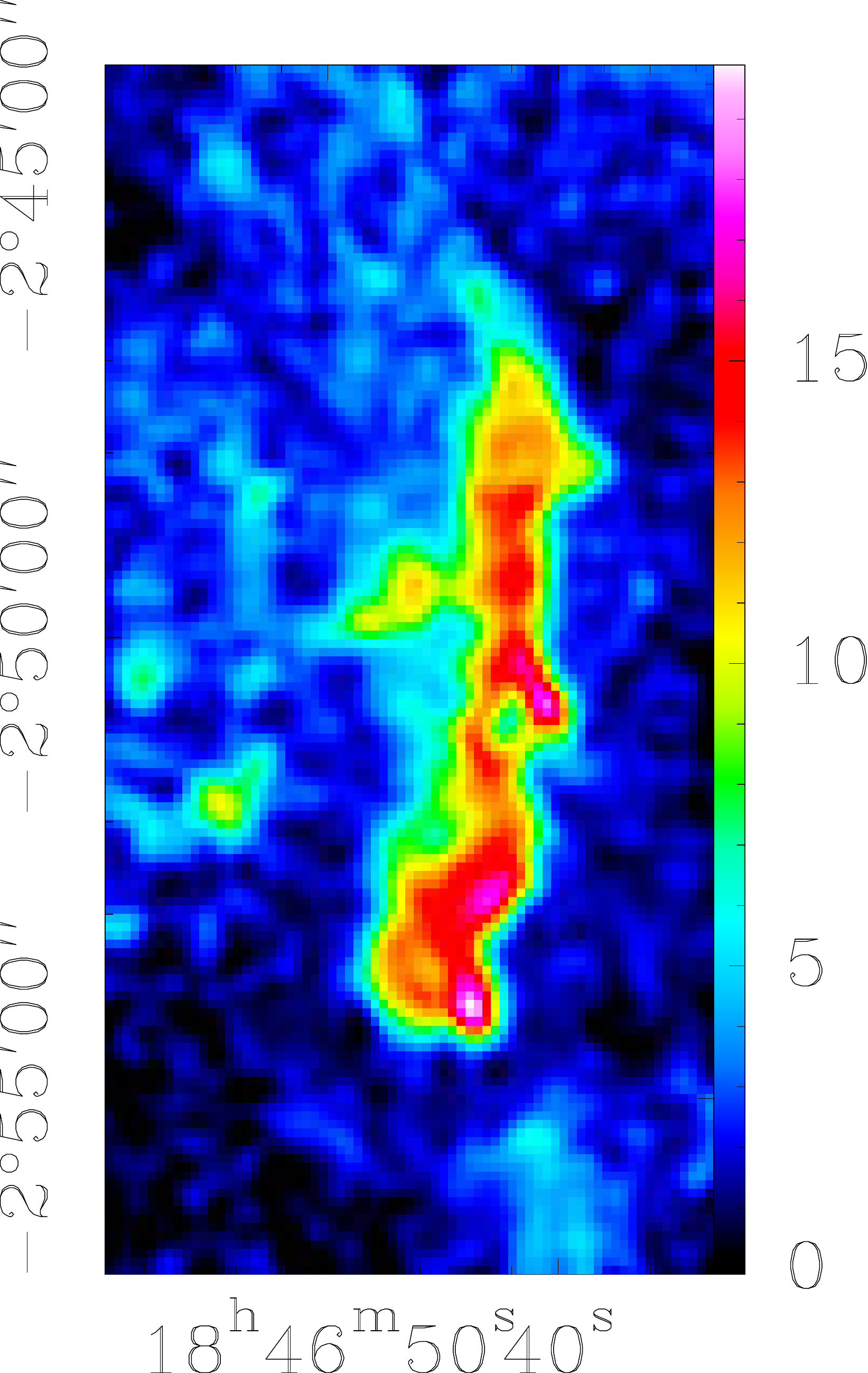}}}\hfill
 \subfloat[(s) Source 19]{\resizebox{!}{0.125\textheight}{\includegraphics[scale=1]{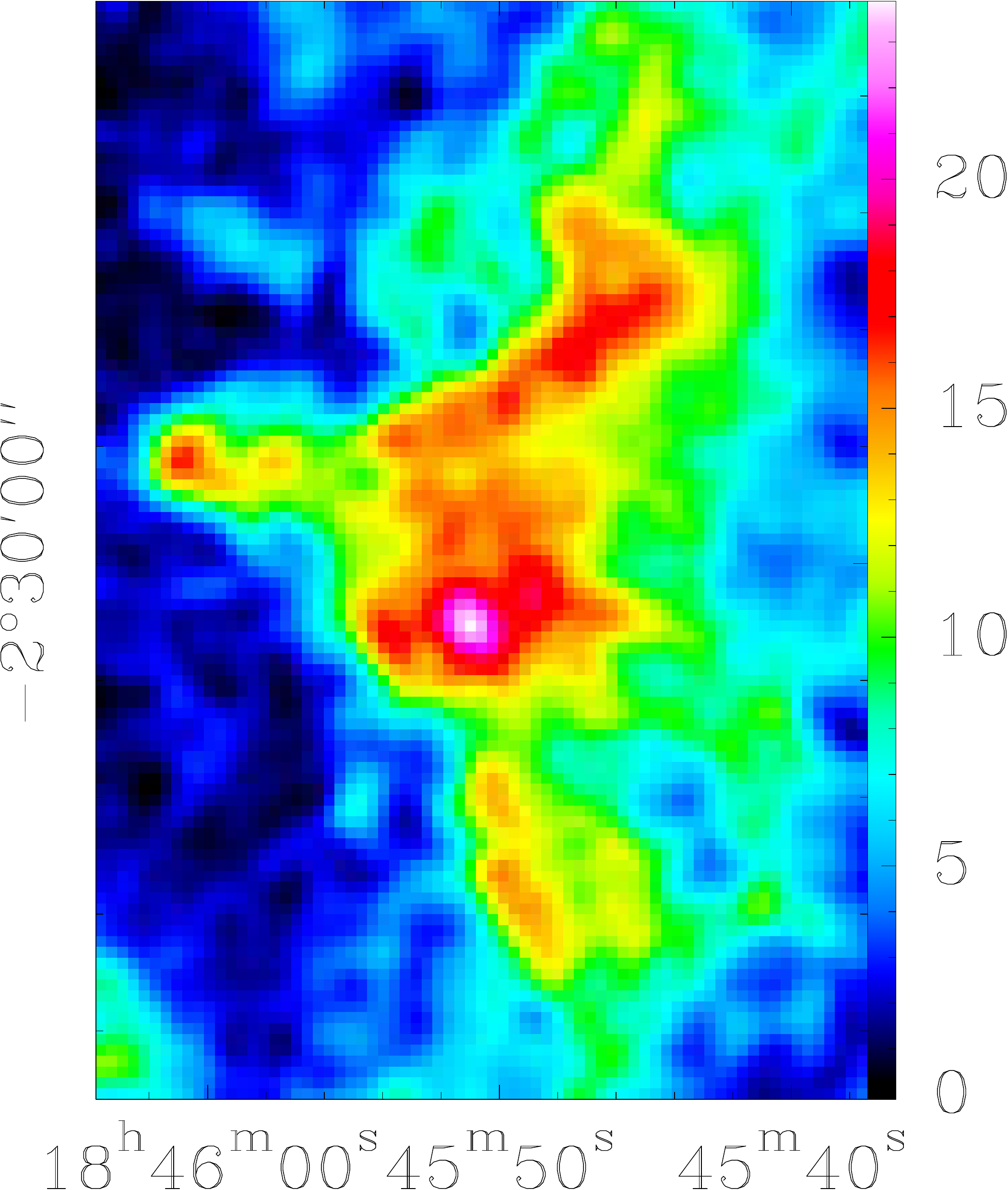}}}\\
 \subfloat[(t) Source 20]{\resizebox{!}{0.125\textheight}{\includegraphics[scale=1]{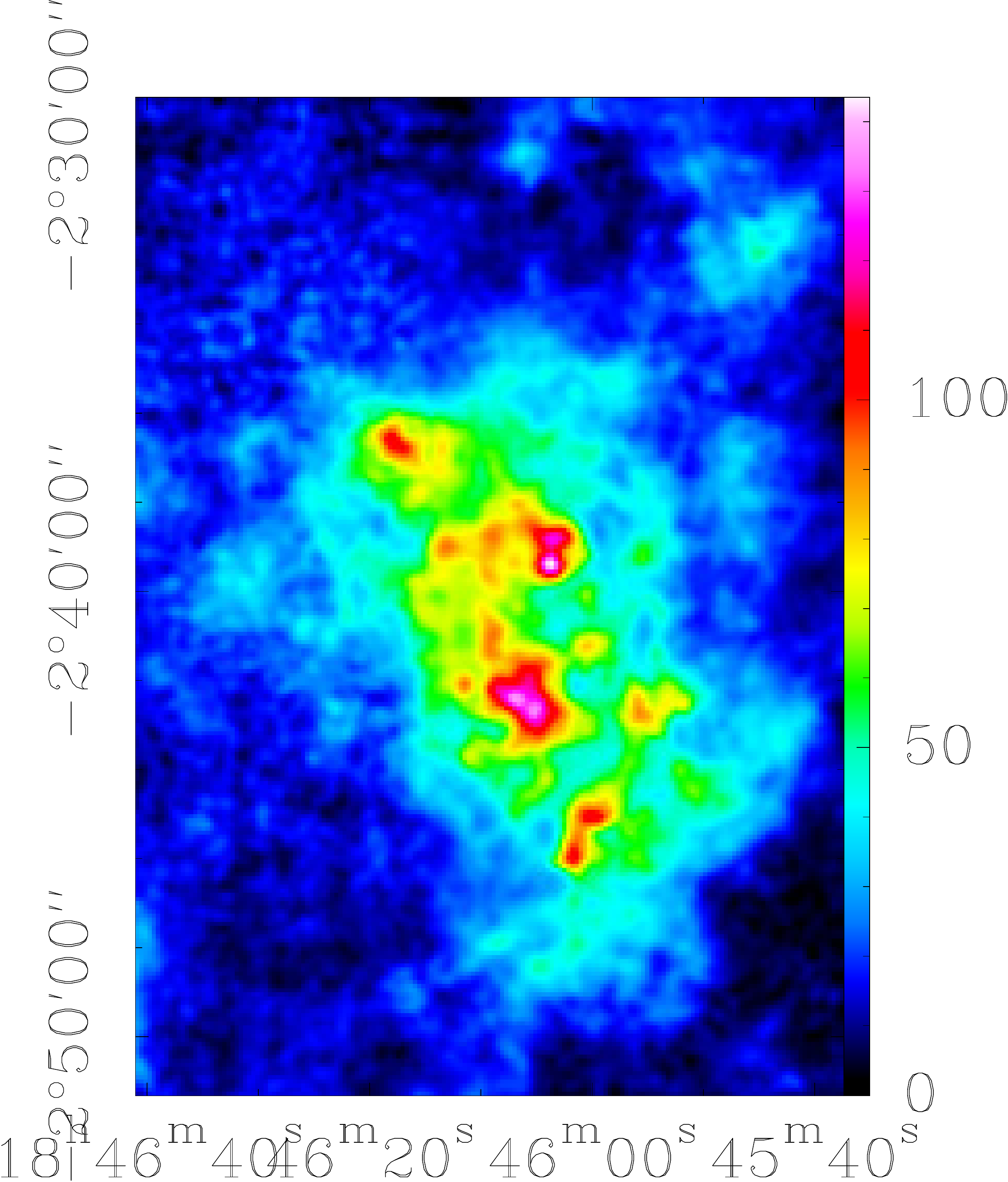}}}\hfill
 \subfloat[(u) Source 21]{\resizebox{!}{0.125\textheight}{\includegraphics[scale=1]{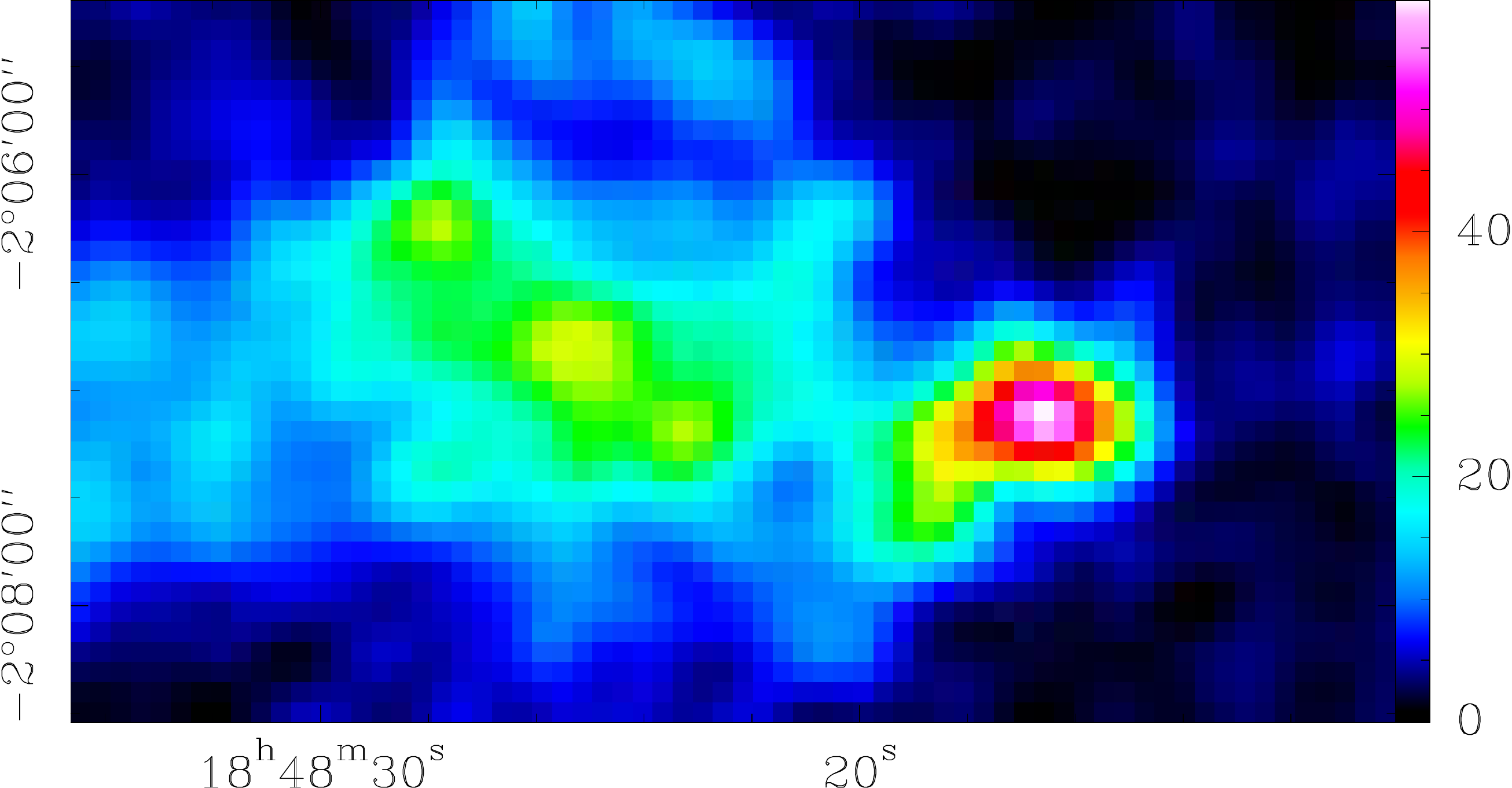}}}\hfill
 \subfloat[(v) Source 22]{\resizebox{!}{0.125\textheight}{\includegraphics[scale=1]{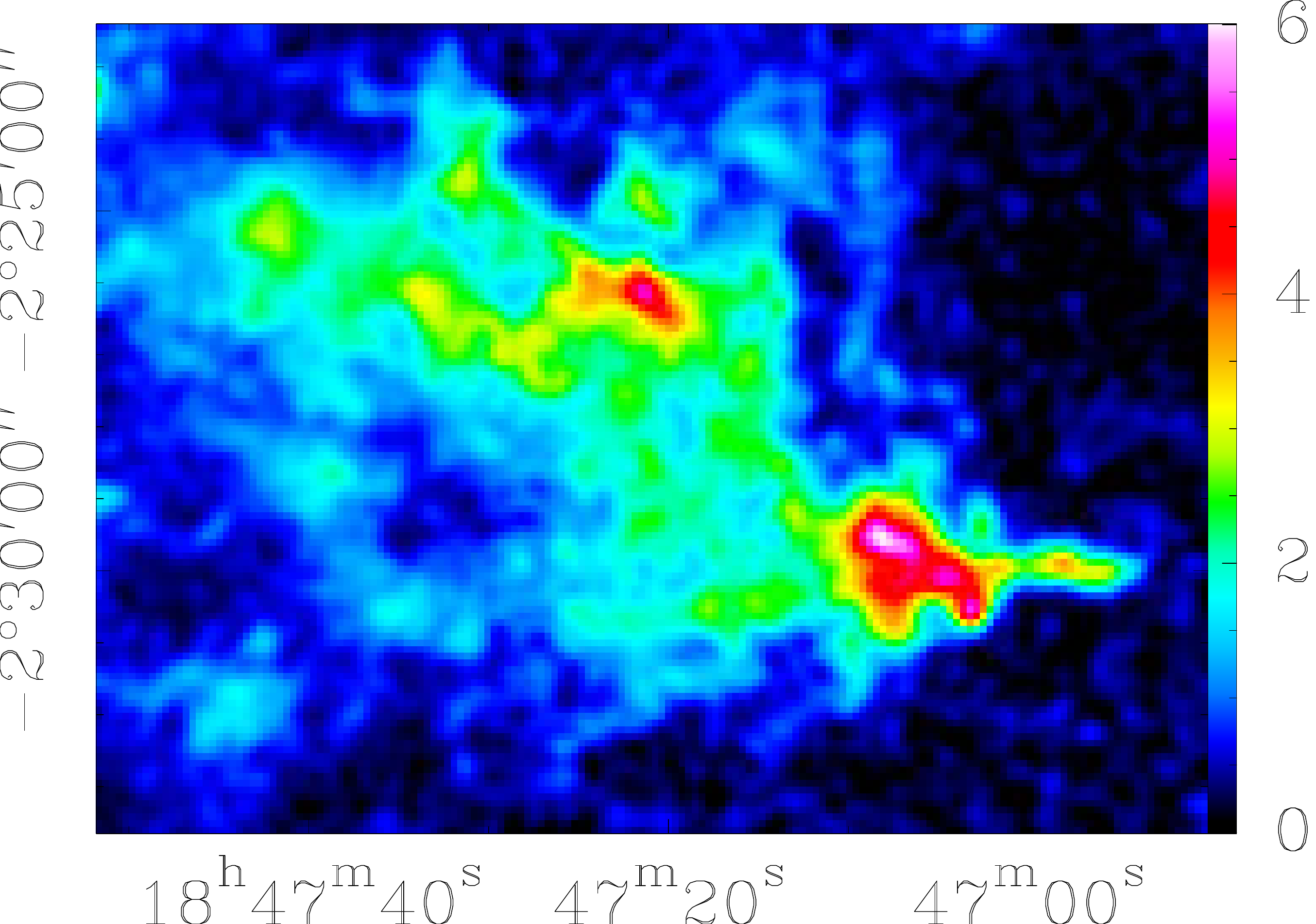}}}\hfill
 \subfloat[(w) Source 23]{\resizebox{!}{0.125\textheight}{\includegraphics[scale=1]{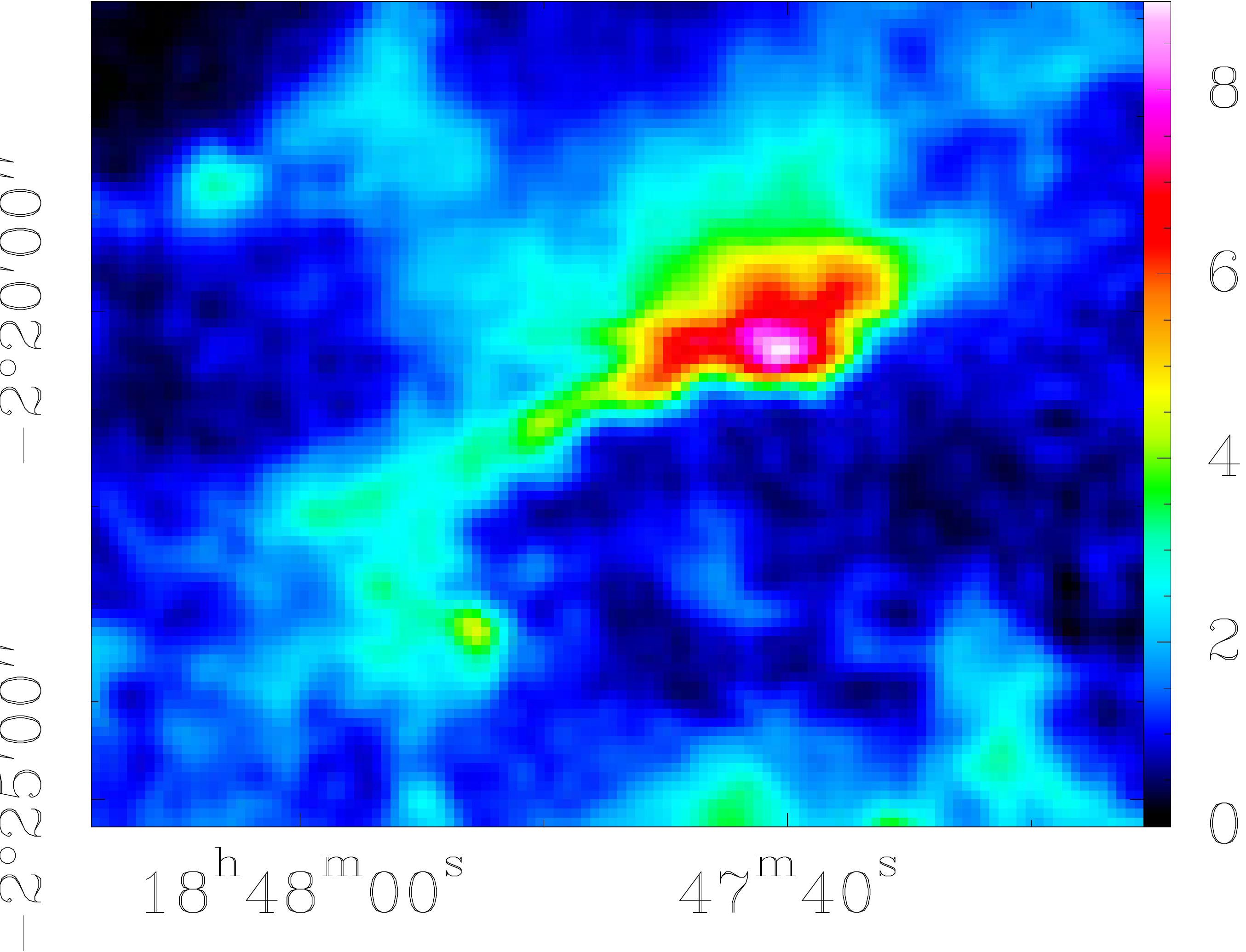}}}\\
 \subfloat[(x) Source 24]{\resizebox{!}{0.125\textheight}{\includegraphics[scale=1]{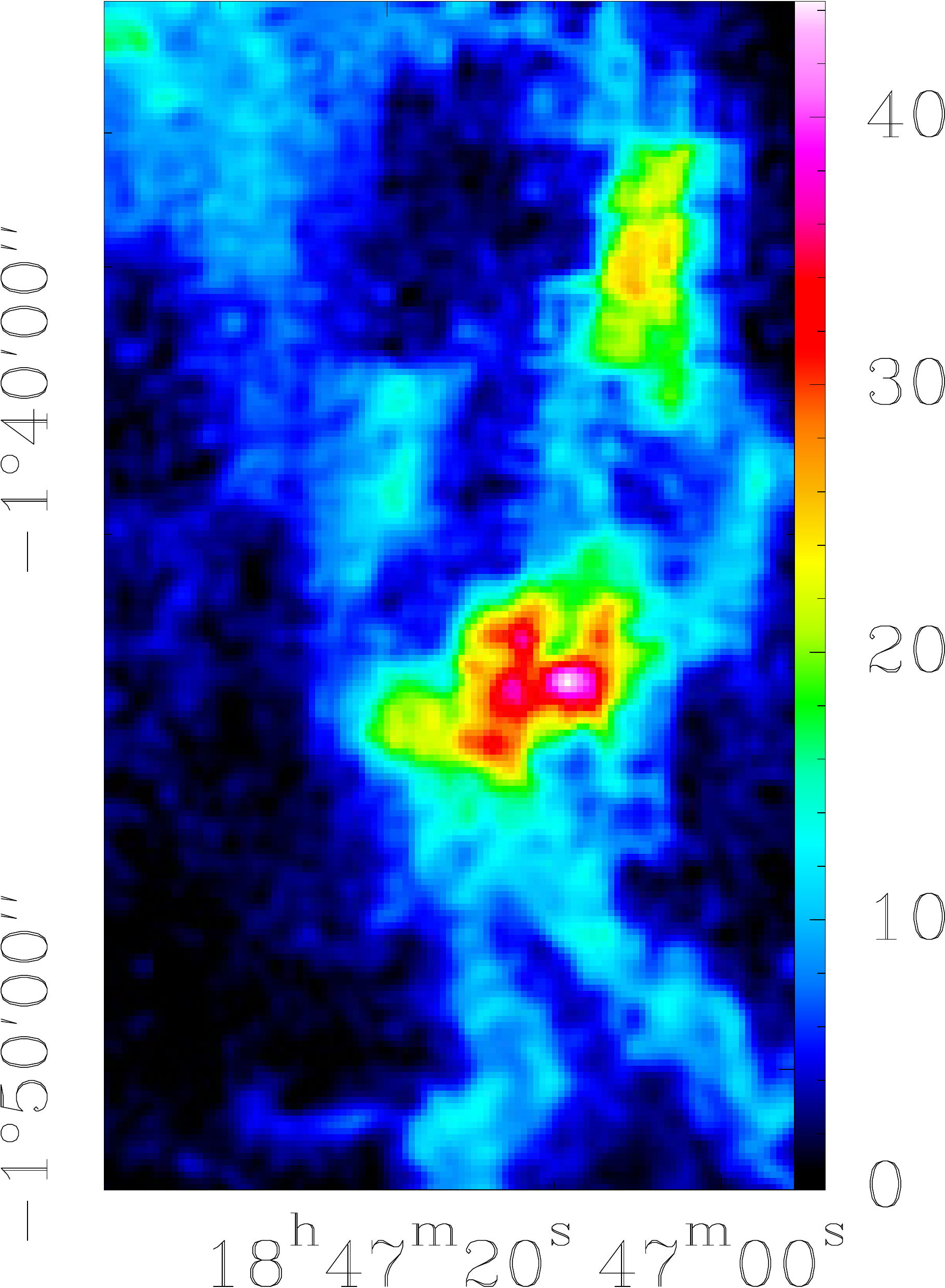}}}\hfill
 \subfloat[(y) Source 25]{\resizebox{!}{0.125\textheight}{\includegraphics[scale=1]{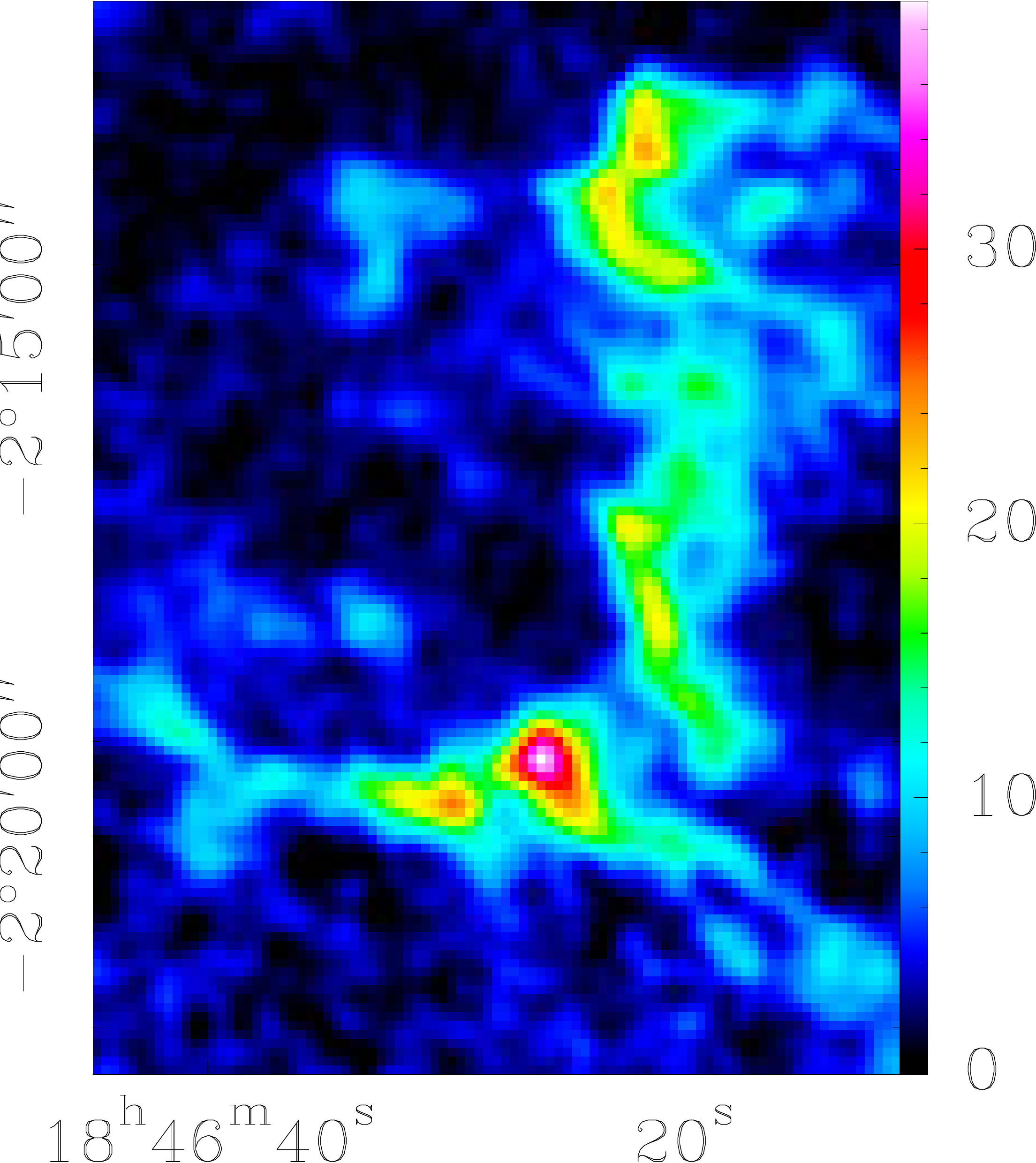}}}\hfill
 \subfloat[(z) Source 26]{\resizebox{!}{0.125\textheight}{\includegraphics[scale=1]{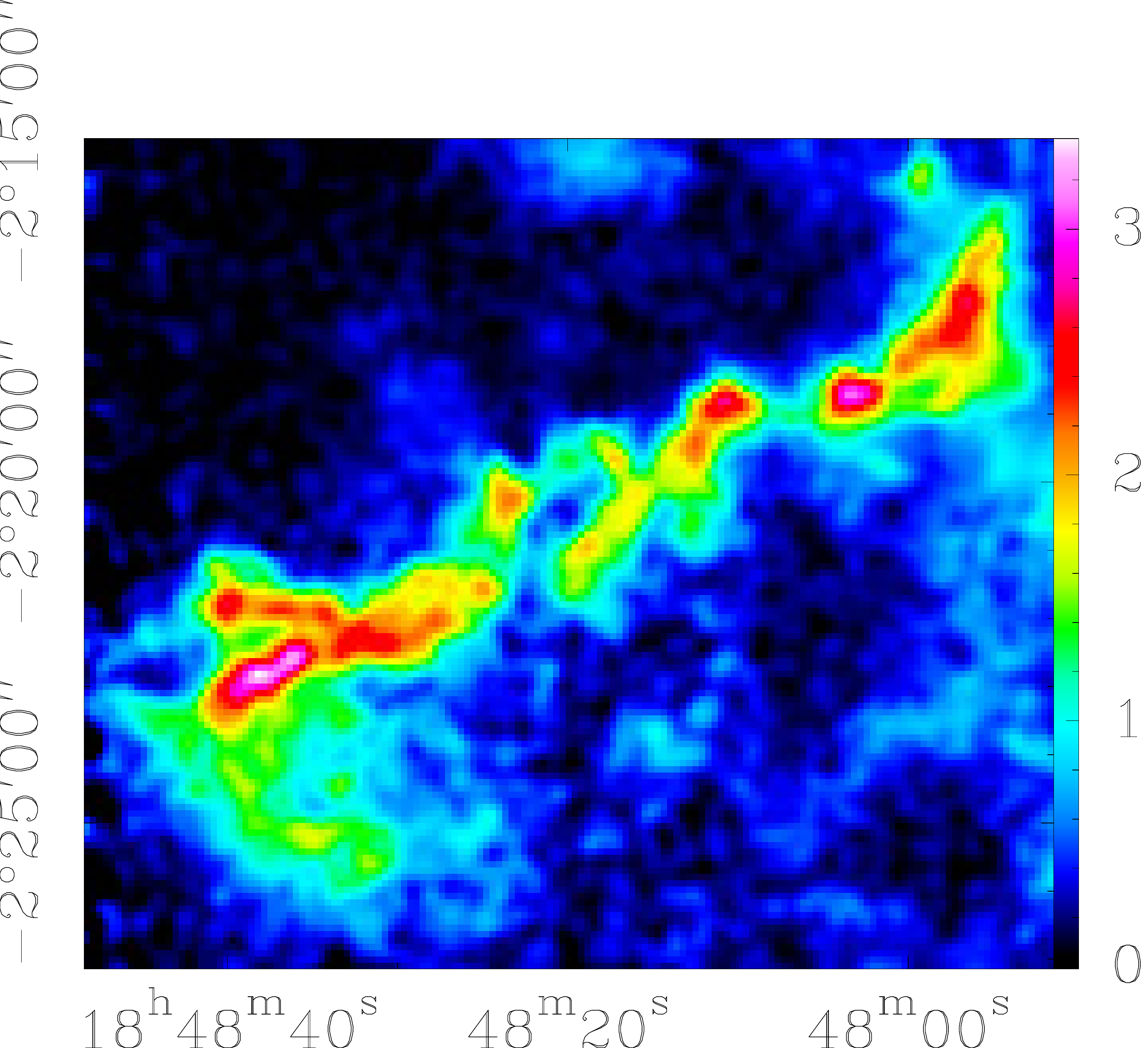}}}\hfill
 \subfloat[(aa) Source 27]{\resizebox{!}{0.125\textheight}{\includegraphics[scale=1]{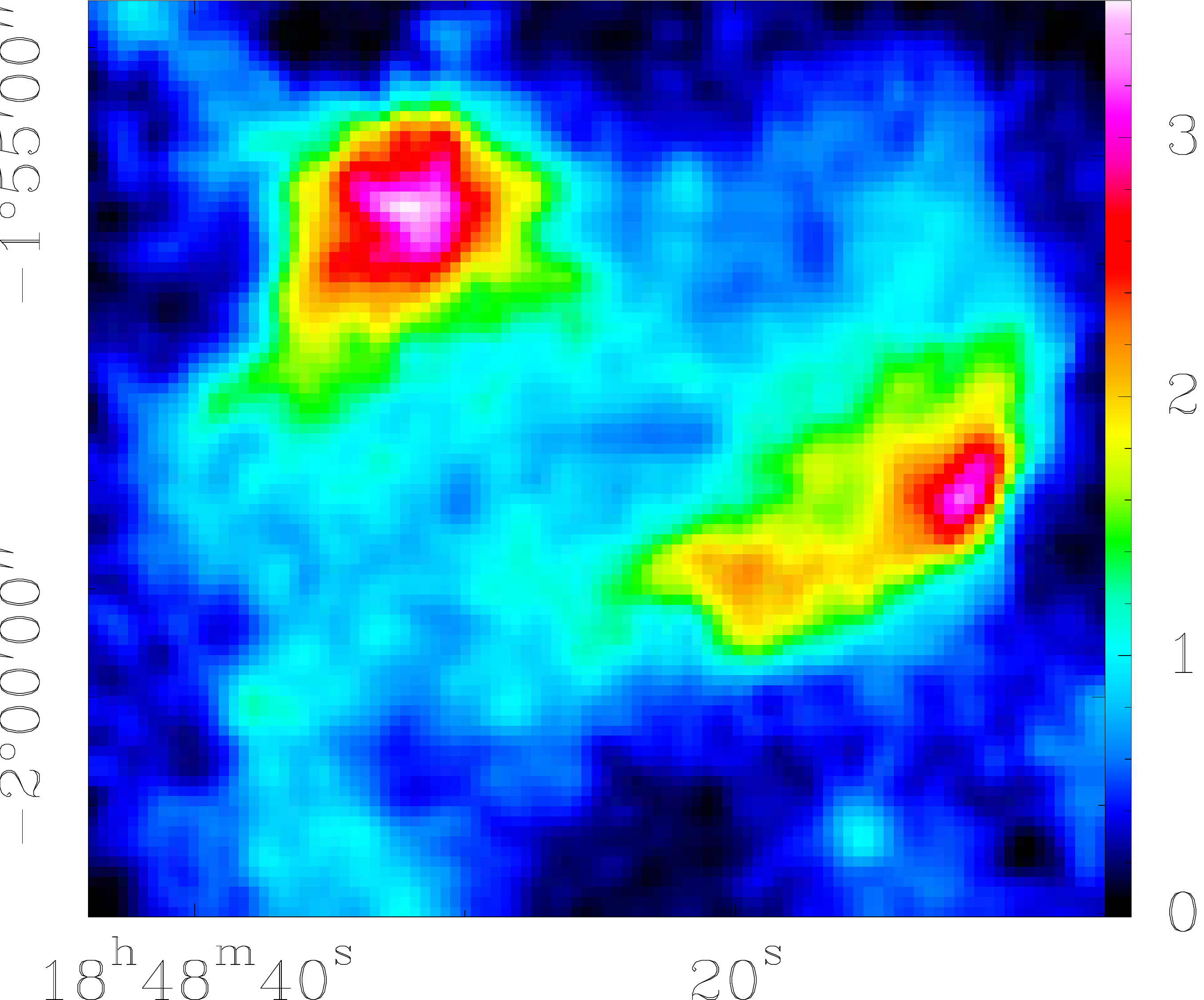}}}\hfill
 \subfloat[(ab) Source 28]{\resizebox{!}{0.125\textheight}{\includegraphics[scale=1]{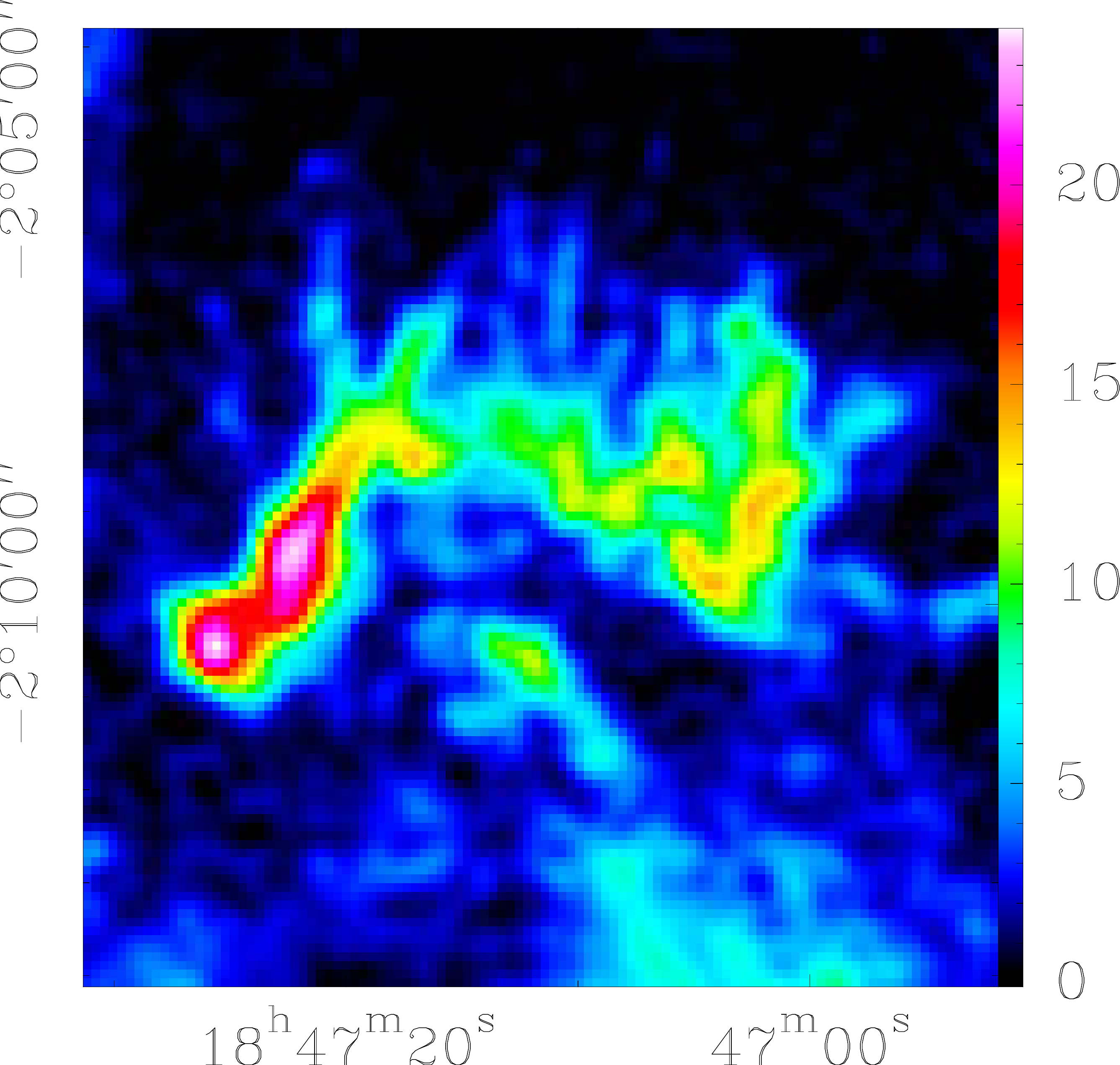}}}\hfill
 \subfloat[(ac) Source 29]{\resizebox{!}{0.125\textheight}{\includegraphics[scale=1]{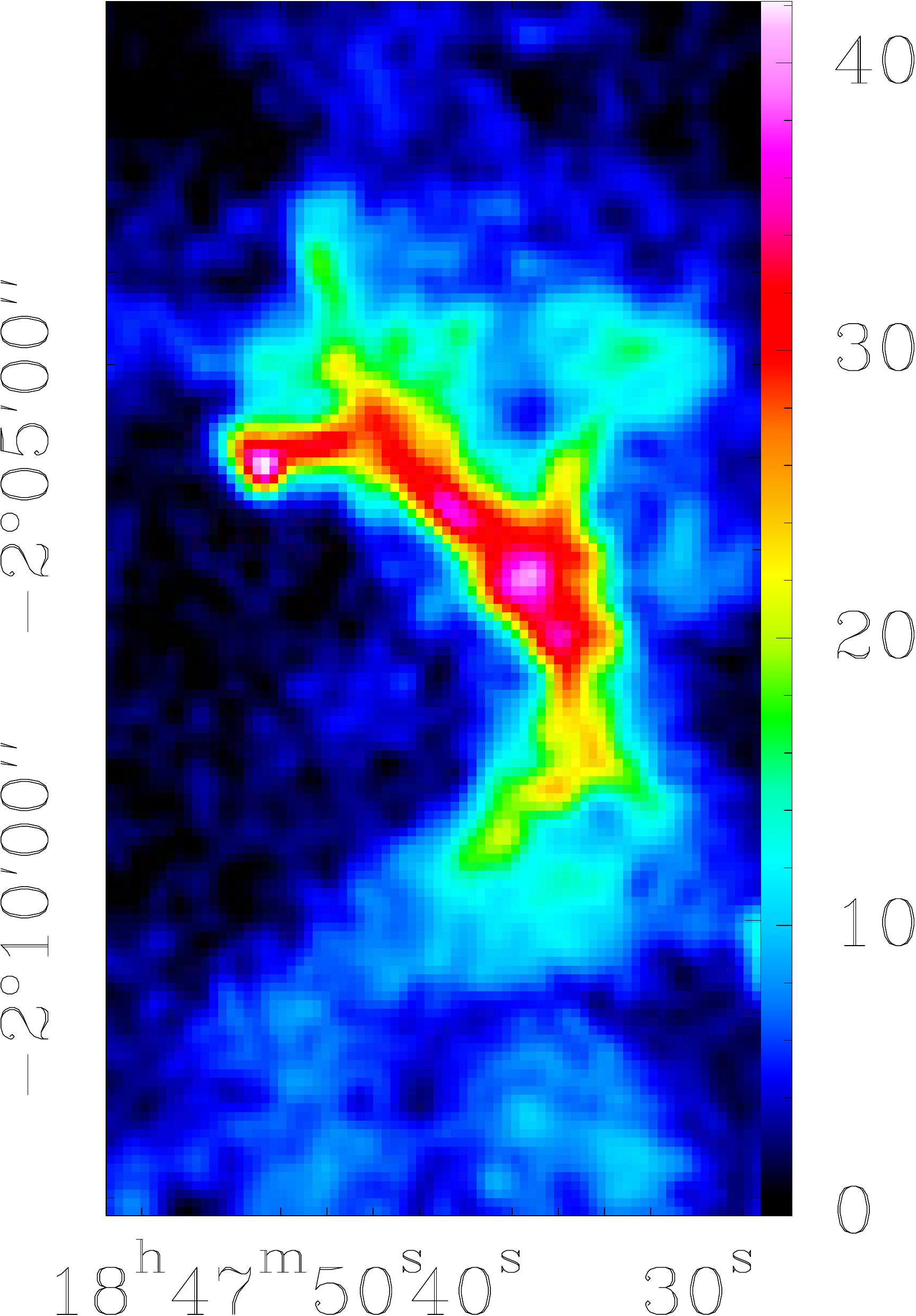}}}
 \end{minipage}
\caption{Integrated $^{13}$CO~(2--1) maps in K\,km\,s$^{-1}$ of all identified sources. The plots are not proportional to their physical size, as source sizes differ strongly.}
\label{fig:appendixsourcemaps}
\end{figure*}

This source is found in the very center of our region maps, directly south of the W43-Main cloud. Its shape resembles a crescent moon, opened toward the southeast, where it is sharply bound. The
outside is more diffuse and shows several outflows away from the center. See Fig.~\ref{fig:appendixsourcemaps}~(ac) for a plot of the $^{13}$CO emission.
The extent of the source is 12~pc from northeast to southwest, and the filament has a typical width of 2 to 3~pc.
Two stronger clumps with an integrated $^{13}$CO intensity of 40~K\,km\,s$^{-1}$ are seen in the center and the northeastern tip.
The strong backbone of this source has still an integrated intensity of $\sim20$~K\,km\,s$^{-1}$, where line peak intensity goes up to 18~K.

The central western part of the source moves with a relative radial velocity of 112~km\,s$^{-1}$ and increases to 118~km\,s$^{-1}$ toward both ends of the crescent. 
The lines show widths of 2 to 3~km\,s$^{-1}$ in the central region, decreasing to 1~km\,s$^{-1}$ in the outer parts.
H$_2$ column densities only reach 
a few $10^{22}$~cm$^{-2}$ across the inner parts of the structure. We find a total mass of $1.2\times10^4$~$M_{\sun}$.

In the dust emission maps of ATLASGAL and Hi-GAL, only the strong backbone of this source can be seen. The weaker outliers are not traced. The Spitzer 8~$\mu$m map shows several bright compact
sources in the center, and some extended emission in the south is most probably related to this source. However, the northern tip that shows strong CO emission is not traced by Spitzer at all.

\section{List of found sources}\label{app:sourcesplots}

\begin{figure*}[htb]
\centering
\begin{minipage}{18cm}
 \subfloat[(a) Source 1]{\resizebox{!}{0.105\textheight}{\includegraphics[scale=1]{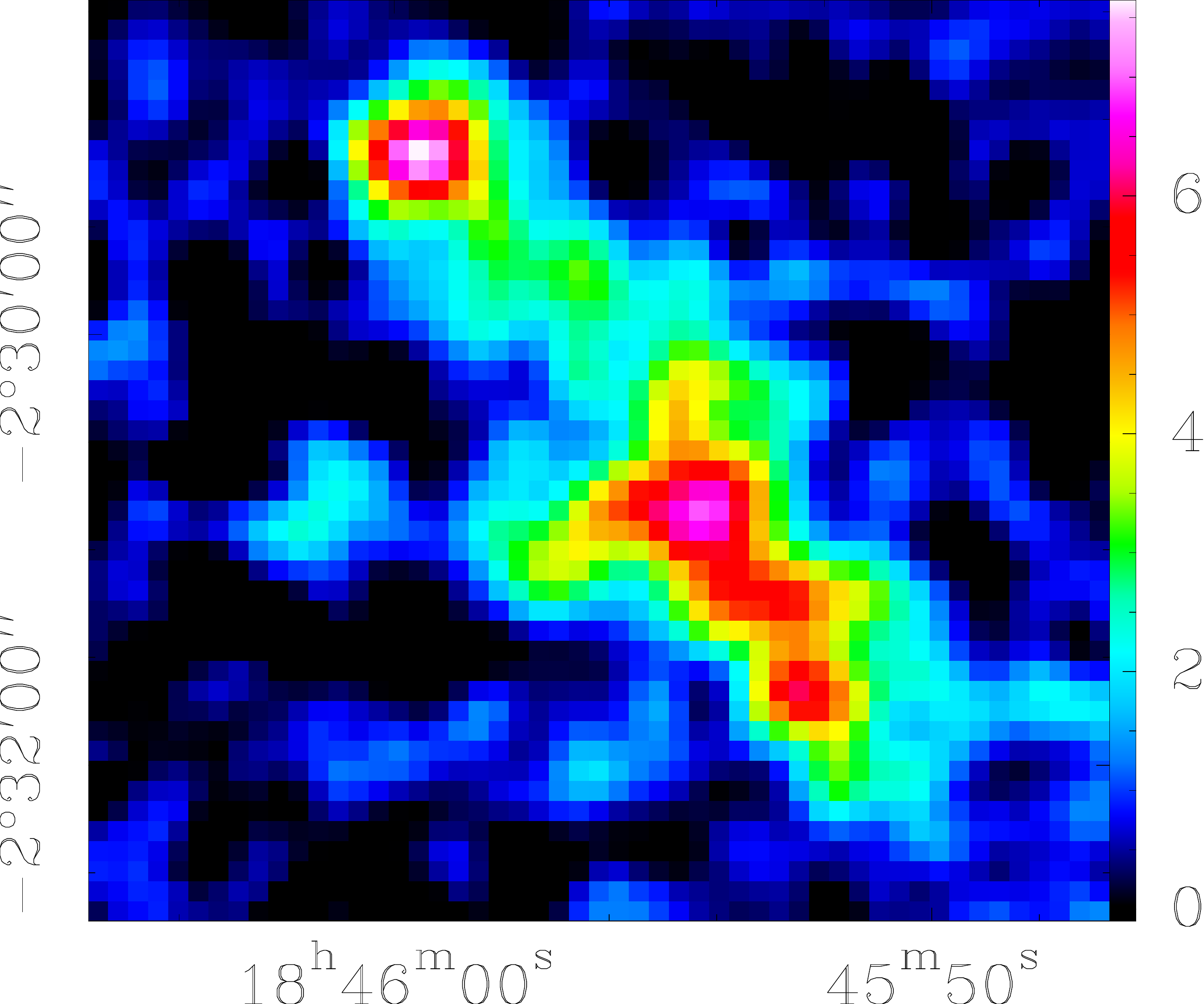}}}\hfill
 \subfloat[(b) Source 2]{\resizebox{!}{0.105\textheight}{\includegraphics[scale=1]{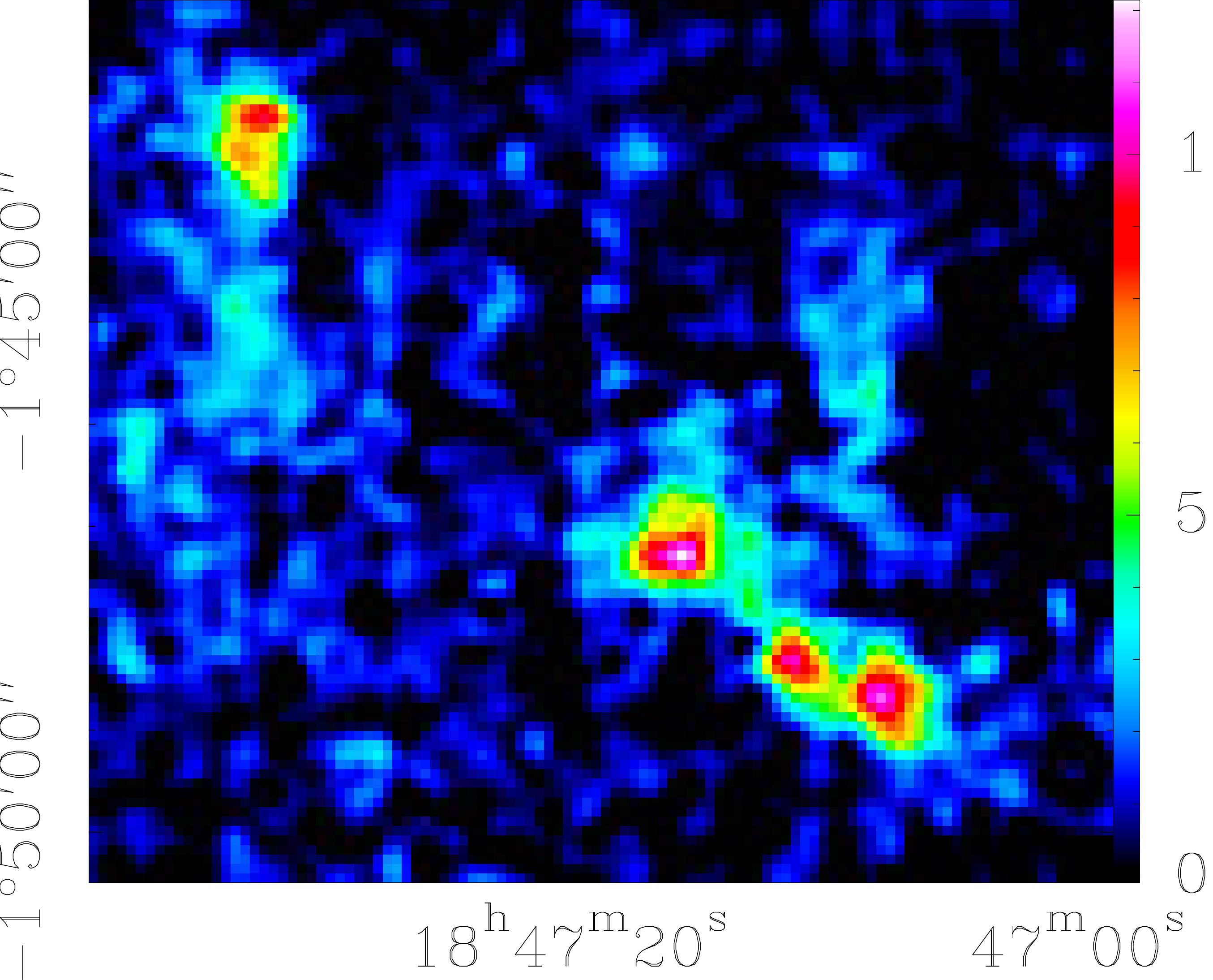}}}\hfill
 \subfloat[(c) Source 3]{\resizebox{!}{0.105\textheight}{\includegraphics[scale=1]{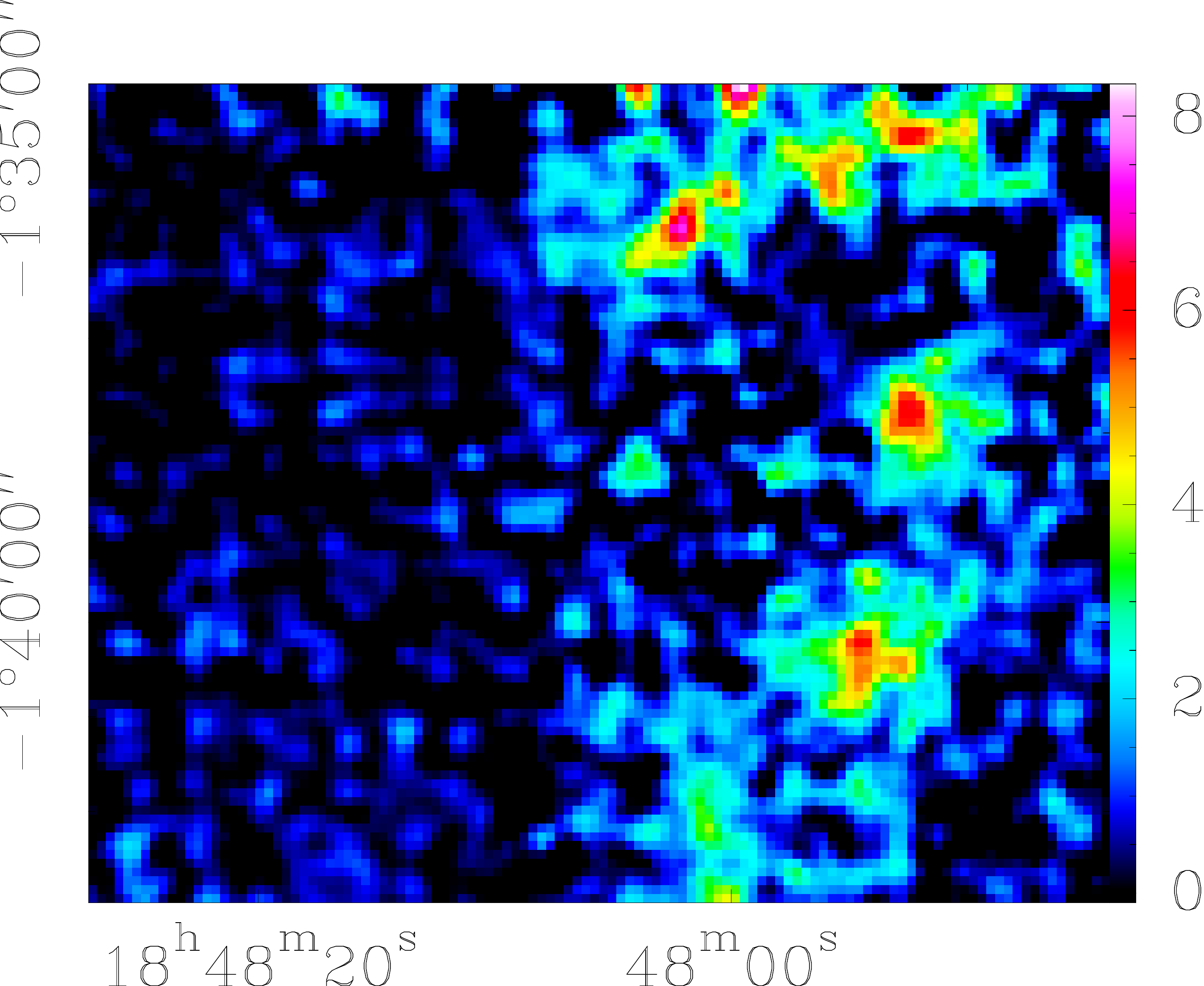}}}\hfill
 \subfloat[(d) Source 4]{\resizebox{!}{0.105\textheight}{\includegraphics[scale=1]{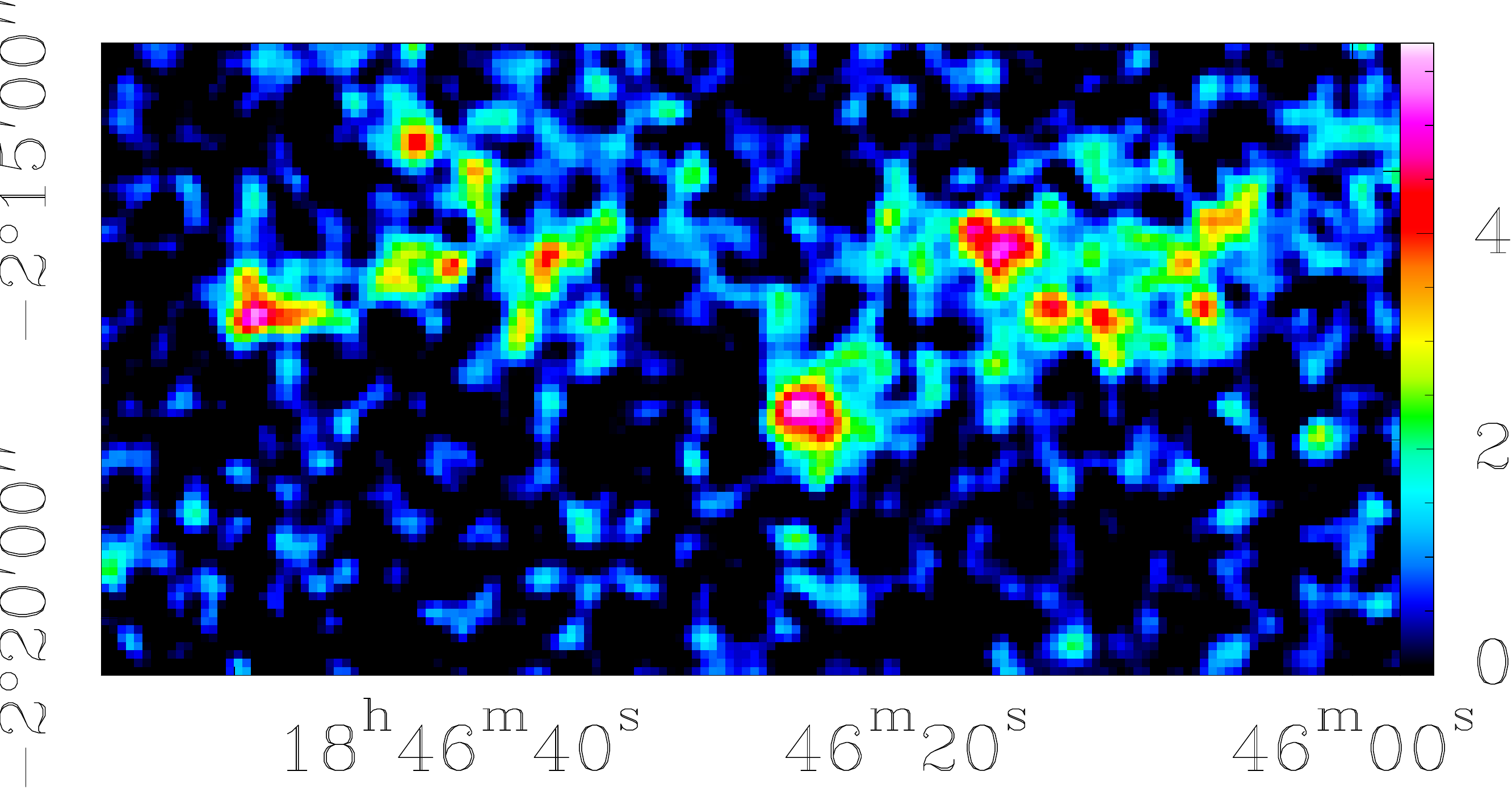}}}\hfill
 \subfloat[(e) Source 5]{\resizebox{!}{0.105\textheight}{\includegraphics[scale=1]{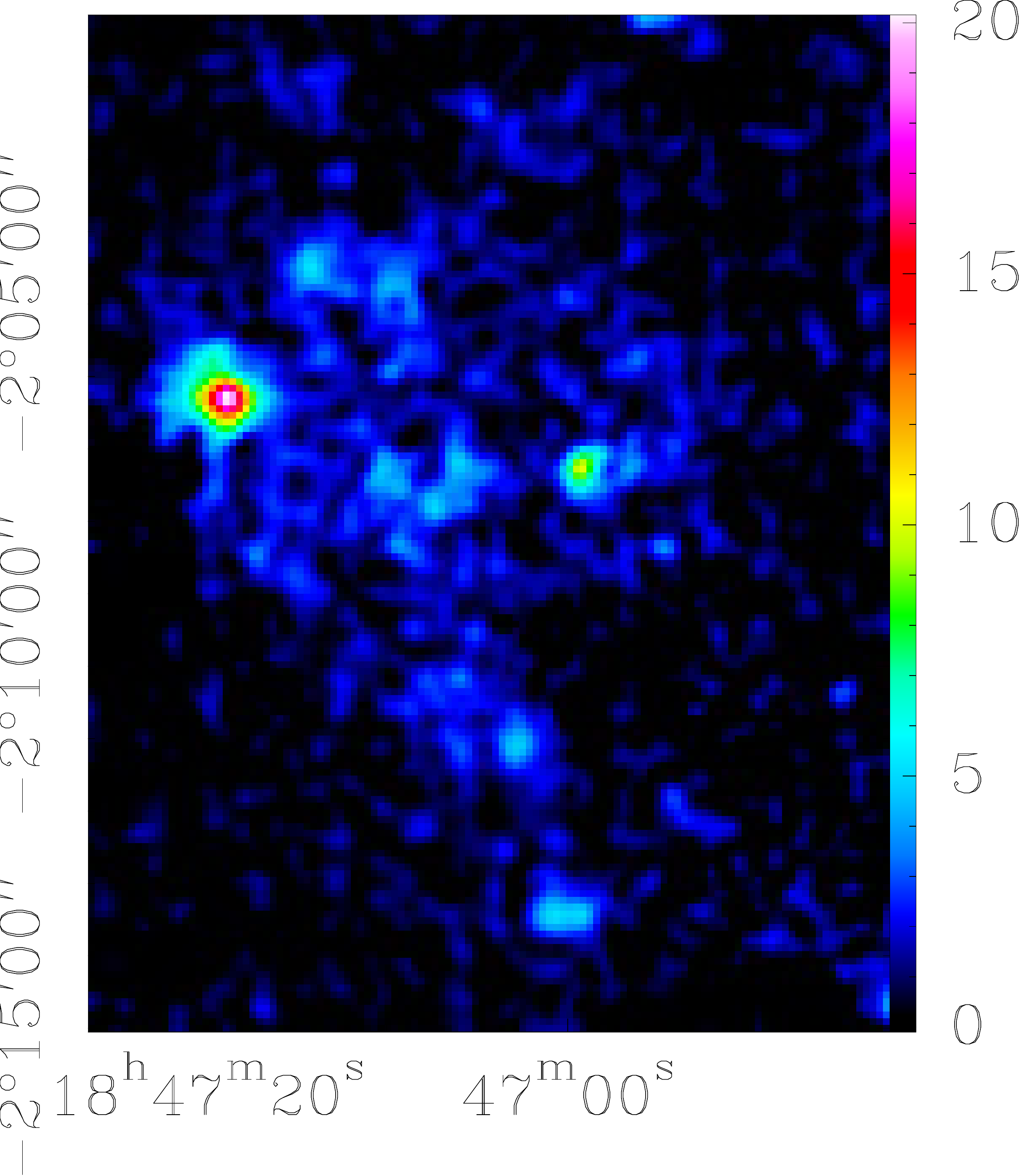}}}\\
 \subfloat[(f) Source 6]{\resizebox{!}{0.125\textheight}{\includegraphics[scale=1]{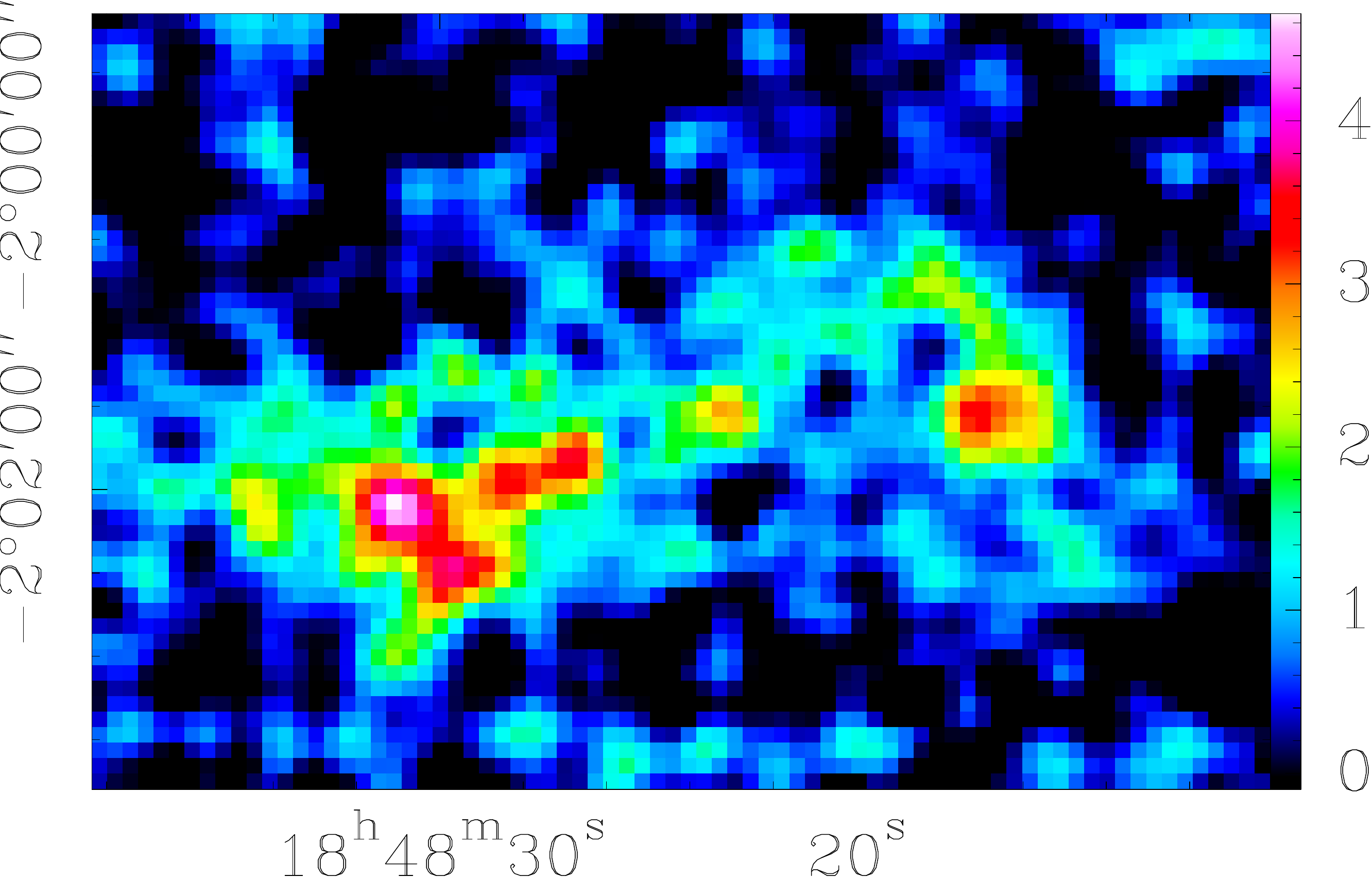}}}\hfill
 \subfloat[(g) Source 7]{\resizebox{!}{0.125\textheight}{\includegraphics[scale=1]{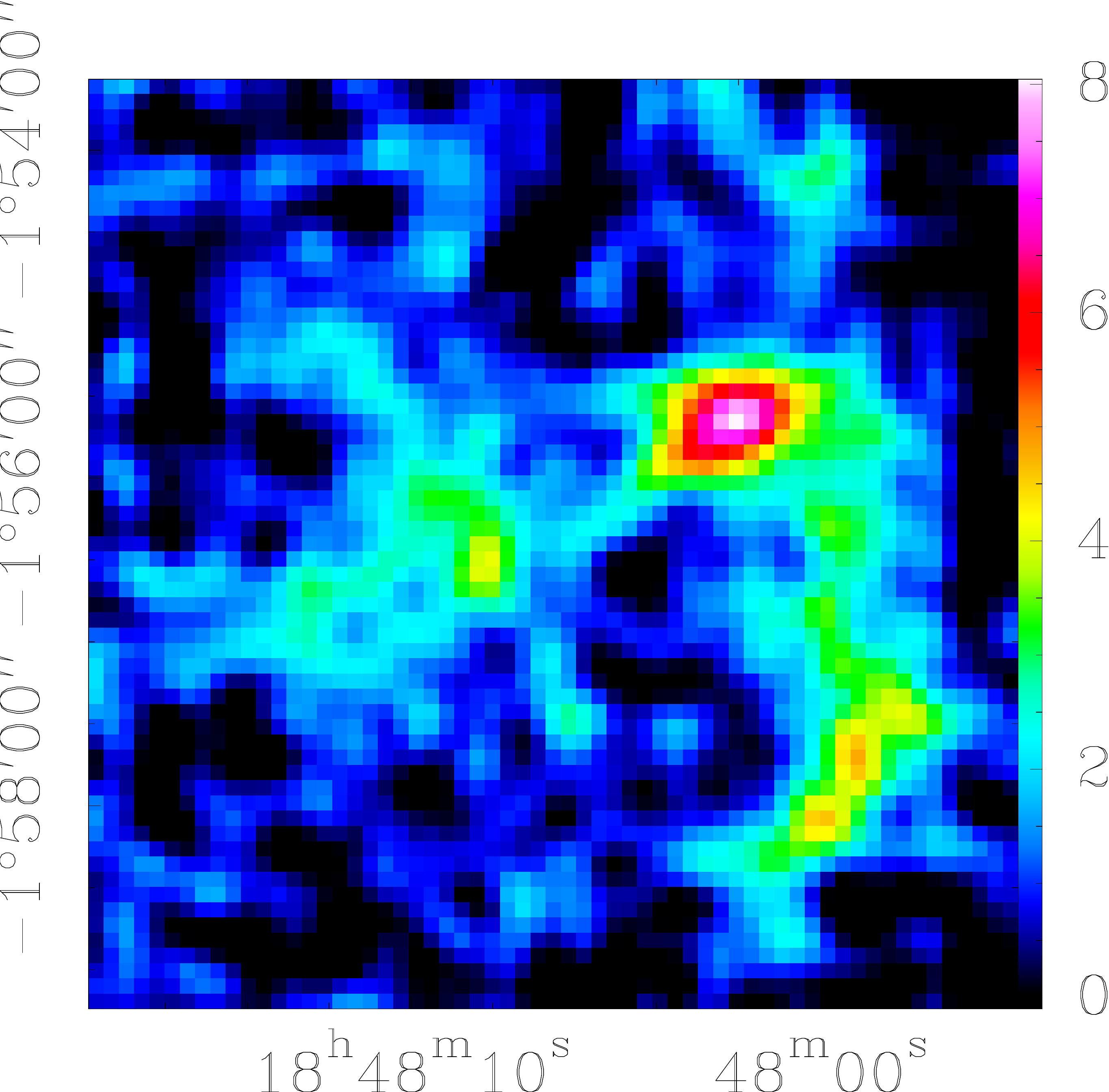}}}\hfill
 \subfloat[(h) Source 8]{\resizebox{!}{0.125\textheight}{\includegraphics[scale=1]{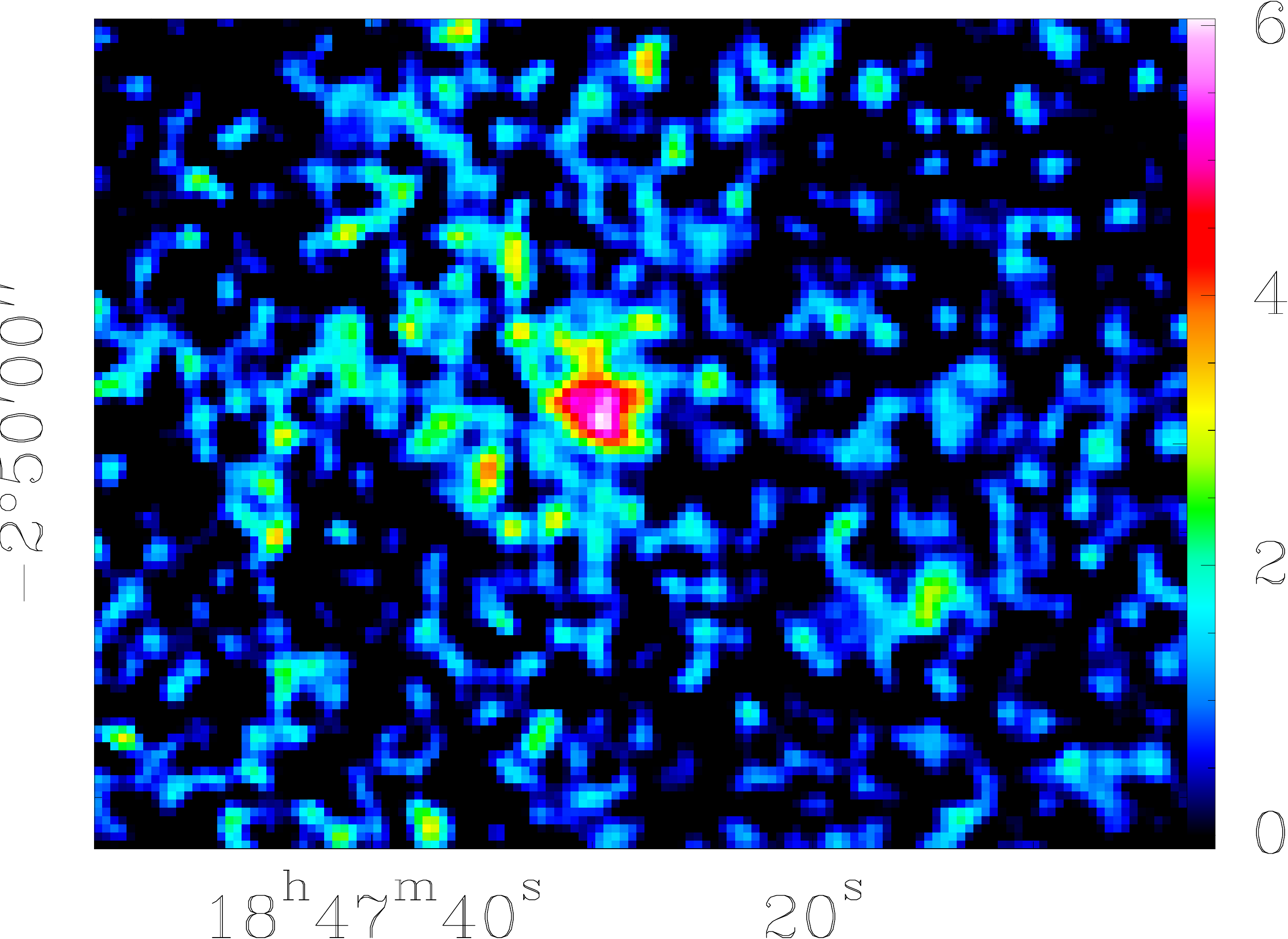}}}\hfill
 \subfloat[(i) Source 9]{\resizebox{!}{0.125\textheight}{\includegraphics[scale=1]{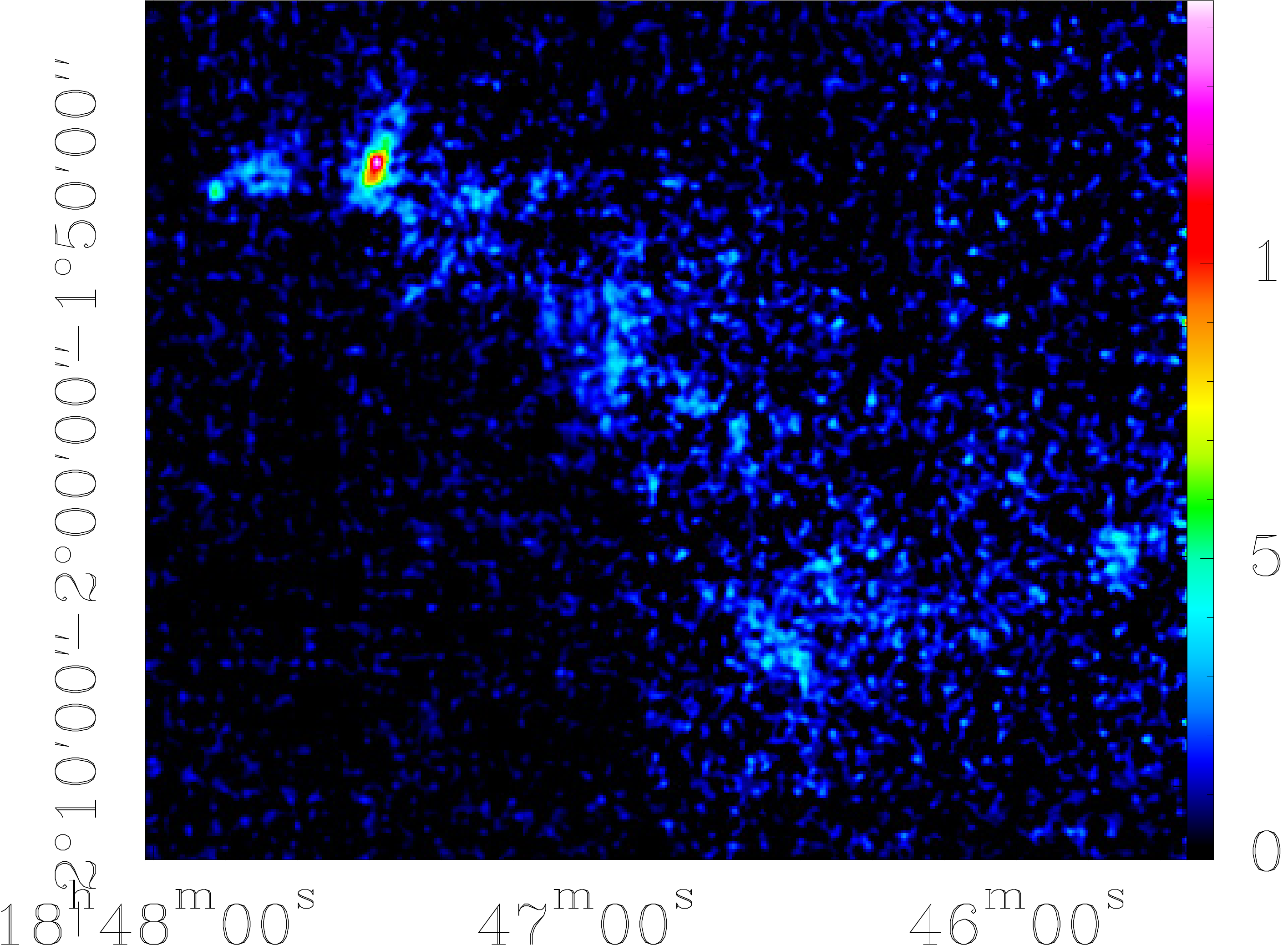}}}\\
 \subfloat[(j) Source 10]{\resizebox{!}{0.125\textheight}{\includegraphics[scale=1]{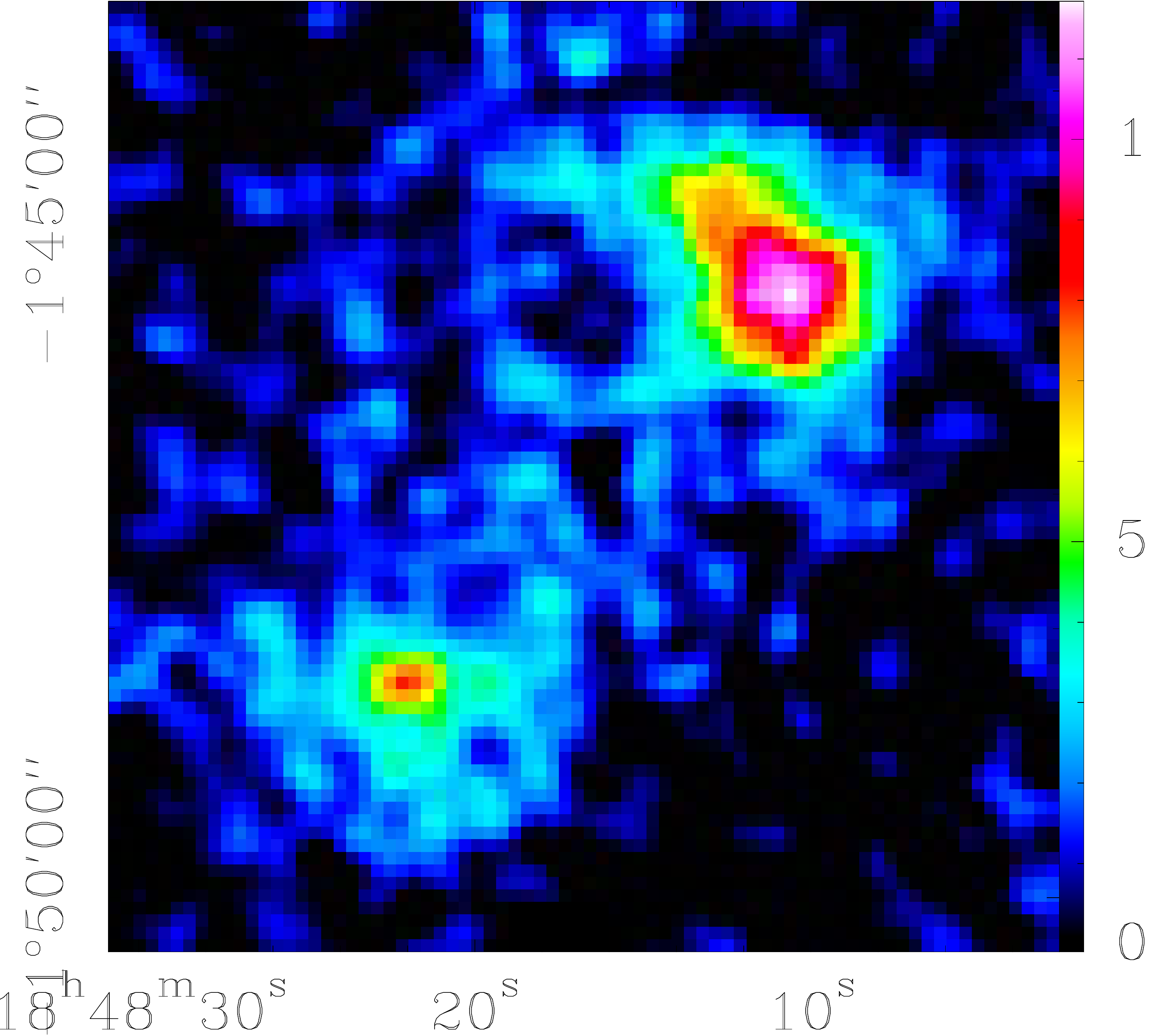}}}\hfill
 \subfloat[(k) Source 11]{\resizebox{!}{0.125\textheight}{\includegraphics[scale=1]{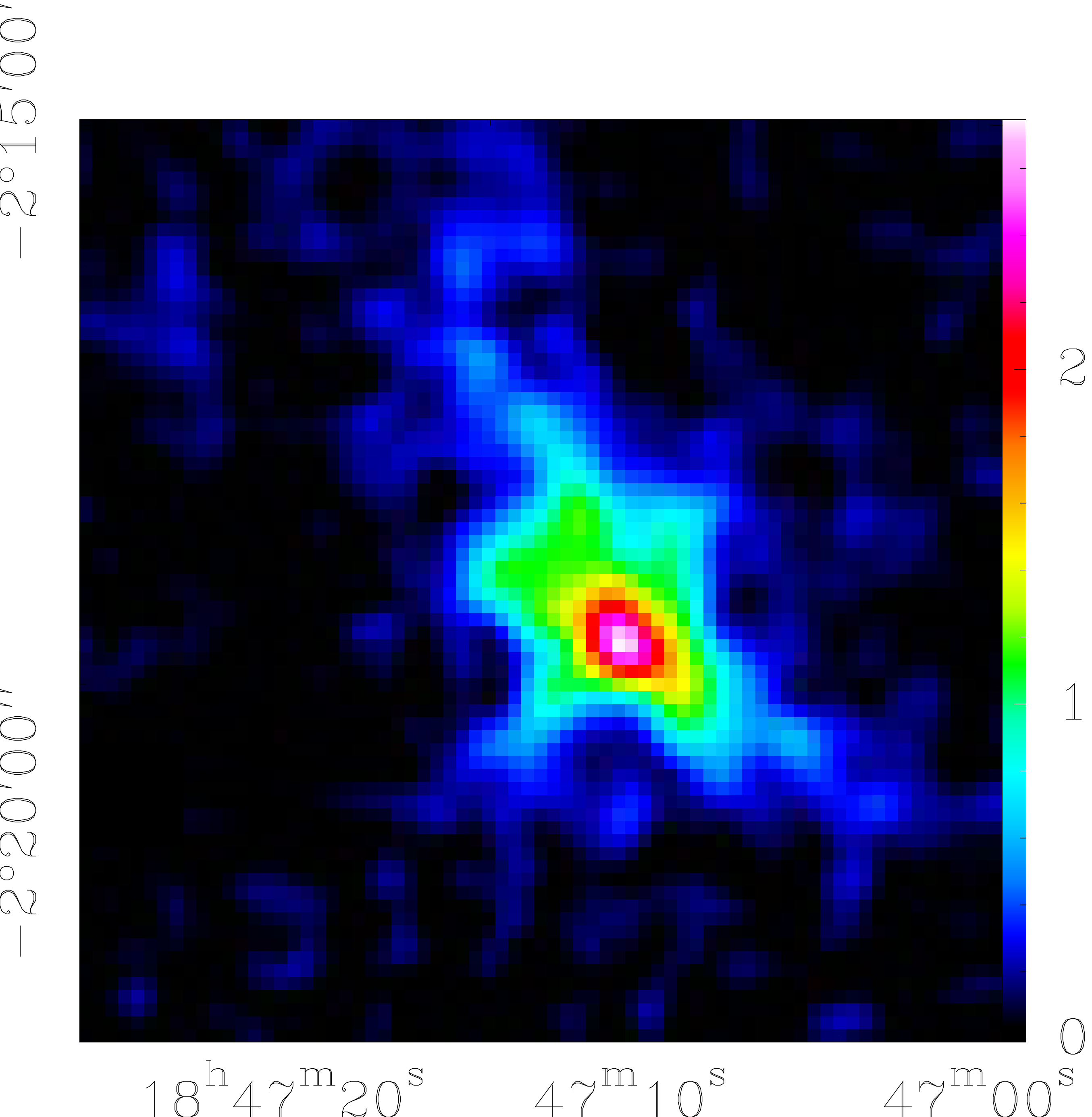}}}\hfill
 \subfloat[(l) Source 12]{\resizebox{!}{0.125\textheight}{\includegraphics[scale=1]{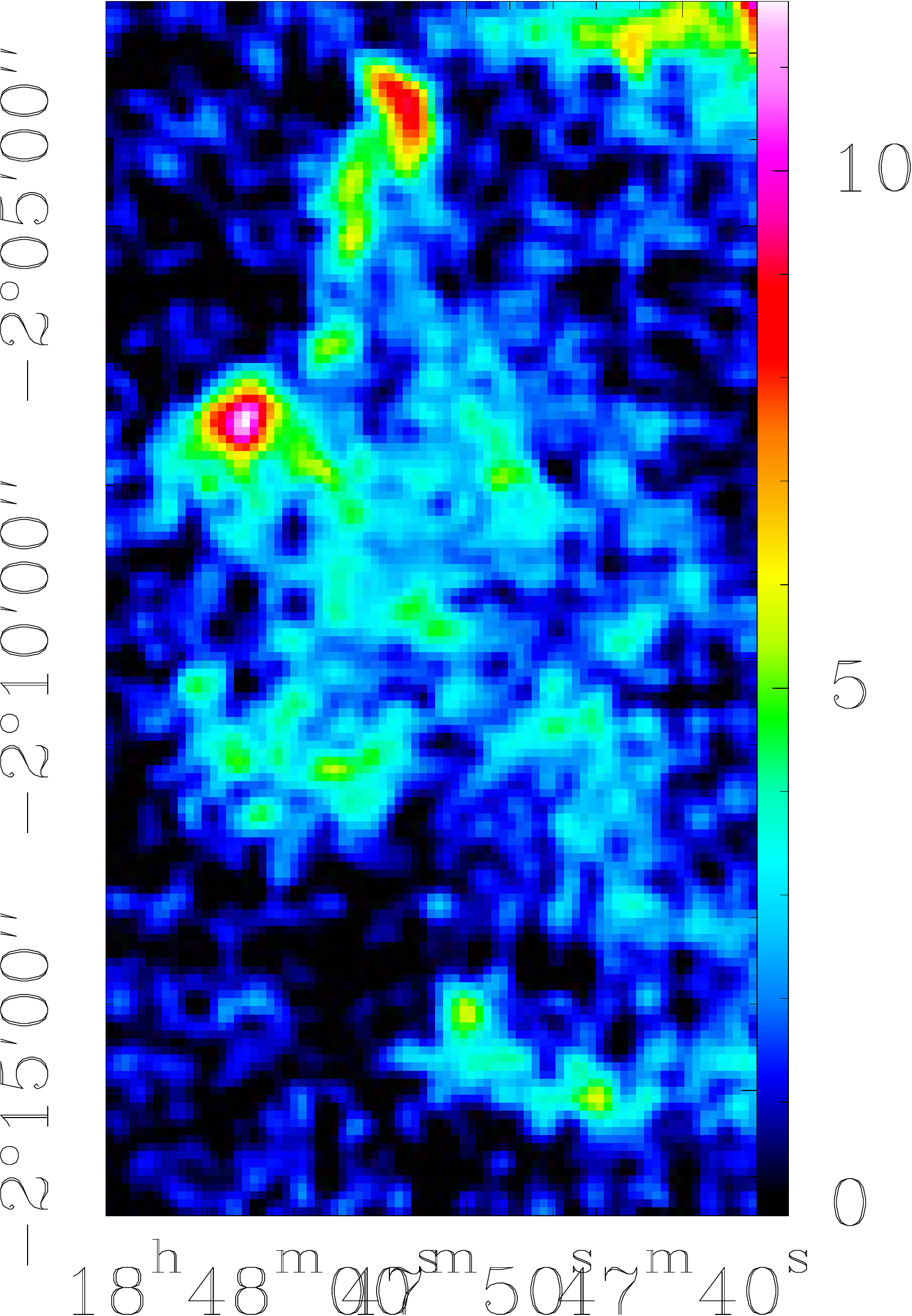}}}\hfill
 \subfloat[(m) Source 13]{\resizebox{!}{0.125\textheight}{\includegraphics[scale=1]{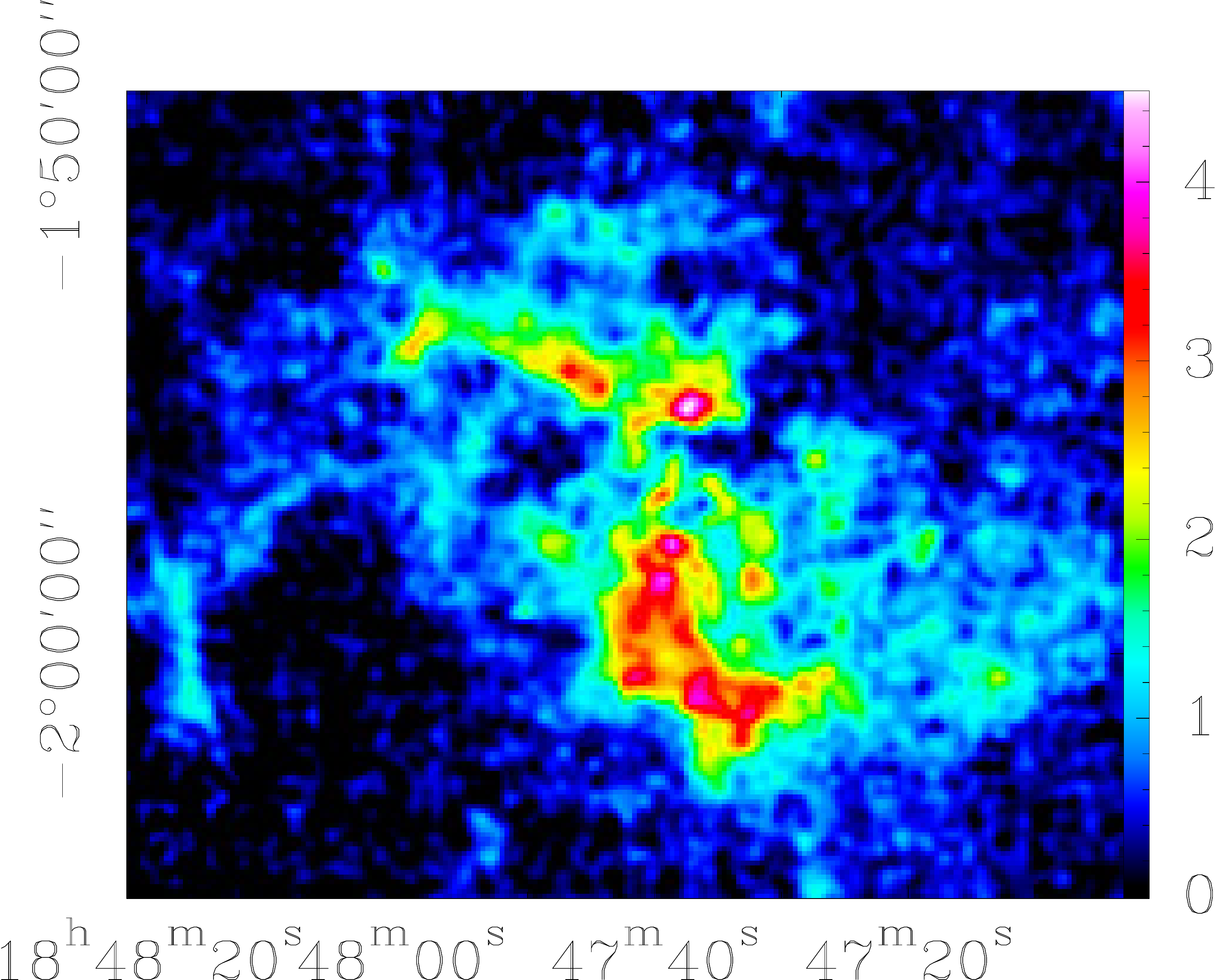}}}\hfill
 \subfloat[(n) Source 14]{\resizebox{!}{0.125\textheight}{\includegraphics[scale=1]{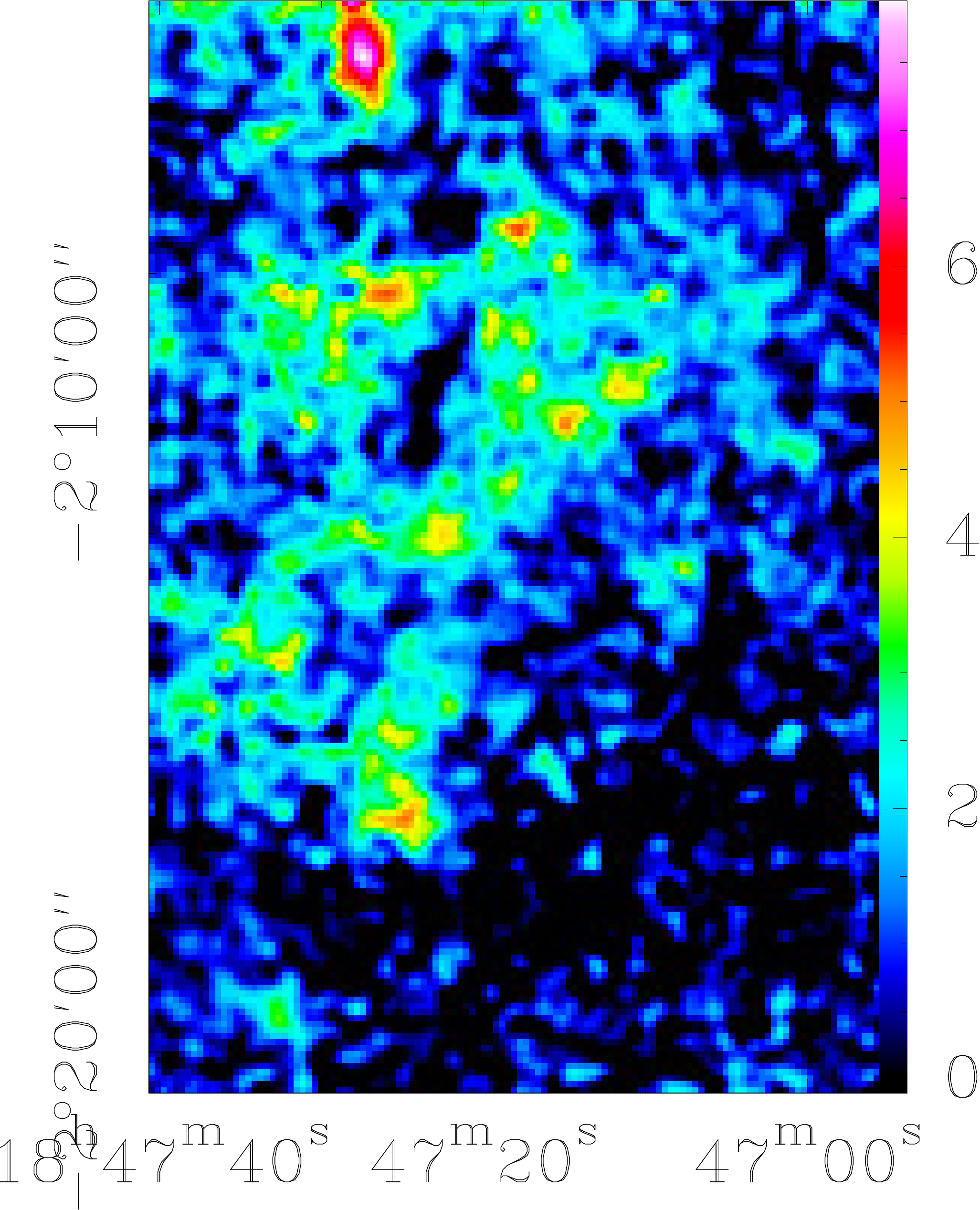}}}\\
 \subfloat[(o) Source 15]{\resizebox{!}{0.125\textheight}{\includegraphics[scale=1]{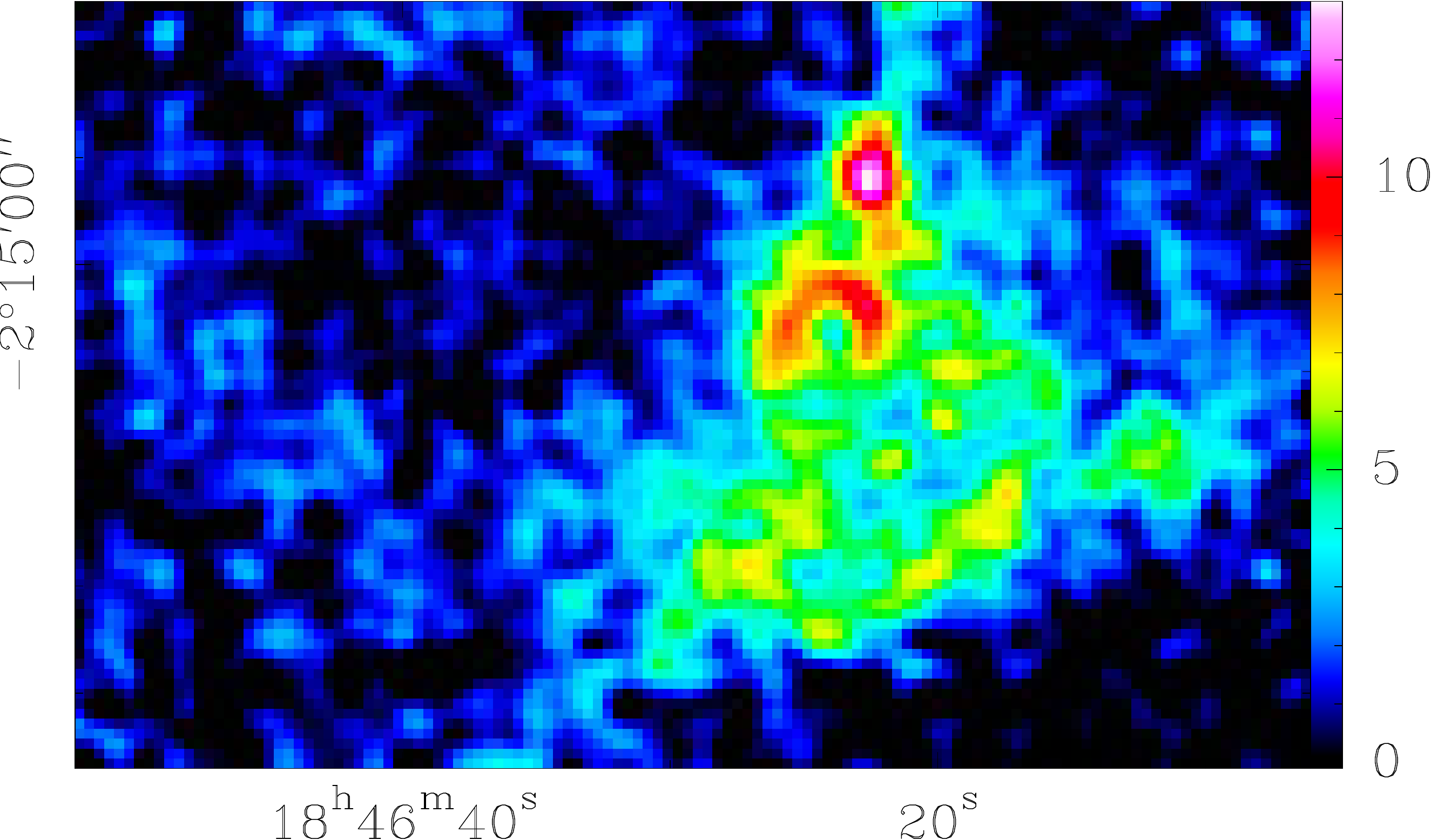}}}\hfill
 \subfloat[(p) Source 16]{\resizebox{!}{0.125\textheight}{\includegraphics[scale=1]{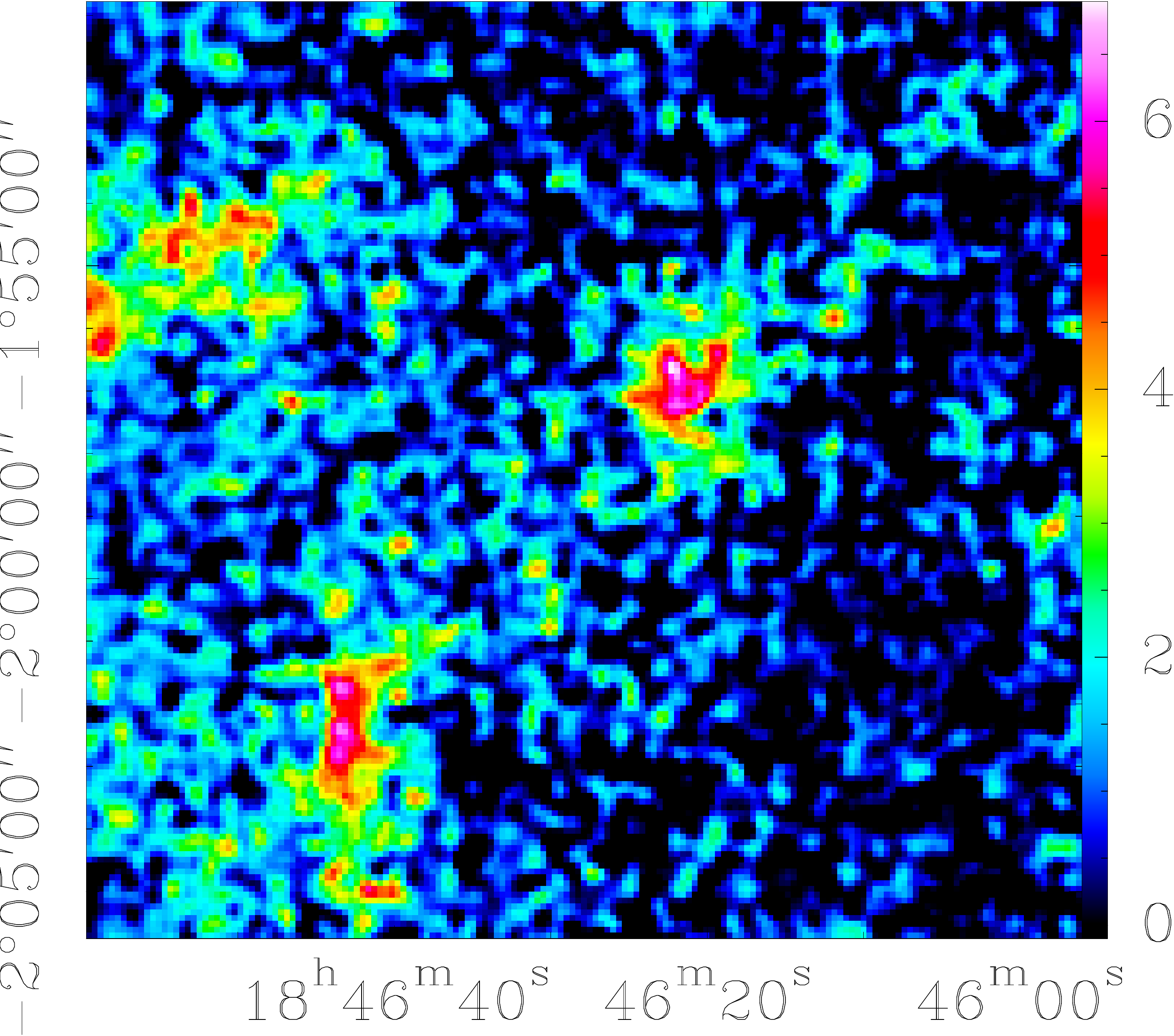}}}\hfill
 \subfloat[(q) Source 17]{\resizebox{!}{0.125\textheight}{\includegraphics[scale=1]{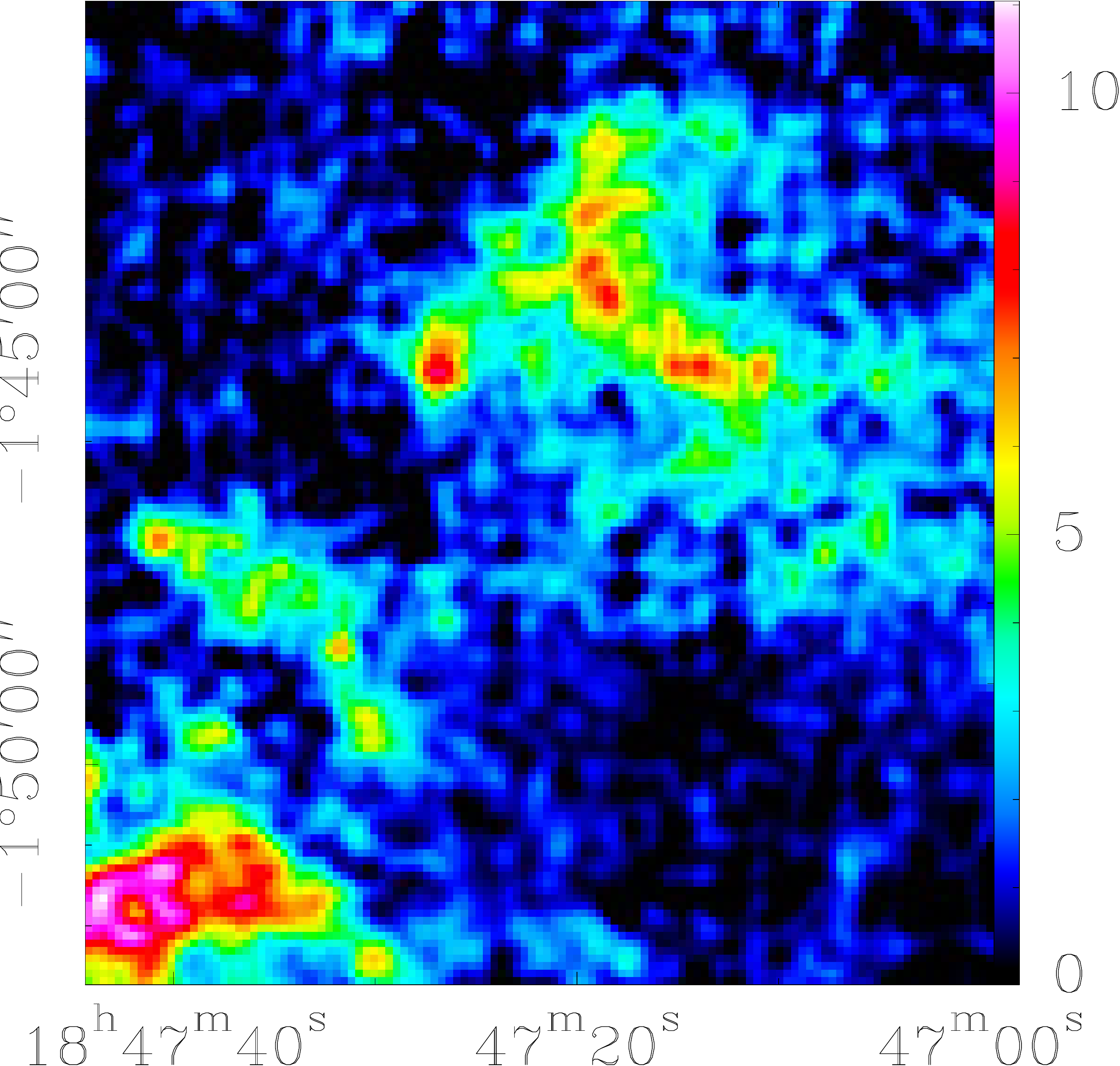}}}\hfill
 \subfloat[(r) Source 18]{\resizebox{!}{0.125\textheight}{\includegraphics[scale=1]{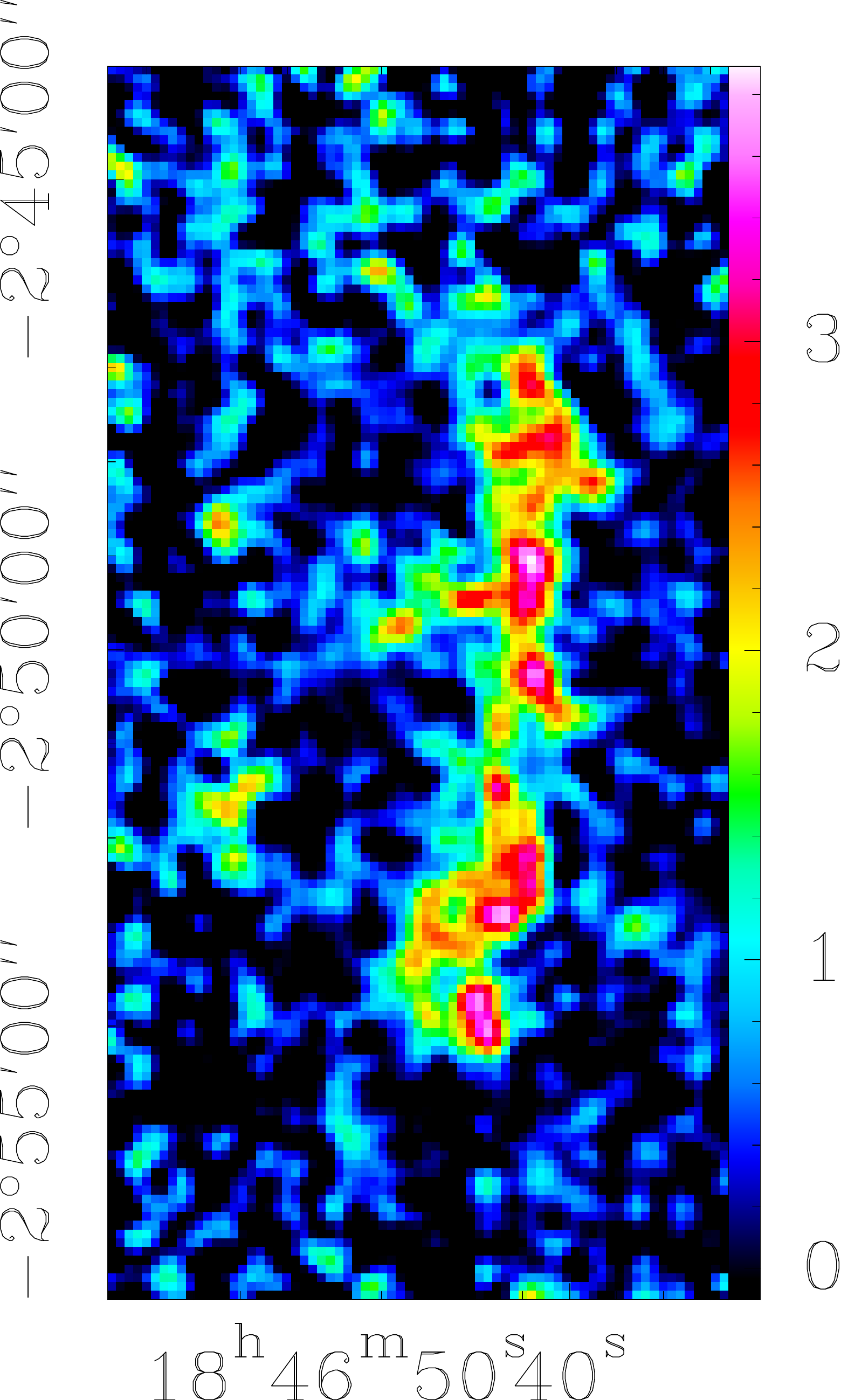}}}\hfill
 \subfloat[(s) Source 19]{\resizebox{!}{0.125\textheight}{\includegraphics[scale=1]{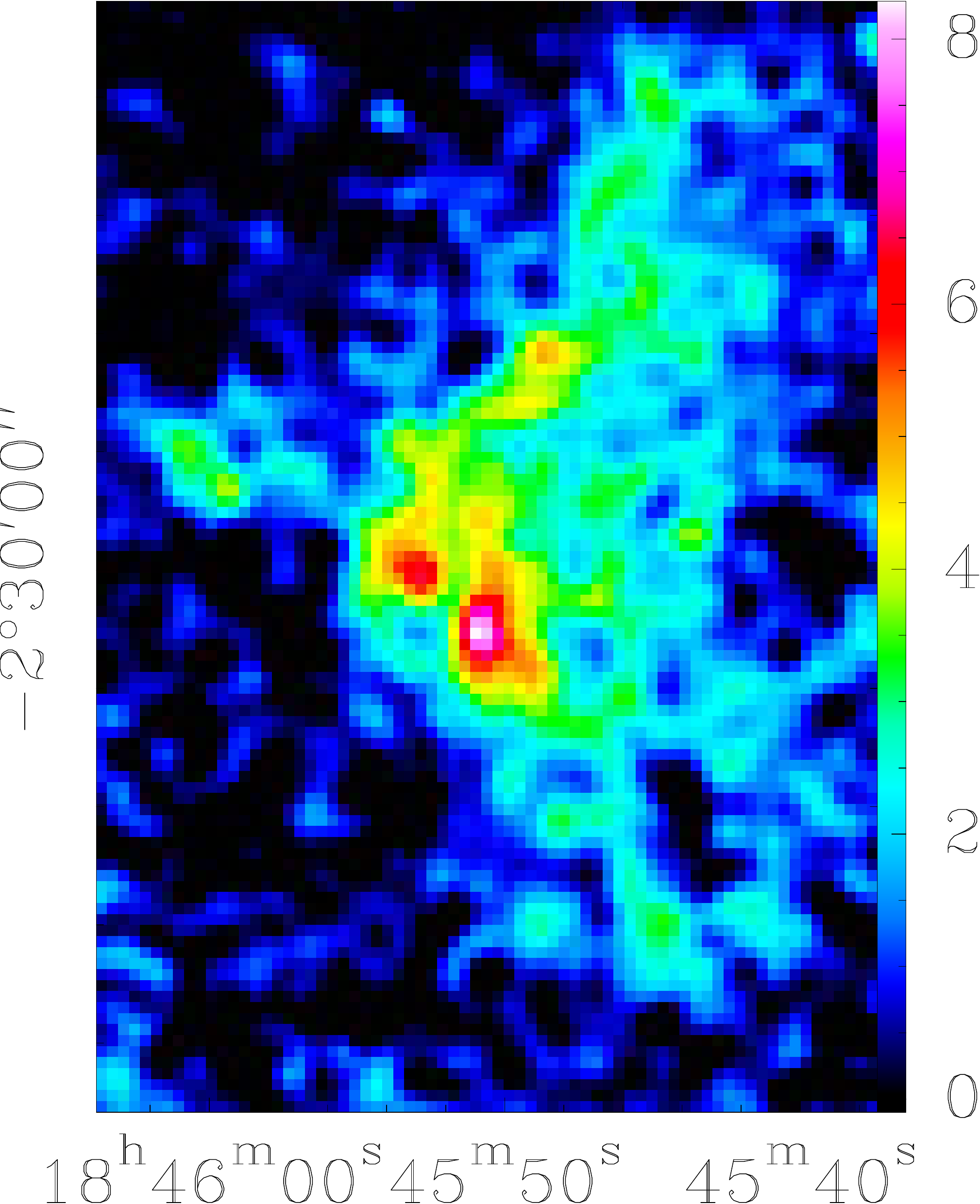}}}\\
 \subfloat[(t) Source 20]{\resizebox{!}{0.125\textheight}{\includegraphics[scale=1]{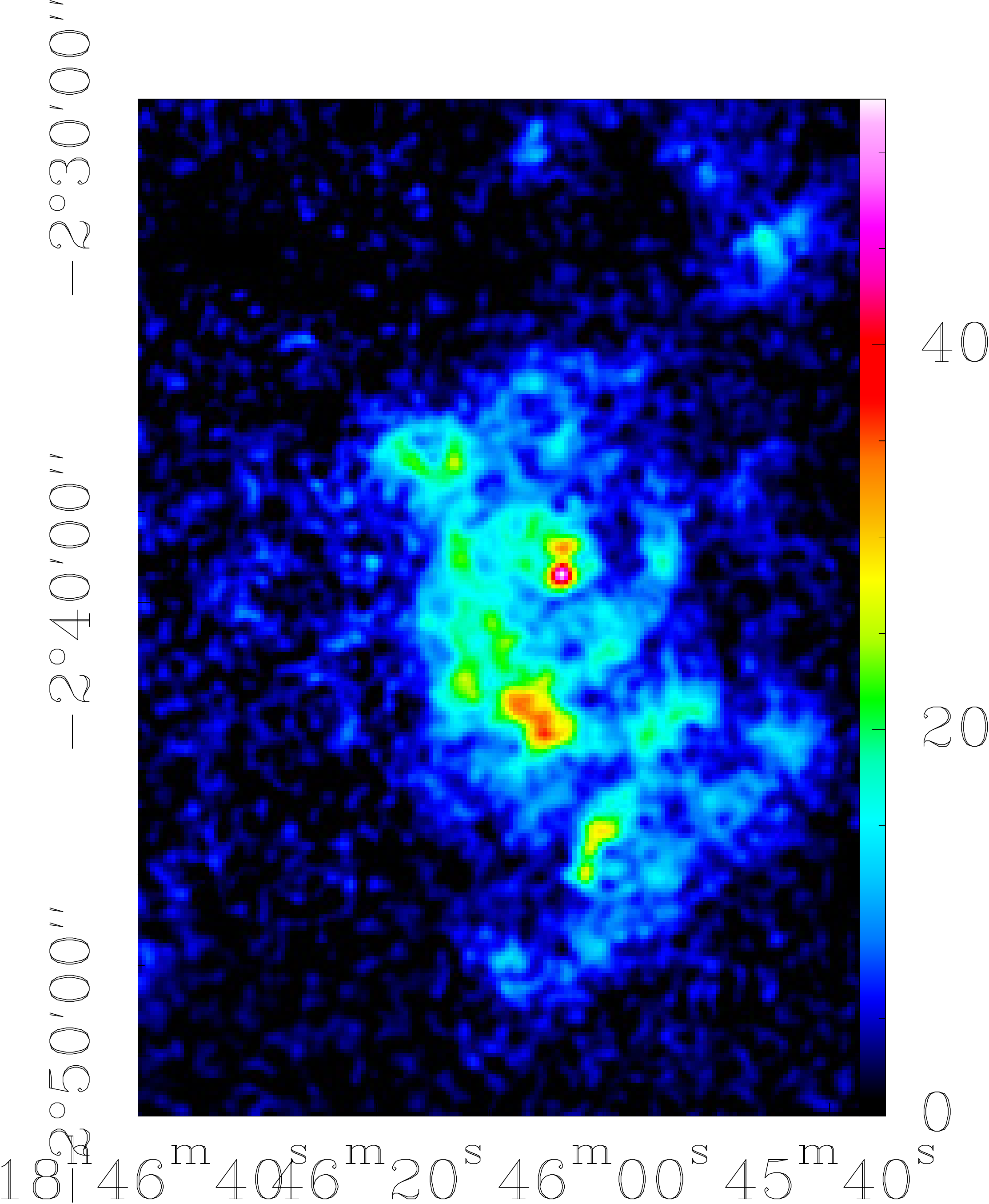}}}\hfill
 \subfloat[(u) Source 21]{\resizebox{!}{0.125\textheight}{\includegraphics[scale=1]{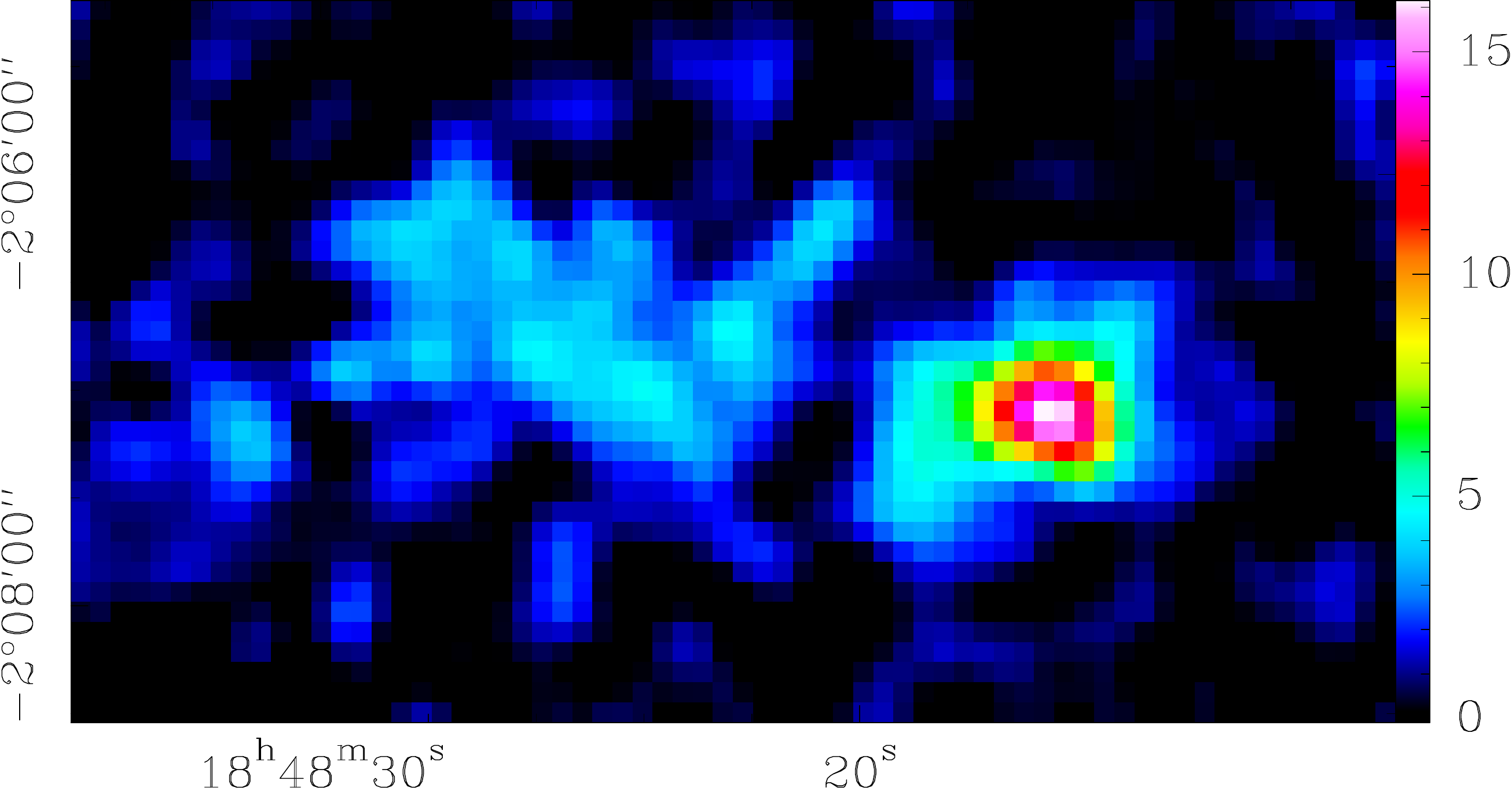}}}\hfill
 \subfloat[(v) Source 22]{\resizebox{!}{0.125\textheight}{\includegraphics[scale=1]{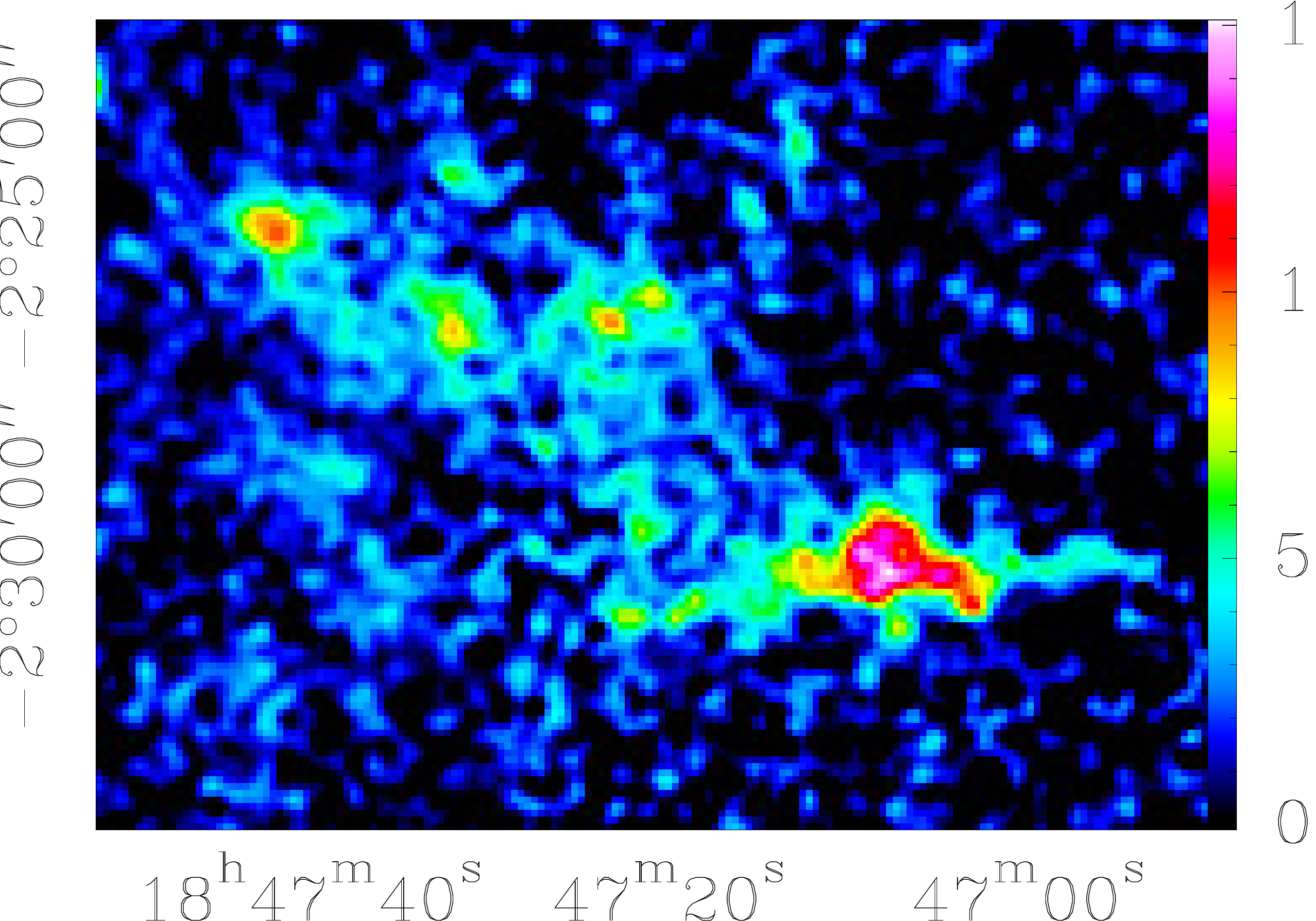}}}\hfill
 \subfloat[(w) Source 23]{\resizebox{!}{0.125\textheight}{\includegraphics[scale=1]{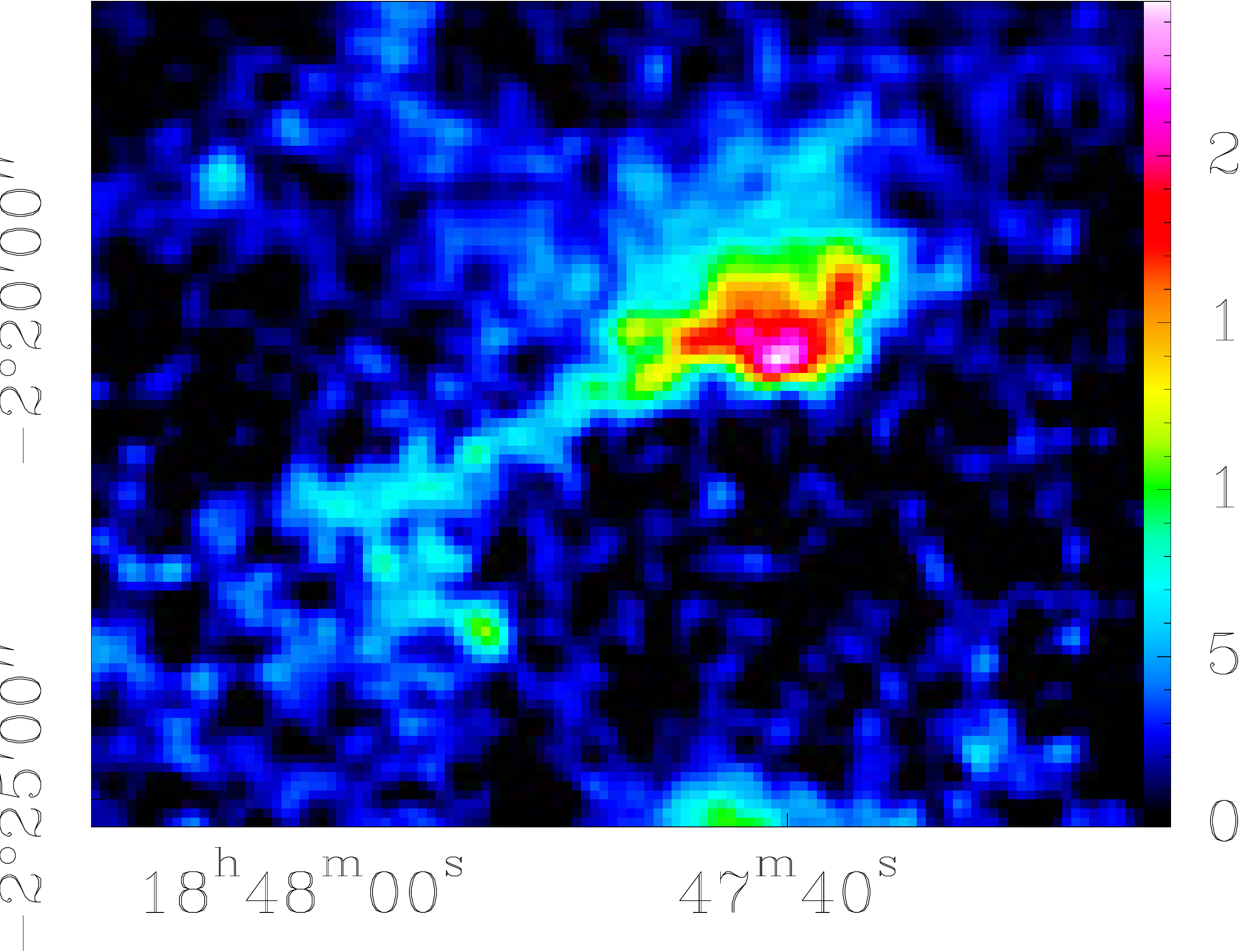}}}\\
 \subfloat[(x) Source 24]{\resizebox{!}{0.125\textheight}{\includegraphics[scale=1]{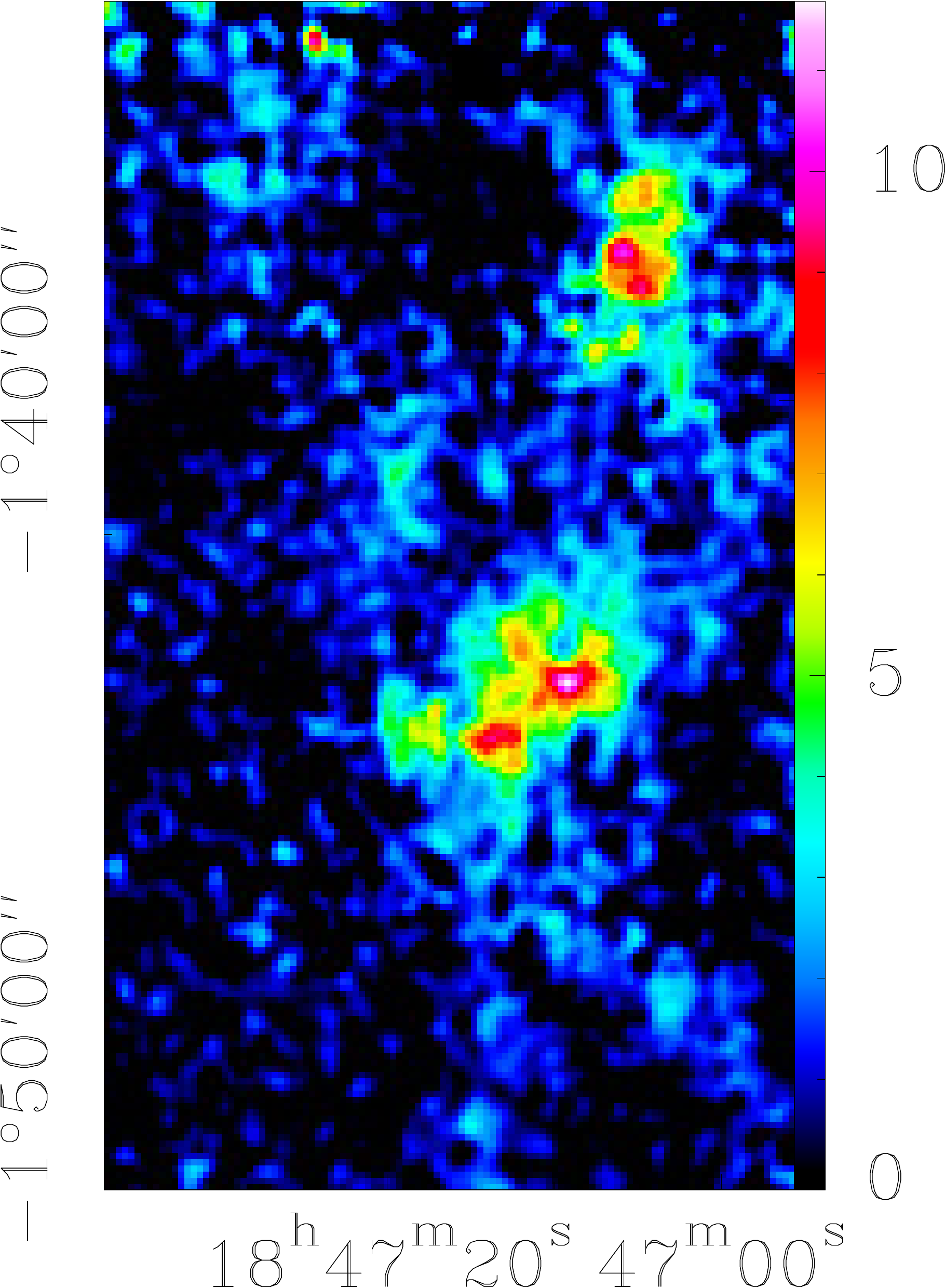}}}\hfill
 \subfloat[(y) Source 25]{\resizebox{!}{0.125\textheight}{\includegraphics[scale=1]{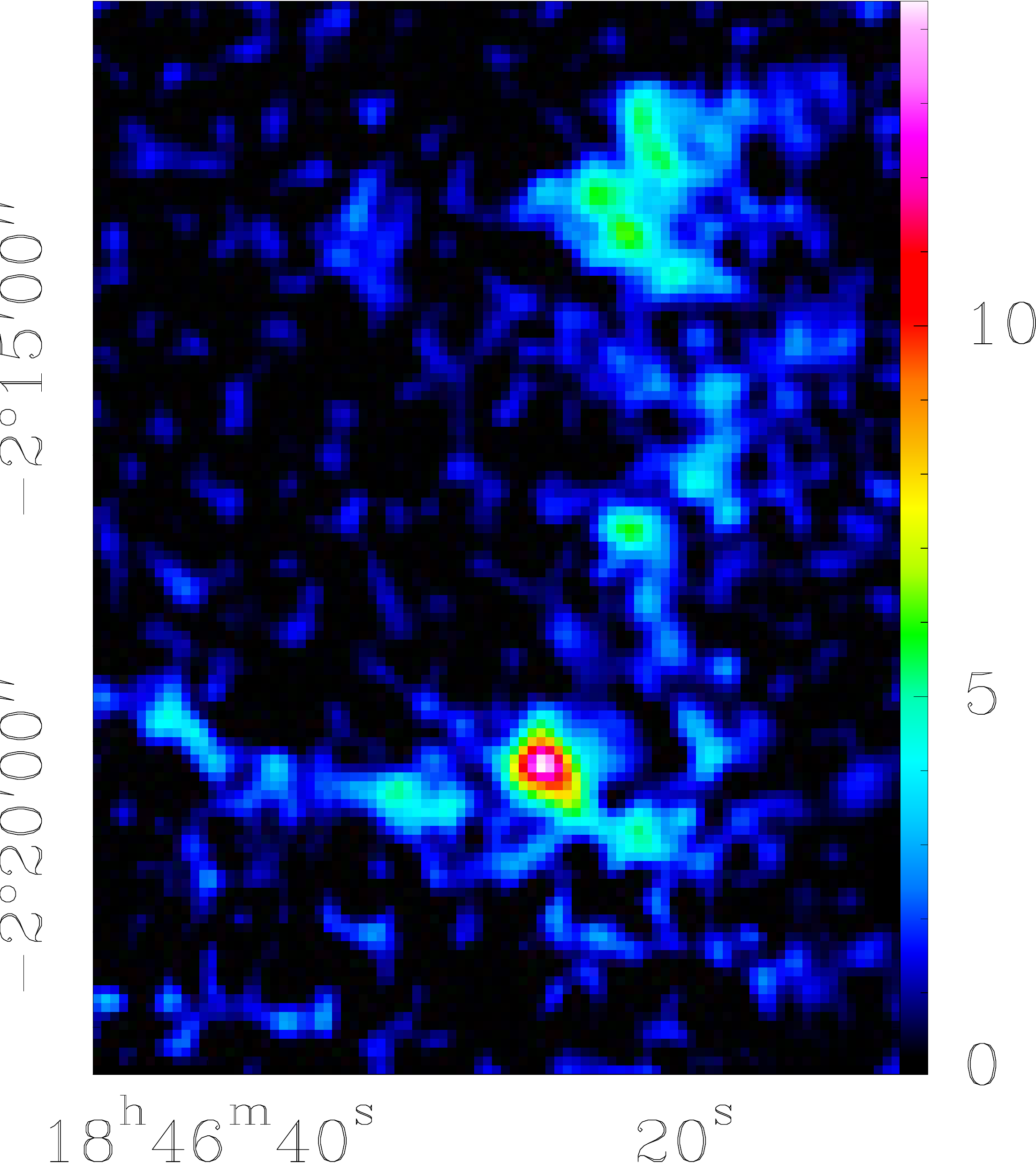}}}\hfill
 \subfloat[(z) Source 26]{\resizebox{!}{0.125\textheight}{\includegraphics[scale=1]{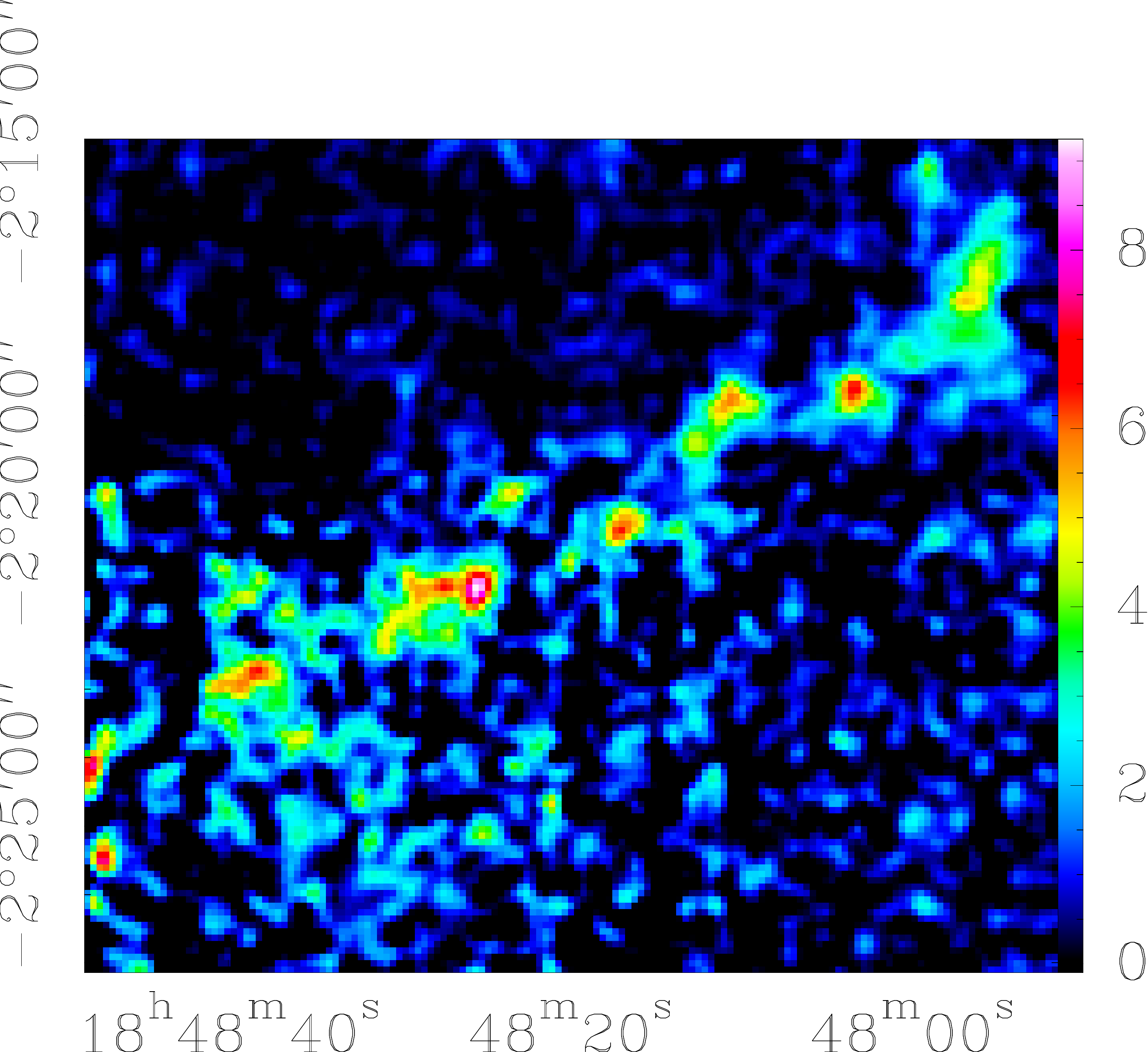}}}\hfill
 \subfloat[(aa) Source 27]{\resizebox{!}{0.125\textheight}{\includegraphics[scale=1]{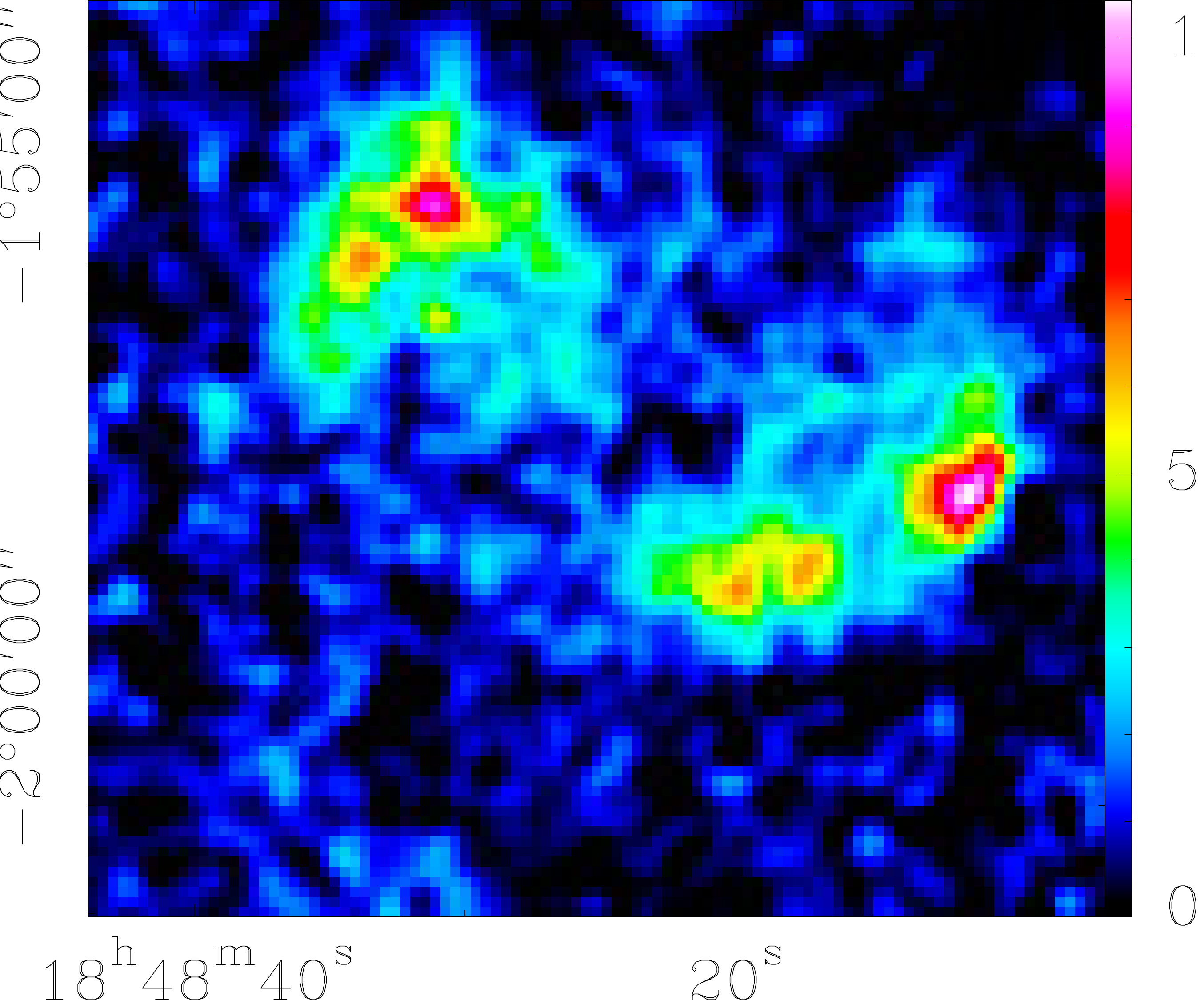}}}\hfill
 \subfloat[(ab) Source 28]{\resizebox{!}{0.125\textheight}{\includegraphics[scale=1]{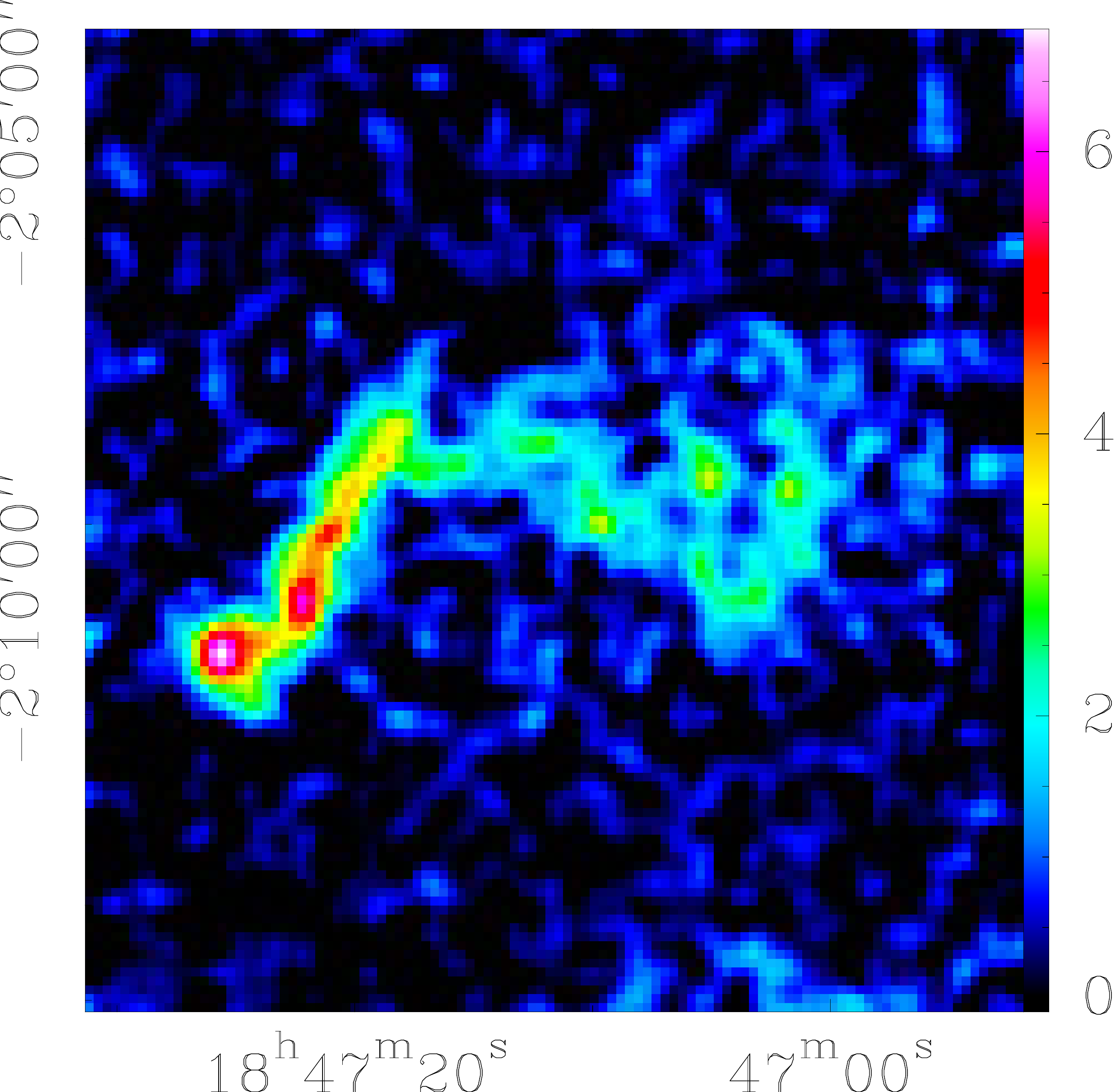}}}\hfill
 \subfloat[(ac) Source 29]{\resizebox{!}{0.125\textheight}{\includegraphics[scale=1]{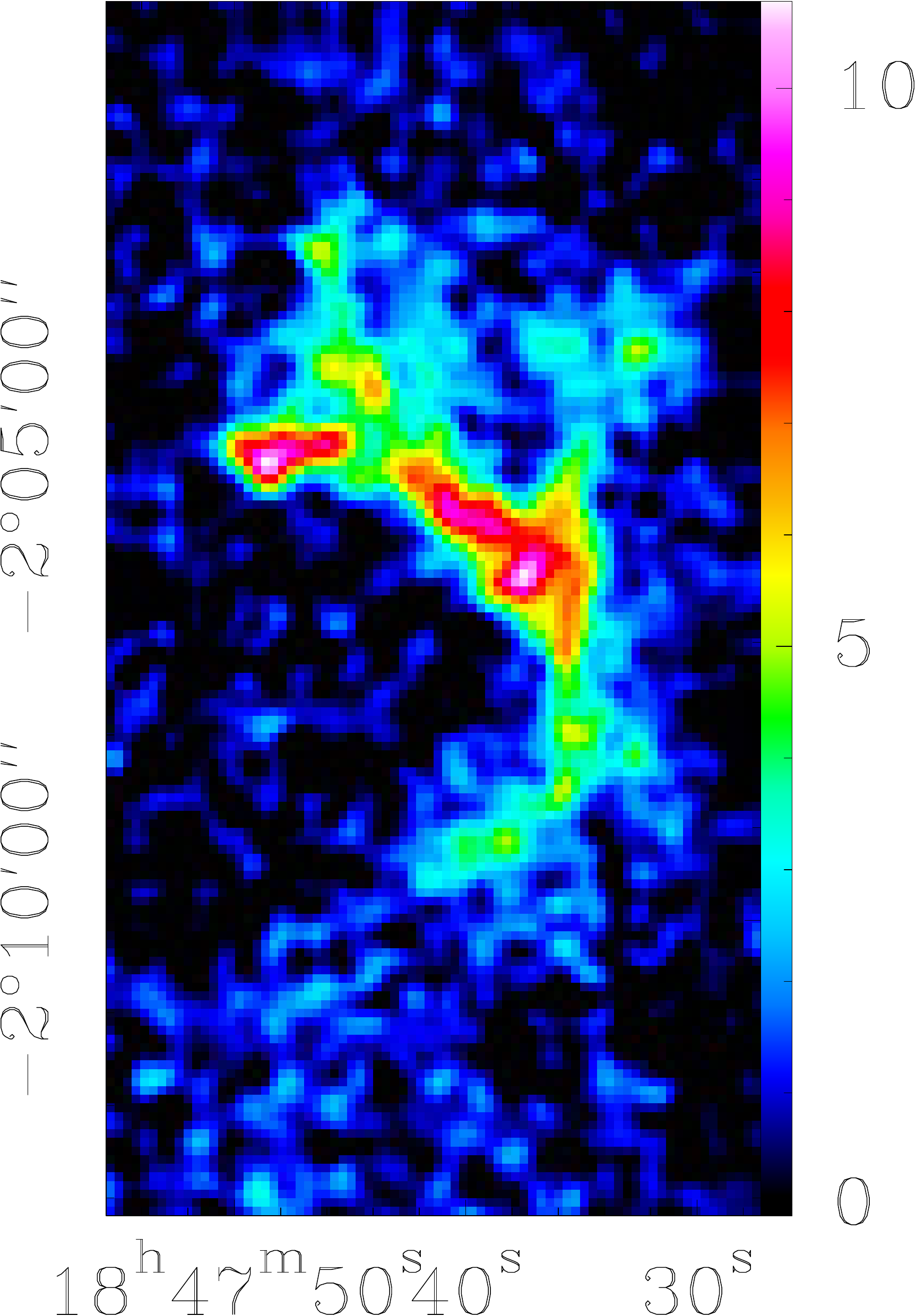}}}
 \end{minipage}
\caption{Integrated C$^{18}$O~(2--1) maps in K\,km\,s$^{-1}$ of all identified sources.}
\label{fig:appendixsourcemapsc18o}
\end{figure*}

\end{document}